\documentclass[12pt,a4paper,twoside,openright]{book}

\usepackage[english]{babel}
\usepackage{fourier} 
\usepackage{CormorantGaramond} 
\usepackage[OT1]{fontenc}
\usepackage{geometry}
\usepackage{ulem}
\usepackage{caption}
\usepackage[square]{natbib}
\usepackage[breaklinks=true]{hyperref} 
\newcommand{\apjs}{ApJS}
\newcommand{\apj}{ApJ}
\newcommand{\apjl}{ApJ}
\newcommand{\mnras}{MNRAS}
\newcommand{\aap}{A\&A}
\newcommand{\aj}{AJ}
\newcommand{\nat}{Nature} 
\newcommand{\prl}{PRL}
\newcommand{\prd}{PRD}
\newcommand{\physrep}{Physics Reports}
\newcommand{\jcap}{JCAP}
\newcommand{\jcp}{JCP}
\newcommand{\araa}{ARA\&A}

\usepackage[dvipsnames]{xcolor}
\usepackage{url}
\usepackage{graphicx}	
\usepackage{wrapfig}
\usepackage{amsfonts}
\usepackage{enumerate} 
\usepackage{colordvi} 
\usepackage{verbatim}
\usepackage{multirow,multicol}
\usepackage{pdfpages}
\usepackage{amsmath,amssymb,amsthm,amscd,textcomp}
\usepackage{mathtools}
\usepackage{amsfonts}
\usepackage{fancyhdr}
\usepackage{booktabs}
\usepackage{array}

\usepackage{listings} 

\usepackage{csquotes}
\usepackage{adjustbox}

\usepackage{xcolor} 
\definecolor{turquoise}{RGB}{64, 224, 208}
\definecolor{RoyalBlue}{RGB}{65, 105, 225}

\usepackage{tikz}

\usepackage{inconsolata} 

\definecolor{codebg}{rgb}{0.95,0.95,0.95}
\definecolor{keywordcolor}{rgb}{0.5,0.0,0.35}
\definecolor{commentcolor}{rgb}{0.25,0.5,0.35}
\definecolor{stringcolor}{rgb}{0.6,0.0,0.0}
\definecolor{numbercolor}{rgb}{0.5,0.5,0.5} 

\lstdefinestyle{mintedstyle}{
    backgroundcolor=\color{codebg},
    basicstyle=\ttfamily\footnotesize\color{black}, 
    breakatwhitespace=false,
    breaklines=true,
    captionpos=b,
    commentstyle=\color{commentcolor},
    frame=single,
    framesep=5pt,
    keepspaces=true,
    keywordstyle=\color{keywordcolor}\bfseries,
    numbers=left,
    numbersep=8pt,
    numberstyle=\scriptsize\color{numbercolor}, 
    rulecolor=\color{purple},
    showspaces=false,
    showstringspaces=false,
    showtabs=false,
    stepnumber=1,
    stringstyle=\color{stringcolor},
    tabsize=4,
    xleftmargin=10pt,
    literate=*{0}{{{\color{black}0}}}1 
        {1}{{{\color{black}1}}}1
        {2}{{{\color{black}2}}}1
        {3}{{{\color{black}3}}}1
        {4}{{{\color{black}4}}}1
        {5}{{{\color{black}5}}}1
        {6}{{{\color{black}6}}}1
        {7}{{{\color{black}7}}}1
        {8}{{{\color{black}8}}}1
        {9}{{{\color{black}9}}}1
        {.}{{{\color{black}.}}}1
        {=}{{{\color{keywordcolor}=}}}1
        {-}{{{\color{keywordcolor}-}}}1
}

\usepackage{cleveref}
\usepackage[caption=false]{subfig}
\crefname{equation}{Eq}{Eqs.} 
\usepackage[labelfont=bf]{caption}
\graphicspath{{./}}
\DeclareGraphicsExtensions{.jpg, .pdf, .png}
\usepackage{bm}
\crefrangelabelformat{equation}{(#3#1#4--#5#2#6)}
\newtheorem*{axiom*}{Cosmological Principle}

\hypersetup{
    colorlinks=true,        
    citecolor=violet,      
    linkcolor=RoyalBlue,         
    urlcolor=RoyalBlue        
}

\bibpunct{\textcolor{violet}{[}}{\textcolor{violet}{]}}{,}{a}{}{;}

\usepackage{fontawesome5} 
\newcommand{\github}[1]{%
   \href{#1}{\faGithub}%
}
\usepackage{bigints}
\usepackage{float}
\usepackage{tikz}
\usetikzlibrary{shapes.geometric, arrows}
\usetikzlibrary{positioning}

\usepackage{tcolorbox}
\tcbuselibrary{breakable,xparse,skins}  

\addto\extrasenglish{%
}

\addto\extrasenglish{%
}

\addto\extrasenglish{%
}

\usepackage[nottoc]{tocbibind} 

\usepackage[titletoc]{appendix}
  
\newcommand{\be}{\begin{equation}}
\newcommand{\ee}{\end{equation}}
\newcommand{\bea}{\begin{eqnarray}}
\newcommand{\eea}{\end{eqnarray}}

\DeclareMathAlphabet{\pazocal}{OMS}{zplm}{m}{n}
\newcommand{\unif}{\pazocal{U}}

\begin{document}
\pagenumbering{gobble}

\begin{minipage}[!ht]{\textwidth}
\begin{center}
\includegraphics[scale=0.075]{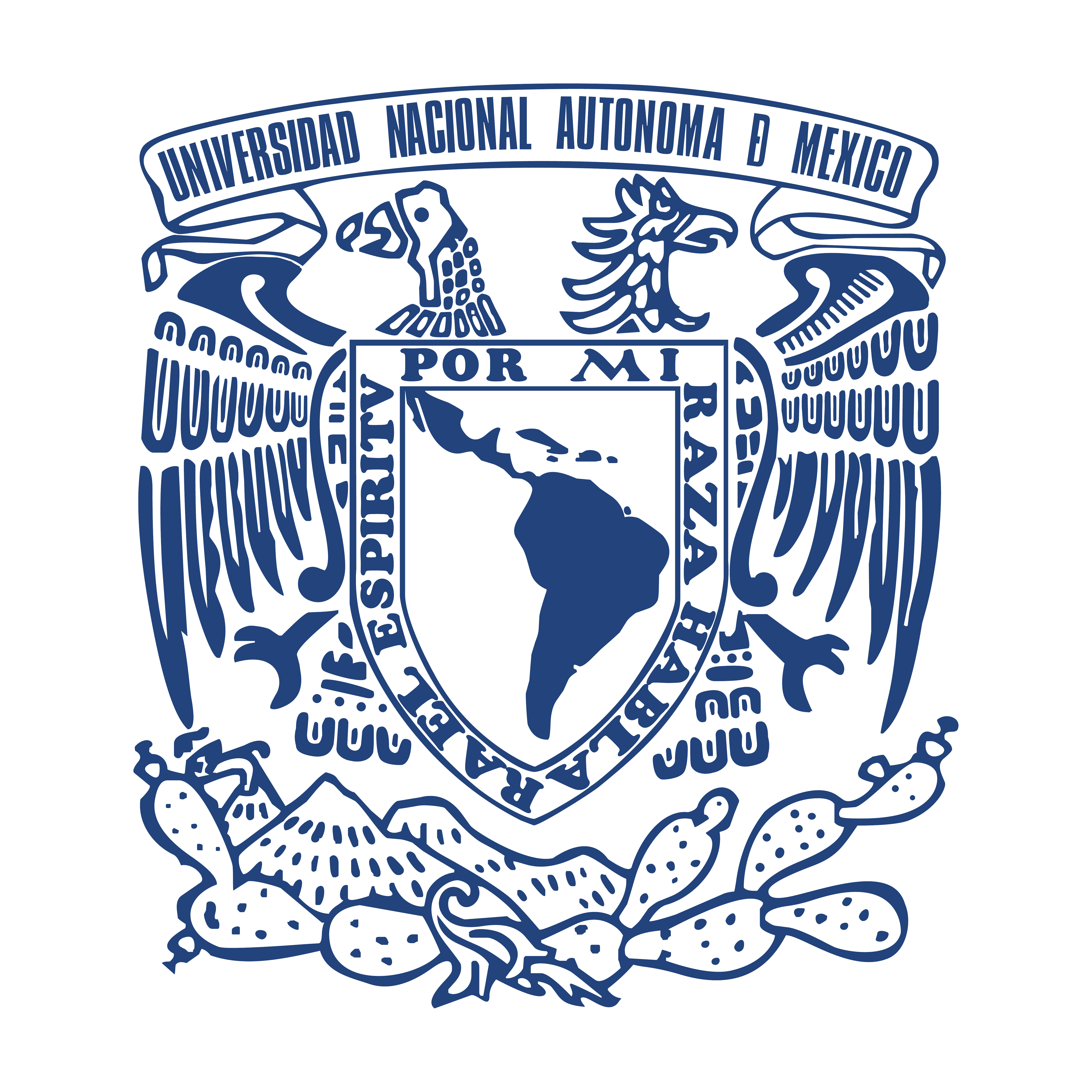}
\includegraphics[width=0.95\textwidth, height=0.04\textheight]{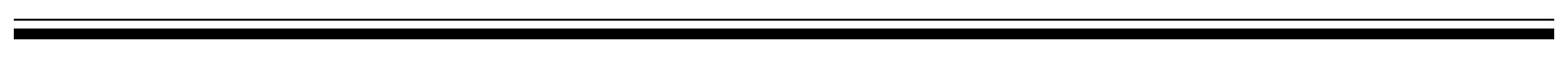}

\vspace{0.25cm}

{\huge\textbf{Universidad Nacional Autónoma de México}}
\\ \vspace{0.5cm}
{\LARGE\textbf{Posgrado en Ciencias Físicas}}
\\ \vspace{0.5cm}
{\Large\textbf{Instituto de Ciencias Físicas}}
\end{center}

\vspace{0.75 cm}

\begin{center}
\Large{\emph{Accelerating cosmological inference of interacting dark energy \\ with neural emulators}}    
\end{center}

\vspace{0.75 cm}
\begin{center}
\huge{T \hspace{1cm} E \hspace{1cm} S \hspace{1cm} I \hspace{1cm} S}\\[1cm]
\end{center}

\begin{center}
  \large{para optar por el grado de}\\[0.75cm]

  \Large{
\textbf{Doctor en Ciencias (Física)} }\\[0.75cm]
\end{center}

\begin{center}
  {\large{PRESENTA:}}\\ \vspace{0.5cm}
  \Large{\textbf{M. C. Gabriel Karim Miranda Carrion}}
\end{center}
\end{minipage}

\newpage

\section*{Supervisores:}
\begin{flushleft}
\textbf{Prof. Juan Carlos Hidalgo Cuéllar}\\
\textit{Instituto de Ciencias Físicas, Universidad Nacional Autónoma de México, Cuernavaca, México.}\\
\vspace{0.5cm}

\textbf{Prof. Alkistis Pourtsidou}\\
\textit{Institute for Astronomy, The University of Edinburgh, Royal Observatory, Edimburgo, Reino Unido.}\\
\end{flushleft}

\vspace{1cm}

\section*{Colaboradores:}

\begin{flushleft}

  \textbf{Dr. Alessio Spurio Mancini}\\
    \textit{Department of Physics, Royal Holloway University of London, Egham, Reino Unido.}\\
    \vspace{0.5cm}
  \textbf{Dr. Pedro Carrilho}\\
    \textit{Institute for Astronomy, The University of Edinburgh, Royal Observatory, Edimburgo, Reino Unido.}\\
  \vspace{0.5cm}
  \textbf{Dr. Davide Piras}\\
    \textit{Centre Universitaire d’Informatique, Université de Genève, Ginebra, Suiza.}\\
  \vspace{0.5cm}
  \textbf{Dr. Benjamin Bose}\\
    \textit{Institute for Astronomy, The University of Edinburgh, Royal Observatory, Edimburgo, Reino Unido.}\\
 \end{flushleft}

\newpage 

\noindent {\Large\textbf{Copyright Notice:}} \\

\noindent This thesis is released under a Creative Commons Attribution (CC BY) license, permitting the copying, distribution, display, and performance of the work, provided the following conditions are met:

\begin{itemize} 
\item Proper attribution is given to the original author.
\item The work must not be used for commercial purposes.
\item No alterations, transformations, or derivative works are allowed unless explicitly permitted.
\item Certain chapters of this thesis, corresponding to published works, remain under the copyright of their respective publishers.
\end{itemize}

\noindent In any case of reuse or distribution, the terms of the license must be clearly communicated. Any exceptions to these conditions can be granted by obtaining permission from the copyright holder. \vspace{1cm}

\noindent {\Large\textbf{Thesis declaration:}} \\

\noindent I hereby declare that this thesis represents the outcome of my research and has not been submitted for any other academic degree or professional qualification. The content of this manuscript is entirely my own, except where appropriately cited and credited, with the exception of sections derived from the following jointly-authored publications:

\begin{enumerate}[I.]
\item Pedro Carrilho, Karim Carrion, Benjamin Bose, Alkistis Pourtsidou, Juan Carlos Hidalgo, Lucas Lombriser, Marco Baldi,  
\textbf{On the road to per cent accuracy VI: the non-linear power spectrum for interacting dark energy with baryonic feedback and massive neutrinos},  
Monthly Notices of the Royal Astronomical Society, Volume 512, Issue 3, Pages 3691–3702 (2022).

\item Karim Carrion, Pedro Carrilho, Alessio Spurio Mancini, Alkistis Pourtsidou, Juan Carlos Hidalgo,  
\textbf{Dark Scattering: accelerated constraints from KiDS-1000 with \texttt{ReACT} and \texttt{CosmoPower}},  
Monthly Notices of the Royal Astronomical Society, Volume 532, Issue 4, Pages 3914–3925 (2024).

\item Karim Carrion, Alessio Spurio Mancini, Davide Piras, Juan Carlos Hidalgo,  
\textbf{Testing interacting dark energy with Stage IV cosmic shear surveys through differentiable neural emulators}, 
Monthly Notices of the Royal Astronomical Society, Volume 539, Issue 4, Pages 3220–3228 (2025).
\end{enumerate}

\chapter*{Dedicatory}

\vspace*{5cm}
\begin{center}

    {\large \textit{
    In loving memory of\\[0.3cm]
    \textbf{Dante} \textcolor{black}{\fontsize{20}{24}\selectfont \faPaw} \textit{and} \textbf{Tita} \textcolor{gray}{\fontsize{16}{20}\selectfont \faPaw},\\[0.75cm]
    though your paws have left this world, your essence still walks beside me.\\ [0.2cm]
    The love you gave will never fade, it remains in my soul forever.
    }}

\end{center}

\vfill

\newpage

\vfill

\begin{center}
\textit{Thesis also dedicated to everyone who shared a path on my journey to this point.}
\end{center}
\chapter*{Acknowledgements}

\vspace{0.5cm}

\noindent \textbf{\large Español:}

\vspace{0.5cm}

\noindent Dedico esta tesis y expreso mi más profundo agradecimiento a mi familia, cuyo respaldo constante ha sido esencial a lo largo de todos mis estudios. En especial, a mi mamá, Marlene Carrion, por su amor incondicional y su apoyo diario. Este logro es tanto tuyo como mío. No encuentro palabras suficientes para agradecerte; tener una madre como tú es un regalo invaluable. Te amo. \\
A mi hermano Ivan Miranda, gracias por tu valentía, resiliencia y determinación, cualidades que me han inspirado durante todo este camino académico.\\
A mi papá Gabriel Miranda, gracias por tu apoyo y por compartir conmigo tus historias. Aunque no siempre estemos cerca, cada gesto y conversación ha dejado una huella imborrable que me acompaña. \\

\noindent No puedo dejar de expresar mi gratitud, una vez más, a mi asesor Juan Carlos Hidalgo. Ha sido un verdadero privilegio contar con su apoyo constante durante este recorrido. Incluso en los momentos de mayor incertidumbre, cuando ninguno de los dos tenía claro por dónde avanzar, siempre hubo disposición, compromiso y confianza mutua. \\
Gracias por motivarme, por impulsarme a seguir adelante, por animarme a participar en congresos, escuelas y charlas, y por compartir conmigo tu experiencia y sabios consejos. Todo ese esfuerzo conjunto ha marcado una diferencia enorme, y siempre te estaré profundamente agradecido. \\ 

\noindent Agradezco a los miembros de mi comité tutoral, Sébastien Fromentau y Octavio Valenzuela, por sus valiosos comentarios y sugerencias durante las evaluaciones, que aportaron perspectivas enriquecedoras para el desarrollo de esta tesis. También agradezco a los miembros del jurado por su participación en mi examen de grado y sus correcciones al manuscrito. \\

\noindent Estoy muy agradecido con el resto de mis familiares, mis amigos/amigas del ICF, mis compañeros de generación, mis amistades de siempre, los del frontenis, y todas las demás personas que, de alguna manera, formaron parte de este trayecto conmigo. Su apoyo, energía y compañia fueron clave para superar los desafíos y celebrar cada logro en este camino. ¡Muchas gracias! \\
\noindent Asimismo, agradezco al ICF por brindarme un espacio de trabajo y una comunidad cálida donde pude llevar a cabo esta tesis. \\

\newpage 

\noindent \textbf{\large English:}

\vspace{0.5cm}

\noindent I would like to express my sincere gratitude to my second supervisor, Alkistis Pourtsidou, for her mentorship throughout the course of this PhD. I'm especially thankful for her insightful feedback and investment of time. Working under her guidance has been both a privilege and a defining part of my academic growth. Thank you for opening the door to work within your research group, and for the trust and openness with which you welcomed me from the very beginning,  for that I'm endlessly grateful. \\

\noindent Special thanks to Alessio Spurio Mancini (he could easily be considered my co-supervisor jeje) for his enormous support and welcoming me to Royal Holloway, University of London, during a research visit. His enthusiasm, expertise, and generosity made that time both productive and truly enjoyable. Thank you for being far more than a collaborator: a mentor and friend. \\
To my friends at RHUL: thank you for making me feel at home. I truly enjoyed being part of the tennis team and sharing great times at the pub. I'm also thankful to RHUL for hospitality. \\

\noindent I'm also very grateful to Pedro Carrilho, whose contributions to this project have been truly valuable. His clear thinking, thoughtful suggestions, and supportive attitude made working together genuinely pleasant. I deeply appreciated both his scientific input and his kindness throughout. \\
I also extend my thanks to Ben Bose and Davide Piras for always being willing to help whenever I had questions or needed feedback. Their support has been greatly appreciated. \\

\noindent Part of the computational work for this thesis was performed on the Cuillin cluster at the Royal Observatory, University of Edinburgh. In which I'm thankful to Eric Tittley for his kind assistance and technical support. I also acknowledge the use of the Chalcaltzingo cluster, part of the computing facilities at ICF-UNAM.

\vfill \noindent \textit{This work was financial supported by a doctoral studentship from CONAHCyT (currently SECIHTI), as well as by the UNAM-PAPIIT grant IG102123: ``Laboratorio de Modelos y Datos (LAMOD) para proyectos de Investigación Científica: Censos Astrofísicos.”}

\section*{Acronyms} 

\begin{table}[H]
    \centering
    \begin{tabular}{>{\centering\arraybackslash}m{3cm} m{8cm}}
        \toprule
        \textbf{Acronym} & \textbf{Definition} \\ 
        \midrule
        \textbf{2PCF} & Two-point Correlation Function \\
        \textbf{AGN} & Active Galactic Nucleus \\
        \textbf{BAO} & Baryon Acoustic Oscillations \\
        \textbf{CMB} & Cosmic Microwave Background \\
        \textbf{CDM} & Cold Dark Matter \\
        \textbf{CPL} & Chevallier-Polarski-Linder \\
        \textbf{CPU/GPU} & Central/Graphics Processing Unit \\ 
        \textbf{CPT} & Cosmological Perturbation Theory \\
        \textbf{DS} & Dark Scattering \\
        \textbf{FLRW} & Friedmann-Lemaître-Robertson-Walker \\
        \textbf{GR} & General Relativity \\
        \textbf{HMC} & Hamiltonian Monte Carlo \\
        \textbf{IDE} & Interacting Dark Energy \\
        \textbf{KiDS} & Kilo-Degree Survey \\ 
        \textbf{LSS} & Large-Scale Structure \\
        \textbf{MCMC} & Markov Chain Monte Carlo \\
        \textbf{ML} & Machine Learning \\ 
        \textbf{NN} & Neural Network \\ 
        \textbf{NS} & Nested Sampling \\
        \textbf{NFW} & Navarro-Frenk-White \\
        \textbf{NUTS} & No-U-Turn Sampler \\ 
        \textbf{PCA} & Principal Component Analysis \\ 
        \bottomrule
    \end{tabular}
    \label{tab:acronyms}
\end{table}

\section*{Notation \& Convention}

Throughout this thesis, certain symbols are used to represent specific concepts and operations under whichever quantity $A$, listed in the following table: \\
\begin{center}
\renewcommand{\arraystretch}{1.5}
\setlength{\tabcolsep}{1.5pt}
\resizebox{0.75\textwidth}{!}{%
\begin{tabular}{>{\centering\arraybackslash}m{3cm} m{8cm}}
\toprule
\textbf{Symbol}  & \textbf{Meaning} \\ 
        \midrule
$\dot{A}$               & Derivative with respect to dynamical time $t$                                      \\ 
$A'$                    & Derivative with respect to conformal time $\eta$                                  \\ 
$\bar{A}$               & Background quantity                                                    \\ 
$A^{(n)}$               & $(n)$-th perturbative order  \\
$\tilde{A}(\boldsymbol{k})$ & Fourier transform of $A(\boldsymbol{x})$                                              \\ 
$\boldsymbol{x}$ & Bold symbols represents $3$-vector in $\mathbb{R}^3$                                                       \\ \hline
\end{tabular}%
}    
\end{center}

\noindent We adopt the following conventions within this work:  

\begin{itemize}
    \item Latin subscripts (e.g., $i$) denote indices in three-dimensional Euclidean space, $\mathbb{R}^3$.
    \item Greek subscripts (e.g., $\mu$) correspond to indices in four-dimensional spacetime, $\mathcal{M}^4$.
    \item The metric signature follows the convention $(-,+,+,+)$.
\end{itemize}

\noindent The Fourier transform of a given field $\Theta$ is defined as:

\begin{equation*}
\widetilde{\Theta}(\boldsymbol{k}, \eta) = \int d^3 x \, e^{-i \boldsymbol{k} \cdot \boldsymbol{x}} \, \Theta(\boldsymbol{x}, \eta),
\quad \Rightarrow \quad 
\Theta(\boldsymbol{x}, \eta) = \int \frac{d^3 k}{(2\pi)^3} \, e^{i \boldsymbol{k} \cdot \boldsymbol{x}} \, \widetilde{\Theta}(\boldsymbol{k}, \eta).
\end{equation*}

\section*{Python libraries}

\begin{center}
\begin{tikzpicture}[
    node distance=1cm and 2cm, 
    every node/.style={draw, circle, text centered, minimum size=2cm, font=\bfseries},
    library/.style={fill=turquoise!30},
    main/.style={fill=Orange!30},
    every path/.style={thick, ->, line width=1.5mm} 
]

\node[main] (python) {\texttt{Python} \par \faPython};

\node[library, above left=of python] (numpy) {\texttt{NumPy}};
\node[library, above right=of python] (scipy) {\texttt{SciPy}};
\node[library, right=of python] (jax) {\texttt{JAX}};
\node[library, below right=of python] (tensorflow) {\texttt{TensorFlow}};
\node[library, below left=of python] (numpyro) {\texttt{NumPyro}};
\node[library, left=of python] (pandas) {\texttt{pandas}};
\node[library, below=of python] (matplotlib) {\texttt{matplotlib}};
\node[library, above=of python] (seaborn) {\texttt{seaborn}};

\foreach \lib in {numpy, scipy, jax, tensorflow, numpyro, pandas, matplotlib, seaborn}
    \draw (python) -- (\lib);

\end{tikzpicture}
\end{center}

\section*{Data Availability and Codes}
\label{codes_data}

The data underlying this thesis will be shared on reasonable request. The public repositories of codes employed in this thesis are listed as follows:

\begin{itemize}
     \item \github{https://github.com/PedroCarrilho/class_public/tree/IDE_DS} \texttt{IDE-CLASS}: This is a modified version of \texttt{CLASS} to obtain the Dark Scattering (DS) linear power spectrum.
    \item \github{https://github.com/alexander-mead/HMcode} \texttt{HMCode}: This computes the pseudo power spectrum.
    \item \github{https://github.com/PedroCarrilho/ReACT/tree/react_with_interact} \texttt{ReACT-IDE}: This computes the halo model reaction spectrum for DS. 
    \item \github{https://github.com/alessiospuriomancini/cosmopower.git} \texttt{CosmoPower}: This code is used for training our emulators.
    \item \github{https://github.com/brinckmann/montepython_public} \texttt{Montepython}: This performed the KiDS-1000 statistical inference.
    \item \github{https://github.com/JohannesBuchner/MultiNest} \texttt{Multinest}: This sampler is chosen for KiDS-1000 analysis.
    \item \github{https://github.com/cmbant/getdist} \texttt{getdist}: This generates the plot contours.\footnote{The plotting style belongs to \texttt{SciencePlots} \citep{SciencePlots}.}
    \item \github{https://github.com/karimpsi22/DS-emulators.git} \texttt{DS-emulators}: This repository has available our neural emulators.
    \item \github{https://github.com/dpiras/cosmopower-jax} \texttt{CosmoPower-JAX}: This loads our emulators in a \texttt{JAX} environment.
    \item \github{https://github.com/DifferentiableUniverseInitiative/jax_cosmo} \texttt{jax-cosmo}: This contains the \texttt{JAX} pipeline for the cosmic shear forecast from a Stage IV galaxy survey.
    \item 
    \github{https://github.com/pyro-ppl/numpyro} \texttt{Numpyro}: This provides the No-U-Turn sampler.
    \item \github{https://github.com/astro-informatics/harmonic} \texttt{harmonic}: This helps to estimate the evidence post-analysis (from chains).
\end{itemize}


\pagenumbering{arabic} \setcounter{page}{1}


\chapter*{Abstract}

The present thesis aims to tackle two critical aspects of present and future cosmological analysis of Large-Scale Structure (LSS). As Stage IV galaxy surveys push the boundaries of precision, the demand for both accurate modelling of the nonlinear matter power spectrum (to capture any deviation from the standard $\Lambda$CDM) and efficient computational techniques for Bayesian parameter estimation, become increasingly important. This goes together with the requirement of thorough tests for alternative cosmological models, essential to avoid spurious results. In this thesis, such challenges are faced by studying the specific case of Dark Scattering (DS) model, which entails the pure momentum transfer between dark matter and dark energy, modulated by the characteristic $A_{\rm ds}$ parameter. \\ 

\noindent In order to capture the DS effects present in the matter fluctuations, we modify the halo model reaction approach and implement these changes into the \texttt{ReACT} code. Subsequently, this allows us to compute the nonlinear DS spectrum and validate the results by comparing them with $N$-body simulations of structure formation. We incorporate further corrections to our predictions by including baryonic feedback and massive neutrinos, studying the degeneracies between these effects and the DS interaction. At $1$ percent accuracy, we find a degeneracy between interaction and baryonic feedback, but not with neutrinos.\\

\noindent In a second stage, we present constraints on the DS model through cosmic shear measurements from the Kilo-Degree Survey (KiDS). An accelerated pipeline, powered by novel neural emulators from the machine learning-based \texttt{CosmoPower} code, speeds up the process by $\mathcal{O}(10^4)$. Our main emulator, for the DS nonlinear matter power spectrum, is trained through predictions from the halo model reaction, preserving its accuracy. Additionally, we include the effects of baryonic feedback from \texttt{HMcode2016}, whose contribution is also emulated. We analyse the complete set of statistics of KiDS-1000, namely Band Powers, COSEBIs and Correlation Functions, for DS in two distinct cases. In the first case, taking into account only KiDS cosmic shear data, we constrain the amplitude of the dark energy -- dark matter interaction to be $\vert A_{\rm ds} \vert \lesssim 20$ b/GeV at $68\%$ C.L. Furthermore, we add information from the cosmic microwave background (CMB) from Planck, along with baryon acoustic oscillations (BAO) from 6dFGS, SDSS and BOSS, approximating a combined weak lensing + CMB + BAO analysis. From this combination, we constrain $A_{\rm ds} = 10.6^{+4.5}_{-7.3}$ b/GeV at $68\%$ C.L. We confirm that with this estimated value of $A_{\rm ds}$ the interacting model considered in this work offers a promising alternative to solve the $S_8$ tension. \\

\noindent Lastly, a more robust analysis to constrain the DS model will be presented, showing forecasts for Stage IV cosmic shear surveys. This analysis is obtained through an automatically differentiable inference pipeline to further accelerate the Bayesian analysis. Using the \texttt{jax-cosmo} library and recent machine learning techniques, we accelerate the exploration of high-dimensional parameter spaces. In which, a hybrid Hamiltonian Monte Carlo sampler, implemented within \texttt{NumPyro}, is employed in order to exploit gradients, achieving accurate results while drastically reducing the computational cost from months on CPU cores to a few days on GPUs. Additionally, we estimate model evidence for each cosmological scenario using the \texttt{harmonic} code, applying three different scale cuts. \\

\noindent To put things into perspective, the methodologies for modelling, accelerating parameter inference, and managing high-dimensional parameters which employ the sophisticated tools presented in this work, hold the potential to contribute to the next wave of cosmological discoveries by upcoming generation of LSS surveys.

\newpage

\chapter*{Resumen}

La presente tesis tiene como objetivo abordar dos aspectos críticos del análisis cosmológico actual y futuro de la Estructura a Gran Escala (LSS, por sus siglas en inglés). A medida que los sondeos de galaxias de la Etapa IV empujan los límites de la precisión, la demanda de una modelización precisa del espectro de potencia de materia no lineal (con el fin de captar cualquier desviación del modelo estándar $\Lambda$CDM) y de técnicas computacionales eficientes para la estimación bayesiana de parámetros se vuelve cada vez más importante. Esto va de la mano con la necesidad de realizar pruebas exhaustivas para modelos cosmológicos alternativos, esenciales para evitar resultados espurios. En esta tesis, tales desafíos se confrontan mediante el estudio del caso específico del modelo de Dark Scattering (DS), que permite el puro intercambio de momento entre la materia oscura y la energía oscura, modulado por el parámetro característico $A_{\rm ds}$. \\ 

\noindent Con el fin de captar los efectos de DS presentes en las fluctuaciones de la materia, modificamos el modelo de halo reaction y implementamos dichos efectos en el código \texttt{ReACT}. Posteriormente, esto nos permite calcular el espectro no lineal del modelo DS y validar los resultados mediante su comparación con simulaciones $N$-cuerpos para formación de estructura. Incorporamos correcciones adicionales a nuestras predicciones mediante la inclusión de contribuciones de bariónes llamado baryonic feedback y neutrinos masivos, estudiando las degeneraciones entre estos efectos y la interacción DS. Con una precisión del $1$ porciento, encontramos una degeneración entre la interacción y el baryonic feedback, pero no con los neutrinos. \\

\noindent En una segunda etapa, presentamos restricciones para el modelo DS a través de mediciones de cosmic shear del Kilo-Degree Survey (KiDS). En un pipeline acelerado, impulsado por novedosos emuladores neuronales del código  \texttt{CosmoPower} basado en aprendizaje automático, acelera el proceso en un factor de $\mathcal{O}(10^4)$. Nuestro emulador principal, es el espectro no lineal de materia del modelo DS, se entrena mediante predicciones del modelo de halo reaction, preservando su precisión. Además, incluimos los efectos de baryonic feedback que provienen del código \texttt{HMcode2016}, cuya contribución también es emulada. Analizamos el conjunto completo de estadísticas de KiDS-1000, que consiste en Band Powers, COSEBIs y Funciones de Correlación, para el modelo DS en dos casos distintos. En el primer caso, considerando solo los datos de cosmic shear de KiDS, restringimos la amplitud de la interacción entre energía oscura y materia oscura a $\vert A_{\rm ds} \vert \lesssim 20$ b/GeV con un $68\%$ C.L. Además, al añadir información del fondo cósmico de microondas (CMB) de Planck, junto con oscilaciones acústicas bariónicas (BAO) de 6dFGS, SDSS y BOSS, aproximando un análisis combinado de Weak lensing + CMB + BAO. De esta combinación, obtenemos la restricción $A_{\rm ds} = 10.6^{+4.5}_{-7.3}$ b/GeV con un $68\%$ C.L. Confirmamos que con este valor estimado de $A_{\rm ds}$, el modelo interactuante considerado en este trabajo ofrece una alternativa prometedora para resolver la tensión en $S_8$. \\

\noindent Finalmente, se presentará un análisis más robusto para restringir el modelo DS, mostrando pronósticos del cosmic shear para los sondeos de galaxia de la Etapa IV. Este análisis se obtiene a través de una pipeline de inferencia automáticamente diferenciable para acelerar aún más el análisis bayesiano. Usando la biblioteca \texttt{jax-cosmo} y técnicas recientes de aprendizaje automático, aceleramos la exploración en espacios de parámetros de alta dimensión. En este enfoque, se emplea un muestreo estadístico híbrido de Monte Carlo Hamiltoniano, implementado dentro de \texttt{NumPyro}, para aprovechar los cálculos de gradientes, logrando resultados precisos mientras se reduce drásticamente el costo computacional, de posibles meses en núcleos de CPU a solo unos pocos días en GPUs. Además, estimamos la evidencia del modelo para cada escenario cosmológico utilizando el código \texttt{harmonic}, aplicando tres escala de cortes diferentes.\\

\noindent En perspectiva, las metodologías del modelado, la aceleración en la inferencia de parámetros y el manejo de parámetros en alta dimensión utilizando las herramientas sofisticadas presentadas en este trabajo tienen el potencial de contribuir a la próxima ola de descubrimientos cosmológicos impulsados por la próxima generación de sondeos de galaxias de gran escala.

\tableofcontents 
\newpage
\listoffigures
\listoftables
\newpage
\thispagestyle{empty}

\newpage
\makeatletter
\newif\ifappendix 

\def\@makechapterhead#1{%
  \vspace*{10\p@}%
  {\parindent \z@ 
        \raggedleft 
        \reset@font\huge\bfseries
        \begin{tabular}{rcl} 
          \ifappendix
            \textbf{Appendix \thechapter} &  {\color{RoyalBlue}\textbf{\Big|}} & 
          \else
            \textbf{Chapter \thechapter} &  {\color{RoyalBlue}\textbf{\Big|}} & 
          \fi
          \Huge #1 
        \end{tabular}
        \par\nobreak
    \vskip 10\p@
  }}
\makeatother

\makeatletter
\renewcommand\subsubsection{\@startsection{subsubsection}{3}{\z@}%
                                     {-5ex\@plus -1ex \@minus .2ex}%
                                     {-1em}%
                                     {\normalfont\normalsize\bfseries}}
\makeatother

\chapter{Introduction}\label{Chapter1}

\vspace{1cm}

\noindent In the vastness of the cosmos, today Universe's energy density budget remains as a profound enigma, and is thought to be governed by two elusive dark components. On one hand, the Dark Matter, an invisible architect that shapes the cosmic structures and plays a crucial role in their formation. On the other hand, impacting on the Large-Scales Structure (LSS), the Dark energy, a mysterious substance that is ripping the Universe apart and accelerating it towards an uncertain future. Both mysterious entities, hidden from direct observation yet, challenge our understanding and ignite our curiosity by standing as one of the most profound puzzles in contemporary cosmology. Surprisingly, almost all cosmological observations are compatible with a simplistic model called $\Lambda$CDM, where $\Lambda$ stands for dark energy in the form of a cosmological constant and CDM refers to cold dark matter, which represents our most effective ideal description of dark matter. However, despite its success, the model presents issues in both fundamental and observational aspects, which do not find a simple satisfactory solution. \\

\noindent Concerning observations, the $\Lambda$CDM model has come into question due to apparent two tensions  (i.e. discrepancies between datasets interpretations in terms of model parameters \citep{2022JHEAp..34...49A}), between unlinked probes estimations. One of the most significant controversies lies in the contrasting measurements of the rate of Universe expansion, quantifying in the Hubble constant, $H_0$, from early-time and late-time probes, which differ at a statistically level of about $\sim$5$\sigma$ \citep{2021CQGra..38o3001D}. Another remarkably tension is present in the amplitude of matter density fluctuations, quantified by the $\sigma_8$ parameter when extrapolated to the present day. This tension is particularly evident in the inferred values of the parameter $S_8 \equiv \sigma_8 \sqrt{\Omega_{\rm m}/0.3}$. Estimation from measurements of the cosmic microwave background (CMB) anisotropies by the Planck satellite in 2018 \citep{Planck:2018vyg} determine $S_8 = 0.825 \pm 0.011$ or The Atacama Cosmology Telescope (ACT) \citep{2020JCAP...12..045C} with $S_8 = 0.816 \pm 0.015 $. In contrast, cosmic shear measurements from weak lensing observations at low redshifts by the Dark Energy Survey (DES) collaboration, yield $S_8 = 0.776 \pm 0.017$ \citep{DESY3:2022}, while the Kilo-Degree Survey (KiDS) reports $S_8 = 0.759^{+0.024}_{-0.021}$ \citep{KiDS:2020suj}. Thus, discrepancies stand at about $2\sigma - 2.5 \sigma$ with respect to Planck expectation. Notably, despite the efforts in combining data from both surveys into a joint analysis, which refines the estimation to $S_8 = 0.790^{+0.018}_{-0.014}$ \citep{2023OJAp....6E..36D}, the tension persists. The final KiDS dataset, KiDS-Legacy \citep{2025arXiv250319441W}, after exhaustive improvement on systematics, delivers their most precise cosmic shear constraints. With a reported value of $S_8 = 0.815^{+0.016}_{-0.021}$, it shows strong consistency ($0.73\sigma$) with Planck results. This highlights the rigorously of data treatment required in current and future weak lensing analyses. Discrepancies of this nature may be attributed to unaccounted systematic effects or complex astrophysical dynamics; however, even after accounting for these factors, if tensions persist, they might indicate the presence of new physics, suggesting that $\Lambda$CDM would be replaced by alternatives \citep{2023MNRAS.526.5494M,2024MNRAS.531L..52C}. \\

\noindent From a theoretical perspective, the cosmology community has been involved for decades in the quest to develop models that would explain the nature of this dark sector. In the pursuit of alternatives to the concordance model, interacting dark energy (IDE) models have gained considerable attention as promising candidates. These models are proposed to accommodate a plethora of interactions, such that, certain couplings serve to mitigate and even solve cosmological tensions after the analysis of a variety of cosmological probes. These include the CMB, galaxy clustering, weak lensing, Baryon Acoustic Oscillations (BAO), supernovae, gravitational waves, other state-of-the-art analysis and future probes (see e.g. \cite{2016PhRvD..94d3518P,2020JCAP...04..008Y,2020PDU....3000666D,2021PDU....3400899L,2021JCAP...10..004C,2023PhRvD.108h4070T,2024arXiv240415232G}). Within this scope, our particular IDE model of interest in this thesis is called Dark Scattering (DS), in which the coupling is characterised by pure momentum transfer between dark energy -- dark matter. The DS interaction is a traditional elastic scattering, akin to Thomson scattering between photons and charged particles, which is allowed to take place between the dark sector without violating any conservation law or fundamental principle. This type of interaction leads to modifications in the growth of cosmic structures, that may help vanishing the observed discrepancies in the $S_8$ parameter. \\

\noindent With advancements on the horizon, the landscape of cosmological observations is preparing for a big leap, driven by the unprecedented data proliferation in size and precision from the Stage IV galaxy surveys. Instruments such as the Dark Energy Spectroscopic Instrument (DESI) \citep{DESI:2016fyo},\footnote{\url{www.desi.lbl.gov}} the \textit{Euclid} satellite,\footnote{\url{www.euclid-ec.org}} \citep{Euclid:2024yrr} the Legacy Survey of Space and Time (LSST) at the Vera C. Rubin Observatory\footnote{\url{www.lsst.org}} \citep{LSSTDarkEnergyScience:2012kar} or the Nancy Grace Roman Space Telescope,\footnote{\href{https://roman.gsfc.nasa.gov}{\texttt{roman.gsfc.nasa.gov}}} are designed to observe millions of galaxies across extensive areas of the sky, thus enabling detailed measurements of the matter density field at very small scales and towards deeper redshifts than their predecessors. \\
\noindent One of such advancements is the recent breakthrough announced with the second Data Release (DR2) of DESI BAO results \citep{2025arXiv250314738D}:  the deviations from $\Lambda$CDM first hinted at in DR1 \citep{2025JCAP...02..021A} have now strengthened. Specifically, when combined with CMB data, the new findings now suggest that a simple cosmological constant may not fully describe the Universe accelerating expansion. Notably, dark energy with a time-evolving equation of state appears to provide a better fit to the data. These results underscore the need to explore beyond $\Lambda$CDM, making alternative scenarios like DS particularly relevant. Every effort from Stage IV galaxy surveys seeks to pin deeper insights down of the properties of dark energy by setting tighter constraints on cosmological parameters and may also contribute to alleviate the aforementioned cosmological tensions. \\

\noindent Achieving these ambitious task, however, will demand to accurately model the nonlinear observables, where physics beyond gravity becomes relevant. This is, in short, baryonic effects. Particularly, at small scales where baryonic processes become relevant (for a review, see \citep{2019OJAp....2E...4C}). Properly taking into account the baryonic feedback is essential for ensuring accurate estimation, as its omission can lead to biased constraints \citep{2018MNRAS.475..676S,2019MNRAS.486.2827D,2019MNRAS.488.1652H,2020MNRAS.491.2424V,2021MNRAS.507.5869A,2022A&A...660A..27T,2023A&A...678A.109A}. Whereas, for beyond $\Lambda$CDM cosmologies present further challenges, as these scenarios involve invoking additional nonlinear effects \citep{Tsedrik:2024cdi}. For example, including the effects of non-zero neutrino mass \citep{2012MNRAS.420.2551B,2014JCAP...12..053M, 2016MNRAS.459.1468M,2018MNRAS.481.1486B,2019JCAP...03..022T,2020MNRAS.491.3101C} modifies the gravitational collapse, thus adding complexity to the modelling. Consequently, these effects must be modelled with sufficient precision, as they may be degenerate with other effects \citep{2012PhRvD..85d3007C,2022MNRAS.512.3691C,2023OJAp....6E..40S}. \\
In hindsight, during the analysis of such extensive datasets we eventually face the task of exploring complex and high-dimensional parameter spaces with the dimensions of the order of hundreds. Traditional methods can be prohibitively slow, demanding extensive processing time, thus high computational power will be required. In this context, sophisticated machine learning methods based on Neural Networks offer to accurately replicate certain demanding cosmological direct computations \citep{SpurioMancini:2021ppk,2023arXiv230404785P,2023JCAP...05..025N,2024A&A...686A..10B,2023MNRAS.518..111D,2024OJAp....7E..73P}, demonstrating excellent agreement with traditional techniques in a fraction of the runtime. In particular, Bayesian neural networks and emulator-based approaches are increasingly employed to enhance parameter inference efficiency. \\
Given that, in the next decade future inference pipelines must efficiently manage high-dimensional parameter spaces through advanced codes and libraries, the computational task calls out for innovative and more efficient approaches to both theoretical modelling and parameter estimation. Furthermore, it is critical to establish novel techniques for rigorous and quantitative model comparison \citep{2008ConPh..49...71T}, ideally decoupling this task from parameter estimation to provide greater flexibility in posterior sampling strategies. \\

\noindent Ultimately, refining our methodologies represents both challenges and opportunities for cosmology. The effort will be rewarding, we may open the door to uncovering new deeper understanding about the fundamental dynamics that govern the Universe, resulting in insights of our quest for knowledge in this vast and mysterious domain. \\

\noindent The present thesis is segmented of four main parts: background material, research work, conclusions and extra material (appendices). We start with a concise yet comprehensive summary on antecedents and topics relevant to this thesis in \autoref{Chapter1}. Then, \autoref{Chapter2} and \autoref{Chapter3} lay the conceptual foundations, which serve as a guide for a clear understanding of the cosmological phenomena discussed to this thesis. In
\autoref{Chapter2}, we describe the nonlinear modelling of the matter distribution, starting from the standard halo model and extending it to the halo model reaction. On the other hand, \autoref{Chapter3} focuses on weak lensing, detailing the formulation of cosmic shear and its observational contributions. \\
The research work is developed across \autoref{Chapter4}, \autoref{Chapter5}, \autoref{Chapter6}, and \autoref{Chapter7}, which constitute the core of this thesis. \autoref{Chapter4} briefly reviews IDE theories, introduces the DS model, and outlines its implementation within the halo model reaction formalism through \texttt{ReACT}. We validate our DS nonlinear matter power spectrum predictions against $N$-body simulations and extend it to get full predictions by including baryonic feedback and massive neutrinos, exploring their potential degeneracies with DS interactions. 
Next, \autoref{Chapter5} mainly introduces \texttt{CosmoPower} for constructing emulators of the DS nonlinear and linear power spectrum, as well as linear and baryonic feedback effects. We evaluate the accuracy of each emulator, benchmark their computational performance, and demonstrate their potential to significantly accelerate cosmological analyses.  \\
With these tools at hand, \autoref{Chapter6} explores constraining the DS model using KiDS-1000 galaxy shear data. We begin by presenting the results from the KiDS-1000 analysis alone, followed by the addition of CMB and BAO information to examine how they refine the DS constraints. The results show that the DS model offers a promising solution to the $S_8$ tension. \\
Subsequently, an equivalent analysis is also presented for a more sophisticated pipeline in \autoref{Chapter7}, showing forecasts of different data vectors from Stage IV cosmic shear survey, obtained through an automatically differentiable inference pipeline powered by \texttt{jax-cosmo} and \texttt{CosmoPower-JAX} to further accelerate the Bayesian analysis. We also highlight our model comparison methodology, which is fully decoupled from the sampling scheme. We summarize and draw conclusions of this thesis in \autoref{Chapter8}. For reference, \autoref{Appendix_a}, \autoref{Appendix_b} and \autoref{Appendix_c}, contain supplementary material that has been omitted from the main text.


\section{Cosmology and gravity}
\label{Sec:GR+cosmo}

The starting point must be General Relativity (GR): the modern theory of gravity first put forth by Albert Einstein in 1915. His theory has revolutionized our comprehension of gravity, which is manifested as a result of how space geometry responds to matter dynamics. Rather than considering three-dimensional space as separate of time, this framework unifies them, treating time as an additional dimension. On this unusual four-dimensional space, better-known as spacetime,\footnote{Spacetime is a Lorentzian manifold that is homogeneous and rigid.} a point is defined as an event, denoted by $x^{\mu}$. The notion of distance between two events is described by the line element:
\bea
ds^2 = g_{\mu \nu} dx^\mu dx^\nu \, . 
\label{Eq:line_element}
\eea

\noindent Here $g_{\mu \nu}$ is the metric tensor, a central dynamical variable in GR that shapes spacetime geometry. Einstein's formulation relies on the mathematical foundations of manifolds and differential geometry \citep{2004sgig.book.....C,2009fcgr.book.....S}, which are elegantly contained into the Einstein field equations:
\bea
G_{\mu \nu} \equiv R_{\mu \nu} - \dfrac{1}{2} R \ g_{\mu \nu} = \dfrac{8 \pi G}{c^4} T_{\mu \nu} \, .
\label{Eq:einstein}
\eea

\noindent In this equation, the \textit{left-hand side} (l.h.s.) $G_{\mu \nu}$ defines the Einstein tensor that provides the curvature of spacetime resulting from massive distributions, where $R_{\mu \nu}$ is the Ricci curvature tensor, $R$ is the Ricci scalar, $G$ is Newton's gravitational constant, and $c$ is the speed of light. On the \textit{right-hand side} (r.h.s.) consists of the total energy-momentum tensor $T_{\mu \nu}$, which quantifies the embodied density and flux of energy and momentum within spacetime as follows:
\[
T^{\mu\nu} = \begin{pmatrix}
    \textcolor{blue}{T^{00}} & \textcolor{teal}{T^{01}} & \textcolor{teal}{T^{02}} & \textcolor{teal}{T^{03}} \\
    \textcolor{teal}{T^{10}} & \textcolor{magenta}{T^{11}} & \textcolor{magenta}{T^{12}} & \textcolor{magenta}{T^{13}} \\
    \textcolor{teal}{T^{20}} & \textcolor{magenta}{T^{21}} & \textcolor{magenta}{T^{22}} & \textcolor{magenta}{T^{23}} \\
    \textcolor{teal}{T^{30}} & \textcolor{magenta}{T^{31}} & \textcolor{magenta}{T^{32}} & \textcolor{magenta}{T^{33}}
\end{pmatrix}
= \begin{pmatrix}
    \textcolor{blue}{\text{Energy density}} & \textcolor{teal}{\text{Energy flux}} \\
    \textcolor{teal}{\text{Momentum density}} & \begin{array}{ccc}
         &  & \\
         & \textcolor{magenta}{\text{Momentum}} &  \\
         & \textcolor{magenta}{\text{flux}} & 
    \end{array}  
\end{pmatrix} \, .
\]

\noindent Moreover, it is necessary that $T_{\mu \nu}$ satisfies the divergence-free condition ($\nabla^\mu T_{\mu \nu} = 0$). Such that, the energy and momentum conservation is preserved, aligning with our intuitive understanding of conservation laws. Moreover, such equations can be also derived via an action principle with the Einstein-Hilbert action,
\bea
S = \dfrac{1}{8 \pi G} \int d^4 \sqrt{-g} \left(R - 2 \Lambda \right) \, .
\label{Eq:einstein_action}
\eea

\noindent Einstein's equations are conformed by ten partial differential equations of the metric tensor $g_{\mu \nu}$. Due to the complexity of these equations, evidently it is infeasible to derive an universal solution. However, by imposing certain symmetries or specifying conditions, it is possible to make them more tractable. As an example, in the weak-field limit where $\Phi$, the gravitational field is relatively weak and velocities are much lower than the speed of light, then we would recover the Newtonian gravity. In this regime, the gravitational field is sourced by the mass density $\rho$, leading to the Poisson equation,
\bea
\nabla^2 \Phi = 4\pi G \rho \, .
\label{Eq:poisson}
\eea

\noindent Nowadays, GR has achieved numerous successful predictions that have been confirmed through both direct and indirect observations. To cite a few instances, it accurately predicts the planetary orbits (improving upon Newtonian mechanics, particularly evident in the precession of Mercury's orbit). Furthermore, GR predicts the existence of compact objects, such as black holes, which has been confirmed through the observation of their shadows by the Event Horizon Telescope (EHT) \citep{2019ApJ...875L...1E}. The signal detection of gravitational waves from merging black holes by the Laser Interferometer Gravitational Wave Observatory (LIGO) \citep{PhysRevLett.116.061102}. \\ 
Exceptionally through these equations, we can derive predictions that offer essential insights into the Universe -- its fundamental components, geometric structure, evolution, and origins. \\
Einstein initially viewed the Universe as static and eternal. However, his solutions were unstable and collapsed due to gravitational interactions. In order to achieve an unchanging Universe, he introduced a constant in his equations, expressed as follows:
\bea
G_{\mu \nu} + \Lambda g_{\mu \nu} = R_{\mu \nu} - \dfrac{1}{2} R g_{\mu \nu} +\Lambda g_{\mu \nu} = \dfrac{8 \pi G}{c^4} T_{\mu \nu} \, .
\label{Eq:einstein_with_constant}
\eea

\noindent $\Lambda$ is the cosmological constant, designed solely to keep the stability of his static model according to his belief. In 1927, Georges Lemaître,\footnote{Historically, Lemaître had also proposed the known Big Bang theory, which he called the hypothesis of the {\it{primeval atom}} as the origin of the Universe.} a pioneering cosmologist, put into a question the notion of absolute Universe in proposing spacetime itself evolves with the passage of time. He provided evidence for an expanding Universe through the observation of redshifted galaxies. Lemaître's findings received little recognition. Years later, Einstein recognized Lemaître's contributions, declaring that the introduction of the cosmological constant had been the worst mistake of his whole life, although perhaps it was not a mistake after all. 

\subsection{Friedmann Equations}
\label{Subsec:Friedmann}

The transition to a dynamic view of the Universe unlocked a new realm to study its evolution. An exact solution to describe the Universe from Einstein's equations was independently constructed by four individuals: Alexander Friedmann (1922), Georges Lemaître (1927), and Howard Robertson (1929), and was later generalized by Arthur Walker (1936). Their work led to the formulation of the so-called Friedmann-Lemaître-Robertson-Walker (FLRW) metric. The line element is expressed as:
\bea
ds^2  = g_{\mu \nu} dx^\mu dx^\nu =   - c^2 dt^2 + a^2(t) \left[\dfrac{d\chi ^2}{1- K \chi^2} + \chi^2 d\theta^2 + \chi^2 \sin^2\theta d\varphi^2 \right] \, ,
\label{Eq:flrw}
\eea

\noindent where $a(t)$ is the scale factor and the intrinsic curvature and denoted by $K$, it tags the geometry of three-dimensional space. When $K = 0$, this points out a spatially flat universe with Euclidean geometry. If $K > 0$, the universe exhibits spherical geometry with finite volume. Otherwise, $K < 0$ corresponds to negative curvature, resulting hyperbolic geometry with infinite volume. It is worth noting that the comoving coordinate in the FLRW, $\chi$, is not a physically observable; instead, the physical coordinate is related as $\boldsymbol{r}_{\rm phys} = \boldsymbol{r} =  a(t) \boldsymbol{\chi}$. The comoving distance is then always equal to the physical distance at the present moment in time.\footnote{From now on, we use natural units where the speed of light is set to unity, $c=1$, unless stated otherwise.} Finally, the scale factor is related to the redshift $z$ via the following expression: $a = 1/(1+z)$.\\

\noindent Yet, at the core of this solution resides a fundamental axiom:
\begin{axiom*}
On sufficiently large scales (at least 100-150 Mpc), the Universe distribution on the LSS ought to be \underline{statistically} isotropic and homogeneous.
\end{axiom*}

\noindent Here, isotropy means that the Universe is the same in all directions: everywhere we look on cosmic scales, no ``direction” looks particularly is favoured from any other. Homogeneity, in contrast, tells us that the Universe has same measurable locations: the same density, temperature, pressure and so on. As far as we can tell, the cosmological principle is commonly accepted in cosmology. Although initially taken as no-evidenced postulate, it have been strongly supported by large-scale observational evidence. \\

\noindent In the simplest scenario of a non-empty Universe uniformly filled with one or more forms of energy, such that, these can be modelled as a perfect fluid associated with a density $\rho$, a velocity $u^\mu$ and pressure $P$. Under these conditions, the energy-momentum tensor is given by:
\bea
T_{\mu \nu} = \left(\rho + P \right) u_\mu u_\nu + P g_{\mu \nu} \, .
\label{Eq:perf_fluid}
\eea

\noindent The cosmological principle forces isotropy on the macroscopic velocity field, meaning that the $4$-velocity of a comoving observer has only a temporal component, $u^\mu = (1,0,0,0)$. Imposing the rules of GR, they will determine the Universe evolution. Thus, the equation derived from Eq.~\eqref{Eq:einstein_with_constant} is obtained as:
\bea
H^2 = \left( \dfrac{\dot{a}}{a} \right)^2 
= \dfrac{8 \pi G}{3} \rho + \dfrac{\Lambda}{3} - \dfrac{K}{a^2}\ ,  \qquad \text{\underline{Friedmann Equation,}}
\label{Eq:Friedman_equation}
\eea

\noindent with $H$ being the Hubble parameter. Its value today is estimated at somewhere around $67 \, \text{km} \ \text{s}^{-1} \text{Mpc}^{-1}$, indicating the rate of cosmic expansion per megaparsec (approximately $3.26$ million light-years). Additionally, taking the trace of Eq.~\eqref{Eq:einstein_with_constant} yields 
\bea
 \dfrac{\ddot{a}}{a} = - \dfrac{4\pi G}{3} \left(\rho + 3P\right) + \dfrac{\Lambda}{3} \, ,
 \hspace{1cm}
\text{\underline{Raychaudhuri Equation.}} 
 \label{Eq:acel_equation}
\eea

\noindent To fully close the system, an additional relation must be specified to ensure that the number of degrees of freedom matches the number of equations. This is commonly done by specifying an equation of state, characterised by the parameter $w$, defined as:
\bea
w = \dfrac{P}{\rho} \, .
\label{Eq:state_equation}
\eea

\noindent The equation of state parameter $w$ marks the influence on how different forms of energy behave over time. We generally assume energy components like radiation, pressure-less matter (CDM), and dark energy. For instance, a value of $w = 0$ corresponds to a matter-dominated era, while $w = \frac{1}{3}$ is for radiation, and $\left(w < -\frac{1}{3}\right)$ is indicative of a dark energy-dominated. Notably, dark energy is often associated with $w=-1$, corresponding to a cosmological constant. Additionally, the density of species $i$ is often expressed as a ratio to the critical density (case of $K=0$), known as the  dimensionless density parameter:
\bea
\Omega_i \equiv \dfrac{\rho_i}{\rho_{\rm crit}} \, .
\label{Eq:density_param}
\eea

\noindent Interestingly, an important aspect of the Friedmann equations is that a solution for a static universe requires fine-tuning. Now, depending on universe composition, it will naturally tend to either expand or contract. But what does it truly mean for space to ``expand" or ``contract"? Unlike physical objects, space cannot be isolated, manipulated, or measured directly. We cannot, for instance, extract a portion of space and conduct experiments on it.  Instead, we infer its properties by observing measurable phenomena, such as galactic motion, the dynamics of cosmic structures, and other astrophysical signals. A key example is the light from distant objects must be redshifted or blueshifted accordingly. \\

\noindent This concludes our discussion; the solutions of Eqs.~\eqref{Eq:Friedman_equation}, \eqref{Eq:acel_equation} and \eqref{Eq:state_equation} delineate a spectrum of possible universes governed by one or more sources of matter. Among the numerous components that can be considered are: baryons, CDM, dark energy, photons, neutrinos, scalar fields, curvature, or any other conceivable entities that could be present. Depending on their energy contributions, a predicted universe may fit with our observations. As an analogy, likely identifying a suspect with his/her fingerprint. 

\subsection{\texorpdfstring{$\Lambda$}CCDM model}
\label{Sec:LCDM}

\begin{figure}[b!]
\centering
\includegraphics[width=0.75\linewidth]{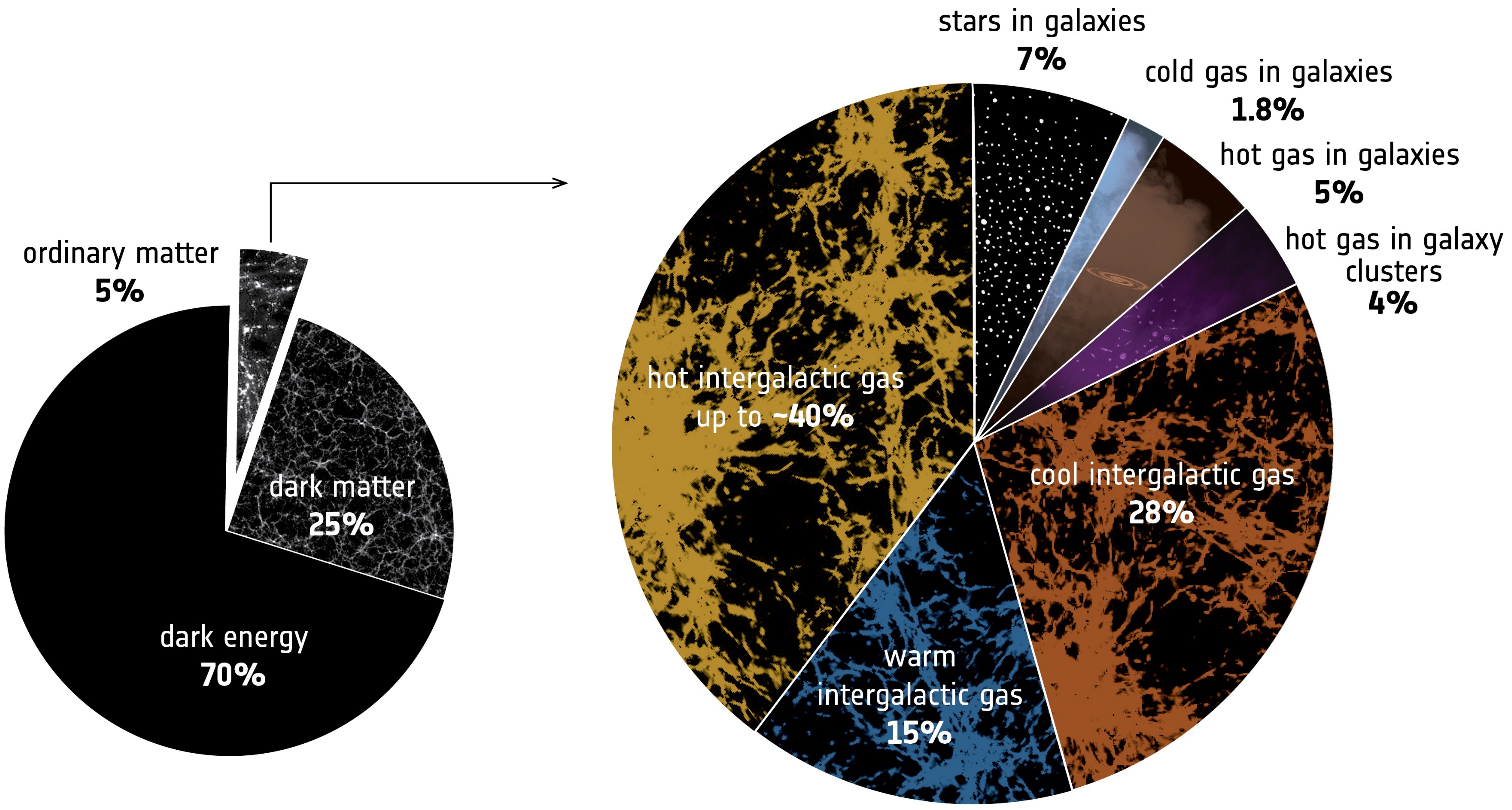}
\caption[Energy density budget of the Universe]{Pie chart of the cosmic energy budget today. The dark sector mostly dominates the Universe, whereas the baryonic density (mostly intergalactic gas) amounts to only about $5\%$. \\ \textbf{Image courtesy:} The European Space Agency (ESA).}
\label{Fig:universe_budget}
\end{figure}

Thanks to precise measurements from various cosmological probes, essentially the anisotropies of the Cosmic Microwave Background (CMB) \citep{Planck:2018vyg} and the Baryon Acoustic Oscillations (BAO) \citep{2018MNRAS.473.4773A}, have provided us with a detailed estimation of the amount from Universe's components. Today, we know that almost all of our observations align exceptionally well with a simplified model known as $\Lambda$CDM, which predicts the Universe's energy budget of Eq.~\eqref{Eq:density_param} mainly constituted by dark matter and, predominantly, dark energy. Remarkably, this dark sector accounts for approximately $95\%$ of today's total energy density budget of the Universe, leaving the rest of components as relatively minor contributors (see \autoref{Fig:universe_budget}).\\

\noindent The present-day energy composition of the Universe can be quantitatively described as follows: radiation, primarily in the form of photons, makes up about $0.005\%$. For massive neutrinos approximately $0.13\%$, while now behave as matter but acted as radiation in the early Universe when their mass was negligible to their kinetic energy. Baryonic matter accounts for roughly $4.9\%$, including stars, gas, plasma, black holes, galaxies, and all matter made of protons, neutrons, and electrons. About $26.8\%$ consists of dark matter, a non-luminous form of matter that is massive, cold, clumps under gravity, pressure-less and does not interact with light. Lastly, $68.3\%$ is attributed to dark energy, a poorly understood component driving the accelerated expansion of the Universe, gradually dominating over and over other forms of energy. \\

\noindent In an attempt to understand the nature of dark energy, various parametrizations have been proposed by extending its equation of state beyond Eq.~\eqref{Eq:state_equation}. The simplest is assuming a constant $w \neq -1$, leading to the $w$CDM model. Further refinement can be achieved through a Taylor expansion of $w$ as a function of redshift at first order,
\bea
w(a) = w_0 + w_{\rm a} \left(1 - a\right) \, ,
\label{Eq:wa}
\eea

\noindent commonly referred to as the Chevallier-Polarski-Linder (also known as CPL) model \citep{2001IJMPD..10..213C,2003PhRvL..90i1301L}. 
Intriguingly, this CPL parametrisation $(w_0, w_{\rm a})$ has recently posed a challenge to the $\Lambda$CDM model, as indicated by the latest DESI BAO DR2 results \citep{2025arXiv250314738D}. This opens up the possibility of probing whether deviations from the standard paradigm suggest valuable insights into new physics underlying cosmic acceleration.

\section{The Perturbed Universe}
\label{Sec:CPT}

\begin{figure}[t!]
\centering
\includegraphics[width=0.5\linewidth]{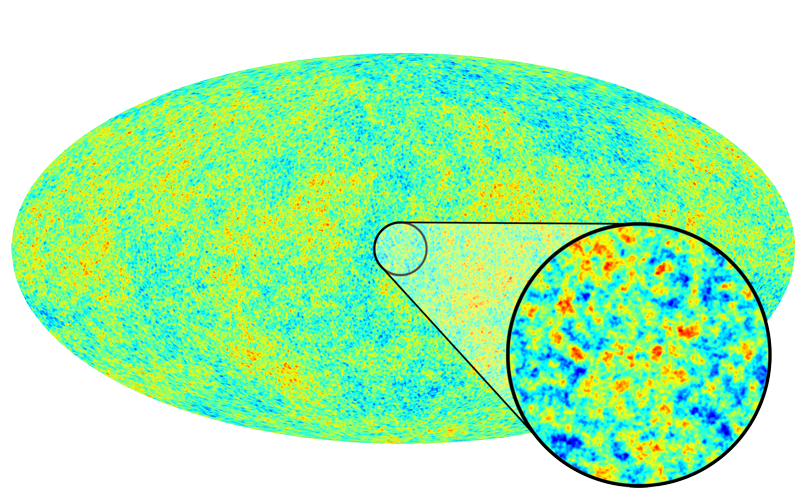}
\caption[Our picture from CMB]{Our picture from CMB measured by Planck satellite \citep{2020A&A...641A...1P} with an average temperature of $T = 2.7$ K over the whole sky. The temperature fluctuates, at the level of just one part in $100,000$. \\ \textbf{Image credit:} Chris North, Cardiff University.}
\label{Fig:cmb}
\end{figure}

So far, we have discussed a statistically homogeneous and isotropic Universe, valid only on large scales, typically beyond $100$-$150$ Mpc, in accordance with the cosmological principle. However, this depiction is insufficient to capture the finer details from inhomogeneities at smaller scales. \\
Perturbation methods have been formidable tools in physics, particularly effective in cosmology, for predicting the evolution of inhomogeneities across various stages. For instance; our understanding of the anisotropies in the CMB (see \autoref{Fig:cmb}) with accuracies increasingly over the years from COBE\footnote{\href{https://lambda.gsfc.nasa.gov/product/cobe/}{\texttt{lambda.gsfc.nasa.gov/product/cobe/}}}(1989) to WMAP\footnote{
\href{https://map.gsfc.nasa.gov}{\texttt{map.gsfc.nasa.gov}}
} (2001) to Planck\footnote{\url{www.esa.int/Enabling_Support/Operations/Planck}} (2009), in which, the use of perturbative linear order has been enough to describe them, since temperature fluctuations are in the order $\delta T/T \sim 10^{-5}$. \\

\noindent On the other hand, regarding LSS, matter perturbations present a distinct situation. As long as the density perturbations satisfy the condition $\delta \rho / \rho \ll 1$ across a wide range of scales, the linear approximation remains effective and accurate. However, as time moves forward, increasingly more structures form, then the linear approximation breaks down at scales below $10$ Mpc, where the dynamics become highly nonlinear due to $\delta \rho / \rho \approx 1$. Therefore, we must either include higher-order terms in the perturbative expansion of the matter density, or resort to computationally expensive numerical simulations. \\

\noindent The central idea in Cosmological Perturbation Theory (CPT) \citep{2002PhR...367....1B,2012reco.book.....E,2020moco.book.....D} is considered a framework for studying small deviations from a flat FLRW background as follows,
\bea
ds^2 = a^2 [-(1+2\Phi)d\eta^2 + (1-2\Psi)\delta_{ij} dx^i dx^j ] \, .
\label{Eq:FLRW_pert}
\eea

\noindent This metric is written in the widely used gauge known as the Newtonian gauge\footnote{Newtonian limit is appropriate describing scales smaller than the Hubble horizon (i.e., sub-horizon scales) and non-relativistic matter components. For a rigorous description one must resort to perturbations in the framework of GR (see \citep{2009PhR...475....1M} for details).} used in CPT. In this gauge, $\Phi$ and $\Psi$ represent the Bardeen potentials, which distinguishes the perturbations of spacetime. \\
On top of this, in a perturbed analysis; each background physical quantity (denoted by $\hat{\Upsilon}$) is adjusted by adding a small linear term, i.e. $\Upsilon(\boldsymbol{x},\eta)  \equiv \bar{\Upsilon} + \delta \Upsilon$, with $\Upsilon \in \{\rho$, $P$, $\boldsymbol{v}$, $\phi \}$ and so forth. \\ 
For now, we focus on density perturbations (for further details, see \autoref{Appendix_a}), and working in Fourier space, so that, the $\boldsymbol{k}$-modes evolve independently and are related to $\boldsymbol{x}$ inversely. Here, we define the density contrast as:
\bea
\delta(\boldsymbol{x}, \eta) \equiv  \dfrac{\bar{\rho} +  \delta \rho}{\bar{\rho}} \, ,
\label{Eq:dens_contrast}
\eea

\noindent with the linear solution expressed as:
\bea
\tilde{\delta}_{\boldsymbol{k}}(\eta) = D_+(\eta) A_{\boldsymbol{k}} + D_{-}(\eta) B_{\boldsymbol{k}} \, .
\label{Eq:delta_lin}
\eea

\noindent where $D_{\pm}$ is the linear growth factor and decay factor respectively, and they depend purely on the background quantities. The decay mode becomes negligible due to the accelerating expansion, so it will not be considered further. We define one more relevant quantity in LSS,
\bea
f = \dfrac{d \ln D_+}{d\ln a} \, ,
\label{Eq:f_growth}
\eea

\noindent where $f$ is referred to as the growth rate function. For example, considering an Einstein-de Sitter (EdS) universe with $\Omega_{\rm m} = 1$ and $\Omega_\Lambda = 0$. Then, we would obtain the solution $\tilde{\delta}^{\rm EdS}_{\bm{k}}(\boldsymbol{x},\eta) = A_{\bm{k}} a(\eta)$, consequently $f = 1$. So, the growth factor scales as $a(\eta)$, with the normalization $D_+(a_0) = 1$. This result is very useful for describing an overdensity through a top-hat model, as described in \autoref{Chapter2}. \\
Furthermore, this method can be extended to higher orders (detailed derivations in \autoref{Appendix_a}) by recurrence formulae. The $n$-th order solution would be written as,
\begin{align}
\tilde{\delta}^{(n)}(\boldsymbol{k}, \eta) = & \ D^n_+   \left[ \prod_{i=1}^{n}  \int  \dfrac{d^3 k_i}{(2\pi)^3} \right] (2\pi)^3 \delta_{\rm D}\left(\boldsymbol{k} - \sum_{i=1}^{n} \boldsymbol{k}_i \right) F_n(\boldsymbol{k}_1, \dots ,\boldsymbol{k}_n) \delta_0(\boldsymbol{k}_1) \cdots \delta_0(\boldsymbol{k}_n) \, .
\label{Eq:delta_n_order}
\end{align}

\noindent This result allows us to explicitly calculate how density perturbation evolves nonlinearly. This however is limited to only dark matter, and also does not take into account, things like shell crossing \citep{Szekeres:1995gy}, halo formation \citep{2002PhR...372....1C}, baryonic effects, e.g. active galactic nucleus (AGN) \citep{2007MNRAS.380..877S}, multi-streaming \citep{2023ApJ...950L..13E}, velocity distortion \citep{1987MNRAS.227....1K}, among others. To address these complexities, $N$-body simulations \citep{2005MNRAS.364.1105S,2013JCAP...06..036T} can be employed, as they solve the Poisson equation numerically by evolving the trajectories of $N$ particles under gravity in a box of size $N^3$. However, such simulations are computationally expensive, especially when aiming for high resolution, large volumes, or exploring wide cosmological parameter spaces. Moreover, they often lack baryonic physics or must rely on subgrid modelling for processes like star formation and feedback, introducing uncertainties.\\
As an alternative, codes such as \texttt{CLASS} (Cosmic Linear Anisotropy Solving System) \citep{2011arXiv1104.2932L} and \texttt{CAMB} (Code for Anisotropies in the Microwave Background) \citep{2011ascl.soft02026L} are Boltzmann solvers designed to compute the evolution of coupled perturbations like photons, neutrinos, baryons, and dark matter from the early radiation-dominated era to the present. These perturbations provide a pathway to deriving quantities that can be directly compared with cosmological observations.

\section{Observables}
\label{sec:observables}

To reconcile observations with our theoretical models, we must identify observable signatures in measurable quantities. 
In an expanding Universe, the simplest evolving quantity to think is its density, which decreases as the expansion increases the volume. \\
To quantify this, we will describe the statistics of the matter distribution. A good starting point is the linear density field, $\delta^{(1)}(\boldsymbol{r})=\delta_0(\boldsymbol{r})$, which helps us determine the matter power spectrum, $P(\boldsymbol{k})$. We begin by considering the probability of finding a particle, galaxy, or halo within a small volume $dV_1$, given by $\mathsf{P}_1 = \bar{n} dV_1$, where $\bar{n}$ represents the mean density. Whereas, the probability of finding another one in $dV_2$, separated by a distance $\boldsymbol{r}$ would be written as:
\bea
\mathsf{P}_{12}(\boldsymbol{r}) = \bar{n} \left[ 1 + \xi(\boldsymbol{r}) \right] dV_1 dV_2 \, ,
\label{Eq:prob_2pcf}
\eea

\noindent where $\xi(\boldsymbol{r})$, known as the two-point correlation function (2PCF), captures the excess probability of finding pairs of objects at separation $\boldsymbol{r}$, beyond what would be expected for a random distribution. \\
According to the predictions of inflation, a period of rapid exponential expansion in the very early Universe in which quantum fluctuations were stretched to cosmological scales, seeding the initial conditions for structure formation. As a result, these primordial density fluctuations are expected to follow a nearly Gaussian distribution characterized by a standard deviation $\sigma$, a statistical feature that is inherited by the matter fields at later times.
\bea
\mathsf{P}(\delta) d\delta  = \dfrac{1}{\sqrt{2\pi}\sigma} \exp(-\delta^2/(2\sigma)^2) d\delta \, .
\label{Eq:prob_gauss}
\eea

\noindent As a result, we can model $\delta(\boldsymbol{r})$ as Gaussian fields, shaped by random processes in the early Universe. Then, the 2PCF\footnote{A limitation is, the 2PCF does not describe non-Gaussian fields, thus requiring higher-order statistics like the three-point correlation function (3PCF).} can be defined in terms of the density contrast, 
\bea
\xi(\boldsymbol{r}) \equiv \langle \delta_0(\boldsymbol{r}') \delta_0(\boldsymbol{r}'+\boldsymbol{r})  \rangle \, .
\label{Eq:2PCF}
\eea

\noindent Notice by ergodicity, $\langle \cdot \rangle$ denotes an ensemble average meaning the average overdensity across several realizations of the random process. Different estimators can be used to calculate the 2PCF, typically involving comparisons to random samples \citep{1999A&A...343..333K}. \\ 
Now, we can cast Eq.~\eqref{Eq:2PCF} into Fourier space in order to define the following:
\bea
P_{\rm L}(\boldsymbol{k}) \equiv  \int d^3 \boldsymbol{r} \ \xi(\boldsymbol{r}) \exp(-i \boldsymbol{k} \cdot \boldsymbol{r}) \, .
\label{Eq:2pcf_fourier}
\eea

\noindent Here, $P_{\rm L}(\boldsymbol{k})$ is the linear matter power spectrum. Assuming isotropy symmetry (i.e. rotation invariance), we can simplify it to $P_{\rm L}(\boldsymbol{k}) = P_{\rm L}(k)$. Additionally, considering homogeneity (translation invariance), the power spectrum is typically expressed as:
\bea
 \langle \delta_0(\boldsymbol{k})\delta_0(\boldsymbol{k}')\rangle = (2\pi)^3   \delta_{\rm D}(\boldsymbol{k} + \boldsymbol{k}') \ P_{\rm L}(k) \, . 
 \label{Eq:Power_spectrum}
\eea

\noindent Conventionally, a dimensionless power spectra is often more employed, given by,
\bea
\Delta(k) = \dfrac{k^3 P(k) }{2\pi^2} \, .  
\label{Eq:Delta_k}
\eea

\noindent Importantly, its statistical meaning of the power spectrum is deduce from Eq.~\eqref{Eq:Power_spectrum}, it is an approach to estimate the variance of density distribution, since the mean $\langle \delta_0 \rangle$ must be zero. To make this variance meaningful on specific physical scales, the density field is smoothed over a region of radius $R$ using a window function. The resulting variance, $\sigma^2(R)$, is given by:
\bea
\sigma^2(R) = \int d \ln k \ |W(kR)|^2 \Delta(k) \, ,
\label{Eq:sigma}
\eea

\noindent where $W(kR)$ is the Fourier transform of the window (or smoothing) function. This function filters the density perturbations, emphasizing modes relevant to the scale $R$ while suppressing others. A common choice is a real-space top-hat filter, which corresponds to averaging the density within a spherical region of radius $R$. One particularly important cosmological parameter, $\sigma_8$, is defined as the root-mean-square fluctuation of the density field smoothed on scales of $R = 8$ Mpc, extrapolated to redshift $z = 0$,
\bea
\sigma_8 \equiv \sigma\left(R=8 \ {\text{Mpc}}, z = 0 \right) \, . 
\label{Eq:sigma_8}
\eea

\noindent The $8$ Mpc scale is chosen as a transitional point, which marks a standardised benchmark from small scales to large-scale. \\
Furthermore, to improve the prediction of matter distribution, we could add all orders in the perturbative expansion of the density contrast, as defined in Eq.~\eqref{Eq:delta_n_order} in order to refine the matter power spectrum, like
\bea
P_{\delta \delta}(k,z) = \langle \delta(k, \eta)\delta(k', \eta)\rangle = \sum^{\infty}_n \sum^{\infty}_m \langle \delta^{(n)}(k, \eta)\delta^{(m)}(k', \eta)\rangle \, .  
\label{Eq:Ps_full}
\eea

\noindent Then, making use of the Wick's theorem in which assuming $\delta$ is Gaussian, terms with $l = n + m$ being odd contribute zero, while even terms are considered. Thus, we obtain:
\bea
\langle \delta(k, \eta)\delta(k', \eta)\rangle = \sum_{n, m} \, \prod_{l \, \in \,\text{even}}  \langle \delta^{(n)}(k, \eta)\delta^{(m)}(k', \eta) \rangle \, .
\label{Eq:Wicks}
\eea

\noindent Depending on where is the order truncated after a finite number of terms, the nonlinear regime can be exploited to provide better accuracy. In this thesis, we only introduce the perturbative expansion at the 1-loop\footnote{The \textit{loop} term involves a closely analogy of diagrammatic representation from quantum field theory.} contribution, by the following expression,
\bea
P_{\delta \delta}(k,z) =  P_{\rm L}(k,z) + P_{1\rm \text{-}loop}(k,z) \, .
\label{Eq:nonlinear_matter_power}
\eea 

\noindent Being the 1-loop term as,
\bea
P_{1\rm \text{-}loop}(k,z) = 2P_{(1,3)}(k, z) + P_{(2,2)}(k, z) \, .
\label{Eq:P_1loop}
\eea 

\noindent This correction will refine the nonlinear regime of a modified halo model introduced later. In respect to initial conditions of the matter power spectrum, the inflation theory suggests that the primordial spectrum shape is determined by the underlying mechanisms of inflation. Predominantly, predicting Gaussian perturbations with a quasi-scale invariant dimensionless power spectrum
\bea
P(k) = A_s \left(\dfrac{k}{k_*}\right)^{n_s-1} \, ,  
\label{Eq:Pk_primordial}
\eea

\noindent where $A_s$ is the primordial scalar amplitude, $k_* = 0.05 ~{\text{Mpc}^{-1}}$ is the pivot scale guarantees homogeneity and isotropy, $n_s \sim 1$ represents the spectral index. If $n_s = 1$ then we would obtain the Harrison-Peebles-Zel’dovich spectrum and has the preference of being scale invariant. To evolve this spectrum, it is typically connected through the transfer function $T(k)$ that contains messy physics, i.e. a range of cosmological processes, including BAO, radiation driving, neutrino free-streaming, and matter-radiation equality that affect the growth of perturbations across scales.\\

\noindent Notably, the standard cosmological model, $\Lambda$CDM, incorporates this evolution and is parametrised by six fundamental quantities:
\bea 
\boldsymbol{\theta}_{\Lambda \rm CDM} = \{ \Omega_{\rm b}, \Omega_{\rm cdm}, h, \tau_{\rm reio}, n_s, A_{\rm s} \} \, ,
\label{Eq:LCDM model}
\eea

\noindent where each parameter plays a specific role in shaping both the primordial spectrum and its evolution through the transfer function.  The takeaway from this section is the possibility to generate observables from theory, this enables us to extract vital information and then derive relevant quantities, as we just illustrated on the matter power spectrum. In this context, \texttt{CLASS} and \texttt{CAMB} are useful for high-precision calculations of temperature and matter power spectra, effectively linking theoretical predictions with observational data through robust inference analysis.

\section{Cosmo-Statistics: Bayesian Inference}
\label{Bayesian_cosmo}

The interplay between observation and theory has pursuit to any natural science that seeks to probe a scientific model. In this process, statistical analysis are designed to extract insights and interpret data. Conventional approaches often rely on Frequentism statistics, however, these lack in effectively incorporating prior knowledge or managing uncertainties inherent in data. Here, the probability $\mathsf{P}$ is fundamentally tied to the frequency of events over many repeated trials. In cosmology, naturally, we face an unique challenge: there is only one observable realisation of the Universe. In contrast, this limitation makes the Bayesianism \citep{2014bmc..book.....H,2017arXiv170101467T} statistics\footnote{
Both the statistics schemes ought to follow the probability rules.} particularly valuable, as it allows us to incorporate our beliefs (inside probability distributions) based on an unique data, treating them as dynamical things. In addition, this simultaneously estimates multiple parameters in cosmological models while quantifying uncertainties. \\

\noindent The fundamental principle of the Bayesian inference relies on the conditional probabilities\footnote{Mathematically expressed as $\mathsf{P}(X \vert Y)$, it is read as the probability of $X$ under the condition that $Y$ has occurred.} applied into the famous Bayesian theorem, which provides a mathematical relation between a distribution of the parameters $\boldsymbol{\theta}$ associated with an assumed model $\mathsf{M}$ and observed data $\mathsf{D}$, as follows:
\bea
\mathsf{P} (\boldsymbol{\theta} \vert \mathsf{D}, \mathsf{M})
= \frac{\mathsf{P} ( \mathsf{D} \vert \boldsymbol{\theta}, \mathsf{M} ) \cdot \mathsf{P} ( \boldsymbol{\theta} \vert \mathsf{M} )}{\mathsf{P} ( \mathsf{D} \vert \mathsf{M} )} \, .
\label{Eq:bayes_def}
\eea

\noindent The term on the l.h.s. stands in for the posterior distribution of the parameters, $\mathsf{P}(\boldsymbol{\theta} \vert \mathsf{D}, \mathsf{M})$, which is updated based on the observed data and the assumed model. On the r.h.s., $\mathsf{P}(\mathsf{D} \vert \boldsymbol{\theta}, \mathsf{M}) \equiv \mathsf{L}$ is known as the likelihood, describing the probability of the data given the parameters and the model. Next to, $\mathsf{P}(\boldsymbol{\theta} \vert \mathsf{M}) \equiv \pi(\boldsymbol{\theta})$ defines the prior distribution of the model parameters. Finally above, the probability of data for a given model is deemed the evidence $Z_\mathsf{M}$ which will be relevant for this thesis and described later, for now acts as a normalization factor for the posterior distribution and is given by:
\bea	
Z_\mathsf{M} \equiv \mathsf{P}  ( D \vert \mathsf{M} ) = \int_{d\Omega_{\boldsymbol{\theta}}} \mathrm{d} \boldsymbol{\theta} \, \mathsf{P}  ( D \vert \boldsymbol{\theta}, \mathsf{M} ) \times \mathsf{P}  ( \boldsymbol{\theta} \vert \mathsf{M}) \, .
\label{Eq:evidence}
\eea

\noindent The likelihood function evaluates the agreement between a theoretical predictions and the observational data. It compares the observed data vector, $\boldsymbol{d}$, to the predicted data vector, $\boldsymbol{t}(\boldsymbol{\theta})$, where $\boldsymbol{\theta}$ represents the parameters of interest. When the data is Gaussian-distributed, the log-likelihood would be given by:
\bea
\log \, \mathsf{L}(\boldsymbol{\theta}, \mathsf{M}) \propto -  \frac{1}{2} \sum_{ij} (d_i - t_i)^\top \mathsf{C}^{-1}_{ij} (d_j - t_j) \, ,
\label{Eq:log_likelihood}
\eea

\noindent where $\mathsf{C}_{ij}$ represents the covariance matrix, which encodes the statistical uncertainties and correlations present in the data. Its inverse is referred to as the precision matrix. Under this assumption, the data errors usually follow a multivariate Gaussian distribution and the exponent of Eq.~\eqref{Eq:log_likelihood} can be seen as a $\chi^2$-like test, 
\bea
\chi^2(\boldsymbol{\theta}) = \sum_{ij} (d_i - t_i)^\top \mathsf{C}^{-1}_{ij} (d_j - t_j) \, .
\label{Eq:chi_square}
\eea

\noindent Additionally, this allows to quantify the goodness of fit between the model and the data. A smaller $\chi^2$ value yield a better fit, suggesting that the model is more consistent with the observed data. Unlike, a larger $\chi^2$ value points to a poorer fit, showing discrepancies between the model and the observations. \\

\noindent The prior knowledge from past observations or theory can greatly influence parameter estimation. Although, subjectivity is one of the most common criticisms over Bayesian inference because it requires specifying a prior before one can actually employ the inference analysis, as well as the choice of prior can strongly shape the results. To counter this neutrality, we often rely upon for the use of non-informative priors, like flat priors.\\

\noindent From Eq.~\eqref{Eq:bayes_def}, Bayesian credible levels (C.L.) can be created as regions (contours) containing probabilities such as $68\%$, $95.4\%$, or $99.7\%$. Whilst for Gaussian distributions, these correspond to familiar $1\sigma$, $2\sigma$, and $3\sigma$ intervals, offering an intuitive summary of parameter uncertainty. In short, Bayesian inference is a solid approach recognized for its reliable parameter estimation. \\

\noindent To perform cosmological parameter estimation in \autoref{Chapter6}, we use the user-friendly code for Bayesian analysis, \texttt{Montepython} \citep{2013JCAP...02..001A}, primarily because it includes the likelihood module we need. Other notable options may include \texttt{Cobaya} \citep{2021JCAP...05..057T},  \texttt{CosmoSIS} \citep{2015A&C....12...45Z}, \texttt{CosmoMC} \citep{2013PhRvD..87j3529L} among others. \\ \texttt{Montepython} enables automatic integration with \texttt{CLASS}, making it easy to use in modified versions, and supports likelihoods from experiments like Planck, SDSS, and DES. It also runs in MPI parallelisation. \\

\subsection{Techniques for Posterior Sampling}
\label{subsec:sampling}

Sampling within the Bayesian paradigm is not a straightforward computational task; it blends both creativity and rigour. Different techniques offers unique perspectives that expand the scope of potential applications to the analysis, making them valuable tools in cosmological studies and other fields. \\  

\noindent The first approach to describe is so-called Markov Chain Monte Carlo (MCMC), being a widely adopted technique in statistical inference. MCMC operates in a probability-based process through sequential iterations, where each step generates a sample depending on the previous one. Over time, this process produces a ``chain" of samples in which is traced-up the path (also called a ``walker") of each iteration. These samples yield an approximation of the target posterior distribution, along with a quantitative uncertainty. 

\subsubsection{Diagnostic of MCMC chains} 

The Gelman-Rubin diagnostic \citep{1992StaSc...7..457G} is a widely recognised test for assessing convergence in MCMC methods. In practice for well-behaved posterior densities without pathologies, we track the variances of multiple independent chains from different starting points, then, it is evaluated whether them have sufficiently explored the parameter space and converged to the distribution of interest. Formally, considering $K$ parallel MCMC chains with $x_1, x_2, \ldots, x_K$ samples respectively, the Gelman-Rubin value, $\hat{R}$, is calculated as:
\bea
\hat{R} = \sqrt{\frac{W}{B} + \frac{1}{N}} \, .
\label{Eq:G-R_value}
\eea

\noindent Here $N$ is the number of samples in each chain. While, $W$ is the average within-chain variance as:
\bea
W = \frac{1}{K} \sum_{j=1}^K s_j^2 \, .
\label{Eq:W_chains}
\eea

\noindent $s_j^2$ is the sample variance of the $j$-th chain and $B$ takes into account the variance between-chain (from different chains), calculated as:
\bea
B = \frac{N}{K-1} \sum_{j=1}^K (\bar{x}_j - \bar{x})^2 \, ,
\label{Eq:B_chains}
\eea

\noindent where $\bar{x}_j$ is the mean of the $j$-th chain, and $\bar{x}$ is the overall mean across all chains. In general, $\hat{R}$ is calculated separately for each parameter. Essentially, with Eq.~\eqref{Eq:G-R_value} in hand, it is possible to quantify the convergence of the MCMC process through comparing the variance between different chains to the variance within each chain.\\

\noindent Running MCMC methods may guarantee convergence as $t \rightarrow \infty$. Infinite sequential runtime is, of course, humanly unattainable; however, we can identify a state where the samples are sufficiently representative of the target distribution, allowing us to truncate the process.\\
Thus, a value of $\hat{R} \approx 1$ indicates that the chains have fully converged and are sampling from the same distribution. When $\hat{R} > 1$, the variance of between-chain is larger than the within-chain, implying that the chains have not yet converged and more iterations are necessary. Typically, a criterion for convergence is $R < 1.03$, though stricter thresholds may be used based on the problem and confidence level required.

\subsubsection{Metropolis-Hastings} 

Based on this principle of sequential sampling, perhaps the commonly employed MCMC algorithm is Metropolis-Hastings \citep{1953JChPh..21.1087M,1970Bimka..57...97H}. This well-established algorithm generates samples by suggesting new points in the parameter space and evaluating them based on a condition of likelihood ratio. The proposed points are either accepted or rejected, then such determines whether to proceed to the new values or stays at the current position. Its ability to adapt to different target distributions allows for broad applicability across various inference tasks by using only the likelihood and prior information.

\noindent The step-by-step of the algorithm can be structured upon a few essential elements: 
First, it begins with a starting point $\boldsymbol{\theta}_0$ in parameter space, randomly grabbed from the prior. Alongside this, the problem requires specifying a likelihood function, $\mathsf{P}(\mathsf{D} \vert \boldsymbol{\theta})$. Secondly, to propose a new candidate point, denoted as $\boldsymbol{\theta}_{\rm new}$, a transition distribution, $\mathsf{Q}(\boldsymbol{\theta}_{\rm new} \vert \boldsymbol{\theta}_0)$, is introduced.
In order to step-forward, a likelihood comparison is evaluated as:
\bea
\alpha = \min \left(\frac{\mathsf{P} ( \mathsf{D} \vert \boldsymbol{\theta}_{\rm new}) \cdot \mathsf{P} ( \boldsymbol{\theta}_{\rm new})}{\mathsf{P} ( \mathsf{D} \vert \boldsymbol{\theta}) \cdot \mathsf{P} ( \boldsymbol{\theta})} \cdot \frac{\mathsf{Q}(\boldsymbol{\theta}_{\rm new} \vert \boldsymbol{\theta})}{\mathsf{Q}(\boldsymbol{\theta} \vert \boldsymbol{\theta}_{\rm new})}\, , \, 1 \right) \, .
\label{Eq:accept_rate}
\eea

\noindent In which the new candidate is accepted using an acceptance criterion based on the probability $\alpha$. Thirdly, a random value $u$ is drawn from a uniform distribution between $0$ and $1$. If the value of $\alpha$ exceeds $u$, the candidate is accepted as the new initial state in the chain, i.e. $\boldsymbol{\theta}_{\rm new} \rightarrow \boldsymbol{\theta}_0$. Otherwise, if $\alpha \leq u $, the current point is retained and it is kept as the starting point for the procedure. 

\noindent The goal is to continue drawing samples in order to get the target distribution for the posterior distribution $P(\boldsymbol{\theta}|D)$ from Eq.~\eqref{Eq:bayes_def} by repeating the previous steps for a large number of iterations to allow the chain to ``burn-in" and reach a steady state. After this, the samples from the chain are considered representative of the posterior distribution. \\

\noindent Let us take a closer look at Eq.~\eqref{Eq:accept_rate}. If the proposal distribution $\mathsf{Q}$ is symmetric (e.g., Gaussian) and the model remains unchanged, the proposal probabilities cancel out, leaving only the relative posterior probabilities. 

\noindent To sum up, this algorithm allows us to generate samples from a probability distribution that lacks a practical expression, providing an estimation when analytical solutions are infeasible.

\subsubsection{Hamiltonian Monte Carlo}

The next algorithm to point out is called Hamiltonian (or Hybrid) Monte Carlo (HMC) \citep{1987PhLB..195..216D,2011hmcm.book..113N}, which represents a robust sampler designed to circumvent the inefficiencies of random walk proposals used in conventional MCMC methods. HMC follows the mathematical principles of Hamiltonian dynamics by proposing new states via distant proposals based on physics-inspired trajectories that are farther apart yet maintain a high acceptance probability, facilitating more efficient exploration of the posterior distribution. \\

\noindent Within this scheme, a fictitious ``momentum" variable, $\boldsymbol{p}$, is introduced and linked to the model parameters $\boldsymbol{\theta}$. These quantities define the phase-space $(\boldsymbol{\theta}, \boldsymbol{p})$ in which Hamiltonian dynamics can be simulated. The Hamiltonian (or total energy) is written down by the following form:
\bea
H(\boldsymbol{\theta}, \boldsymbol{p}) = K(\boldsymbol{p}) + U(\boldsymbol{\theta}) = \dfrac{1}{2} \boldsymbol{p}^T \boldsymbol{M}^{-1} \boldsymbol{p} -\log \mathsf{P}(\boldsymbol{\theta}\vert \mathsf{D}) \, .
\label{Eq:hmc_hamiltonian}
\eea

\noindent Being $K(\boldsymbol{p})$ the kinetic energy, with $\boldsymbol{M}$ standing as a positive definite mass matrix and $U(\boldsymbol{\theta})$ represents the potential energy, derived from the negative log-posterior probability density.
Hence, the system evolution with a fictitious time $\tau$, is governed by the Hamilton equations \citep{2002clme.book.....G}:
\bea
\frac{d\boldsymbol{\theta}}{d\tau} = \frac{\partial H}{\partial \boldsymbol{p}} \, , \quad \frac{d\boldsymbol{p}}{d\tau} = -\frac{\partial H}{\partial \boldsymbol{\theta}} \, .
\label{Eq:hamilton_eqs}
\eea

\noindent The HMC algorithm structure unfolds as follows: Beginning at an initial state $(\boldsymbol{\theta}_0, \boldsymbol{p}_0)$, where $\boldsymbol{\theta}_0$ is drawn from the prior, and $\boldsymbol{p}_0$ is sampled from the momentum distribution, typically a Gaussian. Then, numerically integrate both Eqs.~\eqref{Eq:hamilton_eqs} over a fixed number of steps through employing methods like a modified Euler’s method called leapfrog \citep{Hairer_Lubich_Wanner_2003} that updates a new candidate state $(\boldsymbol{\theta}_{\rm new}, \boldsymbol{p}_{\rm new})$ using a small discretisation step $\epsilon$. This method is reversible and volume-preserving, making it well-suited for simulating Hamiltonian dynamics accurately. Similarly, the acceptance criterion is based on the comparison between the current and proposed states like,
\bea
\alpha = \min \left( \exp\left[-\Delta H \right] \, , 1 \right) \, ,
\label{Eq:accept_rate_hmc}
\eea

\noindent with $\Delta H = H(\boldsymbol{\theta}_{\rm new}, \boldsymbol{p}_{\rm new}) - H(\boldsymbol{\theta}_0, \boldsymbol{p}_0)$. The proposed state is always accepted whether it has lower or equal energy than the current state. Intuitively, this reflects lower energy states correspond to higher probabilities in the target distribution, which we aim to sample from. Otherwise, it is also accepted with probability $\alpha$. Accepting higher-energy proposals prevents the sampler from getting trapped in local modes of the posterior (regions of high probability), thereby
allowing escape of narrow peaks to better explore the parameter space efficiently.\\
Remarkably, HMC offers key benefits, such as; employing the log-posterior gradient (via the potential energy term) to inform trajectories and guide exploration, effectively making it geometry-aware (how steep or flat the distribution is in different directions). \\
Despite its strengths, HMC depends on precise tuning of hyperparameters, including step size and the number of integration steps. Moreover, setting the mass matrix $\boldsymbol{M}$ is non-trivial task for high-dimensional or complex distributions, since there is no a well-established blueprint. Poor choices can lead to unstable trajectories or inefficient sampling. However, various improvements exist, including using state windows for acceptance, fast trajectory approximations, tempering for isolated modes, and short-cuts to minimise useless trajectories from taking much computation time. Noteworthy, adaptive algorithms like the No-U-Turn Sampler (\texttt{NUTS}) \citep{JMLR:v15:hoffman14a}, which will be later applied to the analysis of this thesis, as it adjusts step size and trajectory length dynamically but still requiring fine tuning for an efficiency inference. \\

\noindent In summary, HMC combines the rigour of classical physics-based dynamics with the flexibility of Bayesian inference. Its ability to scales well with high-dimensional parameter spaces makes it a potentially powerful tool for forthcoming cosmological analyses.

\subsubsection{Nested Sampling}

The second approach, Nested Sampling (NS) \citep{2004AIPC..735..395S,2022NRvMP...2...39A}, simultaneously provides inference posteriors for model parameters and estimates the evidence (or marginal likelihood) of the model in a single run. Nevertheless, when facing an analysis where dozens or more parameters are in played, so the process of computing model evidence results quite complicated by the necessity to solve high-dimensional integrals from Eq.~\eqref{Eq:evidence}. \\
In this context, NS transforms such multidimensional integral into an one-dimensional problem, as follows:
\bea
\mathsf{X}(\lambda) = \int_{\mathsf{L} > \kappa} \mathsf{P}(\boldsymbol{\theta}) \, d\boldsymbol{\theta} \, ,
\label{Eq:1d_ns}
\eea

\noindent which is easier to evaluate. Here, $\mathsf{L}$ is the likelihood previously defined in Eq.~\eqref{Eq:log_likelihood}, while $\kappa$, the upper limit of the integral, serves as a threshold. Unlike traditional samplers, NS systematically explores the prior space, focusing on regions of higher likelihood. From Eq.~\eqref{Eq:1d_ns} the evidence can be evaluated by calculating:
\bea
Z(\lambda) = \int^{\max{\mathsf{L}}}_0 \mathsf{X}(\lambda) \, d\lambda 
= \int^1_0 \mathsf{L}(\mathsf{X}) \, d\mathsf{X} \, . 
\label{Eq:z_ns}
\eea

\noindent Noticing, the evidence is given by the area under the likelihood curve. The NS algorithm can be assembled briefly as follows: It starts with a collection of $N$ live points $\{\boldsymbol{\theta}_i\}_{i=1}^N$ drawn uniformly from the prior distribution $\mathsf{P}(\boldsymbol{\theta})$. Then, each live point is ranked based on their likelihood values, $\mathsf{L}(\boldsymbol{\theta})$. Next, we track the lowest likelihood $\mathsf{L}_{\rm{min}}$ (or highest value of $\mathsf{X}$). The identified point is removed from the set (becoming a \textit{dead point}), and a new sample with a higher likelihood value is searched and incorporated to guide the exploration towards the posterior distribution, while being subject to the constraint $\mathsf{L} > \mathsf{L}_{\rm{min}}$. This ensures through iteratively ``killing” the lowest likelihood points and replacing them with new samples, $\boldsymbol{\theta}_0$, that the algorithm focuses on regions of increasing likelihood (more probable regions), with samples identified as \textit{live points}. The iteration stops when the live points no longer significantly affect the evidence, this occurs when the likelihood is sufficiently close to the maximum. Meanwhile, the evidence $Z$ is estimated (with its error bar) as the sum of likelihood values weighted by their respective prior volumes across iterations, as $\Delta Z = \mathsf{L}_\text{min} \Delta \mathsf{X}$.\\

\noindent The process is demanding, especially in high dimensions, requiring techniques like ellipsoidal or slice sampling to improve efficiency and scalability. In particular, there are popular NS algorithms available, each with its own approach (see \citep{2023StSur..17..169B} for an overview). For instance, \texttt{MultiNest} \citep{2009MNRAS.398.1601F} software, a highly successful sampler known for its effectiveness across a broad spectrum of research applications. It offers a fast approach, particularly for models with tens of parameters, and handles multimodal distributions by sampling live points within ellipsoidal regions.
Another well-known code is \texttt{PolyChord} \citep{2015MNRAS.450L..61H}, designed for tackling high-dimensional problems, particularly those with over $100$ parameters. Although it may be slower than others, its precision in estimating evidence using slice sampling \citep{2000physics...9028N} makes it especially reliable. Sophisticated code like \texttt{UltraNest} \citep{2021JOSS....6.3001B} code offers notable improvements in computational speed and memory usage which results in faster performance and optimised memory usage compared to previous samplers.
Recent developments have led to \texttt{JAXNS} \citep{2020arXiv201215286A} software, a cutting-edge library built on \texttt{JAX} \citep{jax2018github} software. This package offers efficiency and scalability, making it a key library presented in \autoref{Chapter7}. NS alongside evidence estimation, stand out as a powerful scheme for model selection. Notably, two external tool -- \texttt{MCEvidence} and \texttt{harmonic} allow for Bayesian evidence computation through post-processing of MCMC chains.

\section{Model comparison}

Cosmological inference analysis typically fall into two main steps: parameter constraints and model comparison. As cosmology advances into a data-driven era, establishing robust methods to test and contrast competing models is essential. A widely accepted method for comparing cosmological models through Bayesian evidence \citep{2008ConPh..49...71T}. In a single model parameter estimation, the evidence is typically negligible unless multiple models are under consideration. For model selection, however, it becomes essential. The key quantity is the evidence ratio between competing models, \textit{e.g.} considering the model $\Lambda$CDM, and an alternative model $M$:
\bea
B = \frac{Z_M}{Z_{\rm \Lambda CDM}} =  \frac{\mathsf{P} ( D  \vert M )}{ \mathsf{P} ( D \vert \Lambda {{\rm CDM}} )} \, .
\label{Eq:bayes_factor}
\eea

\noindent This ratio is well-known as the Bayes factor $B$. According to Jeffreys’ scale  \citep{1939thpr.book.....J,2013JCAP...08..036N} (reported in \autoref{tab:jeff_evidence}), this ratio would also provide a qualitative measure of indicating how strongly the data supports for an alternative model over $\Lambda$CDM model. 

\begin{table*}[t!]
\centering 
\caption[The Jeﬀreys’ scale]{Interpretation of the Jeffreys’ scale for comparing two competing models, $M_i$ and $M_j$, based on the Bayes factor values $B_{ij}$ and its logarithmic scale. Table sourced from \citep{2013JCAP...08..036N}.}
\renewcommand{\arraystretch}{2}
\setlength{\tabcolsep}{3.5pt}
\begin{tabular}{ccc}
\hline\hline
$B_{ij}$ & $ K_{ij} = \ln B_{ij}$ & \textbf{Evidence} \\
\hline
$1 \leq B_{ij} < 3$   & $0 \leq K_{ij} < 1.1$ & Weak \\
$3 \leq B_{ij} < 20$  & $1.1 \leq K_{ij} < 3$ & Definite \\
$20 \leq B_{ij} < 150$ & $3 \leq K_{ij} < 5$ & Strong \\
$150 \leq B_{ij}$      & $5 \leq K_{ij}$      & Very Strong \\
\hline\hline
\end{tabular}
\label{tab:jeff_evidence}
\end{table*}

\noindent Bayes factor acts as a natural discriminator among competing models, inherently penalizing those models with excessive set of parameters. Beyond fitting data, it aims to identify models that reflect our understanding of the Universe. We will apply a model comparison into our analysis in \autoref{Chapter6} and \autoref{Chapter7} of this thesis.  

\subsection{Learnt harmonic mean estimator}
\label{subsec:harmonic}

Computing the evidence is often attached to inference analysis. However, there are some techniques to compute the evidence independently of the inference method (thus we are not limited to NS), while others derive it directly from posterior samples \citep{2017arXiv170403472H,2019arXiv191206073J,2024arXiv240412294S}.\\ 

\noindent One such technique involves using the learnt harmonic mean estimator \citep{2021arXiv211112720M} method, which trains a target distribution $\varphi(\boldsymbol{\theta})$ using a normalising flow. Once the flow can approximate the posterior while having tighter tails, it is used to compute the harmonic mean estimator \citep{Newton94}:
\bea
\hat{\rho} = \frac{1}{N} \sum_{i=1}^{N} \frac{\varphi (\boldsymbol{\theta}_i)}{\mathsf{P}(D \vert \boldsymbol{\theta}_i ) \mathsf{P} (\boldsymbol{\theta}_i)} = \frac{1}{Z} \, , \quad \boldsymbol{\theta}_i \sim \mathsf{P}(\boldsymbol{\theta} \vert D) \, ,
\label{eq:harmonic_estimator}
\eea

\noindent where $N$ is the number of posterior samples $\{\boldsymbol{\theta}_i\}_{i=1}^N$. We refer the reader to \citep{2021arXiv211112720M,2023arXiv230700048P} for more details. In \autoref{Chapter7}, we adopt the \texttt{harmonic} package to estimate Bayesian evidence using the learnt harmonic mean estimator. This sampling-scheme is agnostic and has demonstrated accuracy \citep{2024OJAp....7E..73P} on par with approaches like NS, while offering flexibility on the method by using only posterior samples and their associated probability densities.\\

\section{Cosmological Tensions}

As far as we can interpret, parameter tension in cosmology occurs when different probes independently measure the same quantity (or more), yet their inferred constraints exhibit discrepancies -- often reflected in non-overlapping $\sigma$ uncertainties. \\
In order to capture how much $\sigma$s are separated two different estimation, we basically assume that the posteriors are represented by Gaussians (some times there are not) with associated $\sigma$, then we count how many $\sigma$s are away. For instance; considering the established $\Lambda$CDM model with CMB data yields an estimation of $H_0 = 67.4 \pm 0.5 \ \text{km} \ \text{s}^{-1} \text{Mpc}^{-1}$. Whereas the late Universe, for example the project of SH$0$ES (Supernovae $H_0$ for the Equation of State) \citep{2016ApJ...826...56R} yields an estimation of $H_0 = 73.5 \pm 1.4 \ \text{km} \ \text{s}^{-1} \text{Mpc}^{-1}$, which is in approximately $4.2\sigma$ tension with the early Universe estimation! This seems to depend on whether the observable signatures are based on the early or late Universe. \\

\noindent Modern cosmological inference is increasingly moving towards integrating diverse datasets \citep{2022A&A...660A..27T,2023OJAp....6E..36D,2023JCAP...12..023G,2023PhRvD.107b3531A,2025arXiv250314454C,2025arXiv250314743D}, including those from LSS surveys, CMB measurements, gravitational wave observations, and other sources, in order to refine the estimation of cosmological parameters and their interrelations.

\newpage
\thispagestyle{empty}
\chapter{Structure formation}\label{Chapter2}

\vspace{1cm}

\begin{figure}[t!]
\centering
\includegraphics[width=\linewidth]{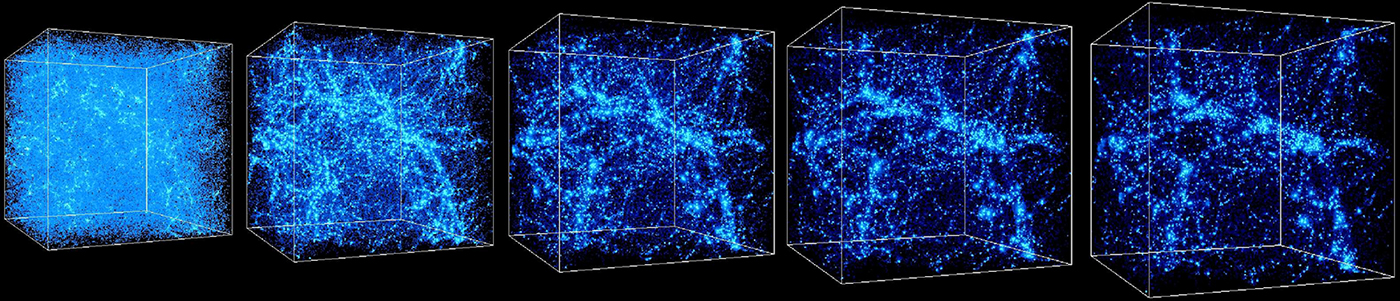}
\caption[Cosmological LSS simulation]{Evolution of cosmic structures in the CDM model with dark energy, illustrating the formation of clusters and large-scale filaments. The frames show the snapshots of structure formation within a $43$ Mpc (comoving) simulation box at redshifts $z = 10.0, 2.0, 1.0, 0.5, 0.0$. \\ \textbf{Image credits:} Imagen taken from \cite{10.1088/978-0-7503-3775-5ch7}. Visualisations by Andrey Kravtsov. Publicly available at: \url{http://cosmicweb.uchicago.edu/filaments.html} (version April 2021).}
\label{Fig:n_body}
\end{figure}

Understanding the Large-Scale Structure (LSS) on the Universe requires an in-depth focus of galaxy formation and clustering. These are usually conducted through $N$-body and hydrodynamical simulations (see \autoref{Fig:n_body}), as they provide the most reliable approach for tracing the nonlinear evolution of cosmic structures, aiming at reproducing observations, from initial conditions provided by CPT.
Nevertheless, these simulations are computationally demanding. Therefore, performing inference analyses poses a computational intensive challenge, since thousands of likelihood evaluations are required. To address this, nonlinear analytical methods are often adopted as ``effective" substitutes. 
In this context, rigorous tests of the nonlinear modelling and validation against outputs of $N$-body simulations are mandatory. Particularly those designed for non-standard cosmologies require rigorous tests in order to ensure that the modelling capture the full scope of structure formation dynamics accurately. \\
It is now known that, there is no complete expression for the matter power spectrum at full nonlinear order in CPT that would includes all eﬀects observed in simulations at very small scales. In order to ease the discrepancies, formalism such as the Effective Field Theory of LSS (EFTofLSS) \citep{2012JCAP...07..051B,2012JHEP...09..082C,2018JCAP...12..035D} extends corrections to the 1-loop order. It systematically incorporates bias terms into the galaxy power spectrum in redshift space to capture nonlinear effects more accurately, improving the modelling of the matter power spectrum.
Alternatively, the Halo model theory has been put forward, and we shall adopt and describe this approach in the present chapter (see \cite{2000MNRAS.318.1144P,2000MNRAS.318..203S,2001ApJ...546...20S,2002PhR...372....1C,2010MNRAS.408..300G} and \cite{2023OJAp....6E..39A} for a comprehensive review). This approach is inspired by products of LSS simulations (see e.g. The Millennium Simulation Project\footnote{\url{https://wwwmpa.mpa-garching.mpg.de/galform/virgo/millennium/\#pictures}} or CAMELS\footnote{\url{https://www.camel-simulations.org/items}}) that exhibit with a high-resolution the gravitational evolution of galaxies clustering and dark matter field, evolving into a cosmic web of filaments and nodes. Such nodes are deemed as dark matter haloes, which represent highly nonlinear structures.

\section{Halo model}
\label{Sec:Halo_Model}

The key assumption of the halo model is that all matter density is confined inside haloes; therefore, the matter density at position $\boldsymbol{x}$ is given by a superposition of each halo contribution as follows,
\bea
\rho_{\rm m}(\boldsymbol{x}) = \sum^{\text{all haloes}}_i \rho_h(\boldsymbol{x} - \boldsymbol{x}_i  | M_i) \, ,  \qquad \text{(Pure Dark Matter)}
\label{Eq:rho_hm}    
\eea

\noindent with $\boldsymbol{x}_i$ located at the centre-of-mass position, and where $M_i$ is the contained mass in the $i$-th halo.\footnote{While the quantities expressed here depend on the redshift, $z$ does not play a role in our calculations, and we shall thus omit it for now.} Following up on our earlier discussions, we seek to establish a connection between matter density and the statistical quantities. Accordingly, in the halo model, the matter power spectrum is represented given by the sum of two terms, 
\bea
P_{\rm m}(k) =  P_{{\rm 1h}}(k) + P_{{\rm 2h}}(k) \, .
\label{Eq:Halo_spectra_model}    
\eea

\begin{figure}[t!]
\centering
\includegraphics[width=0.5\linewidth, angle=270]{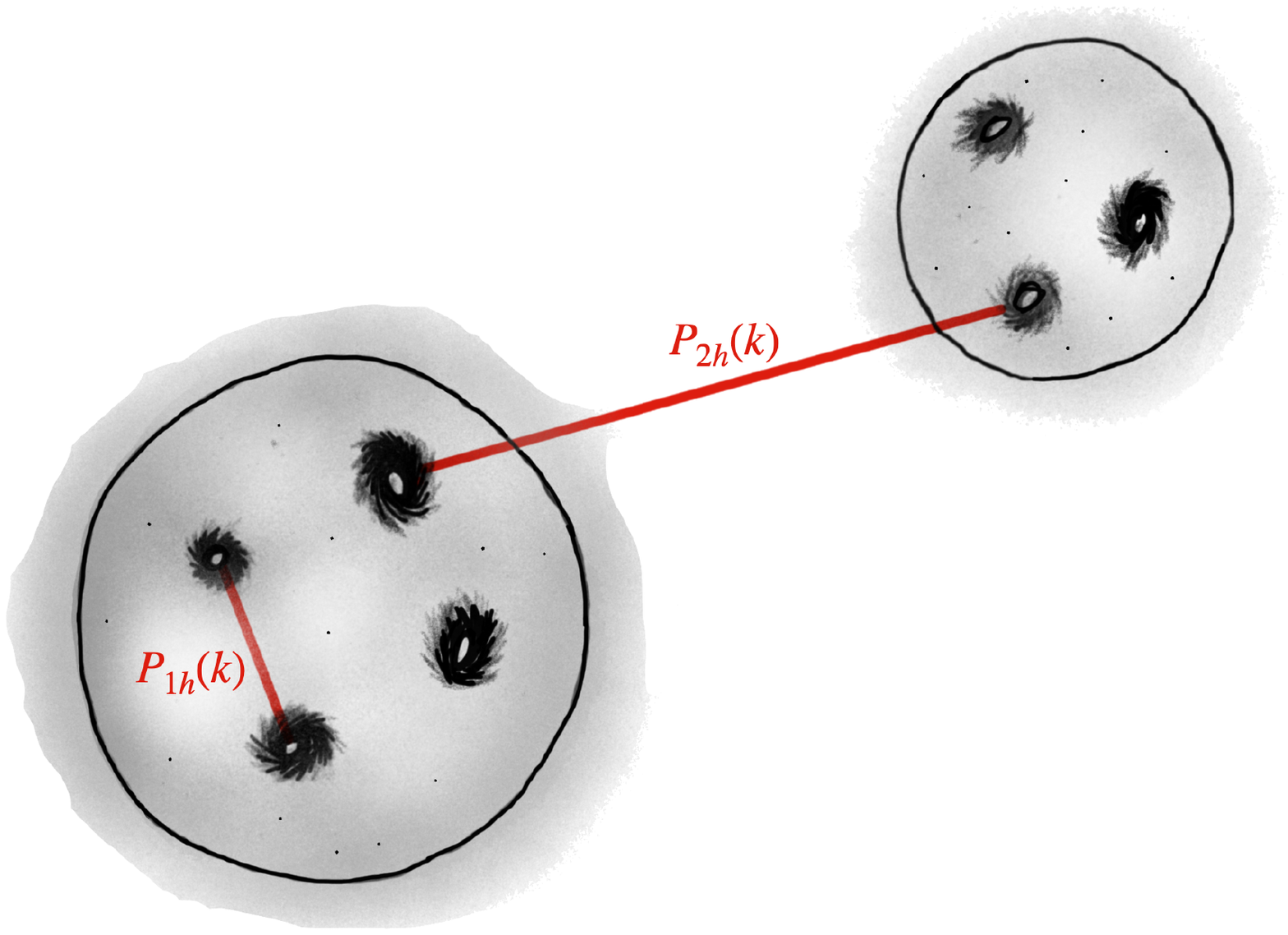}
\caption[Halo model illustration]{Illustration of the contributions to matter power spectrum of the halo model. The $P_{{\rm 1h}}(k)$ term considers correlations in isolated haloes. In contrast, $P_{{\rm 2h}}(k)$ term accounts for correlations between different haloes. \\ \textbf{Image credits:} Ivan Miranda, Morelos, Mexico.}
\label{Fig:1h_and_2h}
\end{figure}

\noindent The 1-halo term, denoted by the subscript ${\rm 1h}$, accounts for the correlation of galaxy pairs within a single halo and is dominant on small scales, whereas the 2-halo term, with subscript ${\rm 2h}$, counts galaxy pairs, each from different haloes, and becomes important on large scales, see schematically \autoref{Fig:1h_and_2h}.

\noindent To derive such expression, it is commonplace to normalise the density profile of the halo, $\rho_h$, as $u(\boldsymbol{x} - \boldsymbol{x}_i  | M_i) \equiv \rho_h(\boldsymbol{x} - \boldsymbol{x}_i  | M_i)/M_i$, then substituting this into Eq.~\eqref{Eq:rho_hm}, we obtain 
\bea
\rho_{\rm m}(\boldsymbol{x}) = \sum^{\text{all haloes}}_i \, N_i \, M_i \, u(\boldsymbol{x} - \boldsymbol{x}_i  | M_i) \, .
\label{Eq:rho_hm_Ni}  
\eea

\noindent The summation extends over small volume elements, $\Delta V_i$, each carefully chosen to contain at most one halo. The indicator variable is $N_i = \{0,1\}$, which specifies whether a halo centre is present in that volume. Under these premises, we can find the mean value of $\rho_{\rm m}(\boldsymbol{x})$ by averaging over all haloes,
\bea
\left\langle \rho_{\rm m}(\boldsymbol{x}) \right\rangle &=& \left\langle  \sum_i \, N_i \, M_i \, u(\boldsymbol{x} - \boldsymbol{x}_i  | M_i) \right\rangle = \sum_i  \int^\infty_0  \mathrm{d}M \, M \, \Delta V_i \, \dfrac{\mathrm{d} n(M)}{\mathrm{d} M} \, u(\boldsymbol{x} - \boldsymbol{x}_i | M) \, \nonumber \\
&=& \int^\infty_0 \mathrm{d}M 
\, M \,\dfrac{\mathrm{d} n}{\mathrm{d} M} \int \mathrm{d}^3 x'  u(\boldsymbol{x} - \boldsymbol{x}_i | M) = \int^\infty_0 \mathrm{d}M 
\, M \,\dfrac{\mathrm{d} n}{\mathrm{d} M} = \bar{\rho}_{\rm m} \, . 
\label{Eq:avg_rho_hm_Ni}  
\eea

\noindent This ensures consistency to the halo model,\footnote{This result in Eq.~\eqref{Eq:avg_rho_hm_Ni} provides a condition that all matter resides within haloes and that, on average, it remains unbiased relative to itself.} with $\bar{\rho}_{\rm m}$ representing the mean comoving matter density. The notable step here is expressing the ensemble average like $\int \mathrm{d}M \, \Delta V_i \, \mathrm{d}n/\mathrm{d} M$, i.e. as an integral over halo masses, where we also introduced the halo mass function, $\mathrm{d}n/\mathrm{d} M$, which represents a counting of the haloes of mass $M$ per volume unit. Lastly, we replaced the discrete sum over $i$ with a continuous integral over the volume $\mathrm{d}^3 x$'. Since our focus is on the statistical properties of haloes, we express the computations in terms of the density contrast from Eq.~\eqref{Eq:dens_contrast}. Then, substituting it into the 2PCF definition in Eq.~\eqref{Eq:2PCF}, we obtain
\bea
\xi_{\rm mm}(\boldsymbol{r}) = \left\langle \delta_{\rm m}(\boldsymbol{x}) \delta_{\rm m} (\boldsymbol{x}+\boldsymbol{r}) \right\rangle = \dfrac{1}{\bar{\rho}^2} \sum_i \sum_j \left\langle N_i \, N_j \, M_i \, M_j \, u(\boldsymbol{x} - \boldsymbol{x}_i | M_i) \, u(\boldsymbol{x}_1 - \boldsymbol{x}_j | M_j) \right\rangle \, ,  
\label{Eq:2PCF_hm}    
\eea

\noindent where we define $\boldsymbol{x}_1 = \boldsymbol{x} +\boldsymbol{r}$. Moreover, we decompose the above expression into two parts: $i=j$ means 1-halo contributions, while $i \neq j$ are 2-halo contributions. Applying the same procedure as in Eq.~\eqref{Eq:avg_rho_hm_Ni}, the 1-halo correlation function is given by:
\bea
\xi_{{\rm 1h}}(\boldsymbol{r}) & = &  \left\langle \delta_{\rm m}(\boldsymbol{x}) \delta_{\rm m} (\boldsymbol{x}+\boldsymbol{r}) \right\rangle_{{\rm 1h}}   = 
 \dfrac{1}{\bar{\rho}^2} \sum_i \left\langle N^2_i \, M^2_i \, u(\boldsymbol{x} - \boldsymbol{x}_i | M_i) \, u(\boldsymbol{x}_1 - \boldsymbol{x}_i | M_i) \right\rangle  \nonumber \\ 
& = & \dfrac{1}{\bar{\rho}^2} \int \mathrm{d}M M^2 \dfrac{\mathrm{d} n}{\mathrm{d} M} \int \mathrm{d}^3 x'  u(\boldsymbol{x} - \boldsymbol{x}' | M) u(\boldsymbol{x}_1 - \boldsymbol{x}' | M) \, .  
\label{Eq:2PCF_1h}    
\eea

\noindent Unlike 1-halo, in 2-halo the correlation comes from different haloes, so this induces a linear halo bias $b(M)$. This means that haloes are biased tracers of the underlying matter density field and are connected by the relation $\delta_h = b(M) \delta_m$. Implying now the correlation function is given by, 
\bea
\xi_{\rm hh}(\boldsymbol{r}|M_1,M_2) = b(M_1) b(M_2) \xi^{\rm L}_{\rm mm}(\boldsymbol{r}) \, ,
\label{Eq:xi_hh}    
\eea

\noindent where $\xi^L_{\rm mm}(\boldsymbol{r})$ is the 2PCF of the linear matter-matter. Substituting this into Eq.~\eqref{Eq:2PCF_hm}, we derive the 2PCF for the 2-halo term as:
\bea
\xi_{{\rm 2h}}(\boldsymbol{r}) = & \dfrac{1}{\bar{\rho}^2}  {\displaystyle \int} \mathrm{d}M_1 M_1 b(M_1) \dfrac{\mathrm{d} n(M_1)}{\mathrm{d} M_1}  {\displaystyle \int} \mathrm{d}M_2 M_2 b(M_2) \dfrac{\mathrm{d} n(M_2)}{\mathrm{d} M_2} \ \times \notag \\ & {\displaystyle \int} \mathrm{d}^3 x' {\displaystyle \int}  \mathrm{d}^3 x'' u(\boldsymbol{x} - \boldsymbol{x}' | M_1) u(\boldsymbol{x}_1 - \boldsymbol{x}'' | M_2) \xi^{\rm L}_{\rm mm}(\boldsymbol{x}' - \boldsymbol{x}'') \, .
\label{Eq:2PCF_2h}    
\eea

\noindent Due to spherical symmetry, the profile $u(\boldsymbol{x} - \boldsymbol{x}_i | M)$ simplifies to $u(r, M)$, where $r = |\boldsymbol{x} - \boldsymbol{x}_i|$. Consequently, the 2PCFs from Eq.~\eqref{Eq:2PCF_1h} and Eq.~\eqref{Eq:2PCF_2h} can be expressed in terms of the power spectrum defined in Eq.~\eqref{Eq:Power_spectrum} through a Fourier transform and we can return their redshift dependency, 
\bea
P_{{\rm 1h}}(k, z) = \int^\infty_0 \mathrm{d}M \dfrac{\mathrm{d} n(M,z)}{\mathrm{d} M} \left( \dfrac{M}{\bar{\rho}} \right)^2 \vert u(k,M) \vert^2 \, , 
\label{Eq:P_1h} \\
P_{{\rm 2h}}(k, z) = P_{\rm L}(k, z) \left[ \int^\infty_0 \mathrm{d}M \dfrac{\mathrm{d} n(M,z)}{\mathrm{d} M}\left( \dfrac{M}{\bar{\rho}} \right) b(M,z) u(k,M) \right]^2 \, .
\label{Eq:P_2h}    
\eea

\noindent Hence, the above expressions are summed to yield Eq.~\eqref{Eq:Halo_spectra_model}. To ensure further consistency to the halo model, the following condition is imposed: 
\bea
\int^\infty_0 \mathrm{d}M \, \dfrac{\mathrm{d} n(M)}{\mathrm{d} M} b(M) M &=& \bar{\rho}_{\rm m} \, .
\label{Eq:bias_halo_condition}    
\eea

\noindent This imposes the following limit: on large scales, the nonlinear dark matter power spectrum must converge to the linear power spectrum, meaning $P_{{\rm 2h}}(k, z) \to P_{\rm L}(k, z)$ as $ k \to 0 $.\\
The normalised density profile in the Fourier space corresponds to,
\bea
u(k,M) = 4 \pi  \int^{r_{\rm vir}}_0 dr \, u(r,M)r^2 \mathrm{sinc}(kr) \, ,
\label{Eq:u_profile_fourier}    
\eea

\noindent with $\mathrm{sinc}(kr) = \sin(kr)/kr$. The cut-off value $r_{\rm vir}$ is called virial radius, which besides marks the halo radius, this prevents divergence at large radii, the profile is typically truncated at the halo radius. Therefore, this establishes a spherical region with a virial mass, 
\bea
M_{\rm vir} = \dfrac{4\pi}{3} r^3_{\rm vir} \Delta_{\rm vir} \bar{\rho}_{\rm m} \, .
\label{Eq:mass_vir}    
\eea

\noindent $\Delta_{\rm vir}$ being the virial overdensity expressed as:
\bea
\Delta_{\rm vir} = [1 + \delta(a_{\rm vir})]\left( \dfrac{a_{\rm col}}{a_{\rm vir}}\right)^{-3} \, ,
\label{Eq:delta_vir}    
\eea

\noindent where $a_{\rm col}$ and $a_{\rm vir}$ denote the scale factors at the collapse and virialisation time, respectively. This quantifies the excess density of the halo with respect to the background matter density (usually either $200$, or $200$ times the critical density, or else the virial definition).

\subsection{Spherical Collapse} 
\label{subsec:spherical_collapse}

Haloes are formed from regions where the initial density field were sufficiently large  to collapse under its own gravity. In order to approximate such process, the evolution of a spherical top-hat overdensity is considered (see \citep{2010MNRAS.406.1865P} for a relativistic approach). This approach simplifies the complexity of gravitational collapse by considering a symmetric overdense region. Assuming initial conditions at $a_i$, the density contrast is given by,
\bea
\delta(r,a) = \left(\frac{r_i}{r}  \right)^{3}(1+\delta_i)  - 1\, ,
\label{Eq:delta_init}
\eea

\noindent where $r_i$ and $\delta_i$ denote the initial radius and density contrast, respectively. To model the nonlinear collapse of a top-hat overdensity, the density profile is described by a piecewise function:
\bea
\rho  = \begin{cases} 
     \bar{\rho}(1+\delta) & r \leq R \, , \\
    \bar{\rho} & R \leq r \, .
     \end{cases}
\label{Eq:top_hat_overdensity}
\eea

\noindent Following Ref. \cite{2020moco.book.....D}, we combine the continuity and Euler equations (see \autoref{Appendix_a} for equations), to derive a second-order differential equation of the nonlinear density contrast. The evolution of the nonlinear density results in,
\bea
\ddot{\delta} + 2 H \dot{\delta} - \dfrac{4}{3}\dfrac{\dot{\delta}^2}{(1+\delta)} = \dfrac{(1+\delta)}{a^2} \nabla^2 \Phi \, .
\label{Eq:nonlineal_eq}    
\eea

\noindent Furthermore, employing Eq.~\eqref{Eq:delta_init}, the equation can be re-expressed in terms of the radial coordinate $r$ as
\bea
\dfrac{\ddot{r}}{r}  =  - \dfrac{4\pi G}{3} \left[\bar{\rho}_{\rm CDM} + (1+3\omega_{\rm DE}
)\bar{\rho}_{\rm DE} ) \right] - \dfrac{4\pi G}{3} \bar{\rho}_{\rm CDM} \delta \, .
\label{Eq:r_ddot3}
\eea

\noindent Here $\bar{\rho}_{\rm CDM}$ and $\bar{\rho}_{\rm DE}$ are the background densities of CDM and dark energy, and $w_{\rm DE}$ is the dark energy equation-of-state parameter. Notice we need to solve the spherical collapse times in order to determinate the virial quantities, as required by the halo model computations.

\subsection{Density profile}
\label{subsec:density_profile}

Customarily, the Navarro-Frenk-White (NFW) density profile \citep{1997ApJ...490..493N} is widely regarded as the standard model for describing the density distribution of dark matter haloes in halo-model codes. Its functional form is given by,  
\bea
u(r, M) = \dfrac{\rho_h(r, M)}{M} = \dfrac{\rho_s}{M} \dfrac{1}{(r/r_s) (1 + r/r_s)^2},
\label{Eq:NFW}    
\eea

\noindent where $r_s$ is the separation between the core and the rest of the halo. The normalization constant $\rho_s$ is obtained from integrating the total halo mass $M = \int dx^3 \rho_h(r, M) $, (with an upper limit at $r_{\rm vir}$), leading to,
\bea
\rho_s = \rho_{\rm crit} \dfrac{\Omega_{\rm m}(z) \Delta_{\rm vir}}{3} \left[ \ln(1+c) - \dfrac{c}{1+c} \right]^{-1}.
\label{Eq:rho_s}    
\eea

\noindent Where we have defined a concentration parameter $c_{\rm vir} \equiv r_{\rm vir}/r_s$. The NFW profile is widely favoured for its strong agreement with numerical simulations \citep{2013MNRAS.432.1103L}, capturing some features, such as the steep central cusp and gradual flattening at larger radii.

\subsection{Halo mass function}
\label{subsec:halo_mass_function}

Previously, we denoted $n(M)$, which estimates the population of virialised dark matter haloes per volume unit. For haloes with masses within the range $[M, M + \mathrm{d}M]$, the comoving halo mass distribution function is given by:
\bea
\dfrac{\mathrm{d} n(M,z)}{\mathrm{d} M} = \dfrac{\bar{\rho}}{M} \, \nu f(\nu) \, \dfrac{\mathrm{d} \ln \nu}{ \mathrm{d} \ln M} \, .
\label{Eq:halo_mass_function}    
\eea

\noindent We have further defined the halo mass function $dn/d\ln M$, where $\nu$ is the peak-height and the function $\nu f(\nu)$ is known as the multiplicity function.\footnote{In case all the mass resides in haloes, then $f(\nu)$ must satisfy: $\int^{\infty}_{0} f(\nu) d\nu = 1$.} While the Press-Schechter formalism initially proposed to calibrate a simple Gaussian distribution for $\nu f(\nu)$, the Sheth-Tormen distribution\footnote{The Sheth-Tormen formalism accounts for the ellipticity of haloes, improving agreement with simulations.} has since proven more accurate calibration of the halo mass function \citep{2001MNRAS.323....1S} with simulations. Its multiplicity function is given by:
\bea
 \nu f(\nu) = A \sqrt{\dfrac{2}{\pi}q\nu^2} \left[ 1 + (q\nu^2)^{-p} \right]\exp(-q\nu^2/2) \, .
\label{Eq:ST_distribution}    
\eea

\noindent The peak-height is $\nu \equiv \delta^2_{\rm crit}/\sigma^2(M, z)$. The parameters $q$ and $p$ are empirical constants derived from fitting to simulations. Here, $\delta_{\rm crit}$ is the critical contrasts density for spherical halo collapse calculated at $z = 0$, which strongly depends on the associated cosmology, and $\sigma^2(M, z)$ corresponds to the mass variance obtained from Eq.~\eqref{Eq:sigma}. Simulation calibrations \citep{1999MNRAS.308..119S} yield the best-fit parameters: $A=0.3222$, $q=0.75$ and $p=0.3$. 
Alternatives for computing the halo mass function from the primordial matter density field include Peaks theory \citep{1986ApJ...304...15B}, in which high-density points are identified as precursors to haloes, or the excursion set formalism \citep{1991ApJ...379..440B} that connects overdense regions to their eventual collapse into bound structures. Another strategy has been the direct emulation of the function via Principal Component Analysis (PCA) in recent studies \cite{2020ApJ...901....5B}.

\subsection{Concentration-mass relation}
\label{subsec:cm_relation}

To determine $r_s$, presented in Eq.~\eqref{Eq:NFW}, the concentration-mass ($c$-$M$) relation is a crucial tool, since it is a semi-analytic approach to link halo concentration to its virial mass $M_{\rm vir}$. This relation often assumes a power-law dependence on such mass, though more complicated models incorporate redshift-dependent and cosmological quantities, as expressed by:
\bea
c_{\rm vir}(M_{\rm vir}, z) = \dfrac{r_{\rm vir}}{r_s}= \dfrac{c_0}{1+z}\left(\dfrac{M_{\rm vir}}{M_{*}} \right)^{-\alpha} \dfrac{f_{\rm DE}(z\to \infty)}{f_{\rm \Lambda CDM}(z\to \infty)} \, .
\label{Eq:c-m_relation}    
\eea

\noindent Here, the values of the fitting parameters are $c_0=0.9$ and $\alpha = 0.13$, while the characteristic mass scale $M_*$ is obtained by the condition $\nu(M_*) = 1$. This mass controls when the halo concentration becomes approximately independent of virial mass. Note that Eq.~\eqref{Eq:c-m_relation} quantifies a trend of smaller haloes being more centrally concentrated, on average, with respect to their more massive counterparts.

\subsection{Halo bias}
\label{subsec:halo_bias}

Haloes act as biased tracers of the underlying matter distribution; therefore, their spatial distribution exhibits a systematic bias relative to the matter density field. To model this bias, the peak-background split formalism \citep{1996MNRAS.282..347M,1999MNRAS.308..119S} is usually adopted to calculate an approximate linear halo bias, particularly in the 2-halo term of Eq.~\eqref{Eq:P_2h}, which describes the influence of the large-scale power spectrum on halo formation. \\
When combined with the Sheth-Tormen formalism, this yields an expression for the linear halo bias as a function of the mass and is given by:
\bea
b(M) = 1  - \dfrac{1}{\delta_{\rm c}} \left( 1 + \dfrac{\mathrm{d} \ln f(\nu)}{\mathrm{d} \ln \nu} \right) = 1 - \dfrac{1}{\delta_{\rm c}} \left( 1 - q 
\nu^2 + \dfrac{2p}{1+(q\nu^2)^p} \right) \, .
\label{Eq:bias_halo}    
\eea

\noindent The halo bias is also expressed as $b =\delta_h/\delta_{\rm m}$. Note that estimating bias via peak-background split has been questioned \citep{2010MNRAS.402..589M} with concerns over inaccuracies, therefore calibrated bias \citep{2010ApJ...724..878T} relations are often preferred. \\

\noindent With the components in place, including halo bias, we can compute the nonlinear dark matter power spectrum of Eq.~\eqref{Eq:Halo_spectra_model} using the standard halo model. 
However, this model has limitations, particularly when accounting for non-zero neutrino masses or modifications to dark energy or gravity. Specifically, the standard halo model assumes a mass function calibrated to a specific cosmology (e.g., $\Lambda$CDM), which may not apply to alternative cosmologies with different dark energy or gravity theories. \\
In this thesis, we opt to resort the halo model reaction which handles these limitations through allowing a more flexible mass function treatment. This reaction framework is well-suited for alternative cosmological models and extends the capacity to fit observational data with greater reliability compared to the standard approach.

\section{Halo model reaction}
\label{sec:halo_reaction}

A promising method for modelling nonlinearities for alternative cosmologies scenarios is the halo model reaction formalism (see e.g. \cite{2019MNRAS.488.2121C,2020MNRAS.491.3101C,2020MNRAS.498.4650B,2021MNRAS.508.2479B,2022MNRAS.512.3691C,2023MNRAS.519.4780B}). This method has proved useful to reach percent-level accuracy against $N$-body simulations in predicting the power spectrum for various models beyond $\Lambda$CDM. Furthermore, the halo model reaction formalism represents the first framework capable of accurately incorporating the nonlinear effects of massive neutrinos in beyond-$\Lambda$CDM cosmologies \citep{2020MNRAS.491.3101C}, as well as strongly suggesting that baryonic feedback can be reliably modelled independently from massive neutrinos and dark energy \citep{2021MNRAS.508.2479B}. The halo model reaction formalism introduces a function, $\mathcal{R}(k,z)$, which quantifies deviations in the nonlinear matter power spectrum for alternative cosmologies. It bridges the gap between a fiducial pseudo-cosmology and the target cosmology through such function defined as follows:
\bea
P^{\rm alt}_{\rm NL}(k,z) \equiv \mathcal{R}(k,z) \times   P_{\rm NL}^{\rm pseudo}(k,z) \, .
\label{eq:Reaction_spectrum_2}
\eea

\noindent The reaction function can account for modifications introduced by the alternative cosmological theory (denoted by the superscript ``${\rm alt}$"), including non-standard dark sector dynamics, modified gravity, or other deviations from $\Lambda$CDM. The ``${\rm pseudo}$” term refers to a quasi-$\Lambda$CDM cosmology because its initial conditions are tempted to exactly match those of the target cosmology, satisfying $P^{\rm alt}_{\mathrm{L}}(k,z) = P^{\rm pseudo}_{\mathrm{L}}(k,z)$ at the desired redshift $z$ in order to target the modified cosmology. The reaction is defined as follows:
\bea
\mathcal{R}(k,z) =  \dfrac{[(1-\mathcal{E})\exp(-k/k_\star) + \mathcal{E}] P^{\rm alt}_{\rm L}(k,z) + P^{\rm alt}_{\rm 1h}(k,z)}{P^{\rm alt}_{\rm L}(k,z)+P^{\rm pseudo}_{\rm 1h}(k,z)}  \, .
\label{Eq:Reaction_def}
\eea

\begin{figure}[t!]
\centering
\includegraphics[width=0.75\textwidth]{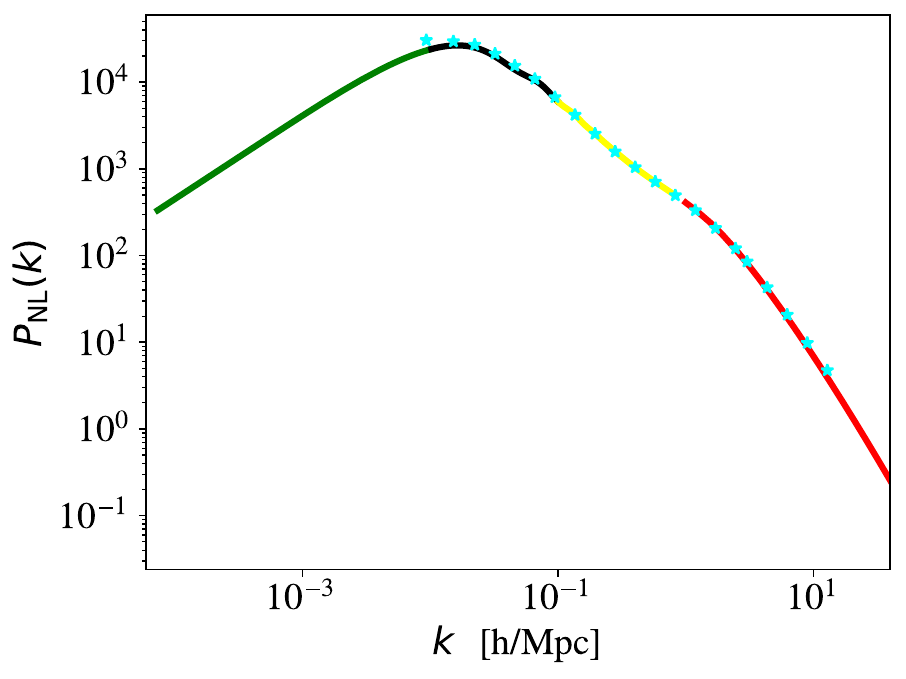}
\caption[Reaction behaviour]{The halo model reaction behaviour on the spectrum (exemplified for a realisation over an IDE model). The green region ($k\leq 0.01~h/{\text{Mpc}}$) is primarily controlled by the linear spectrum, when the reaction simplifies to $\mathcal{R} = 1$ in Eq.~\eqref{Eq:Reaction_def}. The black region ($0.01 ~h/{\text{Mpc}} \leq k\leq 0.1~h/{\text{Mpc}}$) represents to scales which are well described by nonlinear CPT. Whereas the yellow region ($0.1 \ h/$Mpc $\leq k\leq 1\ h/$Mpc) is controlled by the halo mass reaction function ratio. Finally, the rest of the scales in the red region ($k \geq 1~h/{\text{Mpc}}$) is approximated via $\mathcal{R} \approx P^{\rm alt}_{\rm 1h}/P^{\rm pseudo}_{\rm 1h}$. Cyan stars indicate results from simulations, which align closely with the model predictions.}
\label{Fig:reaction_non_linear_spectrum}
\end{figure}

\noindent The subscripts ``${\rm L}$" refer to linear term (as defined in Eq.~\eqref{Eq:Power_spectrum}), while the subscript ``${\rm 1h}$" refers to the 1-halo term\footnote{Notice that the 2-halo term is reduced to the linear spectrum. According to \cite{2019MNRAS.488.2121C}, the improper integral in Eq.~\eqref{Eq:P_2h} can be set to unity without any measurable impact on the halo model reaction} (see Eq.~\eqref{Eq:P_1h}). Here $k_\star$ and $\mathcal{E}$ are unknown constants rather than free parameters. Specifically, $k_\star$ represents the characteristic scale that dictates the transition between linear and nonlinear regimes. Its numerical value is determined at each redshift by solving the following equation:
\bea
\mathcal{R}(k_0,z) =  \dfrac{P^{\rm alt}_{\rm L}(k_0,z) +P^{\rm alt}_{\rm 1-loop}(k_0,z)+P^{\rm alt}_{\rm 1h}(k_0,z)}{P^{\rm pseudo}_{\rm L}(k_0,z) + P^{\rm pseudo}_{\rm 1-loop}(k_0,z)+P^{\rm pseudo}_{\rm 1h}(k_0,z)} \, .
 \label{Eq:Reaction_k_star}
\eea

\noindent Here the scale is set to $k_0 = 0.06~h/{\text{Mpc}}$ where the 1-loop perturbative contribution to the power spectrum of Eq.~\eqref{Eq:P_1loop} remains accurate and reliable. The parameter $\mathcal{E}(z)$ is obtained via the limit,
\bea
\mathcal{E}(z) = \lim_{k\to 0}\dfrac{P^{\rm alt}_{\rm 1h}(k,z)}{P^{\rm pseudo}_{\rm 1h}(k,z)} \, .
\label{Eq:epsilon_reaction}
\eea

\noindent The effect of the reaction of Eq.~\eqref{Eq:Reaction_def}, $\mathcal{R}(k,z)$, on the nonlinear matter power spectrum across a range of scales is illustrated in \autoref{Fig:reaction_non_linear_spectrum}. \\

\begin{figure}[t!]
\centering
\includegraphics[width=\textwidth]{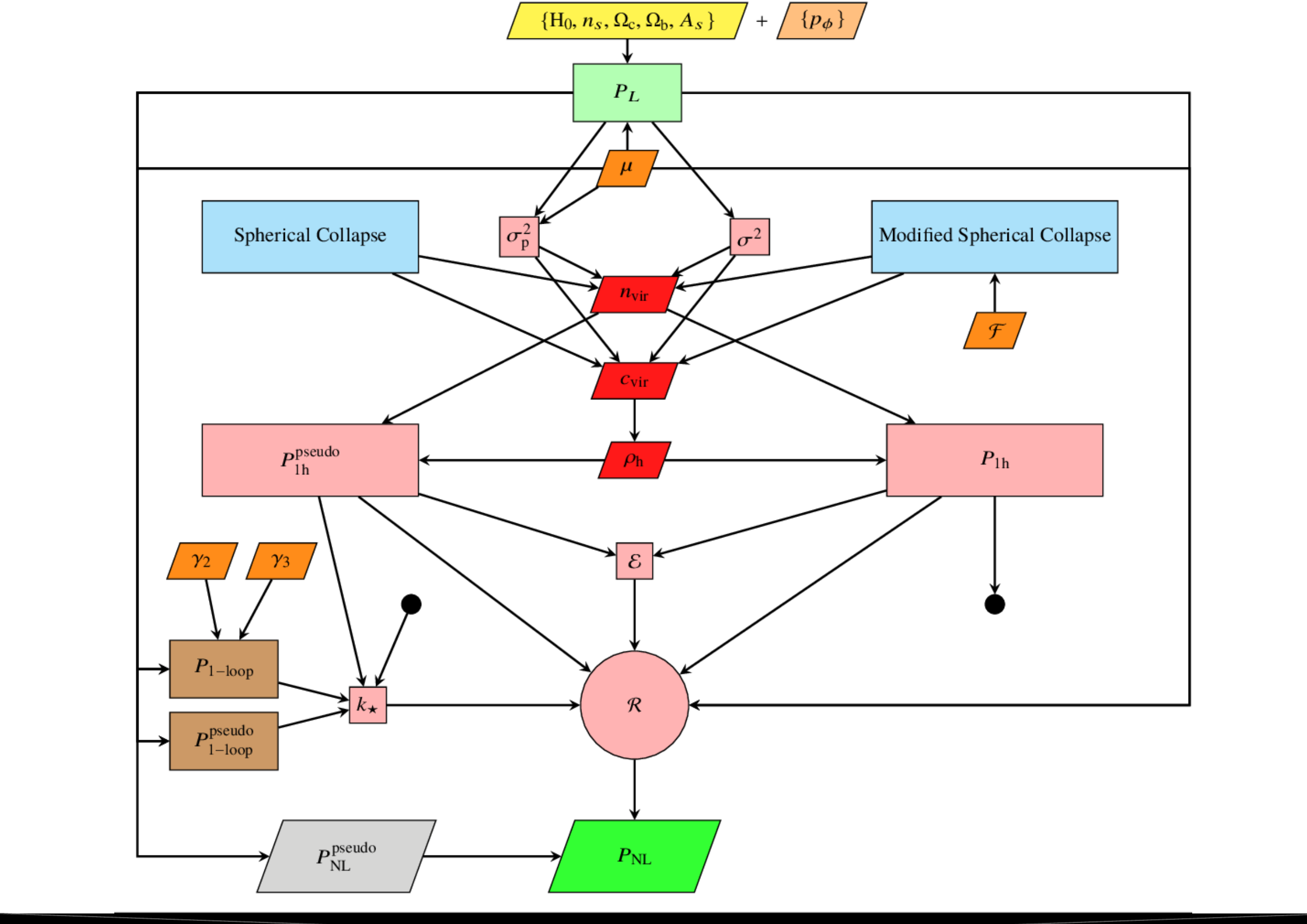}
\caption[\texttt{ReACT} flowchart]{
An overview of the computation of the nonlinear power spectrum $P_{\rm NL}$, within the code \texttt{ReACT}. The process begins with the input of cosmological parameters, indicated by the yellow trapezoid at the top. These parameters are passed to a Boltzmann solver, which computes the linear matter power spectrum. Details of the subsequent steps depicted in the figure are discussed in the text, or readers can refer to \cite{2020MNRAS.498.4650B} for further explanation. Importantly, for the orange trapezoids: the top one, labelled $\rho_\phi$, represents additional parameters that describe modifications to the standard $\Lambda$CDM model. While the trapezoids labelled with $\mathcal{F}$, $\mu$, $\gamma_2$ and $\gamma_3$ are modifications to the gravity.
\\ \textbf{Image credits:} Flowchart taken from \cite{2020MNRAS.498.4650B}.}
\label{Fig:reaction_computation}
\end{figure}

\noindent Furthermore, the modifications on gravity enter directly through the Poisson equation. 
\bea
\nabla^2 \Phi = 4\pi G (1+\mathcal{F}) \bar{\rho}_{\rm m} \delta \, ,
\label{Eq:Poisson_mod}
\eea

\noindent where $\mathcal{F}$ encodes the information of non-standard gravity. In summary, the reaction framework offers an accurate method to map the nonlinear matter power spectrum of $\Lambda$CDM to a desired alternative theory. This formalism has been successfully tested on a variety of models,
such as $w$CDM, CPL, IDE, $f(R)$ gravity and nDGP. \\
\cite{2020MNRAS.498.4650B} have implemented the reaction formalism in a publicly available \texttt{C++} code called \texttt{ReACT}. The code also provides a Python wrapper, which allows it to be run within the Python interpreter, such as in Jupyter notebooks.\\
To compute the reaction, \texttt{ReACT} requires input cosmological model parameters $\boldsymbol{\theta}$. Then, it takes the linear power spectrum generated by either \texttt{CAMB} or \texttt{CLASS}, along with the parameters of the modified theory and its respective additional parameters. Furthermore, the value of $\sigma_8(z=0)$ is saved, as it is essential for calculating the pseudo spectrum, which is based on the $\Lambda$CDM cosmology. Such spectrum can then be computed using \texttt{HMCode} \citep{2021MNRAS.502.1401M} or other similar tool. For further details on steps of the code, refer to the flowchart in \autoref{Fig:reaction_computation}. \\

\section{Reaction in presence of massive neutrinos}
\label{sec:neutrinos_spectra}

Neutrinos, often called ghostly particles, are among the most abundant particle in the Universe, nonetheless, their unique properties continue to intrigue particle physicists, remaining nearly undetectable in direct experiments due to their feeble interactions with matter. Regarding their subtle significant imprints on the evolution of the Universe, offering an opportunity to study their properties through cosmological observations\footnote{Cosmological observations primarily yield constraints on the sum of neutrino masses and provide indirect hints about their mass hierarchy.} by analysing their effects on the structure formation, expansion history or CMB. \\

\noindent A particularly intriguing aspect of neutrino physics \citep{2013neco.book.....L} is their transition from relativistic to non-relativistic behaviour. Initially, in the hot early Universe, neutrinos move at nearly the speed of light due to their high thermal energy. Nevertheless, as the Universe expands and cools, neutrinos momentum decreases, so they eventually shift to a non-relativistic state when the average momentum, $\left\langle p_\nu \right\rangle$, becomes comparable to their rest mass. This transition occurs at a redshift, 
\bea
1+z_{\rm nr} = \dfrac{m_\nu}{3.15 T_{\nu,0}} \, .
\label{Eq:z_nr}
\eea

\noindent In which it is determined by the neutrino mass $m_\nu$ and the neutrino temperature today, $T_{\nu,0} \approx 1.7 \times 10^{-4} \, \text{eV} = 1.9$ K. For example, using the $m_\nu$ mean value from Planck 2018 \citep{Planck:2018vyg} estimation, this implies $1+z_{\rm nr} \sim 100$. \\ 

\noindent When neutrinos become non-relativistic, they reach what is known as the free-streaming scale $k_{\rm fs}$. This free-streaming motion prevents them from clustering on small scales, leaving an additional suppression on the matter power spectrum \citep{2011MNRAS.410.1647A}, since they do not cluster as efficiently as CDM. This interplay between neutrinos and structure formation may play a role in the refinement of the estimation of parameters, such as $H_0$ and $S_8$. For $H_0$, it may be influenced by the impact of neutrinos on the expansion history. Whereas $S_8$ could be affected by differences in the observed amplitude of matter fluctuations due to neutrino free-streaming. \\

\noindent Following \cite{2021MNRAS.508.2479B}, massive neutrinos have been implemented into \texttt{ReACT} through their matter perturbations in the halo model (see also \cite{2014JCAP...12..053M}). The sum of all matter contributions, including from CDM ($\rm c$), baryons ($\rm b$) and neutrinos ($\nu$), is considered as:
\bea
\delta_{\rm m} = (f_{\rm c} \delta_{\rm c} +  f_{\rm b} \delta_{\rm b}) +  f_\nu \delta_\nu \, . 
\label{Eq:delta_m}
\eea

\noindent Here $\delta_{\rm c}$, $\delta_{\rm b}$ and $\delta_\nu$ refer to the perturbations from CDM, baryons, and neutrinos, respectively. The factors $f_{\rm c}$, $f_{\rm b}$, and $f_\nu$ are the relative abundances of these components compared to total matter, with $f_a=\rho_a/\rho_{\rm m}$. The matter power spectrum, which is proportional to the square of the matter perturbation $\delta_{\rm m}^2$, can be written as:
\bea
P^{(\mathrm{m})}(k) = (1-f_\nu)^2 P^{(\mathrm{cb})}(k) + 2 f_\nu (1-f_\nu) P^{(\mathrm{cb}\nu)}(k) + f^2_\nu P^{(\nu)}(k) \, ,
\label{Eq:spectrum_neutrinos}
\eea

\noindent where $P^{(\mathrm{cb})}(k)$ is the power spectrum for CDM plus baryons, $P^{(\mathrm{cb\nu})}(k)$ is the cross-power spectrum between CDM, baryons, and neutrinos, and $P^{(\nu)}(k)$ is the power spectrum of pure neutrinos. From \cite{2020MNRAS.491.3101C}, the reaction function $\mathcal{R}(k)$ is modified to include massive neutrinos\footnote{The effects of massive neutrinos are incorporated at the linear level on the target cosmology.} contributions as follows:
\bea
\mathcal{R}(k,z)=\frac{\bar{f}_{\nu}^{2} P_{\mathrm{HM}}^{(\mathrm{cb})}(k,z)+2 f_{\nu}\bar{f}_{\nu} P_{\mathrm{HM}}^{(\mathrm{cb} \nu)}(k,z)+f_{\nu}^{2} P_{\mathrm{L}}^{(\nu)}(k,z)}{P_{\mathrm{L}}^{(\mathrm{m})}(k,z)+P_{\mathrm{1h}}^{\mathrm{pseudo}}(k,z)}  \, ,
\label{Eq:Reaction_neutrinos}
\eea

\noindent where the superscript $({\rm m}) = ({\rm cb} + \nu )$ accounts for the sum of matter components. The cross component is approximated as $P_{\mathrm{HM}}^{(\mathrm{cb} \nu)} \approx \sqrt{P_{\mathrm{HM}}^{(\mathrm{cb})} P_{\mathrm{L}}^{(\nu)}}$. Additionally, for each ``${\rm HM}$" subscript  $P^{(\cdot)}_{\mathrm{HM}} = [(1 - \mathcal{E})\exp(-k/k_\star) + \mathcal{E}] P^{(\cdot)}_{\mathrm{L}} + P^{(\cdot)}_{\mathrm{1h}}$. Lastly, $\bar{f}_{\nu} = 1 - f_{\nu}$ is the complementary of neutrino fraction. \\

\begin{figure}[t!]
\centering
\includegraphics[width=\linewidth]{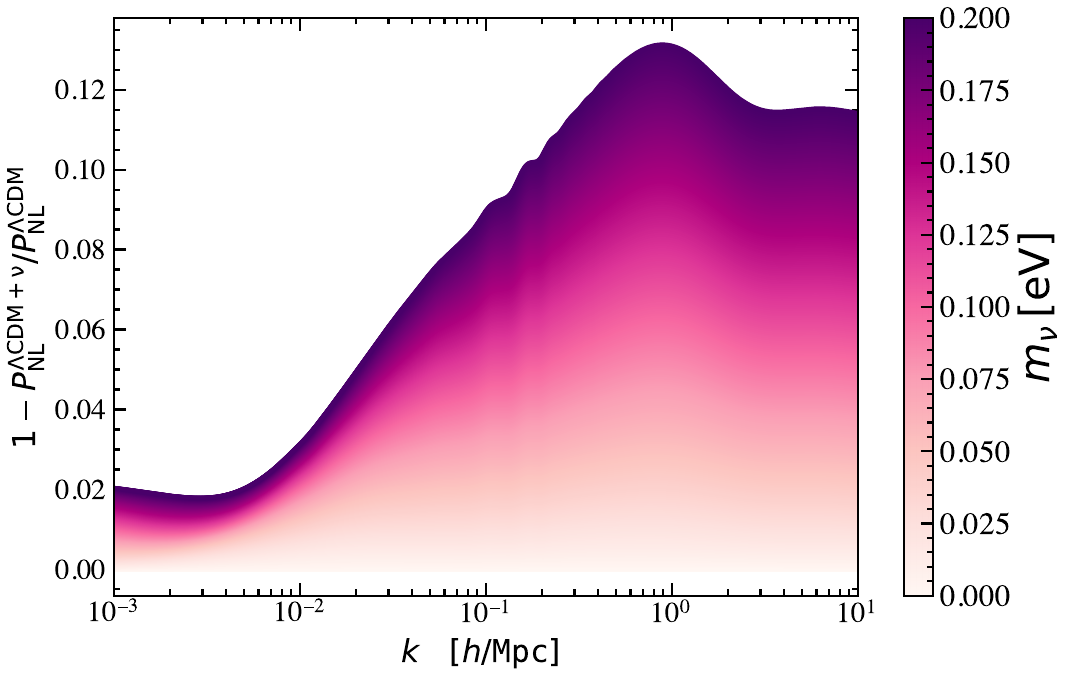}
\caption[Massive neutrinos impact on matter power spectrum]{The impact of massive neutrinos on the  matter power spectrum with $\Omega_\nu = 0$ at $z = 0$, showing the suppression on $\Lambda$CDM through the residual factor $(1 - P^{\Lambda \rm CDM + \nu} / P^{\Lambda \rm CDM})$ across a range of $k$-scales from $10^{-3}$ to $10^1$ in units of $[h/{\text{Mpc}}]$. The colorbar spans the variation of neutrino masses within the range $m_\nu \in [0, 0.2]$ [eV].}
\label{Fig:neutrinos_impact}
\end{figure}

\noindent In \texttt{ReACT}, the effects of massive neutrinos over Eq.~\eqref{Eq:Reaction_neutrinos} are modulated through their mass: $M_\nu = \sum^{N_\nu}_{i} m_{\nu_i}  \sim N_\nu \bar{m}_\nu$. The impact of massive neutrinos on the matter power spectrum is illustrated in \autoref{Fig:neutrinos_impact}. Additionally, their contribution to the energy budget of the Universe is computed like,
\bea
\Omega_\nu = \dfrac{M_\nu}{93.14 h^2} \, .
\label{Eq:Omega_nu}    
\eea

\noindent The reaction method using \texttt{ReACT} is capable of achieving an accuracy of $5\%$ compared to $N$-body simulations at $k < 10~h/{\text{Mpc}}$. While the halo model reaction provides a solid framework for modelling the nonlinear matter power spectrum for dark matter. A natural extension is to incorporate the effects of baryonic feedback, enabling more realistic predictions of the matter power spectrum for weak lensing and related analyses. \\

\section{Baryonic feedback}
\label{sec:baryonic_feedback}

The structure formation on LSS is not solely a playground for dark matter and gravity; it is also profoundly influenced by the dynamic and complex processes driven by baryons (see e.g. \cite{2014JCAP...04..028F,2019OJAp....2E...4C, 2023PhRvD.107b3514S}). As we enter into an era of unprecedented precision, incorporating these effects, dubbed baryonic feedback, is an unavoidable aspect of the study of matter distribution. These baryonic effects can potentially bias the analyses of several cosmological probes, such as weak gravitational lensing \citep{2013MNRAS.434..148S,2023A&A...678A.109A}. Accurate modelling of the matter power spectrum, $P_{\rm NL}(k)$, must be done for probing deviations from the standard $\Lambda$CDM in order to avoid misinterpretations or false detections. Given that the baryonic feedback is an important systematic to consider for ongoing and upcoming surveys through improving the realism of nonlinear matter power spectrum modelling. \\
However, the astrophysical processes underlying baryonic feedback remain an area of active research since it is not fully understood -- hydrodynamical processes that are much more complex to model than gravity such as; gas heating and cooling, the rate of star formation, supernovae explosions, and Active Galactic Nuclei (AGN) feedback, to name a few, can drastically alter the density profiles of haloes of Eq.~\eqref{Eq:NFW}, leading to alterations in the matter power spectrum. \\

\begin{figure}[t!]
\centering
\includegraphics[width=\textwidth]{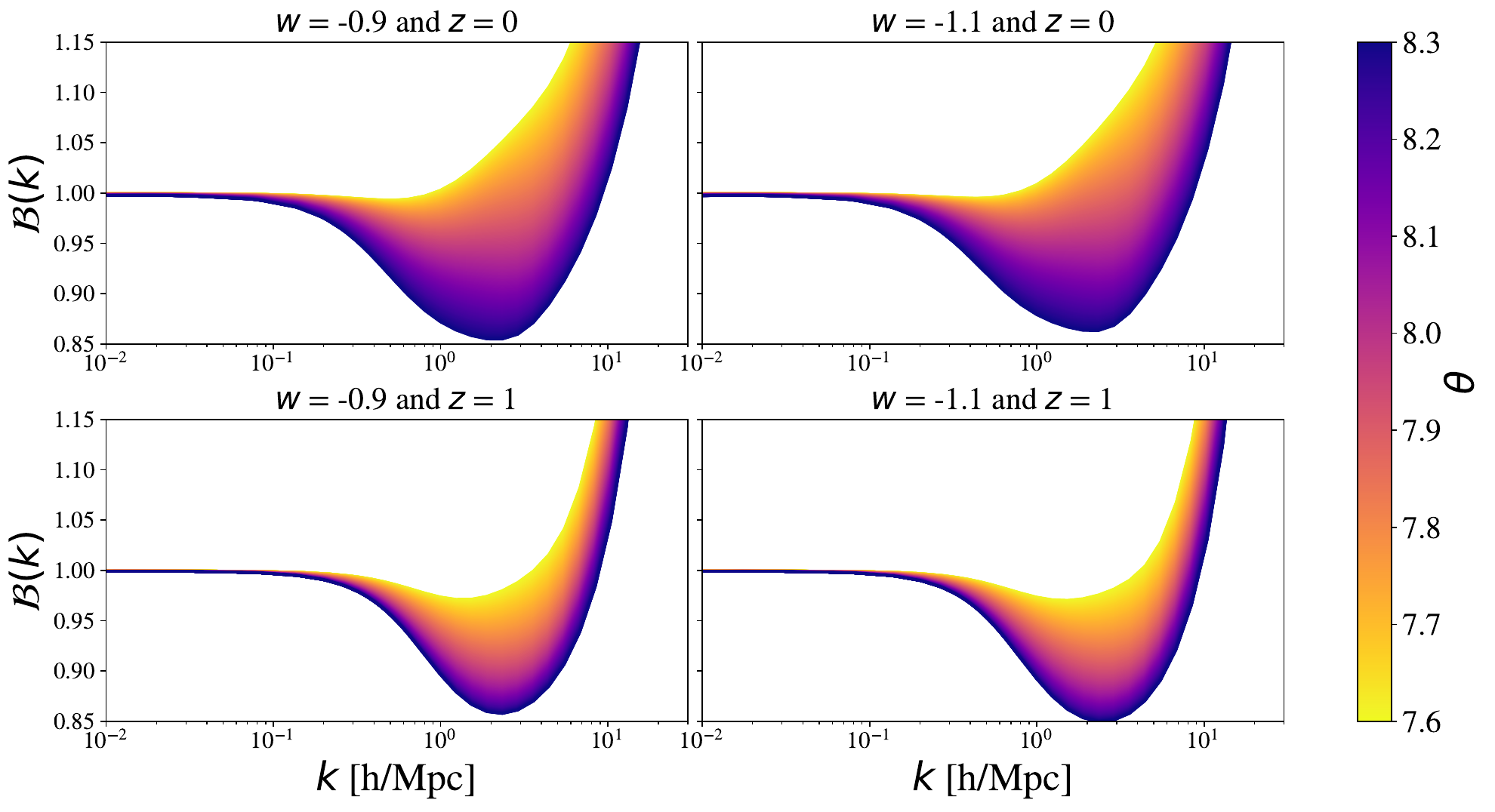}
\caption[Variation of $T_{\rm AGN}$]{The variation of the feedback parameter $\theta \in [7.6, 8.3]$, as shown by the colorbar, and its effects on $\texttt{B}(k)$ for different $w$CDM cosmologies. The top panels show the results of $w=-0.9$ and $w=-1.1$ at $z=0$, while the bottom panels correspond to $z=1$. The $k$-scales in units of $[h/{\text{Mpc}}]$, covering a range from linear to nonlinear scales. At intermediate scales ($k \sim 10^{-1}~h/{\text{Mpc}}$), the feedback parameter $\theta$ suppresses power, reflecting the reduced clustering due to AGN effects. On nonlinear scales $k > 1 ~h/{\text{Mpc}}$, the feedback instead boosts power, indicating enhanced contributions from processes such as gas and stars contributions. The two cosmologies with $w=-0.9$ and $w=-1.1$ highlights the sensitivity of $\texttt{B}(k)$ to variations in dark energy, with subtle differences in the magnitude and scale dependence of the feedback.}
\label{Fig:vary_TAGN}
\end{figure}

\noindent Several strategies have been proposed to incorporate baryonic feedback into structure formation modelling. The approach adopted in this thesis builds on the baryonic halo model, where a baryonic correction factor is parametrised, with the convenience of having it decoupled and marginalised from the pure dark matter power spectrum. Specifically, the baryonic feedback is taken into account as a boost $\texttt{B}(k,z)$ \citep{2021MNRAS.506.4070A, 2021JCAP...12..046G} to the DM-only matter power spectrum as:
\bea
\texttt{B}(k,z) = \dfrac{\hat{P}(k,z)}{P_{\rm DM\text{-}only}(k,z)} \, .
\label{eq:boost_eq}
\eea

\noindent Therefore, to obtain a full power spectrum within the halo model reaction framework, we combine baryonic boost and the reaction from Eq.~\eqref{Eq:Reaction_neutrinos} to obtain,
\bea
P_{\rm NL}(k,z) =   \texttt{B}(k,z) \times \mathcal{R}(k,z) \times P^{\rm pseudo}_{\rm DM\text{-}only} (k,z) \, . 
\label{Eq:full_spectrum}
\eea

\noindent With this prescription we can readily predict the nonlinear matter power spectrum in the presence of baryonic feedback effects. \\ 
In this thesis, we explore two approaches to account for baryonic feedback. In the following paragraphs, their characteristics are described: The first feedback model involves considering its contributions caused by the baryons within the AGN surrounded by dark matter halo. This is built through the standard halo model prescription \citep{2021MNRAS.502.1401M}, in which the baryonic model encodes 6-parameters taking into account effects from AGN feedback and star formation. However, that model was calibrated using the hydrodynamical \texttt{BAHAMAS} simulations \cite{2017MNRAS.465.2936M} to obtain a 1-parameter model, which depends only on the temperature of AGN (in units of [K]), via $\theta \equiv \log_{10} (T_{\rm AGN}/\text{K})$, and was validated in the range $7.6 \leq \theta \leq 8.3$. \\
In this AGN baryonic model, the boost described by Eq.~\eqref{eq:boost_eq} induces a suppression on the power spectrum at intermediate scales, due to gas expulsion from AGN feedback. Conversely, it leads to an enhancement on smaller scales as a result of star formation, as illustrated in \autoref{Fig:vary_TAGN}. This baryonic feedback implementation is accessible through the interface of \texttt{HMCode2020$\_$feedback} \citep{2021MNRAS.502.1401M}, providing a practical tool for modelling these effects within cosmological analyses.

\begin{figure}[t!]
\centering
\includegraphics[width=\textwidth]{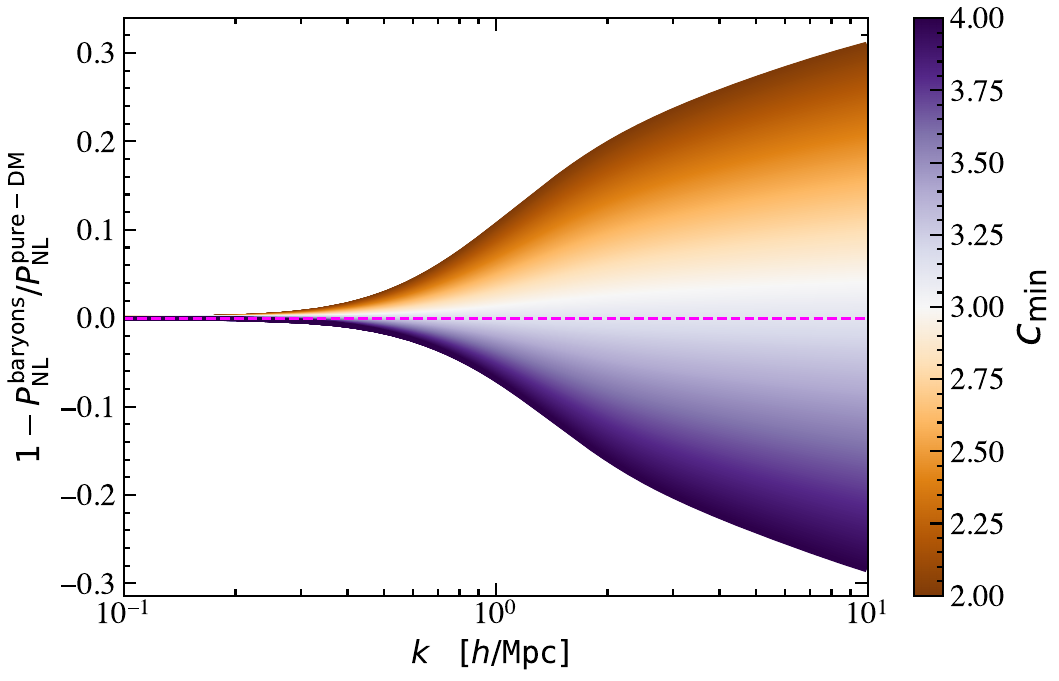}
\caption[Variation of $c_{\rm min}$]{Variation of the residual ratio of $ P^{\rm baryons}_{\rm NL} / P^{\rm pure-DM}_{\rm NL}$, both spectra set within the context of the $\Lambda$CDM cosmology, where one is driven by the feedback parameter $c_{\rm min} \in [2, 4]$, shown in the colorbar. The parameter $c_{\rm min}$ modulates the strength of baryonic feedback relative to the pure-DM case, with smaller values indicating suppression on it. Whereas, the higher ones implies an enhancement. The parameter $\eta_0$ is derived. The magenta dashed line represents the pure-DM case, indicating as a baseline. The results are presented from $k = 10^{-3}$ to $10^1~h/{\text{Mpc}}$.}
\label{Fig:vary_c_min}
\end{figure}

\noindent The second baryonic model is implemented in the older \texttt{HMCode2016} version \citep{2015MNRAS.454.1958M}. In this version, the boost of Eq.~(\ref{eq:boost_eq}) (found it in \texttt{HMCode2016$\_$feedback}) depends on two fitted baryonic parameters, $c_{\rm min}$ and $\eta_0$, which capture the influence of baryons within a halo (see \citep{2015MNRAS.454.1958M} for details), its effect on small scales is also shown in \autoref{Fig:vary_c_min}. Here, $\eta_0$ is determined by the relation $\eta_0 = 0.98 - 0.12 \, c_{\rm min}$. Both are fit to data from the OverWhelmingly Large (OWL) hydrodynamical simulations \citep{2010MNRAS.402.1536S,2011MNRAS.415.3649V,2011MNRAS.417.2020S}. However, for our statistical analysis we will choose \texttt{HMCODE2016} instead of \texttt{HMCODE2020} for baryonic feedback because it has a DM-only limit where baryonic effects vanish -- an advantage not present in the 2020 version. The case where baryons have no influence corresponds to the formula above with a value $c_{\rm min} = 3.13$ (DM-only). \\

\noindent As previously mentioned, this baryonic factor approach provides a simplified way to cover baryonic contributions without requiring full-scale simulations, as these models are generally calibrated using cosmological hydrodynamic simulations. Later, a modification of \texttt{ReACT} will be introduced to study the effects of an interacting dark sector model. In spirit of \cite{2021MNRAS.508.2479B}, we will also aim to include the effects generated from baryon feedback and massive neutrinos. Including both effects in the halo model reaction is vital for achieving accurate predictions that align with weak lensing and other cosmological data.

\newpage
\thispagestyle{empty}

\chapter{Weak Lensing}\label{Chapter3}

\vspace{1cm}

\begin{figure}
\centering
\includegraphics[width=\linewidth]{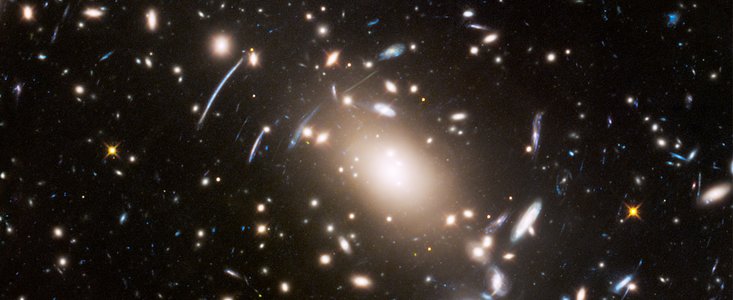}
\caption[Gravitational Lensing in Abell S1063]{This image of Abell S1063 galaxy cluster highlights its large mass, which acts as a cosmic magnifying glass. The lensing effect in this photo shows enlarged light from galaxies aligned behind the cluster, making them bright enough for the NASA/ESA Hubble Space Telescope to observe. This is an example of strong gravitational lensing, where a massive object alters significantly the positions of background galaxies. \\
\textbf{Image credit:} NASA, ESA, and J. Lotz (STScI).} 
\label{fig:lensing}
\end{figure}

As light from distant galaxies travels toward us, it interacts with gravitational fields generated by near massive entities like galaxy clusters, galaxies themselves, their associated dark matter, black holes and so on. This interactions leave observable imprints in the light detected by telescopes, a phenomenon known as gravitational lensing. In a cosmological context, this bending of light gives way to two types of lensing: strong and weak. Strong lensing is defined as light deflections that are prominent enough to be distinguished through direct observations, as illustrated in \autoref{fig:lensing}. This effect occurs when a single massive object along the line of sight to a distant source causes dramatic distortions in the path of lightrays, such as multiple images, arcs, or even Einstein rings and zig-zag signatures. On the other hand, weak lensing distortions on individual scales are visually imperceptible. Instead, it leads to background images of galaxies being distorted in shape and luminosity due to foreground matter distribution, e.g. galaxy clusters or massive dark matter structures. \\

\noindent The statistical correlation of galaxies patterns (their shape or luminosity), is connected to the cosmos energy budget, the map of the dark matter distribution and properties of the LSS of the Universe. 
Moreover, such statistics have proven to be invaluable tools for probing and refining cosmological models. Incidentally, the Stage IV photometric surveys will observe and capture billions of photographs of galaxies, across various scales, including the small, nonlinear ones. \\ 
In this chapter, we explore the required elements for the weak gravitational lensing basis in order to understand its key concepts and mathematical scheme in cosmological studies \citep{2003astro.ph..5089V,2015RPPh...78h6901K,2018ARAA..56..393M,2020moco.book.....D,2025arXiv250107938P} which will be connected to the analysis of this thesis.

\section{Photometric redshifts}
\label{Sec:photo-z}

To start with, it is important to distinguish between two types of galaxy surveys to measure redshifts:

\begin{itemize}

\item[$\bullet$] \textbf{Spectroscopic surveys:} The light from a galaxy goes through a spectrometer, where it is decomposed in the constituent frequencies. This technique, deemed spectro-$z$, enables the determination of redshift with high accuracy. However, it is limited to a few galaxies per observation due to the complexity of the procedure. The redshift is estimated through the difference between the observable $\lambda_0$ and the emitted $\lambda_e$ wavelength as,
\begin{equation}
z \equiv \dfrac{\lambda_0 - \lambda_e}{\lambda_e} \, .
\label{eq:redshift}
\end{equation}

\item[$\bullet$] \textbf{Photometric surveys:} Photographs of a sky region are taken with high resolution, integrating a band of frequencies, and allowing for the capture of a large redshift samples of galaxies, simultaneously. In contrast with the spectroscopic counterpart, there is a lower accuracy in the determination of the photometric redshift (photo-$z$) for single objects.

\end{itemize}

\noindent The data employed in this thesis pertains photometric surveys (both synthetic and observed). Photo-$z$'s are obtained from a probability distribution that is given in terms of the number density of galaxies as a function of redshift. This distribution can be expressed mathematically as:
\bea
n_{g}(z) = \dfrac{1}{N_{g}} \dfrac{\mathrm{d}N_{g}}{\mathrm{d}z} \, ,
\label{Eq:photo-z_dist}
\eea

\noindent where $n_{g}(z)$ is the redshift-dependent galaxy distribution, and $N_{g}$ denotes the total number of galaxies (for a given classification). The type of galaxies associated to specific distributions depends on their individual properties, like brightness and colour. Integrating over all redshifts, it must satisfy the normalization condition:  
\bea
\int_{0}^{\infty} n_{g}(\chi) \, \mathrm{d}\chi = 1 \, .
\label{Eq:norm_photo-z}
\eea

\noindent In which, we re-express Eq.~\eqref{Eq:photo-z_dist} in terms of the comoving distance $\chi(z)$ instead of the redshift $z$. Generally, the comoving distance is related to the physical distance as $r = a(t)\chi$. \\ 

\noindent Photo-$z$'s are usual technique for weak lensing studies. The main advantage of photo-$z$ method lies in the large amount of data captured in each frame. The images of vast sky regions are used to infer the true redshifts distribution for millions of galaxies, thus enabling the determination of statistical properties.\\
In order to obtain photometric redshifts, one calibrates distributions by cross-matching with spectroscopic samples. 
This calibration can be achieved through spectral energy distribution (SED) reconstruction \citep{2000AJ....120.1588B} or machine learning techniques \citep{2018A&A...616A..69B}. \\
However, the trade-off is the reduced precision when compared to spectro-$z$, introducing biases, scatter, and uncertainties that can propagate into considerable systematic effects. Mitigating these issues will enhance the reliability of cosmological parameter constraints derived from weak lensing analysis.

\section{Weak Lensing}
\label{sec:weak_lensing_sect}

Weak gravitational lensing is a cornerstone of modern cosmology, uniquely sensitive to both luminous and dark matter. This is thus a powerful tool to study the nature of dark matter and dark energy. \\
In this section, we describe the behaviour of light geodesics through curved space-time, where its path is bent by the gravitational potential of massive objects. Specifically, we derive the equations needed for the weak lensing formalism.\footnote{The derivations provided are based on \url{https://www.tessabaker.space/images/pdfs/lensing-lecture-tbaker-handout-40742.pdf}.} \\

\noindent Since weak lensing is associated to small matter perturbations, its effects can be modelled using the cosmological perturbation theory (CPT). Specifically, we use Eq.~\eqref{Eq:FLRW_pert}, where the weak fields $\Phi$ and $\Psi$ satisfy $\Psi, \Phi \ll 1$. For convenience, we adopt spherical coordinates. We also consider comoving coordinates for the spatial part of the metric, which remain invariant as the Universe expands. The coordinate, $\chi(z)$ (function of $z$), describes the radial comoving distance from the observer to an object, given by:
\bea
\chi(z) = \int^{t_0}_t \dfrac{\mathrm{d}t}{a(t)} = \int^{z}_0 \dfrac{\mathrm{d}z}{H(z)} \, .
\label{Eq:chi_comoving}
\eea

\begin{figure}[t]
\centering
\includegraphics[width=0.75\linewidth]{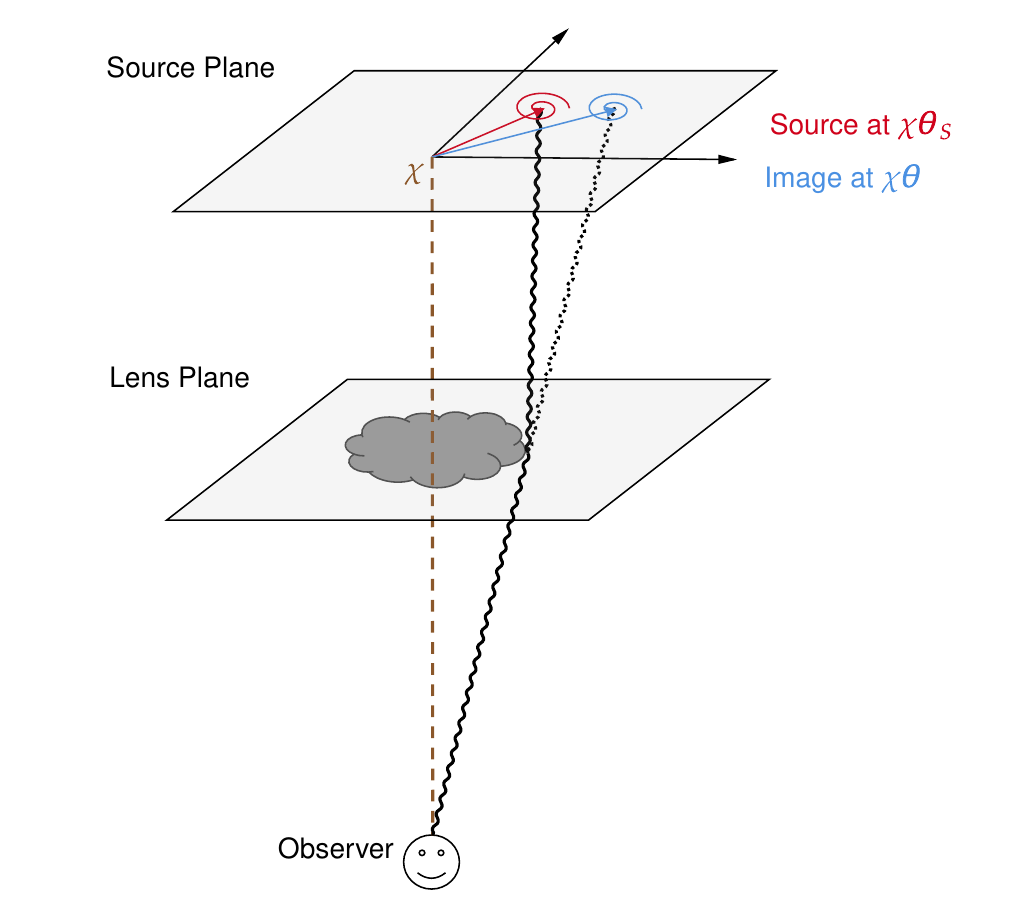}
\caption[Lensing Scheme]{This diagram draws the light path's response to gravity. We consider the positions of the observed image $\chi \boldsymbol{\theta}$ (blue) and the source $\chi \boldsymbol{\theta}_S$ (red): One galaxy  emits a photon at $\chi \boldsymbol{\theta}_S$ ("real") in the source plane, at distance $\chi$. The photon is then deflected by the gravitational field of an object in the lens plane before reaching us, the observer. We thus determine the apparent (``observed") galaxy at $\chi \boldsymbol{\theta}$ in the source plane.
\\
\textbf{Image credit:} Image adapted from \citep{2020moco.book.....D}.} 
\label{fig:lensing_sketch}
\end{figure}

\noindent In this spherical coordinate system, the observer is positioned at the origin. 
As illustrated in \autoref{fig:lensing_sketch}, we locate the source at coordinate $\chi \boldsymbol{\theta}_{S}$, given by:
\bea
\mathbf{x}_{\rm true} = \chi \boldsymbol{\theta}_S = \chi(\theta^{1}_S,\theta^{2}_S, 1) \, .
\label{Eq:true_point}
\eea

\noindent This fully describes the position on the sky. Our plan is to map this true position to its observed (image) counterpart, which at the same distance, is defined as:
\bea
\mathbf{x}_{\rm obs} = \chi \boldsymbol{\theta} = \chi(\theta^{1},\theta^{2}, 1) \, .
\label{Eq:apparent_point}
\eea

\noindent This transformation, $\boldsymbol{\theta}_{S} \to \boldsymbol{\theta}$, will take into account the lensing due to the presences of the gravitational fields.

\noindent Note that the small deviations from the true angle allow us to adopt a small-angle approximation, together with the thin-lens approximation. The latter assumes that the size of the lensing object (located at the lens plane) in comparison with the total distance travelled by the photon is much smaller, thus its dimensions along the line of sight can be neglected. \\

\noindent To describe the trajectory of a photon emitted from a source, we look at the geodesic equation for photons, which satisfies:
\bea
ds^2 = 0  \quad \Rightarrow \quad g_{\mu\nu} p^{\mu} p^{\nu} = 0 \, ,
\label{Eq:geodesic}
\eea

\noindent where $p^{\mu}$ is the four-momentum, defined as:
\bea
p^{\mu} \equiv \dfrac{\mathrm{d}x^{\mu}}{\mathrm{d}\lambda} \, .
\label{Eq:4-momentum}
\eea
 
\noindent With $\lambda$ being the affine parameter. The magnitude of its spatial component is expressed as $\vert p \vert = p = g_{ij} p^{i} p^{j}$. For a photon, the null condition $p^\nu p_\nu = 0$ must be met. Consequently, the temporal component of the photon's momentum can be rewritten as: 
\bea  
p^{0} = \dfrac{p}{\sqrt{1+2\Phi}} \approx p(1-\Phi) \, , 
\label{Eq:p0}
\eea

\noindent where the approximation $(1+2\Phi)^{-\tfrac{1}{2}} \approx 1 - \Phi$ holds for weak fields ($ \Phi \ll 1 $).  
To determine the true position of the source, we now cast the geodesic equation as,
\bea
\dfrac{\mathrm{d}^{2}x^{i}}{\mathrm{d}\lambda^2} = - \Gamma_{\mu\nu}^{i}\dfrac{\mathrm{d}x^{\mu}}{\mathrm{d}\lambda} \, \dfrac{\mathrm{d}x^{\nu}}{\mathrm{d}\lambda} \, .
\label{Eq:geodesic_eq}
\eea

\noindent Where $\Gamma_{\mu\nu}^{i}$ are the Christoffel symbols of the second kind, computed at linear order by neglecting all terms of order $\Phi^2, \Psi^2$, or higher. The respective Christoffel symbols are listed as follows, 
\begin{itemize}
\item[$\star$] $\Gamma^{0}_{00} = H + \dot{\Phi} \, . $
\item[$\star$] $\Gamma^{0}_{0i} = \Gamma^{0}_{i0} = \partial_i \Phi  \, .$
\item[$\star$] $\Gamma^{i}_{00} = \partial^i \Phi \, .$
\item[$\star$] $\Gamma^{0}_{ij} = \Gamma^{0}_{ji} = H \delta_{ij} - a^2[ \dot{\Psi} + 2 H (\Psi + \Phi)]  \delta_{ij} \, .$
\item[$\star$] $\Gamma^{i}_{j0} = \left(H  - \dot{\Psi} \right) \delta^i_j \, .$
\item[$\star$] $\Gamma^{i}_{jk} = \delta_{jk} \delta^{i\ell} \partial_\ell \Psi - 2 \delta^{i}\, _{( j} \partial_{k \, )} \Psi \, .$
\end{itemize}

\noindent Here the symmetrisation is given by, $\partial_{(i} \partial_{j)} E \equiv \tfrac{1}{2}
\left( \partial_i E_j +  \partial_{j}E_i  \right)$. We work on the l.h.s of the Eq.~\eqref{Eq:geodesic_eq} using rule of chain like, 
\bea
\dfrac{\mathrm{d}^{2}x^{i}}{\mathrm{d}\lambda^{2}} & = & \dfrac{\mathrm{d}t}{\mathrm{d}\lambda} \dfrac{\mathrm{d}\chi}{\mathrm{d}t} \dfrac{\mathrm{d}}{\mathrm{d}\lambda} \left[ \dfrac{\mathrm{d}\chi \theta^{i}}{\mathrm{d}\chi} \dfrac{\mathrm{d}t}{\mathrm{d}\lambda} \dfrac{\mathrm{d}\chi}{\mathrm{d}t} \right] \nonumber \\
& = & \dfrac{p(1-\Phi)}{a}\dfrac{\mathrm{d}}{\mathrm{d}\chi} \left[\dfrac{p(1-\Phi)}{a} \dfrac{\mathrm{d}(\chi\theta^{i})}{\mathrm{d}\chi}\right] \, .
\label{Eq:geodesic_eq_lhs}
\eea
 
\noindent In which we have applied both Eq.~\eqref{Eq:chi_comoving} and Eq.~\eqref{Eq:p0}. For the r.h.s. of Eq.~\eqref{Eq:geodesic_eq}, we show the non-zero Christoffel symbols to use,
\bea
-\Gamma_{\mu\nu}^{i}\dfrac{dx^{\mu}}{d\lambda}\dfrac{dx^{\nu}}{d\lambda} & = &
-\Gamma_{\mu\nu}^{i}\dfrac{dx^{\mu}}{d\chi}\dfrac{dx^{\nu}}{d\chi} \left(\dfrac{\mathrm{d}t}{\mathrm{d}\lambda}\right)^2 \left( \dfrac{\mathrm{d}\chi}{\mathrm{d}t} \right)^2 \nonumber \\ & = & 
-\dfrac{p^2(1-\Phi)^2}{a^2}\left[\Gamma_{00}^{i}\left(\dfrac{dx^{0}}{d\chi}\right)^2
+ 2 \Gamma_{0j}^{i}\dfrac{dx^{0}}{d\chi} \dfrac{dx^{j}}{d\chi} 
+ \Gamma_{jk}^{i}\dfrac{dx^{j}}{d\chi}\dfrac{dx^{k}}{d\chi}
\right] \, .
\label{Eq:geodesic_eq_rhs}
\eea

\noindent At this point, we apply the small-angle approximation, thus, we restrict our analysis to first-order terms, whereby the term $\theta \cdot \Phi$ is considered a second-order contribution. Then, matching Eq.~\eqref{Eq:geodesic_eq_lhs} with Eq.~\eqref{Eq:geodesic_eq_rhs}, and substituting the Christoffel symbols, we arrive at:
\bea
\dfrac{\mathrm{d}}{\mathrm{d}\chi} \left[\dfrac{p}{a} \dfrac{\mathrm{d}\chi\theta^{i}}{\mathrm{d}\chi}\right]  \approx 
-\dfrac{p}{a}\left[ a^2 \partial^i \Phi
+ 2 a H \dfrac{d(\chi \theta^i)}{d\chi} + \delta^{i m} \partial_m \Psi \right] \, .
\label{Eq:geodesic_reduced}
\eea

\noindent Due to the fact that the momentum of photons scales as $p \sim a^{-1}$, we can rewrite and expand the geodesic equation as follows, 
\bea
\dfrac{\mathrm{d}}{\mathrm{d}\chi} \left[\dfrac{1}{a^2} \dfrac{\mathrm{d}(\chi\theta^{i})}{\mathrm{d}\chi}\right] & = &
\dfrac{1}{a^2} 
\dfrac{\mathrm{d}^2 (\chi\theta^{i})}{\mathrm{d}^2\chi}
- \dfrac{2}{a^3} \dfrac{\mathrm{d}a}{\mathrm{d}t} \left(\dfrac{\mathrm{d}t}{\mathrm{d}\chi} \right) \dfrac{\mathrm{d}(\chi\theta^{i})}{\mathrm{d}\chi} \nonumber \\ 
& = &
-\left[\partial^i \Phi
+ 2 \dfrac{H}{a} \dfrac{d(\chi \theta^i)}{d\chi} + \dfrac{1}{a^2} \delta^{i m} \partial_m \Psi \right] \, .
\label{Eq:geodesic_reduced_2}
\eea

\noindent Lastly, we simplify some of the expressions to obtain:
\bea
\dfrac{1}{a^2} \dfrac{\mathrm{d}^{2}(\chi\theta^{i})}{\mathrm{d}\chi^{2}} = -\partial^i (\Psi + \Phi) = \dfrac{1}{a^2} \delta^{ij}  \partial_j (\Psi + \Phi) \, .
\label{Eq:geodesic_reduced_3}
\eea

\noindent Now, the above equation is integrated once, yielding 
\bea
\dfrac{\mathrm{d}(\chi\theta^{i})}{\mathrm{d}\chi} =  - \delta^{ij}   \int_{0}^{\chi} \mathrm{d} \hat{\chi}  \, \partial_j \left[\Psi(\mathbf{x}(\hat{\chi})) + \Phi(\mathbf{x}(\hat{\chi}  )) \right] + \text{const} \, .
\label{Eq:geodesic_reduced_4}
\eea

\noindent Integrating once again, we get
\bea
\theta^{i} = - \delta^{ij}  \dfrac{1}{\chi} \int_{0}^{\chi} \mathrm{d} \chi' \int_{0}^{\chi'} \mathrm{d} \hat{\chi}  \, \partial_j \left[\Psi(\mathbf{x}(\hat{\chi})) + \Phi(\mathbf{x}(\hat{\chi}  )) \right] + \text{const} \, .
\label{Eq:geodesic_reduced_5}
\eea

\noindent Since we are interested in the trajectory of a photon emitted from the original source, we label  $\theta^{i}(\chi) = \theta^{i}_{S}(\chi)$. As a necessary condition, in the absence of lensing, the image and the source must coincide, meaning $ \theta^{i} = \theta^{i}_{S}$. This allows us to determine the integration constant and we get,
\bea
\theta^{i}_{S} & = & \theta^{i} - \dfrac{1}{\chi} \delta^{ij}   \int_{0}^{\chi} \mathrm{d} \chi' \int_{0}^{\chi'} \mathrm{d} \hat{\chi}  \, \partial_j \left[\Psi(\mathbf{x}(\hat{\chi})) + \Phi(\mathbf{x}(\hat{\chi}  )) \right] \nonumber \\  & = & 
\theta^{i} - \dfrac{1}{\chi} \delta^{ij}  \int_{0}^{\chi} \mathrm{d} \hat{\chi} \int_{\hat{\chi}}^{\chi} \mathrm{d} \chi'  \, \partial_j \left[\Psi(\mathbf{x}(\hat{\chi})) + \Phi(\mathbf{x}(\hat{\chi}  )) \right] \, .
\label{Eq:geodesic_reduced_6}
\eea

\noindent Given that the region of integration satisfies $ 0 \leq \hat{\chi} \leq \chi' \leq \chi$ (see \citep{2020moco.book.....D}, page 380, for details). We can change the order of integration limits. Finally, we solve the nested integral that yields,
\bea
\theta^{i}_{S} = 
\theta^{i} - \delta^{ij} \int_{0}^{\chi} \mathrm{d} \hat{\chi} \, \left(\dfrac{\chi - \hat{\chi}}{\chi}\right)  \, \partial_j \left[\Psi + \Phi \right] \, .
\label{Eq:geodesic_reduced_7}
\eea

\noindent This expression provides the difference between the angular deflection of an image $\theta^{i}$ and the original position $\theta^{i}_{S}$, incorporating the lensing effects from the gradient of the gravitational potentials\footnote{At late times, the anisotropic stress becomes negligible, allowing us to set $\Phi = \Psi$.} $\Phi$ and $\Psi$. The relative
conformal distances to the lens and the source, encoded in the term $\left(\chi - \hat{\chi} \right)/\chi$, acts as a weighting factor. In case of a cosmology with non-zero spatial curvature, the source position would be given by $\mathbf{x}_{\rm true} = f_K(\chi) \mathbf{\theta}_S$ as well as the observed position. As a result of this, Eq.~\eqref{Eq:geodesic_reduced_7} modifies as follows,
\bea
\theta^{i}_{S} = 
\theta^{i} - \delta^{ij}  \int_{0}^{\chi}   \mathrm{d} \hat{\chi} \, \dfrac{f_K(\chi - \hat{\chi})}{f_K(\chi)} \, \partial_j \left[\Psi + \Phi \right] \, .
\label{Eq:geodesic_K}
\eea

\subsection{Distortion tensor}
\label{subsec:distortion_tensor}

To seek a new representation between the observed and true angular coordinates in terms of lensing distortions, we keep focusing on small deflections. In such case, the relation between them can be linearised. Hence, we express it as:
\bea
\theta^i_{S} \approx \theta^i + \mathcal{A}^i_j \, \delta\theta^j + \ldots \, . 
\label{Eq:coord_1st}
\eea

\noindent This let us to introduce the $2 \times 2$ Jacobian matrix, $\mathcal{A}^i_j$, also known as the deformation matrix, which is defined as:  
\bea
\mathcal{A}^i_j \equiv \dfrac{\partial \theta^i_{S}}{\partial \theta^j} = \chi \dfrac{\partial \theta^i_{S}}{\partial x^j}  \, .
\label{Eq:A_matrix}
\eea

\noindent Note that when we substitute it into Eq.~\eqref{Eq:geodesic_reduced_7},  and for the case in which no lensing is present,  the 
$\mathcal{A}_{ij}$ matrix would reduce to the identity matrix, meaning that the true and observed positions are identical. \\
To capture the lensing-induced distortions, we introduce the distortion tensor:
\bea
\psi^i_j = \mathcal{A}^i_j - \mathrm{I}_2
= \begin{pmatrix}{1-\kappa-\gamma_{1}}&{-\gamma_{2}}\\ {-\gamma_{2}}&{1-\kappa+\gamma_{1}}\end{pmatrix} = - \delta^{ik} \int_{0}^{\chi} \mathrm{d} \hat{\chi} \, \hat{\chi} \, \left(\dfrac{\chi - \hat{\chi}}{\chi}\right) \partial_k \partial_j \left[\Psi + \Phi \right]   \, .
\label{Eq:distortion_matrix}
\eea    

\begin{figure}[t]
\centering
\includegraphics[width=\linewidth]{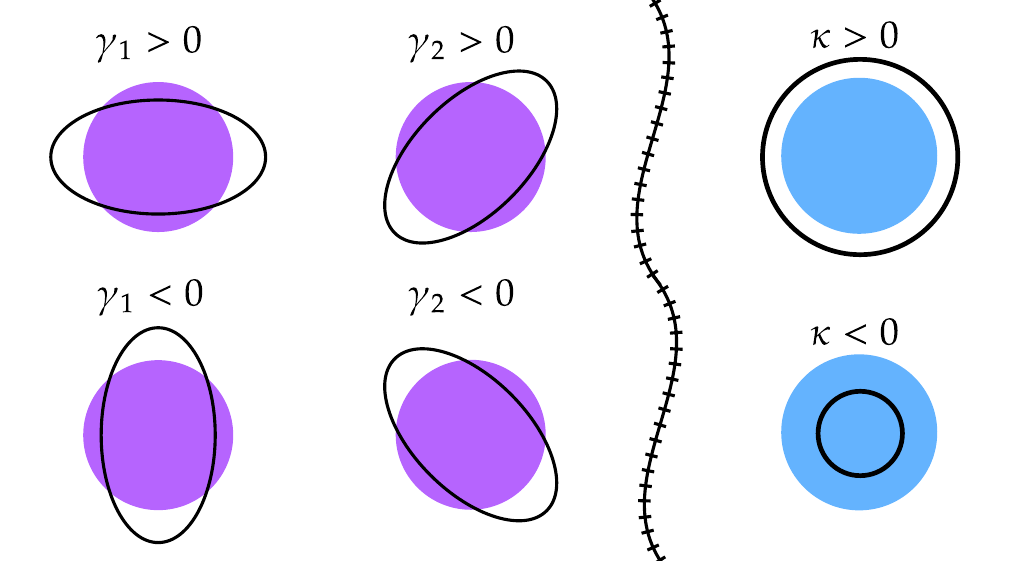}
\caption[The effects of shear ($\gamma_1, \gamma_2$) and convergence ($\kappa$)]{
\textbf{Left:} The effects of shear  components ($\gamma_1, \gamma_2$) of the deformation matrix $\mathcal{A}_{ij}$ on the circular source are illustrated by the transition to an elliptical shape. For instance, shear $\gamma_1$ stretches the purple circle horizontally or vertically, according to its sign, while $\gamma_2$ compresses it along the diagonal and its orientations according also to its sign. \textbf{Right:} The behaviour of convergence ($\kappa$) is displayed through the changes in the size of the blue circle. Positive convergence ($\kappa > 0$) magnifies the image, whereas negative convergence ($\kappa < 0$) reduces its size. However, this effect, representing a change in image size, becomes particularly subtle in the weak lensing regime, where $\kappa \ll 1$.}  
\label{fig:shear_lensing}
\end{figure}

\noindent This tensor captures the deviations from the identity transformation, which encodes key lensing effects: Here, $\kappa$ denotes the \textbf{convergence}, responsible for isotropic magnification of the source (an increase or decrease in the image size), while $\gamma_1$ and $\gamma_2$ correspond to the components of the \textbf{shear}, which account for information on the anisotropic distortions such as stretching, compression (elliptical distortions) and rotation. A schematic representation of these effects illustrated on a circle is \autoref{fig:shear_lensing}.

\subsubsection{Ellipticities}
\label{subsubsec:ellipticities}

Briefly, we now review the connection between galaxy ellipticity and lensing-induced shear. The relation between these quantities comes from the distribution moments of the galaxy image (its brightness in particular). However, the dipole moments cancel out by symmetry, as seen in the left panel of \autoref{fig:shear_lensing}, due to equal shear contributions across all quadrants. This leaves the quadrupole moments as the dominant non-zero terms, defined as:
\bea
q_{ij} \equiv \left\langle  \theta_i \theta_j \right\rangle = \int \mathrm{d} \theta I_{\rm obs}(\boldsymbol{\theta}) \theta_i \theta_j \, .  
\label{Eq:quadrupoles}
\eea

\noindent For a galaxy image with its light profile $I_{\rm obs}$, where the centre is located at $(\theta_1, \theta_2) = (0, 0)$. The elongation and orientation of an elliptical image are represented by 
\bea
\epsilon_1 = \dfrac{q_{xx} - q_{yy}}{q_{xx} + q_{yy}}\, , \quad \epsilon_2 = \dfrac{2q_{xy}}{q_{xx} + q_{yy}} \, .
\label{Eq:epsilons}
\eea

\noindent So using Eq.~\eqref{Eq:quadrupoles}, we can express them as 
\bea
\epsilon_1 = \dfrac{\int \mathrm{d} \theta I_{\rm obs}(\boldsymbol{\theta}) \left(\theta_{xx} - \theta_{yy}\right)}{\int \mathrm{d} \theta I_{\rm obs}(\boldsymbol{\theta}) \left(\theta_{xx} + \theta_{yy}\right)} \, , \quad \epsilon_2 =  \dfrac{2 \int \mathrm{d} \theta I_{\rm obs}(\boldsymbol{\theta}) \, \theta_{xy}}{\int \mathrm{d} \theta I_{\rm obs}(\boldsymbol{\theta}) \left(\theta_{xx} + \theta_{yy}\right)} \, .
\label{Eq:epsilons_2}
\eea

\noindent To avoid lengthy calculations, we proceed by substituting $\theta^i = \left(\mathcal{A}^{-1}\right)^i_{j} \theta^j_S$ into the equations above. In the limit where \(\kappa, \gamma_1, \gamma_2 \ll 1\), this leads to the simplified expressions\footnote{Furthermore, we apply the conservation of surface brightness $I_{\rm obs}(\boldsymbol{\theta}) = I_{\rm true}(\boldsymbol{\theta}_S)$, an important assumption within weak lensing.} as follows:  
\bea
\epsilon_1 \approx \dfrac{4\gamma_1}{2(1-2\kappa)} \, , \quad \epsilon_2 \approx  \dfrac{4\gamma_2}{2(1-2\kappa)} \, .
\label{Eq:epsilons_3}
\eea

\noindent Through extracting the ellipticity components $\epsilon_1$ and $\epsilon_2$ from galaxy images, we can infer the gravitational lensing effects by the shear.\\

\subsection{Power Spectrum of the Distortion Tensor}
\label{subsec:powerspectra_distortion}

We take a step back to reformulate the distortion tensor from Eq.~\eqref{Eq:distortion_matrix} like 
\bea
\psi^i_j = - \delta^{ik} \int_{0}^{\chi_H} \mathrm{d} \chi  g(\chi) \partial_k \partial_j \left[\Psi + \Phi \right]   \, .
\label{Eq:distortion_matrix_new}
\eea    

\noindent Notice that we include new terms, $g(\chi)$ represents the weighting function and it is given by
\bea
g(\chi)  =  \chi \int_{\chi}^{\chi_H} \mathrm{d} \chi' \left(\dfrac{\chi' - \chi}{\chi'}\right) n_g(\chi') \, ,
\label{Eq:window_wl_g}
\eea    

\noindent where the distance $\chi_H$ at the upper limit of the integral defines the Hubble radius $\chi_{H} = 1/H_0$, and $n(\chi)$ is from Eq.~\eqref{Eq:photo-z_dist}. For simplicity, we set the GR limit for the gravitational potentials, i.e. $\Psi = \Phi$. \\ 
We now turn to the calculation of the power spectrum of the distortion tensor. Using the definition of the power spectrum, we write
\bea
P_{\psi_{ijnm}}(\boldsymbol{\ell}) = \int \dfrac{\mathrm{d}^2 \ell'}{(2\pi)^2} \left\langle \tilde{\psi}_{ij}(\boldsymbol{\ell}) \tilde{\psi}^*_{nm}(\boldsymbol{\ell}') \right\rangle \, . 
\label{Eq:distortion_spectrum}
\eea

\noindent The variable $\boldsymbol{\ell}$ is the Fourier counterpart of the two-dimensional angular position $\boldsymbol{\theta}$. The Fourier transform of the distortion tensor from Eq.~\eqref{Eq:distortion_matrix_new}  is computed as follows
\bea
\tilde{\psi}_{ij}(\boldsymbol{\ell}) = \int \mathrm{d}^2 \boldsymbol{\theta} \, \psi_{ij}(\boldsymbol{\theta}) e^{-i \boldsymbol{\ell} \cdot \boldsymbol{\theta}} \, .
\eea  

\noindent Then, we can substitute it into the distortion tensor $\psi_{ij}(\boldsymbol{\theta})$ in real-space, we obtain
\bea
\tilde{\psi}_{ij}(\boldsymbol{\ell}) & = & \int \mathrm{d}^2 \theta  \psi_{ij}(\boldsymbol{\theta}) \exp\left(-i \boldsymbol{\ell} \cdot \boldsymbol{\theta}\right) \nonumber \\ 
& = & - 2 \int \mathrm{d}^2 \theta  \int_{0}^{\chi_H} \mathrm{d} \hat{\chi}  \, g(\chi) \, \partial_i \partial_j \left[\Phi \right] \,  \exp\left(-i \boldsymbol{\ell} \cdot \boldsymbol{\theta}\right) \, . 
\label{Eq:2D_fourier_psi}
\eea

\noindent We apply a 3D Fourier transform to the gravitational potential term and using the Fourier transform of the Laplacian, the expression leads to
\bea
\tilde{\psi}_{ij}(\boldsymbol{\ell}) = 2 \int \mathrm{d}^2 \theta  \int_{0}^{\chi_H} \mathrm{d}\chi  
\int \dfrac{\mathrm{d}^3 k}{(2\pi)^3}  
k_i k_j \, \tilde{\Phi}(\boldsymbol{k}) \, g(\chi) \, \exp\left(-i \boldsymbol{\ell} \cdot \boldsymbol{\theta}\right)  \exp\left(i \boldsymbol{k} \cdot \boldsymbol{x}\right) \, . 
\label{Eq:2D_fourier_psi_2}
\eea

\noindent Using the above expression in Eq.~\eqref{Eq:distortion_spectrum}, we get the following high-dimensional integral
\bea
P_{\psi_{ijnm}}(\boldsymbol{\ell})  &=& 4  \int \dfrac{\mathrm{d}^2 \ell'}{(2\pi)^2} 
\int \mathrm{d}^2 \theta  
\int \mathrm{d}^2 \theta'  
\int_{0}^{\chi_H} \mathrm{d}\chi 
\int_{0}^{\chi_H} \mathrm{d}\chi' 
\int \dfrac{\mathrm{d}^3 k}{(2\pi)^3}  
\int \dfrac{\mathrm{d}^3 k'}{(2\pi)^3}  \nonumber \\
& & \times  k_i k_j k_n k_m \, \left\langle \tilde{\Phi}(\boldsymbol{k}) \tilde{\Phi}^*(\boldsymbol{k}') \right\rangle \, g(\chi) \, g(\chi') \nonumber \\ & &  \times  \exp \left(-i \left[\boldsymbol{\ell} \cdot \boldsymbol{\theta} - \boldsymbol{\ell}' \cdot \boldsymbol{\theta}' \right] \right) \, \exp \left(i \left[\boldsymbol{k} \cdot \boldsymbol{x} - \boldsymbol{k}' \cdot \boldsymbol{x}' \right] \right) \, . 
\label{Eq:distortion_spectrum_full}
\eea

\noindent To manage this integral, we can reduce the integration over $\boldsymbol{k}'$ by substituting the power spectrum definition from Eq.~\eqref{Eq:Power_spectrum} on the $\left\langle \tilde{\Phi}(\boldsymbol{k}) \tilde{\Phi}^*(\boldsymbol{k}') \right\rangle$ term
\bea
P_{\psi_{ijnm}}(\boldsymbol{\ell})  &=& 4  \int \dfrac{\mathrm{d}^2 \ell'}{(2\pi)^2} 
\int \mathrm{d}^2 \theta  
\int \mathrm{d}^2 \theta'  
\int_{0}^{\chi_H} \mathrm{d}\chi 
\int_{0}^{\chi_H} \mathrm{d}\chi' 
\int \dfrac{\mathrm{d}^3 k}{(2\pi)^3}   \nonumber \\
& & \times  k_i k_j k_n k_m \, P_\Phi( \boldsymbol{k}) \, g(\chi) \, g(\chi') \nonumber \\ & &  \times  \exp \left(-i \left[\boldsymbol{\ell} \cdot \boldsymbol{\theta} - \boldsymbol{\ell}' \cdot \boldsymbol{\theta}' \right] \right) \, \exp \left(i \boldsymbol{k} \cdot \left[\boldsymbol{x} - \boldsymbol{x}' \right] \right) \, . 
\label{Eq:distortion_spectrum_full_2}
\eea

\noindent Moreover, the exponential terms can be grouped like 
\bea
P_{\psi_{ijnm}}(\boldsymbol{\ell})  &=& 4  \int \dfrac{\mathrm{d}^2 \ell'}{(2\pi)^2} 
\int \mathrm{d}^2 \theta  
\int \mathrm{d}^2 \theta'  
\int_{0}^{\chi_H} \mathrm{d}\chi 
\int_{0}^{\chi_H} \mathrm{d}\chi' 
\int \dfrac{\mathrm{d}^3 k}{(2\pi)^3}   \nonumber \\
& & \times  k_i k_j k_n k_m \, P_\Phi( \boldsymbol{k}) \, g(\chi) \, g(\chi') \nonumber \\ & &  \times  
\exp \left(-i \left[\ell_1 \theta_1 + \ell_2 \theta_2 - k_1 x_1 - k_2 x_2 \right] \right) 
\nonumber \\ 
& & \times \exp \left(i \left[\ell'_1 \theta'_1 + \ell'_2 \theta'_2 - k_1 x'_1 - k_2 x'_2 \right] \right) \nonumber \\ 
& & \times \exp \left(ik_3 \left[x_3 - x_3' \right]  \right) \, . 
\label{Eq:distortion_spectrum_full_3}
\eea

\noindent Now we recall that the 2D Dirac delta, $\delta_{\rm D}^{(2)}(\boldsymbol{\ell} - \boldsymbol{\ell}')$,  can be cast in the following way
\bea
\delta_{\rm D}^{(2)}(\boldsymbol{\ell} - \boldsymbol{\ell}') = \int \dfrac{\mathrm{d}^2 \theta}{(2\pi)^2} \, \exp\left[-i (\boldsymbol{\ell} - \boldsymbol{\ell}') \cdot \boldsymbol{\theta}\right] \, .
\label{Eq:delta_2d}
\eea

\noindent Changing $\boldsymbol{x} \to \chi \boldsymbol{\theta}$, the integral of Eq.~\eqref{Eq:distortion_spectrum_full_3} is reduced to
\bea
P_{\psi_{ijnm}}(\boldsymbol{\ell})  &=& 4  \int \mathrm{d}^2 \ell' 
\int_{0}^{\chi_H} \mathrm{d}\chi 
\int_{0}^{\chi_H} \mathrm{d}\chi' 
\int \dfrac{\mathrm{d}^3 k}{(2\pi)}   k_i k_j k_n k_m \, P_\Phi( \boldsymbol{k}) \, g(\chi) \, g(\chi')  \nonumber \\
& & \times  \delta_{\rm D}^{(2)}\left(\boldsymbol{\ell} - \chi \boldsymbol{k}^{(2)} \right) \, \delta_{\rm D}^{(2)}\left(\boldsymbol{\ell}' - \chi' \boldsymbol{k}^{(2)}  \right) \, \exp \left(ik_3 \left[\chi - \chi' \right]  \right) \, . 
\label{Eq:distortion_spectrum_full_4}
\eea

\noindent Note that $\boldsymbol{k}^{(2)}$ denotes the first two components of the 3D vector $\boldsymbol{k}$. To simplify the last exponential term, we employ the 1D Fourier transform, which gives $2\pi \delta(\chi-\chi')$, so this is reduced to 
\bea
P_{\psi_{ijnm}}(\boldsymbol{\ell})  &=& 4  \int \mathrm{d}^2 \ell' 
\int_{0}^{\chi_H} \mathrm{d}\chi 
\int \mathrm{d}^2 k \, k_i k_j k_n k_m \, P_\Phi( \boldsymbol{k}) \, g^2(\chi)  \nonumber \\
& & \times  \delta_{\rm D}^{(2)}\left(\boldsymbol{\ell} - \chi \boldsymbol{k}^{(2)} \right) \, \delta_{\rm D}^{(2)}\left(\boldsymbol{\ell}' - \chi \boldsymbol{k}^{(2)}  \right)  \, . 
\label{Eq:distortion_spectrum_full_5}
\eea

\noindent The Dirac delta of $\boldsymbol{\ell}'$ term enforces that $\boldsymbol{k} \to \boldsymbol{\ell}'/\chi$ which is commonly well-known as the Limber approximation. Substituting this condition, the expression simplifies to:
\bea
P_{\psi_{ijnm}}(\boldsymbol{\ell}) = 4  \int_{0}^{\chi_H} \mathrm{d}\chi 
\int \dfrac{\mathrm{d}^2 \ell'}{\chi^2} \, \dfrac{\ell'_i \ell'_j \ell'_n \ell'_m}{\chi^4} \, P_\Phi( \boldsymbol{\ell}'/\chi) \, g^2(\chi) \, \delta_{\rm D}^{(2)}\left(\boldsymbol{\ell} - \boldsymbol{\ell}' \right) \, . 
\label{Eq:distortion_spectrum_full_6}
\eea

\noindent Finally, the remaining Dirac delta eliminates the dependency on $\boldsymbol{\ell}'$, yielding our final expression:
\bea
P_{\psi_{ijnm}}(\boldsymbol{\ell})  = 4  \int_{0}^{\chi_H} \mathrm{d}\chi \dfrac{g^2(\chi)}{\chi^2} \, \dfrac{\ell_i \ell_j \ell_n \ell_m}{\chi^4} \, P_\Phi( \boldsymbol{\ell}/\chi) \, . 
\label{Eq:distortion_spectrum_final}
\eea

\noindent This power spectrum, being a four-index quantity, is statistically connected to the deformation matrix components, as we will now illustrate.
 
\section{Convergence}
\label{sec:convergence}

From Eq.~\eqref{Eq:distortion_matrix}, we realize the convergence $\kappa$ is obtained through
\bea
\kappa = -\dfrac{1}{2}\left( \psi_{11} + \psi_{22}\right) \, . 
\label{Eq:kappa}
\eea

\noindent With this expression at hand, we derive the power spectrum of the convergence: 
\bea
P_\kappa(\boldsymbol{\ell}) = 
\left\langle \kappa \kappa^* \right\rangle  = \dfrac{1}{4} \left\langle \left( \psi_{11} + \psi_{22}\right) \left( \psi^*_{11} + \psi^*_{22}\right) \right\rangle  = \dfrac{1}{4} \left( P_{\psi_{1111}} + P_{\psi_{2222}} + 2 P_{\psi_{1122}}\right)   \, .
\label{Eq:kappa_spectrum}
\eea

\noindent Instead of working in $(\ell_1, \ell_2)$ coordinates, we consider them in polar coordinates, where $\ell_1 = \ell \cos(\varphi)$ and $\ell_2 = \ell \sin(\varphi)$. This leads to simplify the above expressions as
\bea
P_\kappa(\boldsymbol{\ell}) =    \int_{0}^{\chi_H} \mathrm{d}\chi \dfrac{g^2(\chi)}{\chi^2} \, \dfrac{\ell^4}{\chi^4} \, P_\Phi(\boldsymbol{\ell}/\chi) \, .  
\label{Eq:kappa_spectrum_2}
\eea

\noindent Using the Poisson equation in Fourier space, the gravitational potential 
$\Phi$ is connected to matter density perturbations $\delta_{\rm m}$ as follows, 
\bea
\nabla^2_{\boldsymbol{x}} \Phi(\boldsymbol{x}) = 4\pi G \bar{\rho}_{\rm m} a^2 \delta_{\rm m}(\mathbf{x}) \, , 
\quad \Rightarrow \quad
k^2 \tilde{\Phi}(\boldsymbol{k}) = - 4\pi G \bar{\rho}_{\rm m} a^2 \tilde{\delta}_{\rm m}(\mathbf{k}) \, .
\label{Eq:Poisson_fourier_relation}
\eea

\noindent Here, the mean matter density can be written as
\bea
\bar{\rho}_{\rm m} = \dfrac{3 H^2_0}{8\pi G} \Omega_{{\rm m},0} a^{-3} \, .  
\label{Eq:mean_density}
\eea

\noindent Moreover, we redefine the lensing kernel of Eq.~\eqref{Eq:window_wl_g} like,
\bea
W(\chi)  =   \dfrac{3}{2} \Omega_{{\rm m},0} \,  H^2_0  \, \dfrac{\chi}{a} \, \int_{\chi}^{\chi_H} \mathrm{d} \chi' \left(\dfrac{\chi' - \chi}{\chi'}\right) n_g(\chi') \, .
\label{Eq:window_wl_W}
\eea   

\noindent In which $\Omega_{{\rm m},0}$ is the total matter density parameter and $H_0$ is the Hubble constant, both evaluated at the present time. Thus, the convergence power spectrum becomes:
\bea
P_\kappa(\boldsymbol{\ell}) =   \int_{0}^{\chi_H} \mathrm{d}\chi \dfrac{W^2(\chi)}{\chi^2} \, P_{\rm m}(\boldsymbol{\ell}/\chi) \, .  
\label{Eq:kappa_spectrum_3}
\eea

\noindent By recurring to the cosmological principle and adopting a more commonly used notation for the spectrum, we rewrite this as:
\bea
C_{\kappa \kappa}(\ell) = \int_{0}^{\chi_H} \mathrm{d}\chi \dfrac{W^2(\chi)}{\chi^2} \, P_{\rm m}(\ell/\chi) \, .  
\label{Eq:kappa_spectrum_4}
\eea

\noindent This is known as the angular power spectrum for the convergence. Let us remark that this angular spectrum takes into account the total matter distribution, accounting for dark and visible matter, neutrinos and baryonic feedback across the comoving distance between the source objects and the observer. For reference, we finally express its Fourier space 2PCF,
\bea
\left\langle \kappa(\ell)\kappa^*(\ell') \right\rangle = (2\pi)^{2}\delta^{(2)}_{\rm D}(\ell + \ell')C_{\kappa}(\ell) \, .
\label{Eq:kappa_spectrum_other}
\eea

\section{Galaxy Shear}
\label{sec:shear}

Likewise, we can identify the following contributions to the shear components,
\bea
\gamma_{1} = \dfrac{1}{2}\left( \psi_{22} - \psi_{11}\right) \, , \qquad  \gamma_{2} = -\psi_{12} \, . 
\label{Eq:shear_in_distortion}
\eea

\noindent According to which, we proceed to compute the angular power spectra of $\gamma_1$, $\gamma_2$ and their cross-component to get 
\bea
C_{\gamma_1 \gamma_1}(\ell) &=& \cos^2(\varphi) \, C_{\kappa \kappa}(\ell) \, ,  \\
C_{\gamma_2 \gamma_2}(\ell) &=& \sin^2(\varphi) \, C_{\kappa \kappa}(\ell) \, ,  \\
C_{\gamma_1 \gamma_2}(\ell) &=& \sin(2\varphi) \cos(2\varphi) \, C_{\kappa \kappa}(\ell) \, .  
\label{Eq:shear_spectrum}
\eea

\noindent The dependence of $C_{\gamma_i \gamma_j}(\ell)$ on $\varphi$ is evident. This dependence on orientation implies that $\gamma_1$ and $\gamma_2$ are not the most practical variables for analysis. To ease the complexity, shear components are often expressed as a single complex quantity, $\gamma = \gamma_1 + i \gamma_2$, which transforms under a rotation by the angle $\varphi$ by a phase,
\bea
\gamma \to \gamma' = \gamma e^{2 i \phi} \, .
\label{Eq:gamma_transformation}
\eea

\noindent This transformation allows us to define two useful components for the shear, the tangential and cross components, given by:
\bea
\gamma_{t}=-\mathrm{Re}[\gamma e^{-2 i \varphi}]\, ,  \qquad  \gamma_{\times}&=-\mathrm{Im}[\gamma e^{-2 i \varphi}] \, .
\label{Eq:tangential_cross}
\eea

\begin{figure}
\centering
\includegraphics[width=0.75\linewidth]{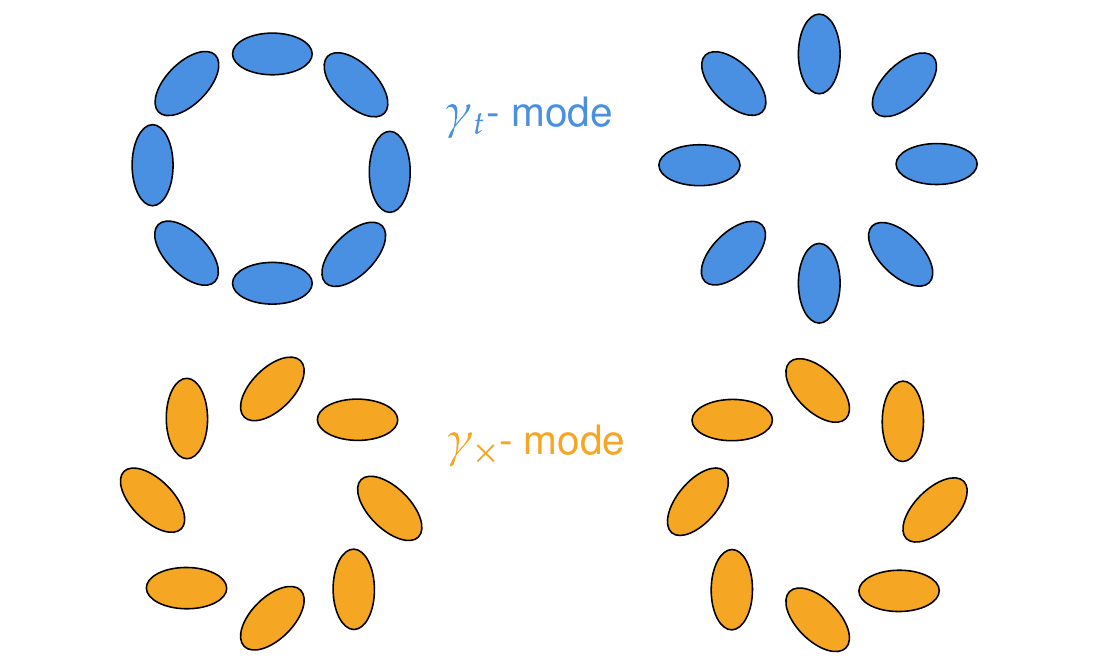}
\caption[Tangential and cross shear effects]{Visualization of the tangential (${\rm E}$-mode) and cross (${\rm B}$-mode) shear modes, respectively. \textbf{Top figures:} These curl-free modes are distortions aligned with, or perpendicular to, a given reference direction. \textbf{Bottom figures:} These curl modes capture distortions oriented at $45^\circ$ degrees to the shear-free directions. \\
\textbf{Image Credit:} Illustration adapted from \cite{2003astro.ph..5089V}.}
\label{Fig:t_x_modes}
\end{figure}

\noindent Such components are deemed as spin-2 modes, see \autoref{Fig:t_x_modes}. Interestingly, the next linear combination between the original and new shear components,
\bea
\gamma_t &=& -\gamma_1 \cos(2\varphi) -\gamma_2 \sin(2\varphi) \, ,  \\
\gamma_\times &=& \gamma_1 \sin(2\varphi) - \gamma_2 \cos(2\varphi)\, .
\label{Eq:shear_new}
\eea

\noindent This transformation results in angular power spectra independent of the angle $\varphi$,
\bea
C_{\gamma_t \gamma_t}(\ell) &=&  C_{\kappa \kappa}(\ell) \, ,  \\
C_{\gamma_\times \gamma_\times}(\ell) &=& 0 \, ,  \\
C_{\gamma_\times \gamma_t}(\ell) &=& 0 \, .  
\label{Eq:shear_spectrum_2spin}
\eea
 
\noindent The fact that the last two spectra are zero suggests two possibilities: either the theoretical model is consistent with data, and any discrepancies in the data are due to noise or unaccounted systematic errors, or the presence of non-zero measurements could indicate missing or incomplete aspects of the model, requiring further refinement. \\ 

\noindent We have carried our derivations assuming a single galaxy population, but real surveys involve galaxy samples at different redshift ranges across the sky. To account for this, we divide these samples into multiple redshift bins, with a certain distribution.  \\
Given $N_{\rm bins}$ tomographic redshift bins, each $(i,j)$ pair of redshift bins will correspond to a term in the angular power spectrum $C^{ab}_{ij}(\ell)$, with $a,b \in \{ \kappa, \gamma_i \}$. For describing it in a non-zero curvature scenario, it is computed the following formula
\bea
C_{i j}^{ab}(\ell) = \int_{0}^{\chi_{H}} \mathrm{d}\chi \, \dfrac{W_i^a \, W_j^b }{f^2_K(\chi)} \, \hat{P}_{\rm NL} \left(k = \dfrac{\ell + 1/2}{\chi}, z \right) \, .
\label{eq:cell_shear_2}
\eea

\noindent Note that in Eq.~\eqref{eq:cell_shear_2} the Fourier scales $k$-modes are differently linked with angular multipoles $\ell$-modes through assuming the extended Limber approximation \citep{2008PhRvD..78l3506L,2017MNRAS.472.2126K}. The sub-index stands for each tomographic bin-$(i,j)$ of redshift, which is associated with the kernel $W_i^a$, such that, given a distribution $n_{s,i}(z)$ of a sourced redshift, it is computed as:
\bea
W_i^a (\chi) = \dfrac{3}{2} \Omega_{\rm m} \dfrac{H_0^2}{c^2} \dfrac{\chi}{a} {\bigintsss\limits_{\chi}}^{\chi_{\rm H}} \mathrm{d} \chi' \, n_{s,i}(\chi') \, \dfrac{\chi'-\chi}{\chi'} \, . 
\label{eq:window_shear_2}
\eea

\noindent While the full analysis of weak lensing often involves a broader range of observables, including the convergence and cross-correlations between various probes, in this thesis, we focus exclusively on the shear as a primary observable, specifically on the $C^{\gamma \gamma}_{ij}(\ell)$ statistics. As previously mentioned, the observed cosmic shear signal is primarily composed of two modes: $\gamma_t$ and $\gamma_\times$, commonly named as ${\rm E}$-mode (curl-free pattern) and ${\rm B}$-mode (divergence-free pattern), respectively (see \autoref{Fig:t_x_modes}). However, under the above definitions, the shear signal is expected to be dominated by ${\rm E}$-mode. On the other hand, we do not expect to measure any significant ${\rm B}$-mode.\footnote{Although its signal can emerge from residual systematics, noise, or other astrophysical effects.} For current surveys, e.g. KiDS, the effects of ${\rm B}$-modes is considered minimal, leaving the analyses using primarily ${\rm E}$-mode measurements. 

\subsection{Intrinsic Alignment}
\label{subsec:IA_NLA}

An important systematic in weak lensing to address is the modelling of Intrinsic Alignments (IA), where galaxies near each other are naturally aligned due to local gravitational fields, regardless of the lensing effects.
Hence, the lensing signal is contaminated by the IA of galaxies. To include it, we add IA terms to Eq.~\eqref{eq:cell_shear_2}. The angular power spectrum (now denoted as $\hat{\gamma}$) is also rewritten as:
\bea
C_{ij}^{\hat{\gamma} \hat{\gamma}}(\ell) = C_{ij}^{\gamma \gamma}(\ell) + C_{ij}^{\gamma \mathrm{I}}(\ell) + C_{ij}^{\mathrm{I}  \gamma}(\ell) + C_{ij}^{\mathrm{I} \mathrm{I}}(\ell) \, .
\label{eq:cell_shear_full}
\eea

\noindent Within the later analysis of this thesis, we consider the window function for the IA contribution from the commonly-used nonlinear alignment (NLA) model \citep{2010PhRvD..82d9901H,2007NJPh....9..444B}, which is modulated by two free parameters, an amplitude $A_{\rm IA}$ and a power law parameter $\eta_{\rm IA}$:
\bea
W_i^{\rm I}(\chi) = - A_{\rm IA} \left( \dfrac{1+z}{1+z_{\rm p}} \right)^{\eta_{\rm IA}} \, n_{i}(\chi) \, \dfrac{C_1 \, \rho_{\rm crit} \, \Omega_{\rm m}}{D_+(\chi)} \, ,
\label{eq:window_NLA_2}
\eea

\noindent where $D_+(\chi)$ represents the growth factor. The constants $C_1$ and the pivot redshift $z_{\rm p}$ are set to $0.64$ and $0.3$, respectively. Ignoring IA in weak lensing analyses can significantly provoke to biases in cosmological parameters such as $S_8$ and $w_0$. While the level of contamination depends on redshift and galaxy sample, several studies have shown that IA can constitute a substantial fraction of the observed shear signal up to tens of percent in some regimes \citep{2010A&A...523A...1J,2015PhR...558....1T}. Proper modelling of IA is therefore essential for robust cosmological inference. Nonetheless, we remark that the modelling of IA and the choice of priors for IA-related free parameters remain open questions in the analyses of weak lensing. 

\section{Kilo-Degree Survey statistics}
\label{sec:KiDS_stats}

Similarly to the methodology outlined in \cite{KiDS:2020suj}, we consider here also the  three statistical sets from the KiDS-1000 analysis of weak lensing \citep{2021A&A...646A.129J}: real-space two-point correlation functions (2PCFs, \cite{2002A&A...396....1S}), Band Powers (BP, \cite{2018MNRAS.476.4662V}), and Complete Orthogonal Sets of ${\rm E/B}$-Integrals (COSEBIs, \cite{2010A&A...520A.116S}). Although each statistical set has its own sensitivity to cosmological parameters and systematic effects, the parameter estimation constraints remain consistent with each other. All these statistics are linear transformations of the cosmic shear angular power spectrum in Eq.~\eqref{eq:cell_shear_full}. Note that, very recently KiDS has released its final dataset, KiDS-Legacy, providing the most precise cosmic shear constraints from the survey. They report $S_8 = 0.815^{+0.016}_{-0.021}$, in close agreement ($0.73 \sigma$) with Planck reported value. The observed increase in $S_8$ is primarily attributed to improved redshift calibration \citep{2025arXiv250319440W}, a larger survey area, and careful image processing. While differences remain between data releases, this underscores the inherent challenges of achieving precision in weak lensing measurements.

\subsection{Shear two-point correlation functions}
\label{sec:2pcfs}

The analysis over the 2PCFs of shear are commonly used as summary statistics in weak lensing probes \citep{KiDS:2020suj,2023PhRvD.108l3518L,2023A&A...678A.109A}. The 2PCFs \citep{1992ApJ...388..272K}, $\xi_\pm$, are formally defined in terms of the tangential $\gamma_\mathrm{t}$ and cross $\gamma_\mathrm{\times}$ shear as follows:
\bea
\xi_\pm(\theta) = \left\langle\gamma_t \gamma_t  \right\rangle (\theta)\pm 
 \left\langle \gamma_\mathrm{\times} \gamma_\mathrm{\times} \right\rangle (\theta) \, .
\label{Eq:xi-pm-def}
\eea 

\noindent Here, 2PCFs depend on the angular separation, $\theta$, between pairs of galaxies. Regarding measurements, the data is binned into $\bar{\theta}$-bins, then the observed tangential and cross ellipticities, $\epsilon^{\rm obs}_t$ and $\epsilon^{\rm obs}_\times$ are used to the following estimator for the signal, 
\bea
\hat\xi^{(ij)}_\pm(\bar{\theta})=\dfrac{\sum_{ab} w_a w_b
 \left[\epsilon^{\rm obs}_{{ t},a}\epsilon^{\rm obs}_{{t},b}
 \pm\epsilon^{\rm obs}_{\times,a}\epsilon^{\rm obs}_{\times,b}\right] \Delta^{(ij)}_{ab}(\bar{\theta}) } 
 {\sum_{ab} w_a w_b (1+m_a)(1+m_b)  \Delta^{(ij)}_{ab}(\bar{\theta}) } \, .
\label{Eq:xipm_meaure}
\eea

\noindent We will not go into details here (we refer to \cite{2021A&A...646A.129J} for further details), but briefly, $\Delta^{(ij)}_{ab}(\bar{\theta})$ limits the sums to galaxy pairs $(a,b)$ and, $w_{a}$ or $w_{b}$ represent their weight. The denominator corrects for biases with an averaged multiplicative bias, $m_a$.  \\

\begin{figure}
\centering
\includegraphics[width=\linewidth]{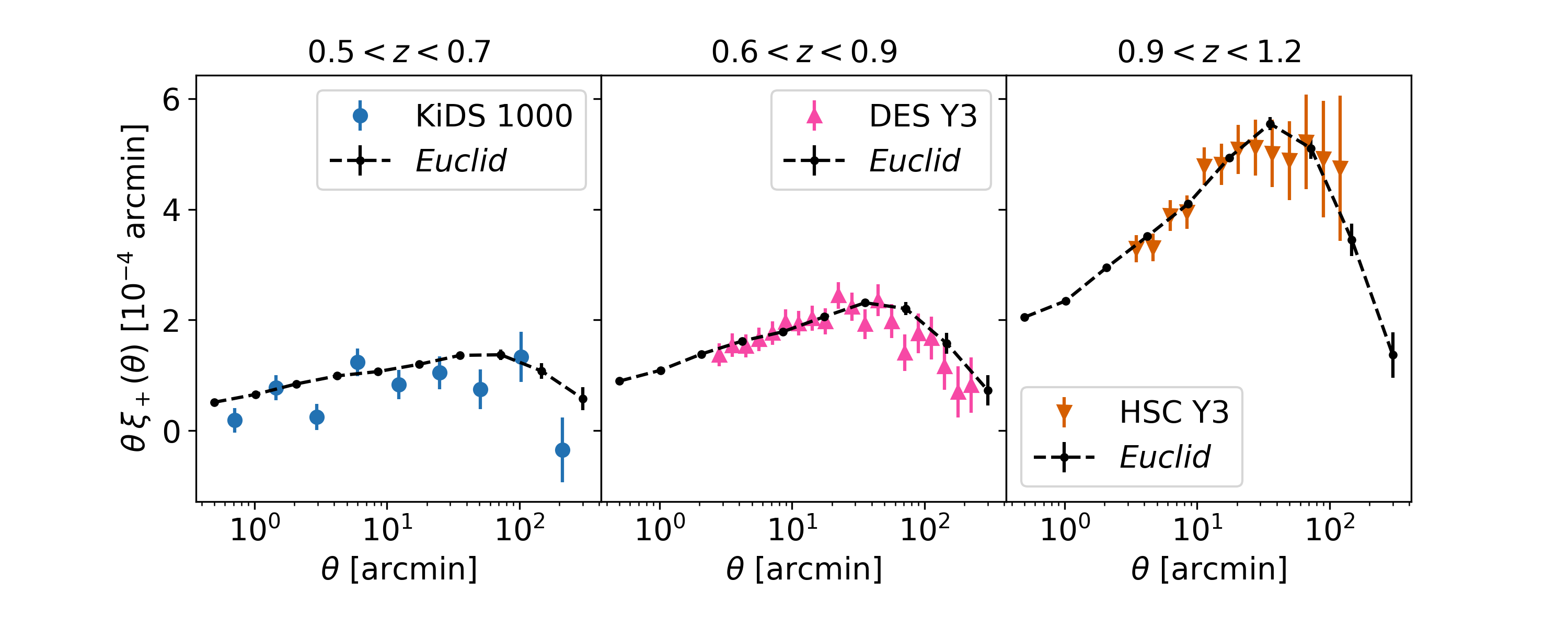}
\caption[Euclid comparison of $\xi_+$ between Stage-III surveys]{Shear correlation function $\xi_{+}(\theta)$ for KiDS-1000 \citep{KiDS:2020suj} on the left plot, then DES-Y3 \citep{2022PhRvD.105b3514A} in the middle, and HSC-Y3 \citep{2023PhRvD.108l3518L} on the right plot, and expected for Euclid satellite in black-dotted lines. Each panel uses data from the tomographic bin with the highest signal-to-noise ratio (S/N) for the respective survey. The S/N for Euclid is notably higher, by a factor of ten, compared to current Stage-III surveys. \\
\textbf{Plots Credits:}  \cite{2024arXiv240513491E}}
\label{Fig:Euclid_shear}
\end{figure}

\noindent Moreover, the 2PCFs can be cast as a linear combination of the ${\rm E}$ and ${\rm B}$-mode angular power spectra,
\bea
\xi_\pm(\theta) =\int_0^{\infty} \frac{ \mathrm{d} \ell\, \ell}{2\pi} {\rm J}_{0/4}(\ell\theta) \left[C_{\rm EE}(\ell)\pm C_{\rm BB}(\ell)\right] \, ,
\label{Eq:xipmPower}
\eea

\noindent where $\mathrm{J}_{0/4}$ are Bessel functions of the first kind. For a comparison of the 2PCFs measurements between Stage IV and Stage-III photometric surveys, refer to \autoref{Fig:Euclid_shear}.

\subsection{Band Powers}
\label{sec:bp}

Band Powers (BP) provide a binned version of the angular power spectrum, typically estimated from another linear transformation of the 2PCFs in Eq.~\eqref{Eq:xipm_meaure}. We can compute BP, denoted as $C_{{\rm E/B},l}$, via
\bea
C_{{\rm E/B},l} = \frac{\pi}{{\mathcal N}_l}\, \int_0^\infty \mathrm{d}  \theta \, \theta\, T(\theta) \left[\xi_+(\theta) \, g_+^l(\theta) \pm \xi_-(\theta)\, g_-^l(\theta)\right] \, , 
\label{Eq:bp_xipm}
\eea

\noindent where $g_\pm^l(\theta)$ is the filter functions and the normalization factor, ${\mathcal N}_l$, is chosen as:
\bea
{\mathcal N}_l = \ln(\ell_{{\rm up},l})-\ln(\ell_{{\rm low},l}) \, .
\label{Eq:bp_normalisation}
\eea 

\noindent With $\ell_{{\rm up},l}$ and $\ell_{{\rm low},l}$ being the upper and lower boundaries of the angular multipole bin, respectively. The integral in Eq.~\eqref{Eq:bp_xipm} must be truncated because the 2PCFs can only be measured over a finite range of angular separations.
The BP are linked to the angular power spectrum through the following expression:
\bea
C_{{\rm E},l} &=&  \frac{1}{2 {\mathcal N}_l} \int_0^\infty \mathrm{d}
\ell\, \ell \left[ W^l_{\rm EE}(\ell)\, C_{\rm EE}(\ell) + W^l_{\rm EB}(\ell)\, C_{\rm BB}(\ell) \right] \, , \\ 
C_{{\rm B},l}  &=& \frac{1}{2 {\mathcal N}_l} \int_0^\infty \mathrm{d}
\ell\, \ell \left[ W^l_{\rm BE}(\ell)\, C_{\rm EE}(\ell) + W^l_{\rm BB}(\ell)\, C_{\rm BB}(\ell)\right] \, .
\label{Eq:bp_formula}
\eea

\noindent These weight functions, which are no longer simple top-hat functions, are expressed as:
\bea
W^l_{\rm EE}(\ell) = W^l_{\rm BB}(\ell)  
= \int_{0}^{\infty}  \mathrm{d}\theta\, \theta\, T(\theta) \left[{ {\rm J}_0(\ell \theta)\, g_+^l(\theta) + {\rm J}_4(\ell \theta)\, g_-^l(\theta) }\right] \, , \\ 
W^l_{\rm EB}(\ell) = W^l_{\rm BE}(\ell)   = \int_{0}^{\infty} \mathrm{d}\theta\, \theta\, T(\theta) \left[{ {\rm J}_0(\ell \theta)\, g_+^l(\theta) - {\rm J}_4(\ell \theta)\, g_-^l(\theta) }\right] \, ,
\label{eq:bpw}
\eea

\noindent where $T(\theta)$ is the selection function. Once again, for further details we refer to \cite{2021A&A...646A.129J}. 
An additional technical aspect (relevant for subsequent analysis) is the set of 8 logarithmically spaced filters within the multipole range $\ell = 100$ to $1500$ in the BP analysis.

\subsection{COSEBIs}
\label{subsec:cosebis}

Unlike BP, COSEBIs are two-point statistics defined over a finite angular range. This framework provides a directly separation of ${\rm E}$-mode and ${\rm B}$-mode contributions. COSEBIs can be also measured through binning the 2PCFs as, 
\bea
E_n = \frac{1}{2} \int_{\theta_{\rm min}}^{\theta_{\rm max}}
\mathrm{d} \theta\,\theta\: 
[F_{+n}(\theta)\
,\xi_+(\theta) +
F_{-n}(\theta)\,\xi_-(\theta)] \, , \\ 
B_n = \frac{1}{2} \int_{\theta_{\rm min}}^{\theta_{\rm
max}} \mathrm{d} \theta\,\theta\: 
[F_{+n}(\theta)\,\xi_+(\theta) -
F_{-n}(\theta)\,\xi_-(\theta)] \, ,
\label{eq:COSEBIsReal}
\eea

\noindent where $F_{\pm n}(\theta)$ are filter functions defined between $\theta_{\rm min}$ and $\theta_{\rm max}$. The $n$-modes are numbered with natural numbers, and their accuracy relies in the number of bins. According to \cite{KiDS:2020suj}, their first few modes efficiently capture nearly all cosmological information. \\
The theoretical expression for COSEBIs can be estimated as,
\bea
E_n = \int_0^{\infty}
\frac{\mathrm{d} \ell\,\ell}{2\pi}C_{\mathrm{EE}}(\ell)\,W_n(\ell) \, ,\\ 
B_n = \int_0^{\infty}
\frac{\mathrm{d} \ell\,\ell}{2\pi}C_{\mathrm{BB}}(\ell)\,W_n(\ell) \, .
\label{eq:En_Bn_Fourier}
\eea

\noindent Being $W_n(\ell)$ the Hankel transforms of $F_{\pm n}(\theta )$. Since COSEBIs focus on specific angular scales, these provide precise control over the $\ell$-scales to be analysed. \\
Each of these statistics probes the weak lensing signal in complementary ways: 2PCFs provide sensitivity to shear correlations in angular configuration space, their Fourier-space counterpart BP offer localized scale-dependent measurements that help isolate systematic effects, and COSEBIs enable rigorous ${\rm E/B}$-mode decomposition, and suppression of ambiguous modes that could bias parameter estimation.

\chapter{Dark Scattering}\label{Chapter4}

\vspace{1cm}

This chapter stems from the first publication of this project \citep{2022MNRAS.512.3691C}. The main difference between that paper and the present chapter lies in the fact that within the paper, we generated the dark matter DM-only pseudo power spectrum using a Python-based code called \texttt{EuclidEmulator2} \citep{2021MNRAS.505.2840E}. This emulator takes as input a scalar amplitude $A_s$, corrected by the growth factor calculated with \texttt{evogrowthpy}, specifically $D^2_{\rm DS}(z)/D^2_{\Lambda\rm CDM}(z)$. Employing such emulator, our results achieve slightly greater accuracy in predicting the pseudo-spectrum on intermediate scales compared to \texttt{HMCODE2020}, the spectrum generator used here. The article has been published in a peer-reviewed international journal. 

\section{Interacting Dark Energy theories}
\label{Sec:IDE}

In the realm of the $\Lambda$CDM model, the dark matter and dark energy always remain uncoupled. Theoretically, however, nothing prevents the gravitationally interaction between these dark components.\footnote{This assumption holds under the condition that the standard model remains decoupled from the dark sector.} This simple appreciation motivates to model dark energy interacting  with dark matter, through models collectively known as Interacting Dark Energy (IDE) \citep{2004ApJ...604....1F,2009JCAP...07..034G,Pourtsidou:2013nha,2016RPPh...79i6901W,2020PDU....3000666D,2020PhRvD.102l3502L,2024PhRvD.110b3529G}. In these scenarios the dark sector exchanges energy and/or momentum depending on the characteristics of the interaction, although there are several choices for the nature of the interaction (e.g. interacting scalars field, interacting fluids, interacting in $f(R)$ gravity and so forth). \\
These models have been the key to explain  some of the irregularities present in the $\Lambda$CDM model, for example, the coincidence problem (see \cite{2016RPPh...79i6901W}), as well as alleviating the previously mentioned data tension, particularly that of $S_8$. This is achieved because the addition of the dark interaction can slow down the growth of dark matter perturbations, thereby reducing their amplitude at late times. 

\noindent The coupling is modelled via a current $Q^\nu$ that represents the energy and momentum exchange between the dark sector. As a result, the energy-momentum tensors of dark matter (labelled with ``${\rm c}$") and dark energy (labelled with ``${\rm DE}$") are no longer separately conserved,
\bea
\nabla_\mu T^{\mu \nu}_{\rm c} = Q^\nu \, , \quad \Longleftrightarrow \quad \nabla_\mu T^{\mu \nu}_{\rm DE} = -Q^\nu \, .
\label{Eq:IDE_coupled}
\eea

\noindent Nonetheless, the total momentum-energy tensor is clearly conserved, $\nabla_\mu T^{\mu \nu}_{\rm c} + \nabla_\mu T^{\mu \nu}_{\rm DE}  = 0$. A more general approach for describing the coupled dark sector is through the Lagrangian,
\bea
\mathcal{L}= - \dfrac{1}{2} g^{\mu \nu} \partial_\mu \phi \partial_\nu \phi - V(\phi) - m\left(\phi\right) \psi \psi^*  + \mathcal{L} \left[ \psi \right] \, , 
\label{Eq:Lagrangian_IDE}    
\eea

\noindent where $\phi$ is an evolving scalar field of the dark energy named \textit{quintessence} and $\psi$ is considered as a scalar field dark matter (SFDM), which has a mass $m(\phi)$.\footnote{If the mass depends only on the SFDM itself, there would be no interaction with the dark energy, i.e. only a self-interaction.} In this representation of the dark sector the coupling current is defined as,
\bea
Q_\nu \equiv \rho_{\rm c} \partial_\nu \phi \dfrac{\partial \ln m\left(\phi\right)}{\partial \phi} \, .
\label{Eq:Coupling_current}    
\eea

\noindent Here $\rho_{\rm c}$ is the dark matter density, and the form of the coupling current is a phenomenological choice. At the background level, the coupling current contribution is: $\bar{Q}^\nu \rightarrow (Q_0(\eta),0,0,0)$, with the form $Q_0$ being fairly diversified (see e.g. \cite{2009PhRvD..79f3518C,2012PhRvD..86j3522B,2013JCAP...02..037L,2013JCAP...12..013T}). From Eq.~\eqref{Eq:IDE_coupled} we obtain the two following equations, the l.h.s. equation we have:
\bea
\dfrac{\partial \bar{\rho}_{\rm{c}}}{\partial \eta} + 3H \bar{\rho}_{\rm c} = Q_0 \, ,
\label{Eq:interaction_cdm}
\eea

\noindent while the on r.h.s. expression for dark energy is,
\bea
\dfrac{\partial \bar{\rho}_\phi}{\partial \eta } + 3H \bar{\rho}_\phi (w_\phi + 1)  = - Q_0 \, .
\label{Eq:interaction_de}
\eea
 
\noindent Now, we embed the IDE theories within the pull-back formalism (see \cite{Pourtsidou:2013nha} for details) in the GR framework coupled to an adiabatic fluid. The advantage of this framework lies in a better perspective at the level of the action, on how such couplings may naturally arise. In addition, we are able to explore whether these models reduce to phenomenological cases under certain limits. However, the most widely studied IDE models are those in which a $\phi$-quintessence is explicitly coupled to pressure-less dark matter, i.e. CDM. The coupling in Eq.~\eqref{Eq:Coupling_current} is often simple, typically involving only one additional parameter. Following this, the total action is given by,
\bea
S =  \int{\rm d}^4 x  \, \sqrt{-g}\, R - \int{\rm d}^4 x \, \sqrt{-g}\, \mathcal{L}(n_{\rm c},\phi,X, Z) \, , 
\label{Eq:Coupling_action}    
\eea

\noindent in which $n_{\rm c}$ is the CDM number density. The terms $X$ and $Z$ are defined as follows:  
\bea
X = \frac{1}{2} g^{\mu \nu} \partial_\mu \phi \partial_\nu \phi \, , 
\label{Eq:X_IDE}
\eea

\noindent representing the kinetic term of the scalar field, and  
\bea
Z = u^\mu \nabla_\mu \phi \, ,
\label{Eq:Z_IDE}
\eea

\noindent which characterizes the coupling between the CDM velocity and the gradient of $\phi$. These Lagrangian, $ \mathcal{L}(n_{\rm c}, \phi, X, Z)$, encodes the dynamics and interactions within the system. So, we can vary the action in Eq.~\eqref{Eq:Coupling_action} by $\delta S = 0$ to obtain
\bea
\delta \mathcal{L} = \dfrac{\partial L}{\partial n_{
\rm c}}\delta n_{
\rm c} + \dfrac{\partial L}{\partial \phi}\delta \phi + \dfrac{\partial L}{\partial X}\delta X + \dfrac{\partial L}{\partial Z}\delta Z \, .
\label{Eq:deltaL}
\eea

\noindent Proceeding this way, we derive several equations through varying the action with respect to its variables, which leads to three sets of equations:

\begin{enumerate}
\item The variation with respect to $\phi$: Requiring $\frac{\delta S}{\delta \phi} = 0$ leads to the Euler-Lagrange equation,
\bea
 \nabla_\mu \left( \dfrac{\partial \mathcal{L}}{\partial X} \nabla_\mu \phi +  \dfrac{\partial \mathcal{L}}{\partial Z} u_\mu \right) = \dfrac{\partial \mathcal{L}}{\partial \phi} \, .
\label{Eq:deltaL_phi}
\eea

\item The variation with respect to $g_{\mu \nu}$: $\frac{\delta S}{\delta g_{\mu \nu}} = 0$, we obtain the Einstein field equations with the total energy-momentum tensor given by,  
\bea
T_{\mu \nu} = \dfrac{\partial \mathcal{L}}{\partial X} \  \partial_\mu \phi \partial_\nu \phi + \left(n_{
\rm c}  \dfrac{\partial \mathcal{L}}{\partial n_{
\rm c}} - Z \dfrac{\partial \mathcal{L}}{\partial Z} \right) u_\mu  u_\nu + \left(n_{
\rm c}  \dfrac{\partial \mathcal{L}}{\partial n_{
\rm c}} - \mathcal{L} \right)g_{\mu \nu} \, . 
\label{Eq:deltaL_tensor_phi}
\eea

\item The variation with respect to $n_{
\rm c}$: For $\frac{\delta S}{\delta n_{
\rm c}} = 0$, we find the conservation law of the fluids,
\bea
\nabla_\mu (n_{
\rm c} u^\mu) = 0 \, .
\label{Eq:deltaL_cons_law}
\eea
\end{enumerate}

\noindent Following \cite{Pourtsidou:2013nha}, with the above equations at hand, the  IDE model families can be classified in three kinds according to their Lagrangian structure $\mathcal{L} = F + f$, where $F$ stands for $\phi$-quintessence and $f$ is for CDM contributions, respectively.

\subsubsection{Type I} 
\label{subsubsec:type1}

The Lagrangian of these type of models does not depend on $Z$ and has the following convenient separated form: 
\bea
\mathcal{L} =  F(X,\phi) + f(n_{
\rm c}, \phi) = X + V(\phi) + f(n_{
\rm c}, \phi) \, .
\label{Eq:type1_L}
\eea    

\noindent In which the CDM density and pressure can be identified by Eq.~\eqref{Eq:perf_fluid} like,
\bea
\rho_{\rm c}= f \, , \qquad P_{\rm c}= n_{
\rm c} \dfrac{\partial f }{\partial n_{
\rm c} } - f = 0 \, .  \label{Eq:type1_rho}
\eea    

\noindent Being set $f(n_{\rm c}, \phi) = n_{\rm c} \exp \left[\alpha (\phi) \right]$, where $\alpha (\phi)$ is a free function. From substituting it into Eq.~\eqref{Eq:deltaL_phi}, then the equation becomes, 
\bea
\nabla_\mu \left( \dfrac{\partial F}{\partial X} \partial_\mu \phi \right) =   \dfrac{\partial \mathcal{L}}{\partial \phi} + \rho \dfrac{\partial \alpha }{\partial \phi} \, . 
\label{Eq:type1_sf_equation}
\eea 

\noindent Subsequently, employing Eq.~\eqref{Eq:deltaL_tensor_phi} to compute the energy-momentum tensor of $\phi$, 
\bea
T^{(\phi)}_{\mu \nu} = \dfrac{\partial F}{\partial X} \partial_\mu \phi \partial_\nu \phi - F g_{\mu \nu} \, . 
\label{Eq:type1_tensor_energy}
\eea 

\noindent Proceeding with this, we now have all the necessary components to find the coupling current for dark energy as defined in Eq.~\eqref{Eq:Coupling_current}. Specifically, we obtain:
\bea
Q^{(\rm I)}_\nu  = - \rho_{\rm c}\dfrac{\partial \alpha }{\partial \phi} \ \nabla_\nu \phi \, , \label{Eq:Coupling_type_1}
\eea
    
\noindent where we have used both, Eq.~\eqref{Eq:type1_sf_equation} and Eq.~\eqref{Eq:type1_tensor_energy}. Note that, this type of coupling describes the transfer of energy and momentum, that impacts the evolution of the Universe and the formation of cosmic structures.

\subsubsection{Type II}
\label{subsubsec:type2}

The Lagrangian is cast as follows,
\bea
\mathcal{L} =  F(X,\phi) + f(n_{
\rm c}, Z)  = X + V(\phi) + f(n_{
\rm c}, Z) \, ,
\label{Eq:type2_L}
\eea 

\noindent where $f(n_{\rm c}, Z) = n_{\rm c} h(Z)$. The coupling of these models introduces a transfer of energy and momentum as well, since the energy-momentum tensors are those of the Type I case. However, unlike before, it is introduced a new coupling function $\beta(Z)$, which is obtained via $h(Z) = \exp \left(\int ds \frac{\beta}{1+\beta s} \right)$.We follow a similar proceed as the previous case. Thus, we find the coupling current to be,
\bea
Q^{(\rm II)}_\nu = \nabla_\mu \left( \rho_{\rm c}\beta u^\mu \right) \nabla_\nu \phi \, .  
\label{Eq:Coupling_type_2}    
\eea

\noindent With density term calculated as $\rho_{\rm c}= f - Z \frac{\partial f }{\partial Z}$.

\subsubsection{Type III}
\label{subsubsec:type3}

Lastly, these coupling is derived through the following Lagrangian: 
\bea
\mathcal{L} =  F(X, Z, \phi) + f(n_{\rm c}) = X + V(\phi) + Z + f(n_{\rm c}) \, .
\label{Eq:type3_L}
\eea

\noindent Likewise, once again we follow the recipe to find the coupling current for this case, resulting in 
\bea
Q^{(\rm III)}_\nu = -\nabla_\mu \left(\dfrac{\partial F }{\partial Z } u^\mu \right) \hat{\phi}_\nu + \dfrac{\partial F }{\partial Z } D_\nu Z + Z \dfrac{\partial F }{\partial Z } u^\mu \nabla_\mu u_\nu \, .
\label{Eq:Coupling_type_3}    
\eea
 
\noindent Here the projector tensor $q_{\mu \nu}$ was used to define $D_\nu = q^\mu_\nu \nabla_\mu$ and $\hat{\phi}_\nu = D_\nu \phi$. Surprisingly, Type III IDE models are special because their coupling involves pure transfer of momentum and thus no coupling $\bar{Q}=0$ at the background level. \\ 
The classification of IDE models presented thus far helps for exploring different interaction mechanisms among dark sector components. Specially, Type III models stand out as a part of our interest for this thesis due to their exclusive momentum transfer properties, particularly found in our model of study. 

\section{Thomson's interaction onto Dark Sector}
\label{sec:DS}

Throughout the present thesis, the model we are interested in studying is \textbf{Dark Scattering} (DS) model (first explored by \cite{2010PhRvD..82h3505S}), consisting of an analogy with Thomson's scattering between electrons and photons by identifying dark sector species like, 
\bea
& \boldsymbol{Q}_{\rm T} =  -(1+w_{\gamma}) \sigma_T  a  \rho_\gamma n_e (\boldsymbol{u}_{e}-\boldsymbol{u}_{\gamma})\, ,  \label{Eq:Thomson_eq} \\
&  \quad  \downarrow  \nonumber \\
& \boldsymbol{Q}_{{\rm DS}}=  -(1+w_{\rm DE}) \sigma_{\rm DS} a \rho_{\rm DE} n_{\rm c}  (\boldsymbol{u}_{{\rm c}}-\boldsymbol{u}_{{\rm DE}}) \, . \label{Eq:DS_eq}
\eea

\noindent In which $w_{\gamma} = 1/3$ for radiation fluid. Here $\boldsymbol{u}_{X}$ denotes the peculiar velocity of species $X \in \{\rm c, \rm DE \}$, where the dark matter and dark energy are assumed to be conformed of fluids, with the latter having density $\rho_{\rm DE}$ and pressure $P_{\rm DE}=w_{\rm DE}\rho_{\rm DE}$, thus defining the equation of state parameter $w_{\rm DE} = w$. The cross-section is effectively between CDM particles and DE, and is given by $\sigma_{\rm DS}$. This interaction is elastic in nature, involving the pure exchange of momentum exclusively between the perturbations of the dark sector. Our analysis assumes the weak field approximation, which assumes non-relativistic velocities and weak gravitational fields. In this regime, the interaction term described by Eq.~\eqref{Eq:DS_eq} does not acquire any nonlinear corrections. Consequently, the linear Euler equations for the interacting dark components are the following:
\bea
\theta_{\rm c}'+\mathcal{H}\theta_{\rm c}+ \nabla^2 \Phi=(1+w) \frac{\rho_{\rm DE}}{\rho_{\rm c}}a n_{\rm c}\sigma_{\rm DS}(\theta_{\rm DE}-\theta_{\rm c}) \, , \\
\theta_{\rm DE}'-2 \mathcal{H}\theta_{\rm DE}- \frac{1}{1+w} 
\nabla^2 \delta_{\rm DE}+ \nabla^2 \Phi = a n_{\rm c}\sigma_{\rm DS}(\theta_{\rm c}-\theta_{\rm DE})\, ,
\eea

\noindent where a prime denotes a derivative with respect to conformal time, $\mathcal{H}=a'/a$ is the conformal Hubble rate, $\theta_X \equiv \nabla \cdot \boldsymbol{u}_X$ are the divergences of the velocities, $\Phi$ is the gravitational potential and $\delta_X = \delta\rho_X/\rho_X$ is the density contrast of species $X$. We assume the sound speed of dark energy fluctuations is $c_s^2=1$. This implies that on sub-horizon scales, the dark energy fluctuations are heavily damped, so that they can be neglected for the evolution of dark matter, resulting in a simplified Euler equation:
\bea
\theta_{\rm c}'+\mathcal{H}(1+\Xi)\theta_{\rm c}+\nabla^2\Phi=0 \, ,
\label{Eq:Euler_simple}
\eea

\noindent where $\Xi$ is a term that encodes the DS interaction term as follows,
\bea
\Xi \equiv   \xi \left( 1+w \right) \dfrac{3\Omega_{\rm DE}}{8\pi G} H \, . 
\label{Eq:Interaction_term}
\eea

\begin{figure}[t]
\centering
{\includegraphics[width=0.75\textwidth]{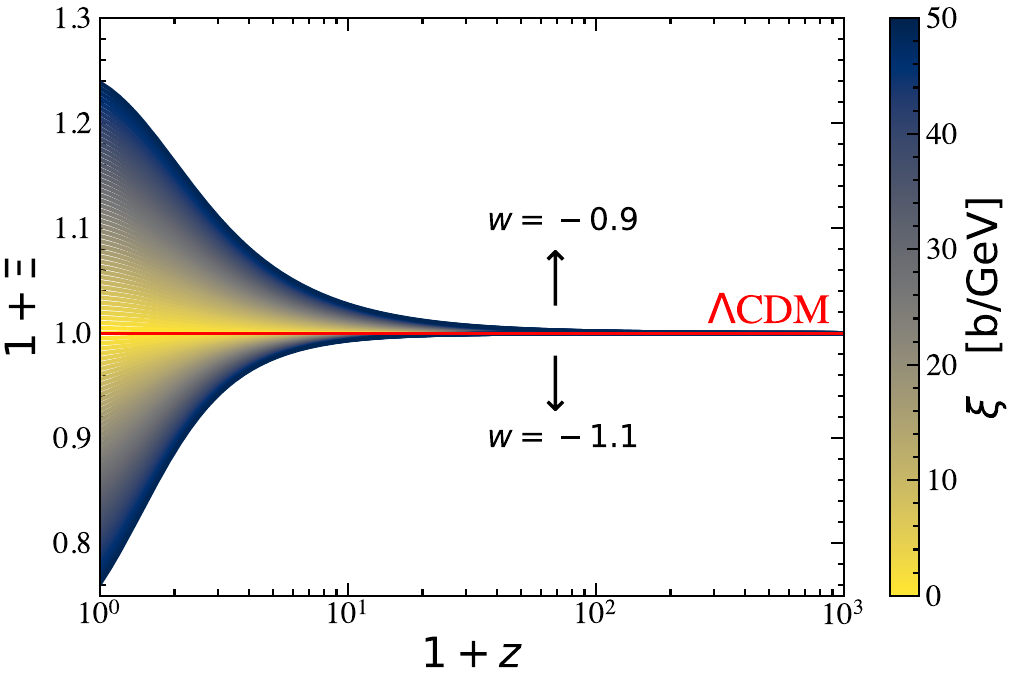}}
\caption[Evolution of the DS interaction term]{Interaction term $\Xi$ as a function of redshift from $0$ to $10^3$, shown for a range of constant equation of state parameter values (indicated by the colorbar). The plot includes two distinct regimes: $w < -1$ and $w > -1$. The horizontal red line represents the $\Lambda$CDM limit.}
\label{Fig:interaction_term}
\end{figure}

\noindent The above term depends only on background quantities and the coupling strength parameter $\xi \equiv \sigma_{\rm DS}/m_{\rm c} \geq 0$ in units of [b/GeV], which encodes information on the CDM particle mass $m_{\rm c}$ and the scattering cross section $\sigma_{\rm DS}$, thereby modulating the strength of the interaction. In addition, the DS model can be considered as an extension of $w$CDM. It has a well-defined $\Lambda$CDM limit when $w \to -1$, which gives $\Xi=0$ (see \autoref{Fig:interaction_term}).\\

\noindent In \cite{2015MNRAS.449.2239B} and \citep{2017MNRAS.465..653B} the DS effects were studied through full $N$-body simulations to evaluate the matter power spectrum for different values of the coupling parameter $\xi$, thereby exploring the nonlinear physics of this cosmology. Under the assumption that the interaction term scales linearly with the dark matter velocity, we can subtract the effect from the collective dynamics and express it for the equation of motion of individual particles as follows,
\bea
\dot{\boldsymbol{u}} = - (1+\Xi) H \boldsymbol{u} - \nabla_{\boldsymbol{r}} \Phi \, .
\label{Eq:interaction_eom}
\eea

\noindent Note that, this Eq.~\eqref{Eq:interaction_eom} reveals that the only effect of the interaction is to introduce additional friction to the evolution of peculiar velocities. There are no other effects relative to a $w$CDM model, as the energy-conservation equation remains unchanged. \\
As mentioned above, this assumption of linearity in the dark matter velocity is also justified under the Newtonian approximation. Such assumption likely extends to many Type III theories resembling DS. Those theories, the nonlinear modelling developed here is expected to apply when expressed in terms of the generic coupling function $\Xi$. While, the DS model may fall under Type III models.

\subsubsection{Parametrised Post-Friedmann}
\label{subsubsec:PPF_IDE}
 
The Parametrised Post-Friedmann (PPF) formalism \citep{2015PhRvD..91h3537S} consists of linearised the perturbations of the coupling current $Q^\nu = \bar{Q}^\nu + \delta Q^\nu$ which are separated in the following variables, 
\bea
q \equiv \delta Q_0 , \quad \text{and} \quad \nabla_i S \equiv \delta Q_i \, .
\label{Eq:interaction_ppf}
\eea

\noindent Turning them now to a linear combination of all other perturbations (fluid and metric variables) with new coefficients $A_n$, $B_n$ and $C_n$. In this context, variables obey gauge transformations and making use of gauge invariants, this can be extended into general description for generalized fields (see \cite{2015PhRvD..91h3537S} for full expressions).
Nonetheless, for our purposes, we restrain ourselves to the common case of the dark energy fluid interacting with CDM, thus the expressions are reduced to,
\bea
q = Q_0 \Psi - 6 A_1 \Phi - 6 A_2 (\dot{\Phi} +\mathcal{H}\Psi ) + A_3 \delta_{\rm DE} + A_4 \delta_{\rm c}+ A_5 \theta_{DE} + A_6 \theta_{\rm c}\, , 
\label{Eq:PPF_q} \\
S = - 6 B_1 \Phi - 6 B_2 (\dot{\Phi} +\mathcal{H}\Psi ) + B_3 \delta_{\rm DE} + B_4 \delta_{\rm c}+ B_5 \theta_{DE} + B_6 \theta_{\rm c}\, .
\label{Eq:PPF_S}
\eea

\noindent For IDE Type III models, by reducing the large parameter space of free functions and coupling types to small set of constants, their coupling current in Eq.~\eqref{Eq:Coupling_type_3} takes the following form, 
\bea
\boldsymbol{Q}^{(\rm III)}= B_3  \nabla \delta_{\rm DE}+B_5 \boldsymbol{u}_{{\rm DE}}+ B_6 \boldsymbol{u}_{{\rm c}}\, .
\label{Eq:PPF_Type3}
\eea

\noindent We impose $A_n = 0$ and $Q^{(\rm III)}_0 = 0$ since the Type III models produce couplings of pure momentum exchange up to linear order in CPT. \\
Regarding the DS coupling current in Eq.~\eqref{Eq:DS_eq}, when we compare it to Eq.~\eqref{Eq:PPF_Type3}, we find that the matching of these two requires $B_3 = 0$. As a consequence DS coupling would not be possible to map exactly to Type III because fixing $B_3 = 0$ translates to set $F$ independent of $Z$ (i.e. $\frac{\partial F }{\partial Z } = 0$, see \cite{2015PhRvD..91h3537S} the role of $B_3$ in Equation 86). In other words from  Eq.~\eqref{Eq:Coupling_type_3} results, it is clearly to see that the model becomes completely uncoupled. \\
Not all is lost, however, as we can still establish an approximate connection between the DS model and Type III models \citep{alessa2025parameterized}. On one hand, \cite{2017MNRAS.465..653B} showed that this can be done approximately, using a Lagrangian of the type $F\propto \exp(-Z)$. On the other hand, if the sound speed $c_s^2=1$, therefore, dark energy perturbations vanished $\delta_{\rm DE} = \boldsymbol{u}_{{\rm DE}} = 0$, so that, making DS model have a similar form to Type III models. Finally, it remains to match the time-dependence of $B_6$ that of the DS interaction as,
\bea
B_6 \to - (1+w_{\rm DE}) \sigma_{\rm DS} a \rho_{\rm DE} n_{\rm c}\, .
\label{Eq:B6}
\eea

\subsubsection{Including non-CDM species}
\label{subsubsec:DS_with_non-CDM-species}

Seeking to add other non-CDM species in our modelling, we split those species into CDM and non-CDM at the linear level, with the latter including baryons and massive neutrinos. We then define the total matter velocity divergence $\theta_{\rm m}\equiv f_{\rm c}\theta_{\rm c}+f_{b\nu}\theta_{b\nu}$ and the velocity difference $\Delta\theta\equiv\theta_{\rm c}-\theta_{b\nu}$. Since $\Delta\theta$ is expected to be small, we can approximate the solution for the $\Delta\theta$ equation by its equilibrium solution (i.e. the solution which gives $\Delta\theta'=0$), that yields into the next equation,
\bea
\theta_{\rm m}'=-\mathcal{H}\left(1+\frac{\Xi f_{\rm{c}}}{1+\Xi f_{b\nu}}\right)\theta_{\rm m}-\nabla^2\Phi \, .
\eea

\noindent This demonstrates that the total matter evolves with an effective coupling function that depends on the relative amount of dark matter in the Universe. While this coupling function has a slightly modified time-dependence relative to the standard coupling function, $\Xi$, it has been verified with a numerical solution from a modified version of \texttt{CLASS} that it is a very good approximation to evaluate the denominator at $z=0$. This introduces an effective coupling constant $\bar\xi$, given by
\bea
\bar\xi=\frac{f_{\rm{c}}}{1+\Xi_0(1-f_{\rm c})} \xi \, ,
\label{Eq:eff_xi}
\eea

\noindent which is modulated by the dark matter fraction, $f_{\rm c}=\rho_{\rm c}/\rho_{\rm m}$ relative to the total matter and $\Xi_0\equiv\Xi(z=0)$. Thus, the interaction term in Eq.~\eqref{Eq:Interaction_term} is reformulated to 
\bea
\Xi(z) \equiv A_{\rm ds} \dfrac{3 \Omega_{\rm DE}}{8 \pi G} H \, ,
\label{eq:Interaction_term}
\eea

\noindent with the effective interaction amplitude, $A_{\rm ds}$, defined as
\bea
A_{\rm ds} \equiv \bar\xi \left(1+w\right)\, . 
\label{eq:Interaction_term_eff}
\eea

\noindent This parameter plays a crucial role in our later Bayesian analysis, as it encodes the strength of the DS interaction and its dependence on the dark energy equation of state. As we commented earlier, 
the DS interaction introduces an additional frictional force that either dragging or pushing on the CDM particles, depending on ${\rm{sign}}(A_{\rm ds})$.
Naturally, this impacts the evolution of matter perturbations, yielding to observable imprints on the matter power spectrum. \\
In the following section, we will extend the halo model reaction framework by incorporating the DS contributions into the code \texttt{ReACT}. Our approximations suggest that we can put together the clustering species when computing the halo model prediction by using an effective coupling. Moreover, we validate our analytical results against simulations products.

\section{Implementation of the Dark Scattering in \texttt{ReACT}}
\label{sec:DS_onto_ReACT}

In the presence of DS interaction, the evolution of the perturbations is modified. Following \cite{2015MNRAS.449.2239B}, the equation of motion in comoving coordinates takes the form
\bea
\dfrac{\partial \boldsymbol{p}}{\partial \eta } = - \Xi \mathcal{H} \boldsymbol{p} - m_{\rm c} a \nabla_{\bm{x}} \Phi \, .
\label{Eq:motion_equation}
\eea

\noindent In which the conformal time is $\eta = \int a \mathrm{d}t$ and $m_{c}$ is the CDM particle mass. Evidently, the DS model assumes there is no background energy exchange, leaving the background evolving as in the uncoupled case. As detailed in \autoref{Appendix_a}, this equation closely seems like Eq.~\eqref{Eq:eom}, with the interaction term acting just as an extra term.
So, we just follow-up the same procedure in order to reacquire the evolution equations for dark sector fluctuations. At the zeroth-order moment, the continuity equation takes the form
\bea
\dfrac{\partial \delta}{\partial \eta} + \nabla_{\boldsymbol{x}} \cdot [(1 + \delta) \boldsymbol{u}] = 0 \, ,
\label{Eq:Continuity_delta_DS}
\eea

\noindent where $\delta = \delta_{\rm m}$ denotes the density contrast of all non-relativistic matter. Surprisingly, this equation is unaltered by the interaction. Followed by the Euler equations,
\bea
\dfrac{\partial \boldsymbol{u}}{\partial \eta} 
+ (\boldsymbol{u} \cdot \nabla_{\bm{x}}) \boldsymbol{u}   
+ \mathcal{H} \boldsymbol{u} + \nabla_{\bm{x}} \Phi + \dfrac{1}{\rho} \nabla_{\bm{x}}(\overleftrightarrow{\sigma}) = - \Xi \mathcal{H} \boldsymbol{u} \, . 
\label{Eq:Euler_DS}
\eea

\noindent The term $\overleftrightarrow{\sigma}$ encodes generalized pressure forces, therefore, in the absence of any pressure perturbation (which is our case for cold pressure-less matter) we obtain $\sigma_{ij} = 0$. To study the nonlinear evolution, we track the evolution of a spherical top-hat overdensity of radius $R$ and mass $M$ in order to approximate the halo formation, 
\bea
M = \dfrac{4}{3}\pi R^3 \bar{\rho} (1+\delta)  = \text{const}\, . 
\label{Eq:mass}
\eea

\noindent We consider the collapse of a top-hat overdensity profile with the following piecewise function, 
\bea
\rho  = \begin{cases} 
\bar{\rho}(1+\delta) & r \leq R \, , \\
\bar{\rho} & R \leq r \, .
\end{cases}
\label{Eq:top_hat_halo_overdensity}
\eea

\noindent Next, we rewrite the continuity and Euler equations in Einstein notation, 
\bea
& \dfrac{\partial \delta}{\partial \eta} + \nabla_{x_i} [(1 + \delta) u_i] = 0 \, , 
\label{Eq:conti_pert} \\
& \dfrac{\partial u_i}{\partial \eta}  + u_j \nabla_{x_j} u_i  + \mathcal{H} u_i + \nabla_{x_i} \Phi = - \Xi \mathcal{H} u_i \, .
\label{Eq:euler_pert}
\eea

\noindent We can now relate the continuity and Euler equations to derive a second-order equation for the evolution of the density. To begin, we take the conformal time derivative of Eq.~\eqref{Eq:conti_pert}, yielding,
\bea
\dfrac{\partial^2 \delta}{\partial \eta^2} + \nabla_{x_i}  \left[ u_i 
\dfrac{\partial \delta}{\partial \eta} + (1 + \delta)
\dfrac{\partial u_i}{\partial \eta}  \right] = 0 \, . 
\label{Eq:conti_pert_prime}
\eea

\noindent Next, we apply the substitutions of Eq.~\eqref{Eq:conti_pert} and Eq.~\eqref{Eq:euler_pert}, to get the following expression:
\bea
\dfrac{\partial^2 \delta}{\partial \eta^2} + \nabla_{x_i} \left[ u_i \left( - \nabla_{x_j} [(1 + \delta) u_j]\right)  + (1 + \delta)
 \left( - u_j \nabla_{x_j} u_i  - \mathcal{H} u_i - \nabla_{x_i} \Phi  - \Xi \mathcal{H} u_i \right) \right] = 0 \, . 
\label{Eq:conti_pert_prime2}
\eea

\noindent After simplifying some terms, we finally obtain an evolution equation for $\delta$,
\bea
\dfrac{\partial^2 \delta}{\partial \eta^2} + \mathcal{H}
\dfrac{\partial \delta}{\partial \eta} -  \dfrac{\partial^2 (1 +\delta) u_i u_j}{ \partial x_i \partial x_j} = \nabla_{x_i}[(1+\delta) \nabla_{x_i} \Phi ]  - \Xi \mathcal{H} \dfrac{\partial \delta}{\partial \eta} \, .
\label{Eq:delta_spherical_prime}
\eea

\noindent Since we adopted the top-hat model in Eq.~\eqref{Eq:top_hat_halo_overdensity}, which its peculiar velocity field in the interior is assumed to be $u_i = h(\eta) r_i = h(\eta) b x_i$. Such that, 
\bea
\dfrac{\partial^2 u_i u_j}{ \partial x_i \partial x_j}  = 12 b^2 h^2 \, .
\label{Eq:amplitude_derivative}
\eea

\noindent Within the overdensity, $b(t)$ represents a different scale factor with respect to the background one. The amplitude can be calculated from Eq.~\eqref{Eq:Continuity_delta_DS} as follows,
\bea
\dfrac{\partial \delta}{\partial \eta} + 3b h(1+\delta) = 0 \, , \quad \Rightarrow \quad h = - \dfrac{1}{3b}\dfrac{\tfrac{\partial \delta}{\partial \eta}}{(1+\delta)} \, . 
\label{Eq:amplitude}
\eea

\noindent Then the Eq.~\eqref{Eq:delta_spherical_prime} can be rewritten as,
\bea
\dfrac{\partial^2 \delta}{\partial \eta^2} + \mathcal{H}
\dfrac{\partial \delta}{\partial \eta} -  \dfrac{4}{3}\dfrac{\left(\tfrac{\partial \delta}{\partial \eta}\right)^2}{(1+\delta)} = (1+\delta) \nabla^2 \Phi - \Xi \mathcal{H} \dfrac{\partial \delta}{\partial \eta} \, . 
\label{Eq:delta_spherical_prime2}
\eea

\noindent Afterwards, we rewrite the equation in terms of cosmic time $t$,
\bea
\mathop{\ddot{\delta}}_{\rm{Acceleration}} + \mathop{(2 H + \Xi) \dot{\delta}}_{\rm{Friction}}  - \dfrac{4}{3}\dfrac{\dot{\delta}^2}{(1+\delta)} = \dfrac{(1+\delta)}{a^2} \mathop{\nabla^2 \Phi}_{\rm{Force}} \, . 
\label{Eq:nonlineal_eq_Xi}    
\eea

\noindent We identify each term representing distinct physical effects in the following bullet-points:

\begin{itemize}

\item \textbf{Acceleration term} $\ddot{\delta}$: This represents the change in the growth rate of the overdensity, analogous to a second derivative describing acceleration in mechanics.  

\item \textbf{Friction term} $(2H + \Xi) \dot{\delta}$: The expansion of the Universe contributes a damping effect on the growth of structures, with $ 2H \dot{\delta}$ being the usual Hubble friction term. The additional contribution from $\Xi \dot{\delta}$ appears by the DS interaction, which either enhances or suppresses structure formation depending on the sign of $\Xi$.  

\item \textbf{Nonlinear term} $-\frac{4}{3} \frac{\dot{\delta}^2}{(1+\delta)}$: This captures deviations from linear growth, specifically the backreaction from the overdensity’s own velocity field, which slows down collapse as densities increase. 

\item \textbf{Force term} $\nabla^2 \Phi$: This represents the gravitational attraction driving the collapse of the overdensity.

\end{itemize}

\noindent Additionally, rather than tracking the evolution of $\delta$, we can describe the evolution of the physical radius $r$ of the overdensity using mass conservation in Eq.~\eqref{Eq:mass}. Taking its first and second time derivative,
\bea
&  \dot{r} = r H - \dfrac{r \dot{\delta}}{3(1+\delta)} \, , \label{Eq:mass_dot} \\
& \ddot{r} = \dot{r} H + r \dot{H} - \dfrac{1}{3}\left[  \dfrac{ (\dot{r} \dot{\delta} + r \ddot{\delta})(1+\delta) - r \dot{\delta}^2}{(1+\delta)^2}\right] \, .
\label{Eq:mass_ddot}
\eea

\noindent Combining both expressions and arranging terms, the following form is found,
\bea
\dfrac{\ddot{r}}{r} =  H^2 +  \dot{H} - \dfrac{1}{3(1+\delta)}\left[  \ddot{\delta} + 2H \dot{\delta} - \dfrac{4}{3}\dfrac{\dot{\delta}^2}{(1+\delta)} \right].
\label{Eq:r_ddot}
\eea

\noindent We substitute the term in the squared-parentheses as Eq.~\eqref{Eq:nonlineal_eq_Xi}, 
\bea
\dfrac{\ddot{r}}{r} =  H^2 +  \dot{H} - \dfrac{1}{3a^2} \nabla^2 \Phi + \dfrac{1}{3}AH\dfrac{\dot{\delta}}{(1+\delta)}.
\label{Eq:r_ddot2}
\eea

\noindent Lastly, incorporating the Poisson equation in comoving coordinates, 
\bea
\nabla^2 \Phi(\boldsymbol{x}, \eta) = \frac{3}{2} \Omega_{\rm m} \mathcal{H}^2 \delta(\boldsymbol{x}, \eta) \, . 
\label{Eq:Poisson_comoving}
\eea

\noindent We arrive at the expression:
\bea
\dfrac{\ddot{r}}{r} + \Xi H \dfrac{\dot{r}}{r} =  - \dfrac{4\pi G}{3} \left[\bar{\rho}_{\rm c} + (1+3 w
)\bar{\rho}_Q ) \right] + \Xi H^2 - \dfrac{4\pi G}{3} \bar{\rho}_{\rm c} \delta \, .
\label{Eq:r_ddot_final}
\eea

\noindent To express the system in a form compatible with the \texttt{ReACT} code, we introduce the normalised comoving radius,
\bea
y \equiv \frac{r}{r_i} - \frac{a}{a_i} \, , 
\label{Eq:React_variable}
\eea

\noindent which measures the deviation of the collapse trajectory from the background expansion. The evolution equation for $y$ then takes the form
\bea
y'' + \frac{H'}{H}y' -  \left(1 + \frac{H'}{H} \right)y + \frac{H_0^2}{H^2} \frac{\Omega_{\rm c, 0}}{2a^3} \, \delta \left( y + \frac{a}{a_i} \right) + \Xi (y'-y) = 0  \, . 
\label{Eq:y_react_evolution}
\eea

\noindent Only for above expression, we apply a slight abuse of notation by using primes to denote derivatives with respect to $\ln a$, i.e. $X' = dX/d \ln a$. \\
The matter density contrast within the collapsing region, $\delta(y,a)$, is given by
\bea
\delta(y,a) = (1+\delta_i) \left(\frac{a_i}{a}y + a \right)^{-3} - 1 \, .
\label{Eq:delta_react}
\eea

\noindent This captures the nonlinear evolution of the overdensity as it deviates from the homogeneous background expansion. As discussed previously, the effect of the dark sector interaction is to generate an additional friction force on dark matter particles. Within the approximations used in our computation, this friction force is clearly not conservative and it cannot be included in the traditional potential term of the virial theorem. Therefore, we must add a non-conservative force $\boldsymbol{F}^{\rm fric} = -m \Xi H \boldsymbol{x}'_i$ to the standard expression
\bea
2\langle T\rangle+\langle W\rangle+\sum^{\text{all particles}}_i\langle \boldsymbol{F}^{\rm fric}_i\cdot \boldsymbol{r}_i\rangle=0 \, , 
\label{Eq:fric_force}
\eea

\noindent where $T$ is the total kinetic energy and $W$ is the potential term of the system. According to \cite{2019MNRAS.488.2121C}, we write the contributions to the Virial theorem in terms of Eq.~\eqref{Eq:React_variable} and in units of $E_0 \equiv \frac{3}{10}M \left(H_0 r_i \right)^2$. The expressions in this case are: 
\bea
 & \dfrac{W_{\rm N}}{E_0} = - \Omega_{\rm m} \left( \dfrac{a^{-1}}{a^2_i}\right) y^2 (1+\delta) \, , \label{Eq:W_Newton} \\
 & \dfrac{W_{\rm DE}}{E_0} = - \dfrac{H^2}{H^2_0}(1+3 w_{\rm DE}) \Omega_{\rm DE}\left(\dfrac{a}{a_i} \right)^2 y^2 \, , \label{Eq:W_DE} \\
 & \dfrac{W_{\rm DS}}{E_0} = - 2 \, \Xi \, \dfrac{H^2}{H^2_0} \left( \dfrac{a}{a_i} \right)^2  y \, \dfrac{d y}{d \ln a }   \,  . \label{Eq:W_DS}
\eea

\noindent For completeness, we present also the total kinetic energy of the top hat, $T$,  
\bea
\dfrac{T}{E_0}=\dfrac{H^2}{H^2_0}\left[\dfrac{a}{a_i} \left(\dfrac{d y}{d \ln a }  + y\right) \right]^2.
\label{Eq:kinetic_T}
\eea

\noindent To summarize, the equation that must be solved to find the virialisation time $a_{\rm vir}$, as well as the corresponding overdensity $\Delta_{\rm vir}$, is 
\bea
2T + W_{\rm N} + W_{\rm DE} + W_{\rm DS} = 0 \, .
\label{y:virial_theorem}
\eea

\begin{figure}[htp!]
\centering
\includegraphics[width=0.75\textwidth]{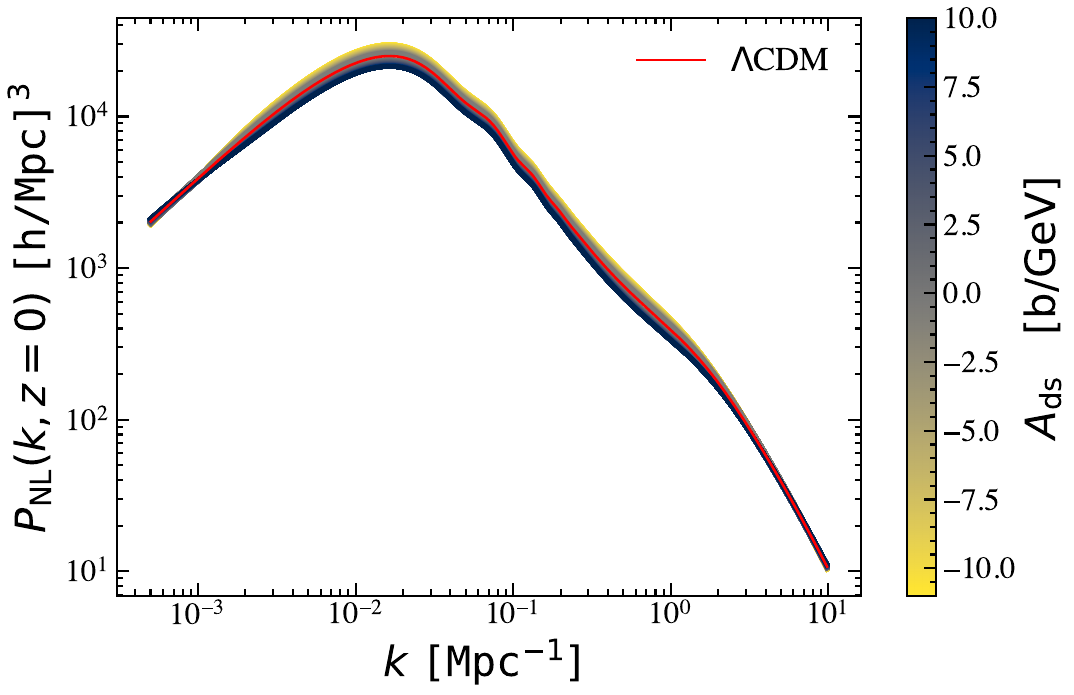}
\caption[Impact of the DS interaction term on matter power spectrum]{Depending on the sign of the interaction parameter $A_{\rm ds}$, we get an enhancement or suppression of the matter power spectrum amplitude.} 
\label{Fig:interaction_term_spectrum}
\end{figure}

\noindent These main modifications from Eq.~\eqref{Eq:motion_equation}, Eq.~\eqref{Eq:y_react_evolution} and Eq.~\eqref{Eq:W_DS} are then incorporated into the calculation of the halo model reaction using \texttt{ReACT}, with the impact of the $A_{\rm ds}$ parameter on the matter power spectrum shown in \autoref{Fig:interaction_term_spectrum}. As a result, we obtain predictions for the DS nonlinear power spectrum, which can be compared to simulations. This validation is carried out in the following section.

\newpage

\section{Modeling Validation}
\label{sec:validation_react}

Theoretical models often face challenges in accurately describing the nonlinear evolution of cosmic structures. As a result, validation through $N$-body cosmological simulations\footnote{Typically, $N$-body simulations, evolve a vast volume (tens of Mpc) with gravitationally interacting $N$-particles over cosmological timescales, with tunable elements like particle count, initial conditions and governing equations.} becomes essential. \\
Motivated by this, one main methodology within this thesis is to test the accuracy of \texttt{ReACT} predictions for the DS matter power spectrum against $N$-body simulations. We use the CPL parametrisation in Eq.~\eqref{Eq:wa} for the evolution of the dark energy equation of state, with $w_0$ and $w_{\rm a}$ being constant parameters. In the case of $w_{\rm a}=0$, we use $w_0 = w$ as a parameter, for simplicity. 

\begin{table}[b]
\centering
\caption[Baseline cosmological parameters of $N$-body simulations]{Baseline cosmological parameters of $N$-body simulations.}
\begin{tabular}{| c | c | } 
\hline  
 Parameter & Value \\ \hline
 $h$ &  $0.678$    \\ 
 $\Omega_{\rm c}$ &   $0.2598$  \\ 
 $\Omega_{\rm b}$ &  $0.0482$   \\ 
 $A_s$ & $2.115 \times 10^{-9}$   \\ 
 $n_s$ & $0.966$ \\ 
\hline
 \end{tabular}
 \label{tab:cosmoparmtab}
\end{table}

\noindent To measure DS power spectra, we have worked with available DS $N$-body simulations products.\footnote{We thank Marco Baldi for providing them, they were used in \cite{2017MNRAS.465..653B}.} They were performed by a modified version of the \texttt{GADGET-2} \citep{2005MNRAS.364.1105S}. 
Moreover, all simulations share the base cosmological parameters given in \autoref{tab:cosmoparmtab} and were set up to $1024^3$ CDM particles evaluated at $z_{\rm ini}=99$ and trace it up to $z=0$ within a box of $1$ Gpc$/h$ per side. Additionally, they all share the same initial seeds so that we can divide-out cosmic variance by taking ratios of power spectra. The resulting CDM particle mass is  $m_{\rm c} = 8\times 10^{10}$ M$_{\odot}/h$ and the spatial resolution is $\epsilon = 24$ kpc$/h$ (equivalent to $k_{\epsilon} = 261~h/\text{Mpc}$). We have power spectrum measurements up to $k=12~h/{\text{Mpc}}$ for $z=0$, $k=9~h/{\text{Mpc}}$ for $z=0.5$ and $k=6~h/{\text{Mpc}}$ for $z=1$. We refer the interested reader to \cite{2015MNRAS.449.2239B,2017MNRAS.465..653B} for a more extended description of the simulations and of the modified $N$-body code.

\noindent In \autoref{Table:modelstab} is summarised a subset of the simulations presented in \cite{2017MNRAS.465..653B}, focusing on three models: $\Lambda$CDM, $w$CDM and CPL with a different $\xi$ interaction strength. We use these simulations for validation and looking for deviations from $\Lambda$CDM model in order to ensure that alternatives are well tested, to avoid false detections. These three different interacting models we consider are only a sample of all the different sets of parameters and time-dependencies for $w$ that are possible in this theory. However, they cover different interaction strengths and values of $w$, two of which ($w$CDM+ and CPL) are relevant in the context of the $\sigma_8$ tension as they reduce its value by approximately $5\%$, matching the discrepancy between CMB and weak lensing data. Additionally, given that the reaction formalism has been shown by \cite{2019MNRAS.488.2121C} to be accurate for many different functions $w(z)$ for the non-interacting case, we expect that the worsening of accuracy will be a function of the interaction strength, effectively given by $(1+w)\xi$. For this reason, analysing different values of that combination is expected to be relevant for general settings. In spite of this, further cases with different time-dependence would be useful to better understand how the accuracy of our predictions varies with parameter choices. \\

\noindent As the $N$-body simulation accounts exclusively for gravitationally interacting particles, the uncorrected coupling, $\xi$, is used in place of the effective coupling $\bar\xi$ defined in Eq.~\eqref{Eq:eff_xi}, effectively assuming $f_{\rm c} = 1$.\footnote{It would be interesting to test the effective coupling approximation in detail, using simulations with more than one type of particle, similar to the ones produced in the work by \cite{2022MNRAS.512.1885F}.} We require to calculate the halo model reaction an accurate pseudo power spectrum -- a fully nonlinear spectrum in a $\Lambda$CDM cosmology, whose corresponding linear spectrum is equal to that of the cosmology with interaction, at the requested redshift.

\begin{table}[htp!]
\centering
\caption[Simulations used for validation]{A summary of the simulations tested for validation}
\begin{tabular}{| c | c | c | c | c | }
\hline  
 Model & $w_0$ &  $w_{a}$  & $\xi$ [b/GeV]  & $\sigma_8(z=0)$  \\ \hline
 $\Lambda$CDM & -1.0 & 0.0 & 0 & 0.8261  \\ 
 $w$CDM$+$  & -0.9 & 0.0 & 10 & 0.7939 \\ 
 $w$CDM$-$ & -1.1 & 0.0 & 10 & 0.8512  \\ 
 CPL  & -1.1 & 0.3 & 50 &  0.7898 \\ 
 \hline
 \end{tabular}
\label{Table:modelstab}
\end{table}

\subsection{\texorpdfstring{$w$}CCDM + DS case} 
\label{subsec:wCDM+DS}

\begin{figure}[t]
\centering
\includegraphics[width=\textwidth]{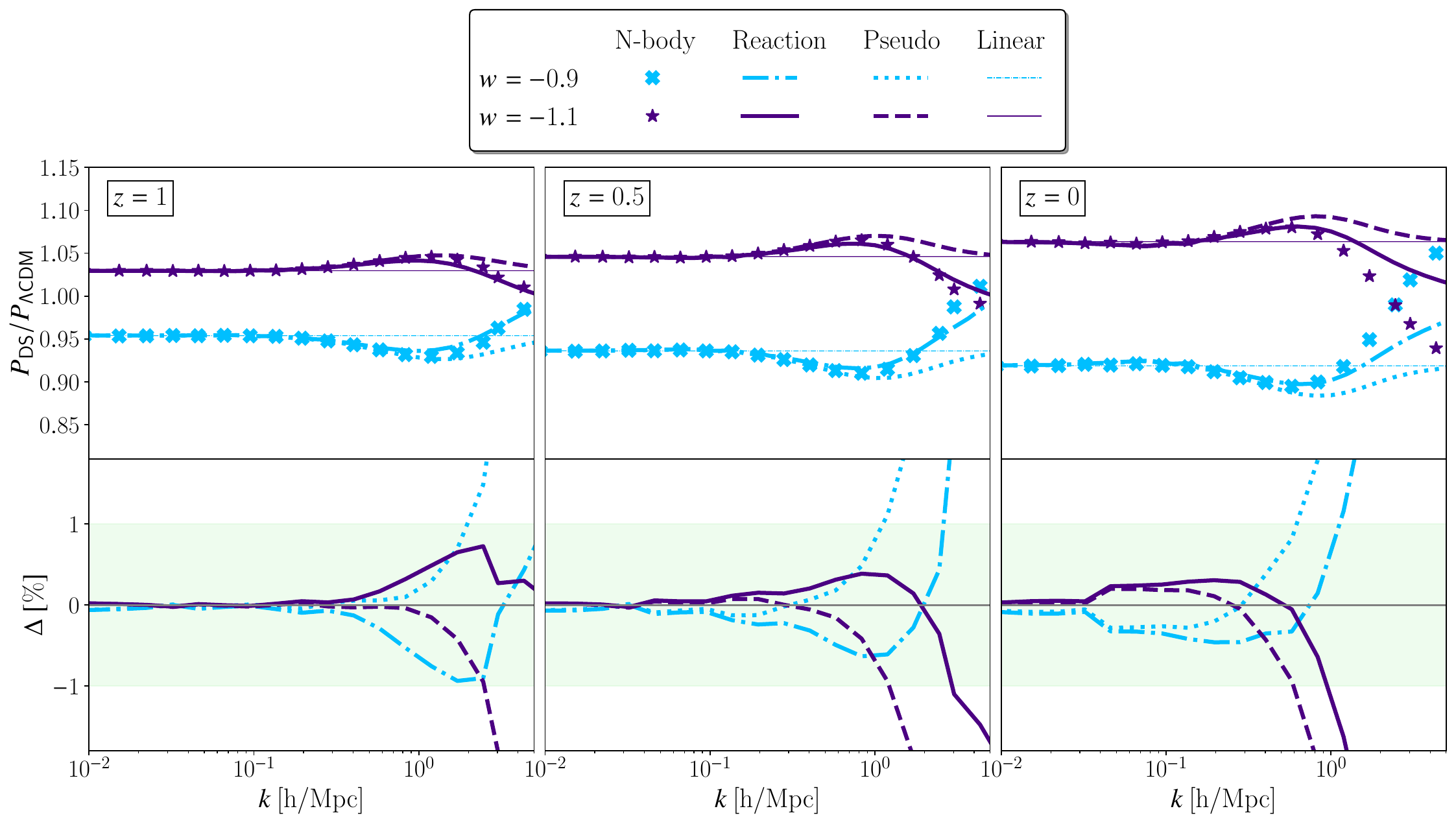}
\caption[$w$CDM+DS in \texttt{ReACT} vs simulations]{{\bf Top:} Ratios of DS spectra to $\Lambda$CDM, for a coupling strength of $\xi = 10$ b/GeV. The blue curves are for $w_0 = -0.9$, where crosses, dash-dotted lines, dotted lines and thin dashed-lines are measurements from simulations, the halo model reaction prediction, the pseudo spectrum prediction and the linear theory prediction, respectively. In purple, we show the results for $w_0 = -1.1$ where the same quantities as before are represented. {\bf Bottom:} The residuals in percentage, $\Delta = 100\,\% \cdot \left( 1 - \hat{P}_{\rm prediction}/\hat{P}_{{\rm N\text{-}body}}\right) $, for the reaction and pseudo spectrum predictions, where $\hat{P}=P_{\rm DS}/P_{\Lambda{\rm CDM}}$ is the ratio shown in the top plot.}
\label{Fig:react_w0}
\end{figure}

We begin by validating our results for the cases of $w$ constant in \autoref{Fig:react_w0}, where we show the ratio between the power spectra for the DS model and the corresponding $\Lambda$CDM model at three redshifts in comparison with simulation measurements. Evidently, we place our reaction predictions within a modified version of \texttt{ReACT} in which the DS model is implemented for cosmologies with $w_0 = -0.9$ and $w_0=-1.1$, the cosmological parameters of \autoref{tab:cosmoparmtab} and a coupling parameter value of $\xi=10$ b/GeV. While the pseudo power spectra were obtained by \texttt{HMCode2020} that takes as input the linear power spectrum provided by \texttt{CAMB}. 

\noindent As it can be inferred from Eq.~\eqref{Eq:Interaction_term} the interaction parameter $\Xi$ depends on whether the value of $w$ lies above or below $-1$. Provoking that these two cases have opposite effects; the case $w_0 = -0.9$ ($\Xi>0$) impacts in a suppression over $\Lambda$CDM spectrum, whereas $w_0 = -1.1$ ($\Xi<0$) results in an enhancement at intermediate scales. Unlike on highly nonlinear scales, in which presents much stronger and opposite effect, as the additional friction for $w>-1$ causes structures to lose energy and collapse to deeper potential wells, thus forming denser structures, with the inverse happening for $w<-1$. \\
In addition as seen in \autoref{Fig:react_w0}, at linear scales the effects of the DS are appreciable, on the other hand the nonlinear are fairly small within the scales that we model accurately. This is particularly true at $z=0$, where the size of the effect is sub-percent at $k<1 ~h/{\text{Mpc}}$. Ignoring the interaction in the calculation of the reaction typically doubles the errors relative to simulations around $k\sim 1~h/{\text{Mpc}}$. Due to the nonlinear effects arise both from the pseudo spectrum and from the reaction. With the large linear effects, the pseudo spectrum shows enhanced nonlinear effects on intermediate scales, which are then compensated by the reaction. Thus, even when the interaction produces nonlinear effects that are small, their modelling is only accurate when the coupling is fully taken into account.

\begin{figure}[t]
\centering
\includegraphics[width=0.8\textwidth]{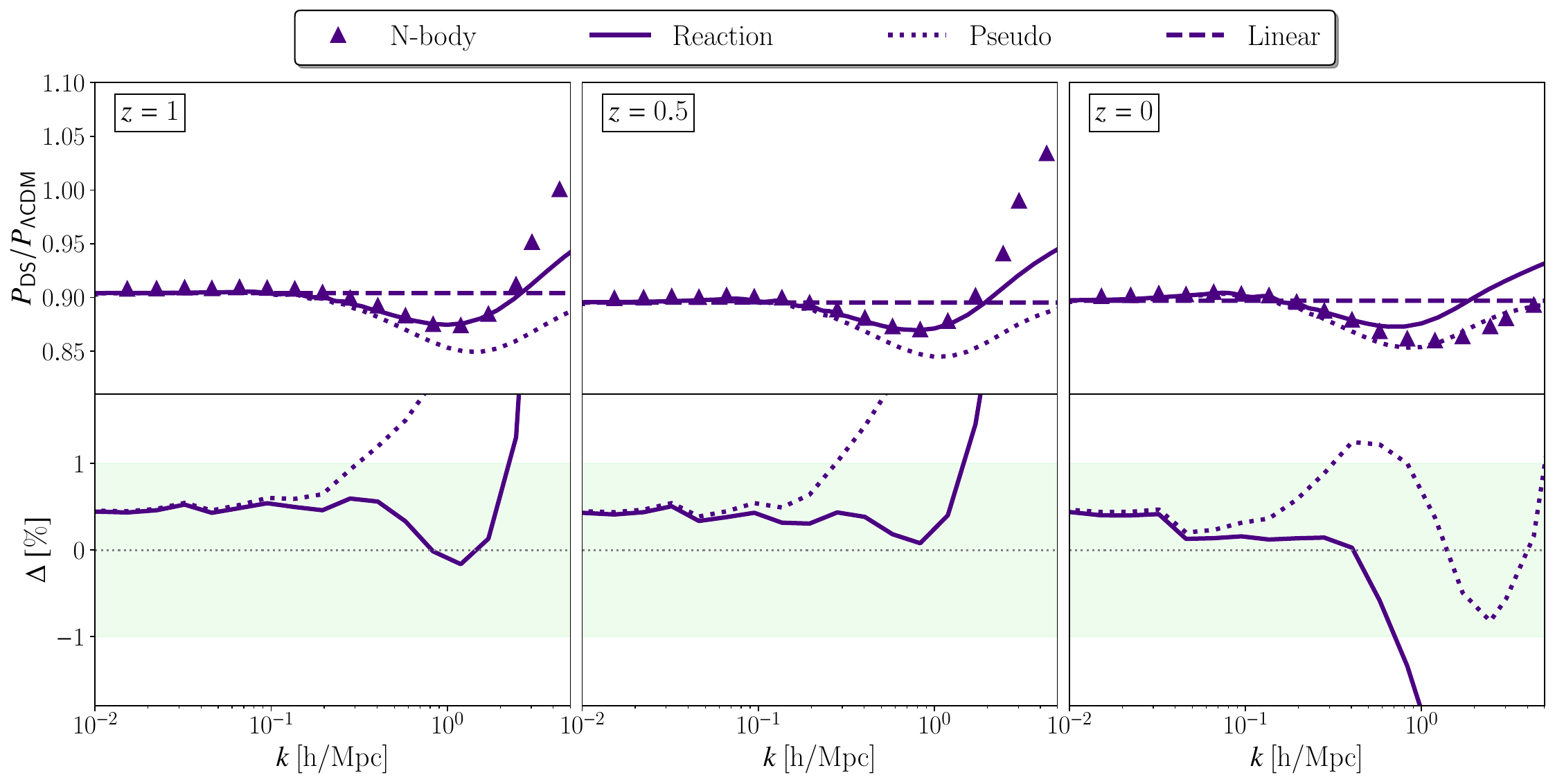}
\caption[CPL+DS in \texttt{ReACT} vs simulations]{{\bf Top:} Ratio of DS spectrum (CPL: $w_0=-1.1, w_{\rm a} = 0.3$) with a value of $\xi = 50$ b/GeV to $\Lambda$CDM. Points are measurements from the simulations whereas solid lines are the halo model reaction prediction. Dotted lines are the pseudo spectrum and dashed lines are the linear theory prediction. {\bf Bottom:} The residuals, as defined in \autoref{Fig:react_w0}, for the reaction and pseudo spectrum predictions. Note that there is a sustained $0.5\%$ discrepancy on large scales. The reason for this is that the $N$-body simulations have been found to capture linear growth with differing accuracy depending on the value of $\xi$, so that the ratio between DS and $\Lambda$CDM is slightly modified with respect to the prediction. This is noticeable here and not in \autoref{Fig:react_w0}, because of the much larger coupling.}
\label{Fig:react_cpl}
\end{figure}

\subsection{CPL + DS case} 
\label{subsec:CPL+DS}

Afterwards, we validate the CPL parametrisation, i.e. with a varying equation of state. We show our results for that case with $w_0 = -1.1,\, w_{\rm a} = 0.3$ and $\xi=50$ b/GeV at $z = 0,\, 0.5,\, 1$ in \autoref{Fig:react_cpl}. This case is interesting because the effective coupling, $A_{\rm ds}$, changes sign at $z=0.5$, first being positive and suppressing linear growth at high redshift, and later enhancing it as redshift goes to zero. The biggest difference comes in the form of a change of shape on the smallest scales, visible in the results at $z=0$, at which the interaction dampened most of the previous nonlinear amplification. All of these nonlinear effects are enhanced here because we study the much larger interaction strength of $\xi=50$ b/GeV. In spite of this, the prediction of the halo model reaction is $1\%$ accurate up to scales of $k \approx 0.8~h/{\text{Mpc}}$ at $z=0$, reaching $k\approx 1.5~h/{\text{Mpc}}$ at higher redshift. In addition, at $z=0$, the errors are never larger than $4\%$ for the entire range of scales available from the simulations. While this is likely due to the compensation between the effects of the interaction from high and low redshift, it shows that our results capture the correct qualitative behaviour in all cases. In addition, contrary to what happened with the cases with $\xi=10$ b/GeV, here the interaction is responsible for larger nonlinear contributions at all redshifts already on intermediate scales, showing that our modelling is robust in this case too.\\

\noindent In summary, the fact that our predictions are accurate for a substantial range of scales and redshifts, particularly for large interaction strengths, demonstrates that the reaction formalism is effective at modelling the nonlinear effects of DS and can be used for the analysis of real data to constrain the interaction strength. For that to be fully realised, however, we must account for all contributions to the power spectrum on small scales, including baryon feedback, massive neutrinos, and their potential degeneracies \cite{2012PhRvD..85d3007C}. At the level of the full spectrum in Eq.~\eqref{Eq:full_spectrum}, now we focus on determining whether a degeneracy exists between DS and baryonic feedback parameters. Likewise, we seek whether the neutrino effects can mimic the interaction contribution. To that end, we attempted to fit a non-interacting model with varying baryonic/neutrinos feedback to an interacting model with fixed baryonic/neutrinos feedback. We explore these effects in the next section.

\section{Interaction degeneracy to baryons and neutrinos}
\label{sec:Degeneracy_baryons_neutrinos}

Throughout this section, we employ the effective coupling approximation of Eq.~\eqref{Eq:eff_xi}, as there would be no realistic scenario for which only CDM is present.

\subsubsection{Neutrinos degeneracy} Massive neutrinos also induce a suppression of matter power spectrum (see \autoref{Fig:neutrinos_impact}), since they do not cluster as efficiently as cold matter on sufficiently small scales. While this effect is less similar to that of the DS interaction, it could also be somewhat degenerate, particularly on intermediate scales. We now aim to investigate the degeneracy between massive neutrinos and the dark sector interaction. The effect of massive neutrinos is fully included in the reaction formalism in Eq.~\eqref{Eq:Reaction_neutrinos}, and available implemented in \texttt{ReACT}. Here, we isolate the scale-dependent nonlinear effects that we are interested in, instead of comparing the full power spectra, we match instead spectra normalized to their large scale value as, 
\bea
Q_{\rm NL}\equiv \frac{P_{\rm NL}}{P_{\rm NL}(k_*)} \, ,
\label{Eq:Q_nl}
\eea

\noindent where $k_* \ll k_{\rm NL}$, so that $P_{\rm NL}(k_*)\approx P_{\rm L}(k_*)$.

\begin{figure}[t!]
\centering
\includegraphics[width=0.5\textwidth]{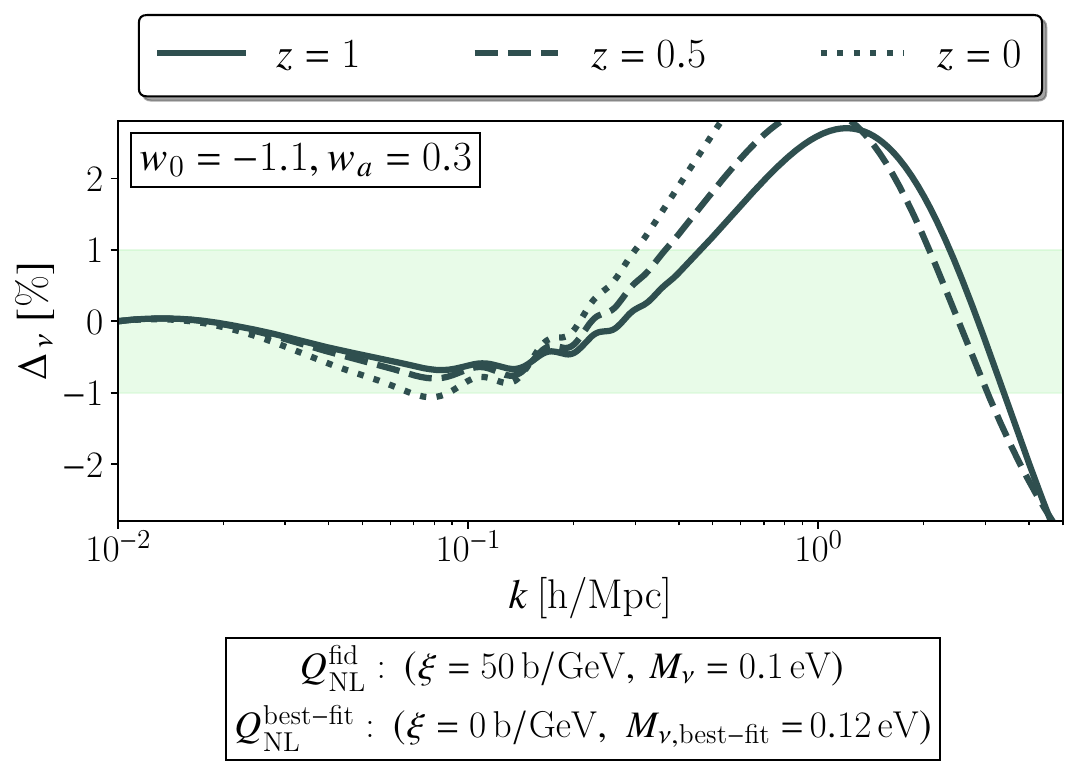}
\caption[Degeneracy between DS and massive neutrinos]{Residuals of the comparison between the DS predictions for $\bar\xi=50$ b/GeV ($\xi=60$ b/GeV), $M_\nu = 0.1$ eV and the non-interacting case with a suitable neutrino mass (best-fit case: $M_{\nu,{\rm best-fit}} = 0.12$ eV), at three different redshifts $z = 1$ (solid), $z = 0.5$ (dashed), and $z = 0$ (dotted). All plots pertain the CPL case: $w_0=-1.1$, $w_{\rm a} = 0.3$. The residual is defined by $\Delta_\nu = 100\,\% \cdot \left( 1 - Q^{\rm fid}_{\rm NL}/Q_{\rm NL}^{\rm best-fit}\right)$.}
\label{Fig:react_neutrinos}
\end{figure}

\noindent We start considering a fiducial spectrum with $\bar\xi = 50$ b/GeV ($\xi=60$ b/GeV), $w_0 = -1.1$, $w_{\rm a}=0.3$, we then generate results with no-interaction ($\xi = 0$), while varying neutrino mass until we are able to find a value $M_{\nu, \rm best-fit}$ which minimises the residuals between this case and the fiducial spectra. We fit all three redshift bins together and we consider only the scales for which our predictions from the halo model reaction have an accuracy of $\Delta \leq 1 \%$ (displayed in \autoref{Fig:react_cpl}), ensuring that the neutrino contribution is not mimicking incorrect effects, which is shown in \autoref{Fig:react_neutrinos}. We found a neutrino mass of $M_\nu = 0.12$ eV with absence of interaction in the dark sector, that best fits the fiducial. We clearly see that a dark sector momentum exchange at the level of $\bar\xi = 50$ b/GeV cannot be mimicked by massive neutrinos alone, even if a single redshift slice is considered in isolation, therefore the case of massive neutrinos, no significant degeneracy is found. However, it should be remarked that we use a single set of dark energy parameters to test both degeneracies, and it is conceivable that a very different dark energy evolution could potentially impact these findings.

\subsubsection{Baryonic feedback degeneracy} 
\begin{figure}[t!]
\centering
\includegraphics[width=0.5\textwidth]{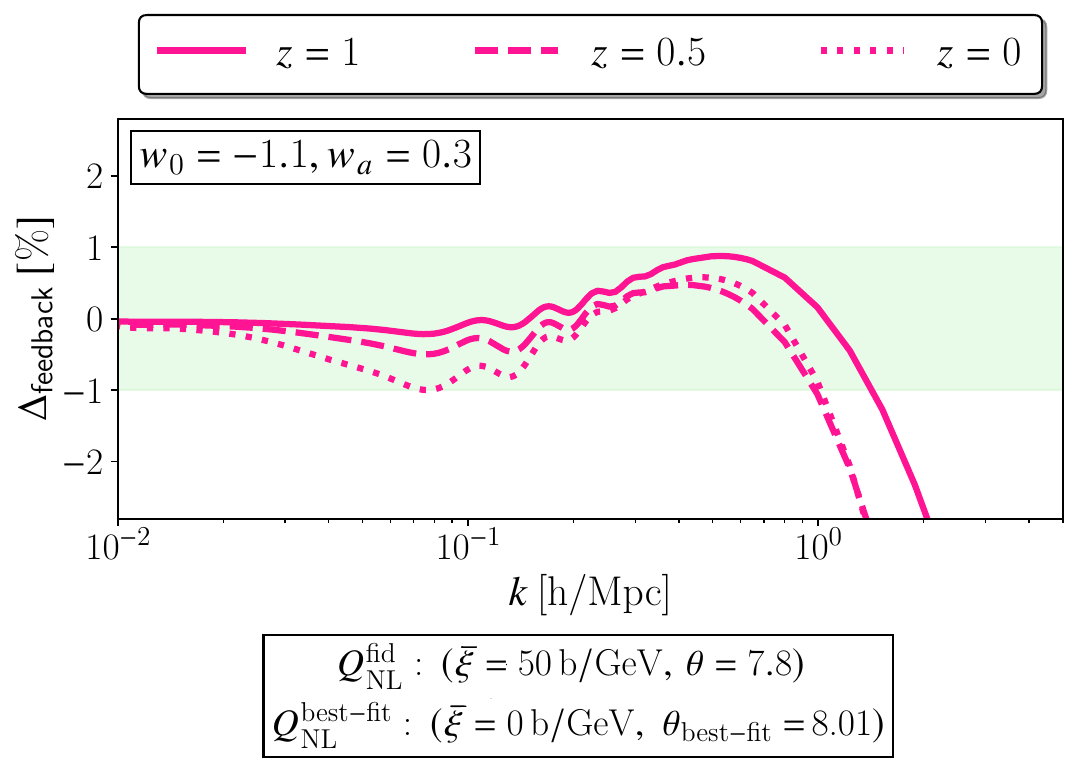}
\caption[Degeneracy between DS and baryonic feedback]{Residuals of the comparison between the Dark Scattering predictions for $\bar\xi=50$ b/GeV ($\xi=60$ b/GeV), $\theta = 7.8$ and the non-interacting case with the best-fit baryonic feedback ($\theta_{\rm best-fit} = 8.01$), at three different redshifts $z = 1$ (solid), $z = 0.5$ (dashed), and $z = 0$ (dotted). In all cases we consider the CPL case: $w_0=-1.1$, $w_{\rm a} = 0.3$. The residual is defined by $\Delta_{\rm feedback} = 100\,\% \cdot \left( 1 - Q^{\rm fid}_{\rm NL}/Q_{\rm NL}^{\rm best-fit}\right)$.}
\label{Fig:react_baryons}
\end{figure}

By using the baryonic feedback from \texttt{HMCODE2020} \citep{2021MNRAS.502.1401M} presents a similar situation, which induces suppression of power on intermediate scales, due to gas expulsion from AGN feedback, and an enhancement on smaller scales due to star formation (see \autoref{Fig:vary_TAGN}). It is therefore likely that there is a degeneracy between the two effects. We proceed in the same way as before, attempting to find a value for the baryonic feedback for which $Q_{\rm NL}$ mimics that of the interacting cosmology. We consider again our fiducial spectrum with $\bar\xi = 50$ b/GeV ($\xi=60$ b/GeV), $w_0 = -1.1$, $w_{\rm a}=0.3$, adding a baryon boost with $\theta = 7.8$ and thus computing $Q_{\rm NL}^{\rm fid}$.  \\
The results from this procedure are illustrated in \autoref{Fig:react_baryons}, where we show the residuals between the fiducial and best fit cases. As seen there the best-fit value is $\theta_{\rm best-fit} = 8.01$, which is substantially different from the fiducial value of $7.8$. It is also clear in the figure that modulating the feedback strength can mimic the effect of the interaction, as it reduced the residuals to below $1\%$ on scales up to $k=1~h/{\text{Mpc}}$, in which we trust our modelling fully. 

\noindent While these simplified tests already reveal some potential degeneracies between DS effects and both, baryons feedback and massive neutrinos, a more thorough MCMC analysis would enable us to exactly pin them down, as well as allowing us to fully validate our nonlinear modelling. The presence of these potentially degeneracy highlights the importance of using a combined analysis with spectroscopic clustering to constrain IDE (as analysed by \cite{2021JCAP...10..004C,2025arXiv250203390T}), 
as their relative independence of baryonic feedback would help in breaking this degeneracy. We aim to fully explore such degeneracy with photometric surveys (e.g. KiDS-1000 dataset) latter in this thesis, where we will also be able to precisely define the range of applicability of our modelling as well as forecast the sensibility of the DS effects shown here.
\chapter{Emulation}\label{Chapter5}

\vspace{1cm}

Before proceeding to the statistical analysis to constrain the DS model, in this chapter we introduce a sophisticated methodology based on Machine Learning (ML) in order to accelerate the inference pipeline. \\
Nowadays, the use of ML\footnote{It is worth mentioning that ML is just a branch of the broader field of artificial intelligence (AI).} has become an essential tool across various scientific disciplines, and its use on astrophysics and cosmology is not lagging behind. Its notable applications are reflected in the rising number of papers utilising ML techniques, as seen in daily submissions on open-access preprint repositories over recent decades.
Initially focused on object classification, ML has since expanded its scope to data science, driven by large-scale experiments and enormous collaborations collecting vast amounts of data, and advancements in computational resources and techniques.\\
For the purposes of this thesis, the focus is on artificial Neural Networks (ANNs), a common method among ML techniques. In light of a substantial validation of the DS halo model predictions against simulations in previous chapter, throughout of this chapter, we proceed forward to create accurate and fast neural emulators of the DS matter power spectra and baryonic feedback. These incorporate newly trained emulators with the aid of \texttt{CosmoPower}, extending the original work of \cite{SpurioMancini:2021ppk} (see also \cite{2021MNRAS.506.4070A,2022MNRAS.512L..44S,2022A&C....3800508M,2022JCAP...11..035G,2023JCAP...05..025N,2023OJAp....6E..20P,2024MNRAS.531.4203B} for similar efforts). From a computational standpoint, emulators have consistently demonstrated promising results to accelerate the forward model, while providing excellent agreement with the more traditional techniques encompassing Boltzmann codes like \texttt{CLASS} or \texttt{CAMB}. This results in essentially the same parameter constraints, but at a fraction of the time. The growing motivation behind this shift is that, in cosmology, accurate parameter estimation has become increasingly expensive. For instance, in a typical MCMC analysis, the power spectra provided by Boltzmann codes are evaluated over roughly $10^5 - 10^6$ likelihood calls, leading to long runtimes (i.e. several days or even months). Consequently, emulators are being used to replace Boltzmann codes, significantly speeding up the inference pipeline. \\ A brief overview of key concepts on Neural Networks and their application to emulation is discussed further in the next section.

\section{Neural networks}
\label{sec:NN}

Neural Networks (NNs) are computational models \citep{bishop2006pattern} designed to identify patterns in data and perform analyses to make predictions, all of that by learning from data. Inspired by the structure and function of biological neurons, they consist of interconnected layers of artificial neurons that process information through weighted connections. Depending on the task\footnote{NNs are excellent in mapping things like complex data, images, audio, and text.} there are different types of NNs like: Artificial Neural Networks (ANNs) for general pattern recognition, Convolutional Neural Networks (CNNs) for image processing, Recurrent Neural Networks (RNNs) for sequential data, and Generative Adversarial Networks (GANs) for data generation, each with architectures tailored to different problems. In this thesis, we will focus on ANNs.  
Mainly, their architecture is composed by multiple neurons to form a layer, then multiple layers stacked yield basically a NN. Those layers consist of an input layer, one or more hidden layers, and an output layer, described as:

\begin{enumerate}
\item  \textbf{Input Layer}: This layer is responsible for receiving the raw input data. Each neuron in this layer represents a specific feature of the input data.

\item \textbf{Hidden Layers}: These layers perform the core computations of the NN. Each neuron applies a transformation to all received inputs and using an activation function $f$, which introduces nonlinearity into the network, subsequently passing the information to the next layer.

\item \textbf{Output Layer:} The final layer generates the desired output by processing transformations from the hidden layers. It is optimised by minimising a target output to identify patterns in the data, producing a classification label, numerical prediction, or other specific outputs.

\end{enumerate}

\noindent Every neuron updates the input signals $\theta$ that are combined through weighted connections. Let $W_{ij}$ represent the weight associated with the connection between the $j$-th input and the $i$-th neuron. The neuron processes these inputs by computing a weighted sum:  
\bea
z_i = \sum_{j} W_{ij} \theta_j + b_i \, ,
\label{Eq:weights_nn}
\eea
\noindent where $b_i$ is a bias term that allows additional flexibility in the transformation. This transformation is then passed through an activation function, which must be a nonlinear function.  \\
NNs learn through an iterative process akin to ``trial and error", where the coefficients in Eq.~\eqref{Eq:weights_nn} are continuously adjusted based on detected patterns of the target output. This cycle repeats until the model reaches an optimal solution or convergence, gradually improving its accuracy with each iteration. In addition, a careful monitoring of the model performance must be taken into account. This requires an internal validation test that results in a learning curve, which evaluates the improvement of the model over training iterations by comparing the training and validation losses. The learning curve is typically defined as:  
\bea
L = \frac{1}{N} \sum_{i=1}^{N} \mathcal{L} (y_i, \hat{y}_i) \, ,
\label{Eq:learning_curve}
\eea

\noindent where $L$ represents the loss, $N$ is the number of data points, $y_i$ are the true values, $\hat{y}_i$ are the predicted values, and $\mathcal{L}$ is a chosen loss function (e.g., mean squared error for regression or cross-entropy for classification). Monitoring this curve helps to detect underfitting, overfitting, or stagnation, so that the network achieves satisfactory performance while maintaining computational efficiency.

\noindent One notable tool in cosmology is a Python-based code called \texttt{CosmoPower}. This code uses NNs to create fast and accurate neural emulators for power spectra, for both the CMB (including temperature, polarisation, and lensing spectra) and matter power spectra or angular galaxy spectra. Since it has proven to greatly improve the efficiency of Bayesian inference, it has been adopted in various analyses where the data is prominent from international collaborations, such as the KiDS, DES, Euclid and Planck. For instance, aiding in accelerating the statistical analysis of the CMB \citep{2023PhRvD.108b3510B,2024EPJWC.29300008B, 2024A&A...686A..10B} or weak lensing analyses \citep{2022MNRAS.512L..44S, 2023OJAp....6E..40S}. 

\section{\tt{CosmoPower}}
\label{sec:cosmopower}

\begin{wrapfigure}{l}{0.25\textwidth}
    \centering
    \includegraphics[width=0.25\textwidth]{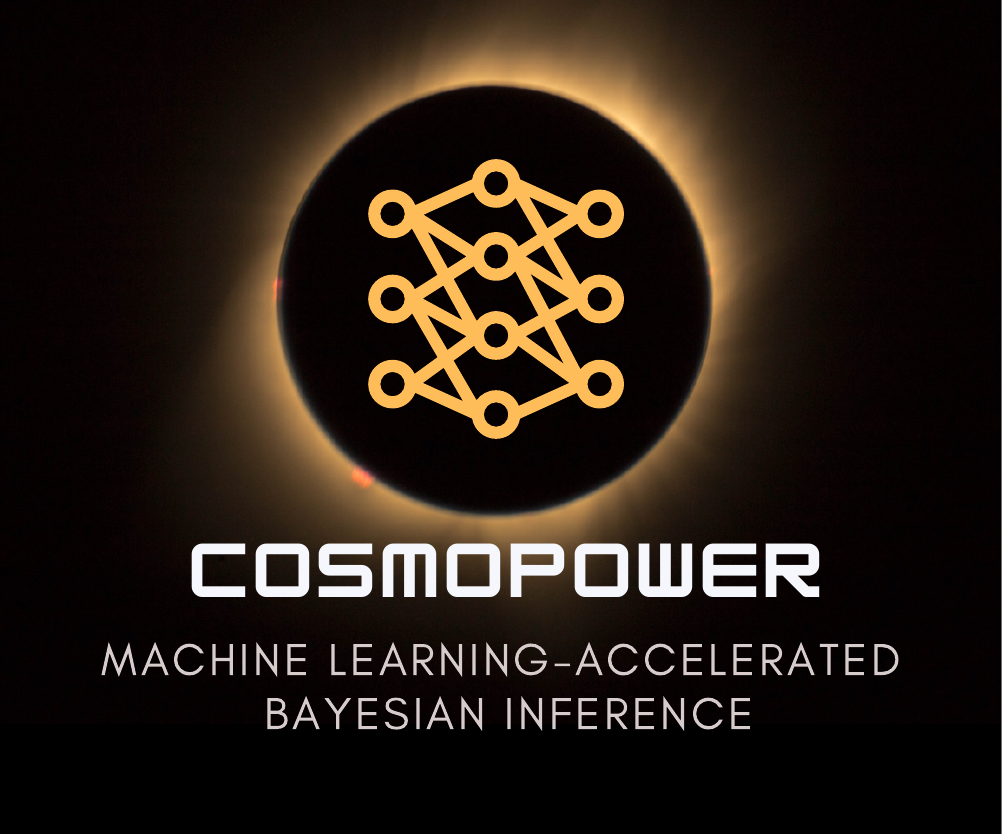}
\end{wrapfigure}

 \texttt{CosmoPower} represents a key code in this thesis. This code produces emulations of the power spectra through a trainings NN on a very high-dimensional parameter spaces, powered by \texttt{TensorFlow} \citep{2016arXiv160304467A}. 

\begin{figure}[t!]
\centering
\includegraphics[width=0.75\textwidth]{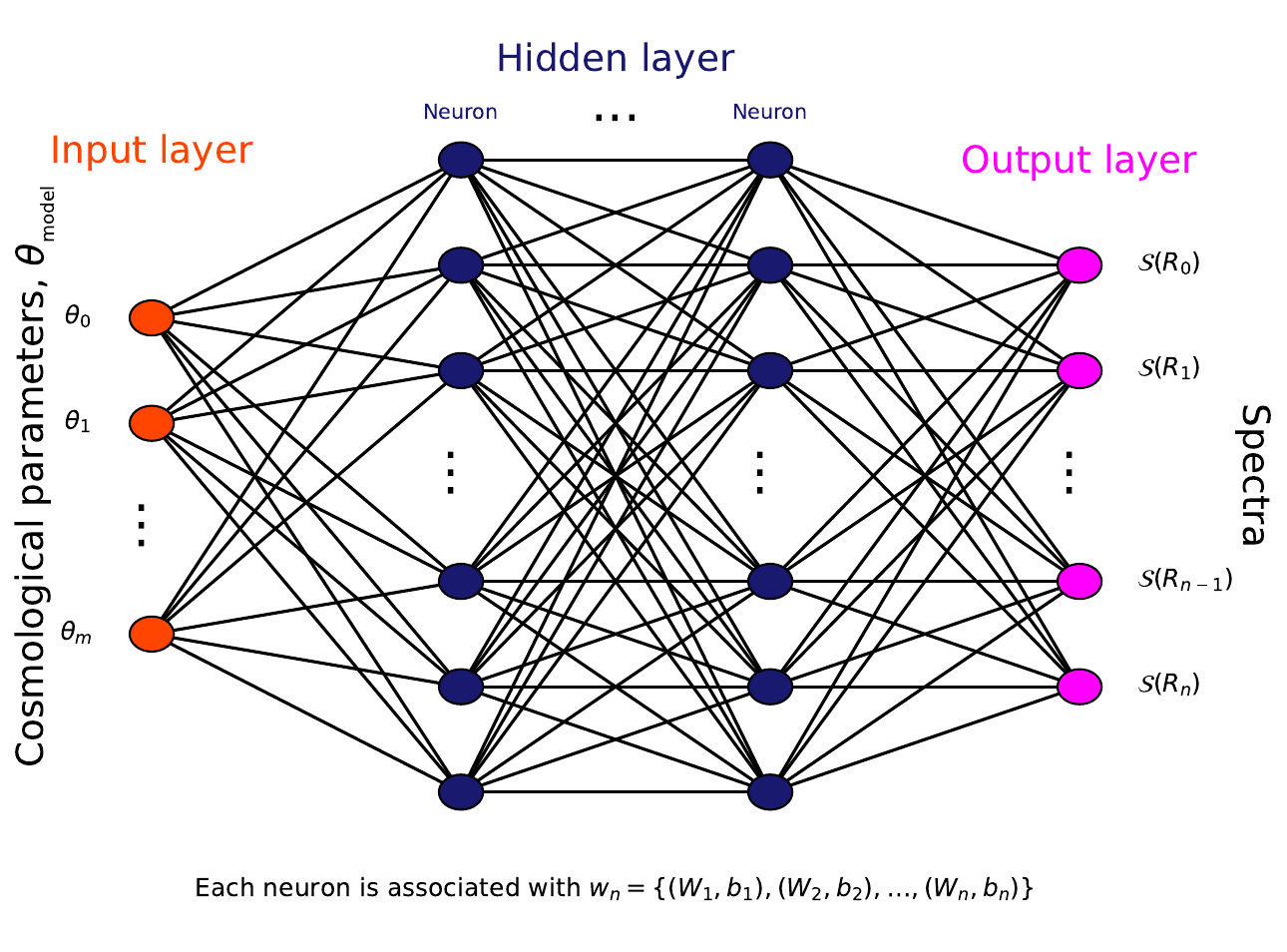}
\caption[Neuronal Network architecture in \texttt{CosmoPower}]{An scheme of the routine of \texttt{cosmopower$\_$NN} module. The NN consists of an input layer (orange circles) filled with cosmological parameters, followed by a set of stacked  hidden layers (blue circles) that process the information through weighted connections. The output layer (magenta circles) corresponds to the predicted power spectra $S(R)$, where $R \in \{k, \ell\}$ represents the modes, and $S \in \{P(k), C_\ell\}$ denotes the spectra. The NN is trained to learn the mapping between cosmological parameters and the power spectra. \\
\textbf{Image Credits:} Scheme adapted from \texttt{CosmoPower}.}
\label{Fig:NN}
\end{figure}

\noindent In essence, constructing emulators requires selecting the target spectrum to be analysed, such as matter or CMB spectra, along with key cosmological parameters. At present, \texttt{CosmoPower} includes only two methods for training. Firstly \texttt{cosmopower$\_$NN} (see \autoref{Fig:NN} for schematically details), in which the input consists of $\boldsymbol{\theta}$ cosmological parameters, each fed into a corresponding $m$-neuron, then pass them through a sequential hidden layers to learn how to map between cosmological parameters and power spectra (similar to classify images). As mentioned above, each neuron is associated with a weight $W_n$ and a bias $b_n$, forming a linear combination of the input that is subsequently passed through a nonlinear activation function. Inspired by the approach in \cite{2020ApJS..249....5A}, \texttt{CosmoPower} implementation adopts the following activation function for all hidden layers of the NNs:
\bea
f(\boldsymbol{x}) = \left ( \boldsymbol{\gamma} + \left ( 1 + e^{-\boldsymbol{\beta}\odot\boldsymbol{x}} \right )^{-1} \odot (1 - \boldsymbol{\gamma}) \right ) \odot \boldsymbol{x} \, ,
\label{eq:act_func}
\eea

\noindent  where $\boldsymbol{\beta}$ and $\boldsymbol{\gamma}$ are learnable parameters optimised jointly with the network weights and biases, and $\odot$ denotes element-wise multiplication. The activation function in Eq.~\eqref{eq:act_func} can be interpreted as a set of element-wise scalar functions applied to each neuron, with independent parameters $\beta_j$ and $\gamma_j$ for each node $j$. \\
This process involves a direct mapping between the cosmological parameters and power spectra for training the NN. A second training method is  \texttt{cosmopower$\_$PCAplusNN}, which incorporates the Principal Component Analysis (PCA) method to match coefficients of the power spectra. As discussed in the \texttt{CosmoPower} paper, both emulation methods were tested on the relevant cosmological power spectra, with the first generally performing better. However, the second method is more effective for the CMB cross temperature-polarisation and lensing potential spectra. For our purposes, we adopt only the first method.

\noindent In a nutshell, \texttt{CosmoPower} basically requires a training set of the cosmological parameters $\boldsymbol{\theta}$ and the associated training and testing set of their log-power spectra. Once trained, the outcome trained emulator can be validated against the testing set, with its accuracy depending on the number of training samples. \\
In this thesis, we develop three emulators: 

\begin{itemize}
    \item[$\star$] \textbf{DS linear matter power spectrum emulator.} 
    \item[$\star$] \textbf{DS nonlinear matter spectrum emulator based on the halo model reaction.}
    \item[$\star$] \textbf{Baryonic feedback emulator.}
\end{itemize}

\noindent Due to the \texttt{ReACT} validation, the $k$-modes in the emulators are restricted to the range $[10^{-3}, 10]$ in units of [$1/$Mpc], while the redshift $z$ is traced up from $0$ to $5$ and treated as an additional input parameter for all emulators. A step-by-step guideline for emulating the linear power spectrum is shown in the next subsection. All emulators are publicly available at \texttt{DS-emulators} \github{https://github.com/karimpsi22/DS-emulators}. 

\subsection{Emulating the linear matter power spectrum}
\label{subsec:accuracy_ds_linear}

At early step it is essential to define the emulator validity range in alignment with the original code validity and the dataset being analysed. This step is crucial to avoid drawbacks on inputs from affecting the training process, reducing the risk of errors and the need for retraining.

\subsubsection{Step I: Training and testing data} In order to generate the training set of nonlinear and linear spectra, we produce a set of spectra for a range of values of 8 cosmological parameters plus a given redshift $z$, as follows: 
\bea 
\boldsymbol{\theta}_{\rm DS} = \{\omega_{\rm b}, \omega_{\rm cdm}, h, n_s, S_8, m_\nu, w, A_{\rm ds}, z \},
\label{Eq:DS_parameters_training}
\eea

\noindent For our statistical analysis, it is an advantage to employ $S_8$ as an input parameter of the emulators instead of $\ln(10^{10} A_s)$. This is motivated by the fact that weak lensing measurements are more sensitive to $S_8$, and also because our reaction-based modelling is directly dependent on the late-time amplitude. So, we linearly generate $N$ samples for each cosmological parameters, like:
\begin{lstlisting}[language=Python]
import numpy as np
# Number of parameters and samples for each.
n_params = 9
n_samples = 200000

# Parameter ranges
obh2 = np.linspace(0.01865, 0.02625, n_samples) # \omega_b
omch2 = np.linspace(0.1, 0.255, n_samples) # \omega_cdm
h = np.linspace(0.64, 0.82, n_samples) # h
ns = np.linspace(0.84, 1.1, n_samples) # n_s
S8 = np.linspace(0.6, 0.9, n_samples) # S_8
mnu = np.linspace(0.0, 0.2, n_samples) # m_\nu
w = np.linspace(-0.7, -1.3, n_samples) # w_0
A_abs = np.linspace(0, 30, n_samples) # |A_ds|
z = np.linspace(0, 5, n_samples) # z
\end{lstlisting}

\noindent Instead of sampling linearly the training  parameters, we employ Latin Hypercube Sampling (LHS) to ensure a fully exploration of the parameter space.
\begin{lstlisting}[language=Python]
import pyDOE as pyDOE
# LHS grid
AllParams = np.vstack([obh2, omch2, h, ns, S8, mnu, w, A_abs, z])
lhd = pyDOE.lhs(n_params, samples=n_samples, criterion=None)
idx = (lhd * n_samples).astype(int)
AllCombinations = np.zeros((n_samples, n_params))
for i in range(n_params):
    AllCombinations[:,i] = AllParams[i][idx[:,i]]

# Saving
params = {'omega_b': AllCombinations[:,0],
          'omega_cdm': AllCombinations[:,1],
          'h': AllCombinations[:,2],
          'n_s': AllCombinations[:,3],
          'S_8': AllCombinations[:,4],
          'm_nu': AllCombinations[:,5],
          'w': AllCombinations[:,6],
          'A': (AllCombinations[:,7])*np.sign(1.0+AllCombinations[:,6]),
          'z': AllCombinations[:,8]}
          
np.savez('your_params_lhs_training.npz', **params)
\end{lstlisting}

\noindent Following the established pipeline, we rescale $A_s$ given the inputted $S_8$ value through the following computation:
\begin{lstlisting}[language=Python]
import classy
cosmo = classy.Class()

def rescaling_As(i):
#   Interaction
    if w_arr[i] == -1.0:
        xi_value = 0.0
    else:
        xi_value = A_arr[i]/(1.0+w_arr[i])
    
    cosmo_As = classy.Class()
    cosmo_As.set({'output':'mPk',
                  'P_k_max_1/Mpc':50.,
                  'z_max_pk':5.0, 'h':h_arr[i],
                  'N_ur':2.0308,
                  'N_ncdm':1,
                  'm_ncdm':mnu_arr[i],
                  'cs2_fld':1.0,
                  'w0_fld':w_arr[i],
                  'wa_fld':0.0,
                  'xi_ds':xi_value,    
                  'Omega_Lambda':0.,
                  'gauge':'Newtonian', 'use_ppf':'yes',
                  'dark_scattering':'yes',
                  'omega_b':omega_b_arr[i],
                  'omega_cdm':omega_cdm_arr[i],
                  'A_s':2.1e-9,
                  'n_s':ns_arr[i]})
    
    cosmo_As.compute()
    target_S8 = S8_arr[i]
    target_sigma8 = target_S8/np.sqrt(cosmo_As.Omega0_m()/0.3)
    new_As = cosmo_As.pars['A_s']*(target_sigma8/cosmo_As.sigma8())**2
    return new_As
\end{lstlisting}

\noindent Subsequently, using the rescaled value of $A_s$ and the corresponding parameter set from Eq.~\eqref{Eq:DS_parameters_training}, we compute the linear power spectrum at the specific $z_i$ with the modified version of \texttt{CLASS} to include the DS model described in \cite{2022MNRAS.512.3691C}, as follows:
\begin{lstlisting}[language=Python]
def linear_generation(i, As_new):
    #Define your cosmology 
    #(what is not specified will be set to CLASS default parameters)
    cosmo.set({'output':'mPk',
               'P_k_max_1/Mpc':50.,
               'z_max_pk':5.0,
               'h':h_arr[i],
               'N_ur':2.0308,
               'N_ncdm':1,
               'm_ncdm':mnu_arr[i],
               'cs2_fld':1.0,
               'w0_fld':w_arr[i],
               'wa_fld':0.0,
               'xi_ds':xi_value,
               'Omega_Lambda':0.,
               'gauge':'Newtonian',
               'use_ppf':'yes',
               'dark_scattering':'yes',
               'omega_b':omega_b_arr[i],
               'omega_cdm':omega_cdm_arr[i],
               'A_s':As_new,
               'n_s':ns_arr[i]})

    try:
        cosmo.compute()
        z = np.linspace(0.0, 5.0, 50)
        
        if z_arr[i] not in z:
            index_remove = np.where(z_arr[i] < z)[0][0] 
            #The nearest redshift to the one will be inserted
            z = np.delete(z, index_remove)
            z = np.sort(np.insert(z, 1, z_arr[i]))

        z_pos = np.where(z == z_arr[i])[0][0] #Index of z_i in z array.

        #Linear DS
        P_lin = np.array([[cosmo.pk(ki, zi)*h_arr[i]**3.0 
                         for ki in k*h_arr[i]] for zi in z])
        sigma_8 = cosmo.sigma8()
        return P_lin, z_pos
\end{lstlisting}

\noindent This process iterates until all $N$ samples\footnote{To enhance efficiency, parallelisation methods should be employed at this stage to accelerate data generation. More extensive training data leads to improve the accuracy of the emulator.} are completed. The resulting data is stored\footnote{It is highly recommended to store spectrum data in logarithmic form, as it accelerates processing during the training stage.} and compressed for training the emulator. Afterwards, we repeat the procedure to generate $20\%$ of the samples for testing. This testing set is kept aside to later validate the accuracy of the emulator.

\subsubsection{Step II: Training} Once the training and testing datasets are prepared, we use \texttt{CosmoPower} to create the DS linear matter power spectrum emulator. So, we load our training set and a list of parameters names,
\begin{lstlisting}[language=Python]
import numpy as np
# Training parameters
training_parameters = np.load('.../your_params_training.npz')

# Training features (= log-spectra, in this case)
training_features = np.load('.../your_logpower_training.npz')
training_log_spectra = training_features['features']

# List of parameter names, in arbitrary order
model_parameters = ['h',
                    'm_nu',
                    'omega_b',
                    'A',
                    'omega_cdm',
                    'n_s',
                    'S_8',
                    'z',
                    'w']
\end{lstlisting}

\noindent Next, \texttt{CosmoPower} with the \texttt{cosmopower$\_$NN} module is initialized, with the desired number of hidden layers and neurons in each layer.
\begin{lstlisting}[language=Python]
import tensorflow as tf
from cosmopower import cosmopower_NN
# Checking that we are using a GPU
device = 'gpu:0' if tf.test.is_gpu_available() else 'cpu'
print('using', device, 'device \n')

# Instantiate NN class
cp_nn = cosmopower_NN(parameters=model_parameters,
                      modes=ell_range,
                      n_hidden = [512, 512, 512, 512], 
                      #4 hidden layers, each with 512 nodes
                      verbose=True)
\end{lstlisting}

\noindent After defining the architecture, the training set is fed into the model, a name is assigned to the emulator, and the hyperparameters are configured accordingly. This training step for the linear matter power spectrum takes around $5-6$ hours, depending on the size of features and the number of training set. This stage is greatly accelerated by using a graphics processing unit (GPU).
\begin{lstlisting}[language=Python]
with tf.device(device):
    # Training stage
    cp_nn.train(training_parameters=training_parameters,
                training_features=training_log_spectra,
                filename_saved_model='Pk_lin_cp_NN_emulator',
                # Cooling schedule
                validation_split=0.1,
                learning_rates=[1e-2, 1e-3, 1e-4, 1e-5, 1e-6],
                batch_sizes=[1024, 1024, 1024, 1024, 1024],
                gradient_accumulation_steps = [1, 1, 1, 1, 1],
                # Early stopping set up
                patience_values = [100,100,100,100,100],
                max_epochs = [1000,1000,1000,1000,1000])
\end{lstlisting}

\noindent The NN architecture and values of hyperparameters are preserved from the original paper. Below are the explanations for each parameter:
\begin{itemize}

    \item[(i)] \textbf{Validation split:} Specifies the portion of spectra set aside for validation.
    
    \item[(ii)] \textbf{Learning rate:} Controls the size of steps taken at each learning epoch.

    \item[(iii)] \textbf{Batch size:} Determines the size of the batch over which the learning step is averaged.

    \item[(iv)] \textbf{Gradient accumulation steps:} Refers to the number of times the model will accumulate gradients before performing an update to the weights.

    \item[(v)] \textbf{Patience value:} Sets how many iterations the network will tolerate being ``stuck" before moving on to the next learning iteration.

    \item[(vi)] \textbf{Maximum number of epochs:} Defines the maximum number of training iterations (epochs). Training stops earlier if the model converges or the early stopping criterion is met.
\end{itemize}

\subsubsection*{Step III: Usage}  The emulator is saved in a file named as \texttt{Pk\_lin\_cp\_NN\_emulator.pkl}, and using it is straightforward. We make a dictionary with the parameter names and their respective values, while we ensure that they fall within the emulator validity range. Then, we can now pass the orderly-arranged dictionary as input to the emulator. For example:
\begin{lstlisting}[language=Python]
import cosmopower as cp
# Load pre-trained NN model
cp_nn = cp.cosmopower_NN(restore=True,
                         restore_filename='.../Pk_lin_cp_NN_emulator')
                         
# Create a dict of cosmological parameters
params = {'omega_b': [0.02242],
          'omega_cdm': [0.11933],
          'h': [0.67],
          'n_s': [0.9665],
          'S_8': [0.8102],
          'm_nu': [0.0],
          'w': [-1.1],
          'A': [-2.0],
          'z': [0]}

# Predictions (= forward pass through the network) -> 10^predictions
spectra = cp_nn.ten_to_predictions_np(params) 
\end{lstlisting}

\subsubsection*{Step IV: Validation} 
\begin{figure}[t!]
\centering
\includegraphics[width=0.75\linewidth]{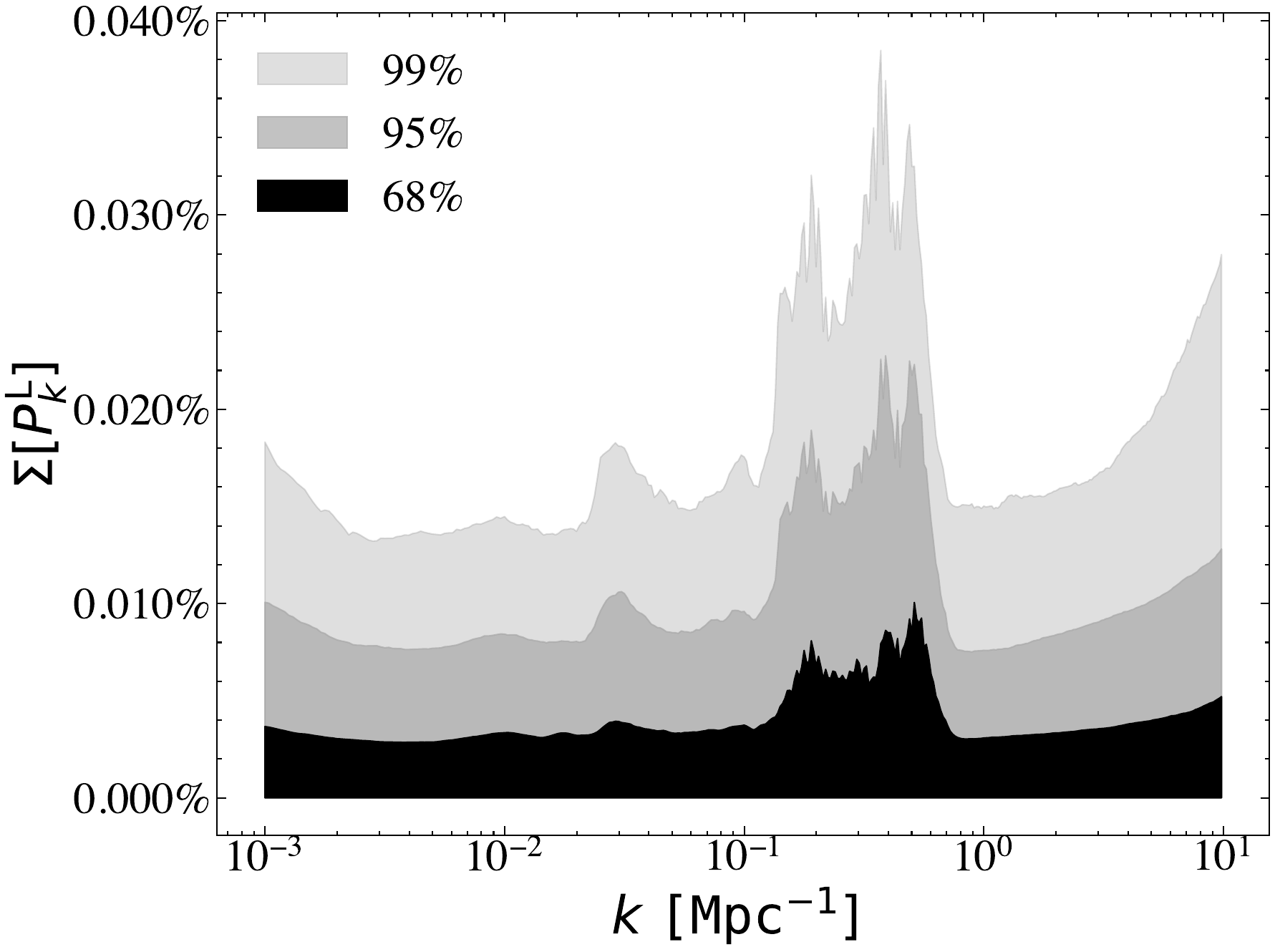}
\caption[Emulator accuracy of the DS linear matter power spectrum]{This plot shows the accuracy of the DS linear power spectrum emulator with $99\%$ of the testing set producing differences smaller than $0.05\%$ to the real value across the entire $k$-range considered, with a slight decrease in accuracy in the region corresponding to the BAO wiggles, which is more difficult to compute for the Boltzmann code.}
\label{fig:acc_linear_ds}
\end{figure}

At this point, the emulator accuracy is evaluated by comparing its predictions to the true values using the stored testing dataset, as shown in \autoref{fig:acc_linear_ds}. All subsequent comparison plots of our emulators are made against $\sim 10^4$ spectra from the testing set, dubbed ``real''. The percentage absolute emulator error is calculated as:
\bea
\Sigma[F_k] = 100\,\% \cdot \left| \frac{ F_{k, \rm{emulated}} - F_{k, \rm{real}}} {F_{k, \rm{real}}}\right| \, ,
\label{Eq:acc_formula}
\eea

\noindent with $F_{k} = \{P_{k}^{\rm{L}}, P_{k}^{\rm{NL}}, \texttt{B}_{k}\}$, and each case has enclosed the areas of the $68$, $95$ and $99$ percentiles.

\subsection{Emulating the nonlinear matter power spectrum}
\label{subsec:accuracy_ds_nonlinear}

\begin{figure}[t]
\centering
\includegraphics[width=0.75\linewidth]{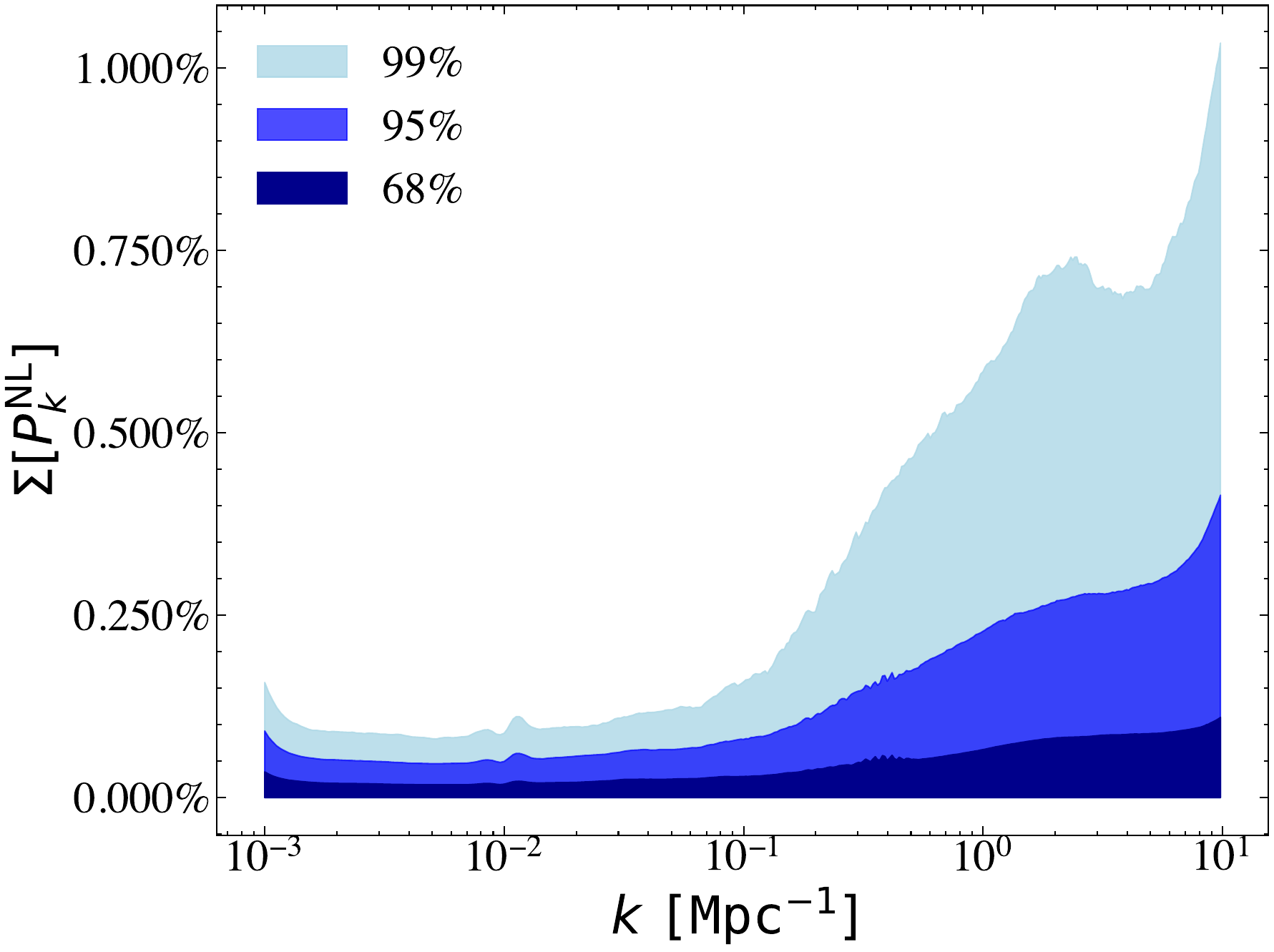}
\caption[Emulator accuracy of the DS nonlinear matter power spectrum]{This plot illustrates that the accuracy of the DS nonlinear power spectrum emulator is better than $1\%$ up to $k = 10~h/{\text{Mpc}}$, thus we are reproducing the output from \texttt{ReACT} with high precision. As already seen in \cite{SpurioMancini:2021ppk}, 
the differences between the emulated and the real predictions increase at highly nonlinear scales. This reflects the intrinsic scatter in the real predictions arising from the numerical complexity of the computation performed by the modelling in that region.}
\label{fig:acc_nonlinear_ds}
\end{figure}

In a similar fashion, we create an emulator for the DS nonlinear matter power spectrum by applying the reaction outlined in Eq.~\eqref{Eq:Reaction_def} using \texttt{ReACT} with the linear spectrum as an input. Since the pseudo spectrum adheres to the standard halo model approach, we opt to use \texttt{HMcode2020}. This choice is motivated by its capability to cover a wide cosmological parameter range, in contrast to alternatives such as \texttt{EuclidEmulator2}, which are bound by more restricted parameter ranges. Finally, from  Eq~\eqref{eq:Reaction_spectrum_2} 
we take the product of the pseudo spectrum times the reaction in order to compute the DS nonlinear power spectrum. \\
As we suggested earlier, we reiterate to carefully define the emulator validity range. In our case, the parameter range is limited to those values where \texttt{ReACT} can resolve the halo model spherical collapse. Specifically, we are free to choose values for a set of cosmological parameters which yield $\sigma_8(z=0)$ values between $0.55$ and $1.4$. In addition, the validity range must be consistent with KiDS-1000 official $\Lambda$CDM analysis.
The accuracy achieved is sub-percent with respect to the predictions of \texttt{ReACT}, as we illustrate in \autoref{fig:acc_nonlinear_ds}.

\subsection{Emulating baryonic feedback}
\label{subsec:accuracy_baryons}

Lastly, we also produce roughly $10^5$ training samples for the contribution of baryonic effects on the matter power spectrum, which is taken into account through the baryonic factor $\mathtt{B}(k,z)$, defined in Eq.~\eqref{eq:boost_eq} using \texttt{HMcode2016}. 
As we discussed previously in \autoref{Chapter2}, here we choose \texttt{HMcode2016} for baryonic feedback because it has a DM-only limit where baryonic effects vanish, unlike the 2020 version. However, we opt instead \texttt{HMcode2020} for computing the DM-only nonlinear power spectrum due to its improved modelling of BAO damping and enhanced treatment of massive neutrinos \citep{2021MNRAS.502.1401M}. \\
Then, we store in a training set and keeping once again $20\%$ of those for testing. This baryonic feedback emulator is parametrised by the following quantities:
\bea
\boldsymbol{\theta}_{\rm feedback} = \{\omega_{\rm b}, \omega_{\rm cdm}, h, n_s, S_8, c_{\rm min}, \eta_0 ,z \} \, .
\label{eq:boost_parameters_training}
\eea

\begin{figure}[t!]
\centering
\includegraphics[width=0.75\linewidth]{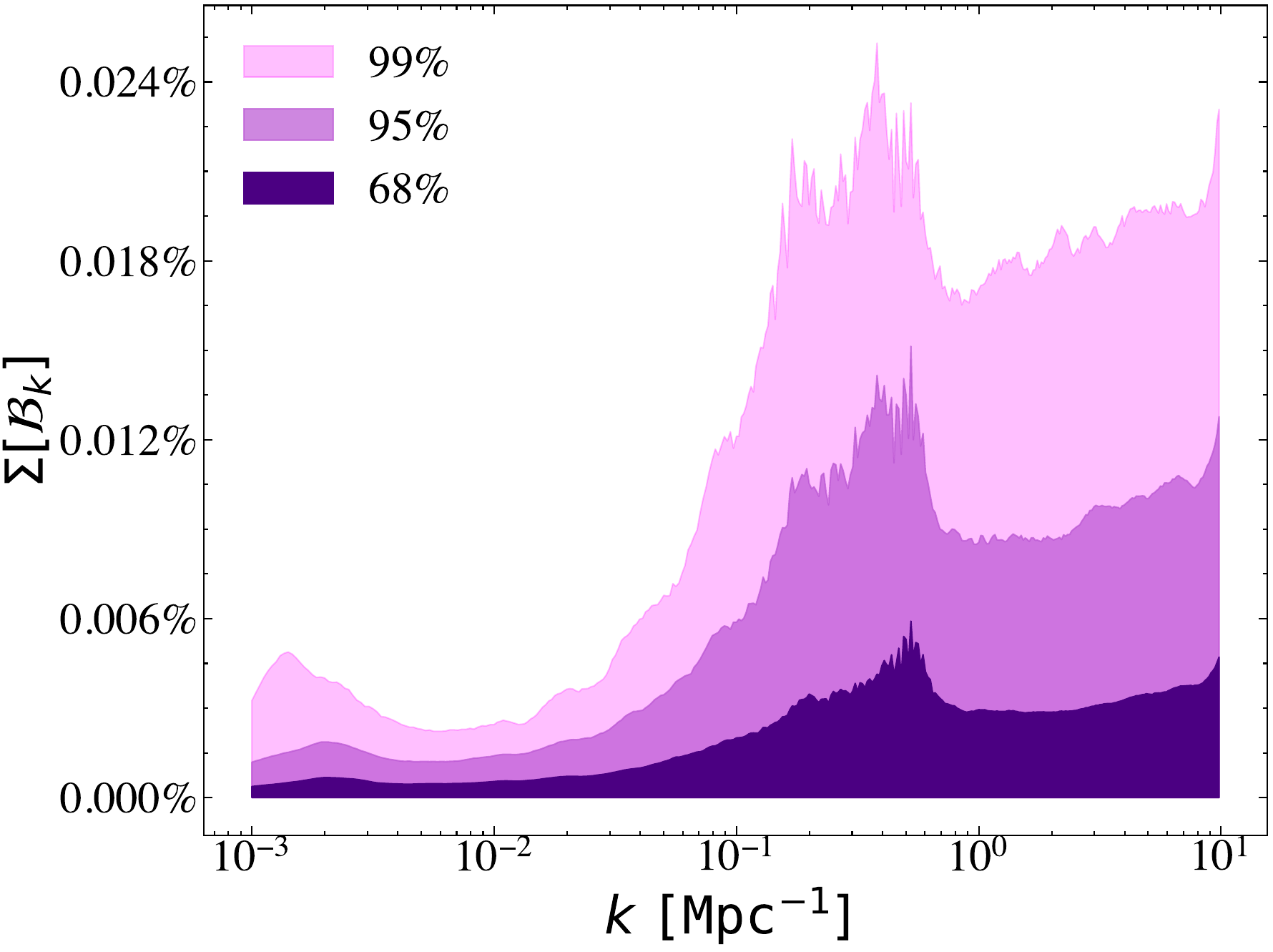}
\caption[Emulator accuracy of the baryonic feedback]{In this plot we display the accuracy of the third emulator, with baryonic effects. In short, $99\%$ are emulated with an error smaller than less than $0.03\%$.}
\label{fig:acc_baryon}
\end{figure}

\noindent Notice this baryonic feedback emulator incorporates fewer input parameters (specifically, only $\Lambda$CDM ones) than the DS emulator. For alternatives to $\Lambda$CDM, baryonic feedback is expected to be accurate enough and is largely unaffected by most cosmological parameters, except for its dependence on the baryon fraction \citep{2021MNRAS.506.4070A, 2021JCAP...12..046G, 2021MNRAS.507.5869A}. Likewise, we report the accuracy of the trained baryonic feedback emulator in \autoref{fig:acc_baryon}. \\

\noindent To recap this chapter, we have developed fast and precise emulators that cover the DS matter power spectra, both linear and nonlinear regimes, as well as baryonic feedback. The parameter ranges for these emulators are presented in \autoref{tab:range_of_emulators}. 
In terms of computational speed, there is already a major improvement at this stage. While generating $\sim 2 \cdot 10^5$ samples with \texttt{ReACT} took nearly two days using $100$ CPU cores, the DS emulator accomplishes the same task in around $\sim 30$ seconds, on a single CPU core. For this reason, we replace the Boltzmann code calls for computing the matter power spectra within the KiDS-1000 pipeline with our emulators in the next chapter.
\newpage 
\phantom{text}

\begin{table}[ht!]
\centering 
\caption[Validity range of input parameters  for neural emulators]{Input cosmological parameters and their validity range for both emulators used in this work. The redshift $z$ is incorporated as an additional input parameter with the range of $[0,5]$. The units of $m_\nu$ are [eV], while $A_{\rm ds}$ has units of [b/GeV].}
  \renewcommand{\arraystretch}{1.45}
  \setlength{\tabcolsep}{3.0pt}
  \begin{tabular}{c c c c}
    &\textbf{Input Parameter}  &  \textbf{Min. value}  & \textbf{Max. value} \\
    \hline
    \hline
     \parbox[c]{5mm}{\multirow{8}{*}{\rotatebox[origin=c]{90}{\textbf{Cosmology}}}} 
     &$\omega_{\mathrm{b}} = \Omega_{\mathrm{b}} h^2$            & 0.01875    & 0.02625 \\
    &$\omega_{\mathrm{cdm}}= \Omega_{\mathrm{cdm}} h^2$          & 0.05       &  0.255\\
    &$h$       & 0.64   &  0.82\\
    &$n_{\textrm{s}}$   & 0.84  & 1.1\\
    &$S_8$          &  0.6    & 0.9\\
    &$m_\nu$          &  0    & 0.2\\
    &$w$          & -1.3    & -0.7 \\ 
    &$|A_{\rm ds}|$    &  0      & 30\\
    \hline
    \hline
    \\[-4ex] \parbox[c]{5mm}{\multirow{2}{*}{\rotatebox[origin=c]{90}{\textbf{Baryons} }}} 
     &$c_{\mathrm{min}}$               & 2   & 4\\
    &$\eta_0$                         & 0.5   & 1 \\[1.8ex]
    \hline
    \hline
    \end{tabular}
\label{tab:range_of_emulators}
\end{table}

\newpage
\thispagestyle{empty}

\chapter{KiDS-1000: DS constraints}\label{Chapter6}

\vspace{1cm}

This chapter showcases the results from my second publication \citep{2024MNRAS.532.3914C}, which represents a core part of this thesis. Here, we apply the computational tools developed up to this point to pin down our specific objectives in placing the first constraints on the DS model through using photometric data. \\
Throughout this chapter, we analyse data from the fourth release of the Kilo-Degree Survey,\footnote{Based on observations made with ESO Telescopes at the La Silla Paranal Observatory under programme IDs 177.A-3016, 177.A-3017, 177.A-3018 and 179.A-2004, and on data products produced by the KiDS consortium. The KiDS production team acknowledges support from: Deutsche Forschungsgemeinschaft, ERC, NOVA and NWO-M grants; Target; the University of Padova, and the University Federico II (Naples).} \citep{2019A&A...625A...2K,2020A&A...637A.100W,2021A&A...647A.124H,2021A&A...645A.105G} known as KiDS-1000. This dataset provides cosmic shear measurements across roughly $1000 \ \text{deg}^2$. Note that we maintain the original cosmic shear and photometric redshift measurements, as well as the data modelling presented in preceding KiDS-1000 analyses \citep{2021A&A...646A.140H,2021A&A...649A..88T,2021A&A...646A.129J}. The main KiDS characteristics are: a sky fraction $f_{\rm sky} = 0.024$, a surface density of galaxies $n_{\rm g} = 6.2$~galaxies/arcmin$^2$ and an observed ellipticity dispersion $\sigma_{\epsilon} = 0.265$. Moreover, KiDS-1000 contains redshift distributions with $5$ tomographic bins over $0 \leq z \leq 2.5$. So, we introduce useful tools and results that will be valuable for future cosmological pipelines and surveys.

\section{KiDS-1000 alone}
\label{sec:k1k} 

We perform the Bayesian analysis by using \texttt{Montepython} \citep{2013JCAP...02..001A}, where our emulators are implemented internally into the pipeline, replacing the usual calls to \texttt{CLASS}. In addition, we selected the sampler \texttt{Multinest} \citep{2009MNRAS.398.1601F}
in order to also obtain the Bayes-factor values using Eq.~\eqref{Eq:bayes_def}, with $\Lambda$CDM as the reference model for comparison against $w$CDM and DS cosmologies.
In the following bullet-points, we share the main set-up of \texttt{Multinest} for our runs: 
\begin{itemize}
    \item \texttt{n\_live\_points = 1000}
    \item \texttt{sampling\_efficiency = 0.3}
    \item \texttt{n\_iter\_before\_update = 200}
    \item \texttt{evidence\_tolerance = 0.01}
    \item \texttt{boost\_posteriors = 10.0}
\end{itemize}

\noindent Those values are taken from the ones called ``optimised" runs from the official KiDS results. While the remaining parameters are set with default values from \texttt{MultiNest} itself. All contour plots in this chapter are obtained using \texttt{GetDist} \citep{2019arXiv191013970L}. \\

\noindent The setup of priors on the cosmological parameters are limited to the validity range of the emulators, previously shown in \autoref{tab:range_of_emulators}. 
Furthermore, baryonic feedback (it is encoded in our baryonic emulator) is incorporated into the pipeline, as it plays a major systematic in weak lensing, which is sensitive to such effects \citep{2023A&A...678A.109A,2024PhRvD.110j3539G,2024arXiv241021980S}. We assume flat distributions on these priors. Before producing posteriors for alternative models, and as a crosscheck, we used our nonlinear spectrum + baryonic feedback emulators to reproduce the $\Lambda$CDM constraints obtained from the KiDS-1000 official results presented in \cite{KiDS:2020suj}. The comparison is presented in \autoref{Appendix_b}, where the contours are produced in around 10 minutes with \texttt{CosmoPower}, as opposed to the few days required with a Boltzmann solver.\\

\begin{table*}
\centering
\caption[Mean and marginalised $68\%$ constraints on key parameters from the KiDS-1000 analysis]{Mean and marginalised $68\%$ constraints on key weak lensing parameters from the KiDS-1000 analysis. We report the log-Bayes factors of each model with respect to $\Lambda$CDM with $B = \frac{\mathcal{Z}_{\rm{model}}}{\mathcal{Z}_{\Lambda\mathrm{CDM}}}$. According to Jeffreys' scale from \autoref{tab:jeff_evidence}, a value of $|\log_{10} B|$ below $0.5$ implies an indecisive advantage over $\Lambda$CDM. Note that the constraints on $w$ are prior dominated for all probes.}
\renewcommand{\arraystretch}{1.75}
\setlength{\tabcolsep}{3.5pt}
\resizebox{\textwidth}{!}{%
\begin{tabular}{|c|ccc|ccc|ccc|}
\hline  
& \multicolumn{3}{c}{\textbf{Band Powers}}                       & \multicolumn{3}{c}{\textbf{COSEBIs}}                           & \multicolumn{3}{c|}{\textbf{2PCFs}} \\
\hline 
& $\Lambda$CDM              & wCDM                       & DS
& $\Lambda$CDM              & wCDM                       & DS
& $\Lambda$CDM              & wCDM                       & DS \\
\hline 
$\Omega_{\mathrm{m}}$ & $0.328^{+0.073}_{-0.31}$ & $0.335^{+0.082}_{-0.11}$ &
$0.353^{+0.092}_{-0.11}$ &  
$0.292^{+0.06}_{-0.11}$ &  
$0.293^{+0.064}_{-0.11}$ & 
$0.293^{+0.066}_{-0.11}$ & 
$0.228^{+0.035}_ {-0.06}$ & 
$0.228^{+0.039}_{-0.063}$ &
$0.226^{+0.038}_{-0.064}$ \\
\hline
$\sigma_8$ & $0.74^{+0.11}_{-0.15}$ & $0.74^{+0.1}_{-0.15}$ & 
$0.708^{+0.087}_{-0.16}$ & 
$0.790^{+0.13}_{-0.15}$  & 
$0.792^{+0.12}_{-0.15}$ & 
$0.79^{+0.13}_{-0.16}$ & 
$0.90 \pm 0.1$ & 
$0.903^{+0.098}_{-0.12}$ &
$0.9^{+0.11}_{-0.12}$ \\
\hline
$S_{8}$ & 
$0.752^{+0.031}_{-0.023}$ & 
$0.754^{+0.034}_{-0.031}$ & 
$0.739 \pm 0.036$ & 
$0.751^{+0.026}_{-0.019}$ & 
$0.753 \pm 0.029$ & 
$0.750 \pm 0.031$ & 
$0.766 \pm 0.019$ & 
$0.770^{+0.025}_{-0.028}$ &
$0.767^{+0.026}_{-0.031}$ \\
\hline
$w$ & 
--- & 
$-0.96^{+0.24}_{-0.13}$ & 
$-0.99^{+0.2}_{-0.15}$ & 
--- & 
$-0.98^{+0.22}_{-0.14}$ & 
$-1.05^{+0.1}_{-0.2}$ & 
--- & 
$-0.99^{+0.22}_{-0.14}$ &
$-1.07^{+0.082}_{-0.19}$ \\ 
\hline
$A_{\rm ds}$ & 
--- & 
--- & 
$-0.3^{+13}_{-8.5}$ & 
--- & 
--- & 
$-3.6^{+7.8}_{-9.9}$ & 
--- & 
--- &
$-4.8^{+7.1}_{-10}$ \\
\hline
$\log_{10} B$   &
--- &
$-0.1295 \pm 0.0011$ &
$0.0787 \pm 0.0018$ &
--- &
$-0.1512 \pm 0.0003$  &
$-0.3780 \pm 0.0021$    &
--- &
$-0.0761 \pm 0.0002 $  &
$-0.4504 \pm 0.0027 $ \\
  \hline 
\end{tabular}
}
\label{tab:bestfit_K1K_params}
\end{table*}

\begin{figure}[t!]
\centering  
\includegraphics[width=0.5\textwidth]{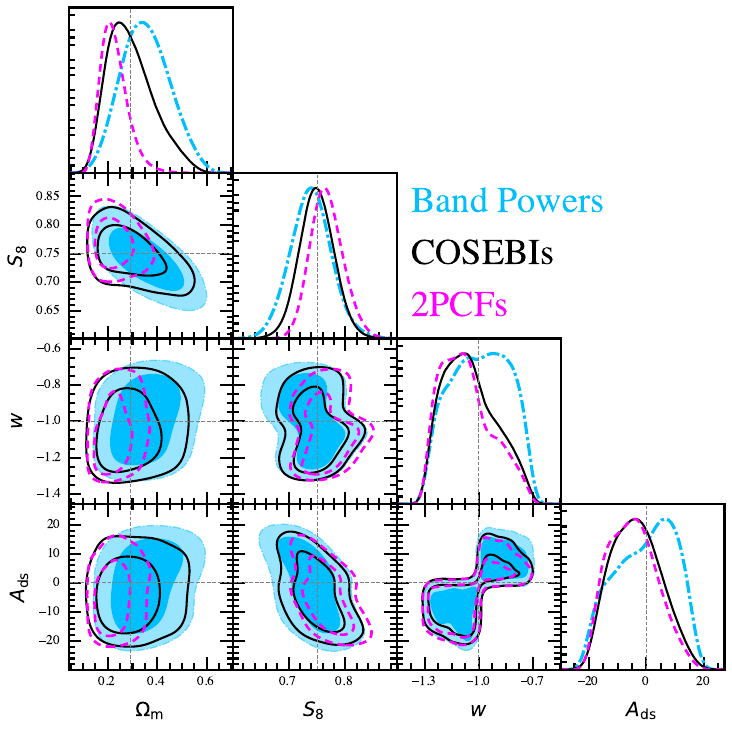}
\caption[Pure KiDS-1000 constraints on DS]{Constraints ($68 \%$ and $95 \%$ marginalised contours) on the key parameters $\Omega_{\rm m}$, $S_8$, $w$, and $A_{\rm ds}$ from all KiDS-1000 statistics sets: Contours for Band Powers (blue), COSEBIs (black) and Correlation Functions (magenta). The dashed lines represent COSEBIs mean values from the $\Lambda$CDM case. The DS parameter $A_{\rm ds}$ has units of [b/GeV].}
\label{Fig:K1K_DS_free_CMB}
\end{figure}

\noindent As we presented in \autoref{Chapter3}, we thus analyse the BPs, COSEBIs and 2PCFs statistics and produce the posterior distribution for several parameters, including $\Omega_{\rm m}$, $S_8$, $A_{\rm ds}$, and $w$. Remarkably, Eq.~\eqref{eq:Interaction_term} depends only on background quantities. Although the model is applicable to a general evolution of the equation of state parameter $w(z)$, we will take it to be constant in this analysis. Throughout our analysis, we assume two massless neutrinos and one massive neutrino with a mass fixed to $0.06$ eV. For the DS parameter $A_{\rm ds}$ and $w$, we assume flat prior distributions $|A_{\rm ds}| \rightarrow \unif [0.0, 30.0]$ b/GeV and $w \rightarrow \unif [-1.3, -0.7]$. Regarding the justification of the nuisance parameters (including the IA parameter) priors we refer to \cite{KiDS:2020suj}. Hence the posteriors are shown in \autoref{Fig:K1K_DS_free_CMB}. 
Note that although a key parameter, $w$, is largely prior-dominated, the KiDS-1000 data alone constrains the other parameter of the model to $\vert A_{\rm ds} \vert \lesssim 20$ b/GeV (at $68\%$ C.L.). This implies that the lensing data alone exhibits sensitivity to the nonlinear effects of the interaction. Furthermore, we report the mean (and the marginalised $68\%$ confidence values) of the several parameters and log-Bayes factor values for DS and $w$CDM ($A_{\rm ds} = 0$) in reference to $\Lambda$CDM, ($w = -1, A_{\rm ds} = 0$). \autoref{tab:bestfit_K1K_params} summarizes those results for the complete KiDS-1000 sets of statistics. As seen by the obtained log-Bayes factor values from Eq.~\eqref{Eq:bayes_factor}, none of the cosmological models exhibits a definitive advantage over the rest. 
Despite this, the current KiDS photometric galaxy catalogue may lack the sample size or precision required to place robust constraints on the DS model. Forthcoming surveys with higher-quality data might provide more robust constraints.

\section{KiDS-1000 with CMB+BAO information}
\label{sec:with_cmb}

As shown above, the KiDS-1000 data provides only an upper bound on the DS amplitude of $\vert A_{\rm ds} \vert \lesssim 20$ b/GeV.
Aiming to constrain the model further, we supplement our analysis with additional information from the Planck measurements of the CMB temperature and polarisation \citep{2020A&A...641A...1P}, as well as from BAO measurements from 6dFGS \citep{2011MNRAS.416.3017B}, SDSS-MGS \citep{2015MNRAS.449..835R} and BOSS \citep{2017MNRAS.470.2617A}. \\
In practice, we apply a prior on cosmological parameters derived from the posterior of the Planck TT+TE+EE+lowE+BAO analysis\footnote{From \url{https://pla.esac.esa.int/pla/}.} of the $w$CDM model \citep{Planck:2018vyg}, as reported in \autoref{tab:priors_k1k_cmb_bao}. Despite this previous CMB+BAO analysis not including the effects of the dark sector interaction, it is a good approximation to use it for our combination. This is because we expect the CMB to be insensitive to the effects of $A_{\rm ds}$ \citep{2016PhRvD..94d3518P}, since those only occur at late time; while the BAO is only sensitive to the expansion history, which is unaltered by DS from $w$CDM. Still, to be conservative, we use instead flat priors\footnote{We compared this setup against using Gaussian priors of the same width, but this did not result in differences in our posteriors.} on cosmological parameters taken from the 1D $2\sigma$ constraints of the CMB+BAO analysis. This approximate method allows us to obtain a robust constraint of the $A_{\rm ds}$ parameter from the combination of KiDS-1000 with CMB and BAO data.

\renewcommand{\arraystretch}{1.75}
\setlength{\tabcolsep}{3.5pt}
\begin{table}
  \centering 
\caption[Priors on key parameters for the combined analysis of KiDS-1000 with CMB+BAO measurements]{We report the setup of the priors considered to cosmological parameters, which are sourced from the Planck TT,TE,EE+lowE+BAO analysis of the $w$CDM model with extended bounds to $2\sigma$.}
  \begin{tabular}{c c}
    \textbf{Parameter} & \textbf{Prior}  \\ \hline \hline $\omega_{\mathrm{b}}$   &  $\unif [0.022,  0.0226]$ \\ \hline 
    $\omega_{\mathrm{cdm}}$ &  $\unif [0.1174, 0.1223]$ \\ \hline
    $h$  & $\unif [0.6594, 0.7163]$ \\ \hline
    $n_s$ & $\unif [0.9571, 0.9736]$ \\ \hline
    $\ln(10^{10} A_s)$ & $\unif [3.0131, 3.0765]$\\ \hline
    $w$  & $\unif [-1.1591, -0.9347]$  \\ \hline
    $m_\nu$  & Fixed   \\ \hline
    $|A_{\rm ds}|$   &  $\unif [0.0, 30.0]$ [b/GeV] \\ \hline
    $c_{\rm min}$ & $\unif [2, 4]$  \\ \hline
    $\eta_0$ & Derived \\ 
    \hline
    \end{tabular}
\label{tab:priors_k1k_cmb_bao}
\end{table}

\begin{figure} 
\centering
\includegraphics[width=0.5\textwidth]{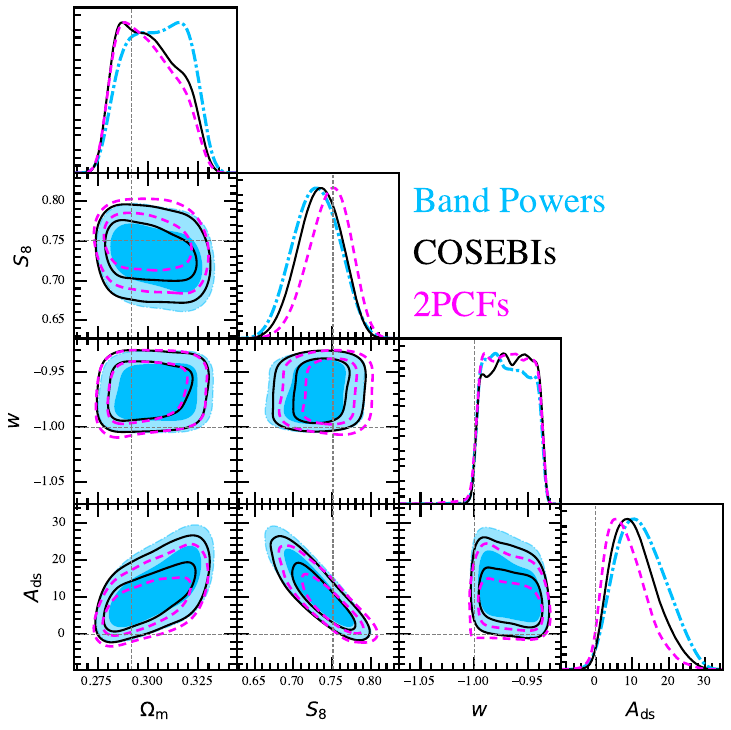}
\caption[KiDS-1000 with CMB+BAO constraints on DS]{Constraints ($68 \%$ and $95 \%$ marginalised contours) on the key parameters $\Omega_{\rm m}$, $S_8$, $w$ and $A_{\rm ds}$ from all KiDS-1000 statistics sets combined with CMB+BAO: Contours for Band Powers (blue), COSEBIs (black) and Correlation Functions (magenta). The dashed lines represent COSEBIs mean values from the $\Lambda$CDM case. The units of $A_{\rm ds}$ are b/GeV.}
\label{Fig:K1K_DS_CMB}
\end{figure}

\noindent The results of this combined analysis are shown in \autoref{Fig:K1K_DS_CMB} and \autoref{tab:bestfit_K1K_params_CMB_priors}. As reference, in \autoref{Appendix_b} we further show the full contour plot of such analysis.
The posteriors exhibit a clear preference for values of $A_{\rm ds} > 0$ over all KiDS-1000 statistics, and consequently $w > -1$. In particular, we see in the $w$-$A_{\rm ds}$ contour, an approximately $2\sigma$ deviation from $\Lambda$CDM, with the COSEBIs analysis giving $A_{\rm ds} = 10.6^{+4.5}_{-7.3}$ b/GeV. This contrasts with the previous KiDS-only result for which there was no preference for a non-zero value of the interaction strength. Note also that while the $w$ constraint appears prior-dominated, the CMB+BAO information is only enforcing it to be in the range $w\in [-1.159, -0.935]$, which only accounts for the upper bound on $w$, and then it is representing a data-driven constraint. The lower bound on $w$ is instead given by the physical prior enforcing the equal signs of $A_{\rm ds}$ and $(1+w)$. It is therefore the substantial preference for a positive interaction amplitude that is driving $w$ to the region of $w>-1$.

\begin{figure}[t!]
\centering 
\includegraphics[width=\textwidth]{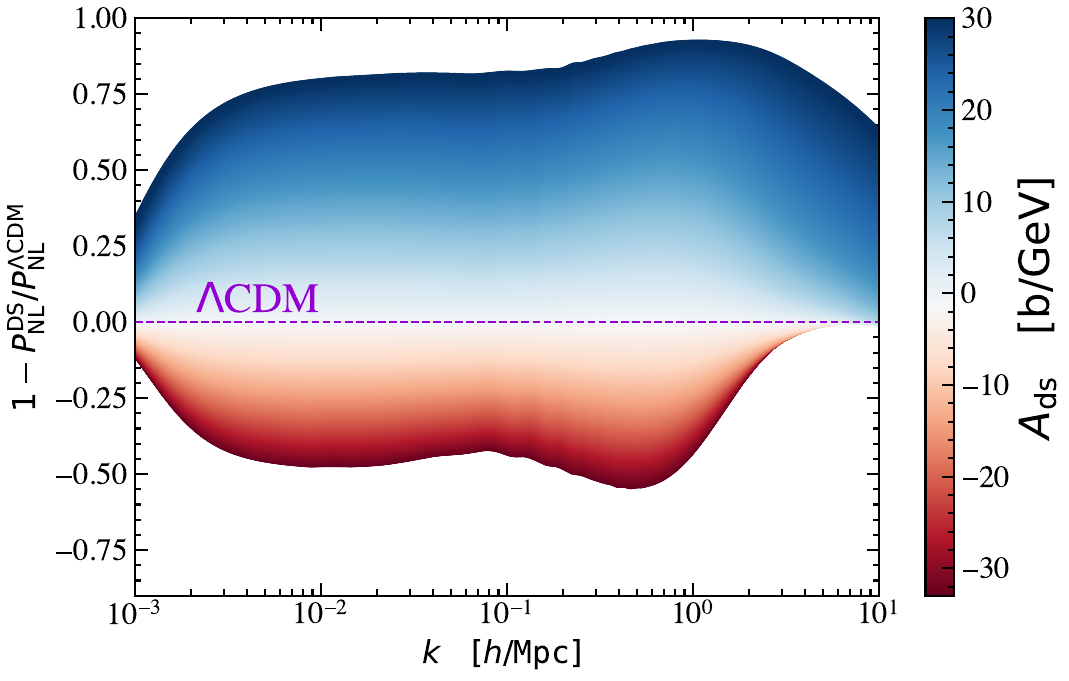}
\caption[$\Lambda$CDM matter power spectrum deviation due to Dark Scattering]{Deviation of from that of a $\Lambda$CDM model (represented in purple dashed-line) in the presence of DS interaction at $z=0$. The colour code on the right indicates that values of $A_{\rm ds}<0$ (warmer colours) enhance the matter power spectrum, while $A_{\rm ds}>0$ values (colder colours) suppress it.}
\label{fig:spectra_impact}
\end{figure}

\noindent The physical explanation for this result is interpreted as follows: \autoref{fig:spectra_impact} shows deviation from $\Lambda$CDM trend due to DS effects, a positive $A_{\rm ds}$ value represents a suppression in amplitude of the matter power spectrum due to an additional frictional force (see Eq.~\eqref{Eq:interaction_eom}). This in turn decelerates the collapse of dark matter density fluctuations, reducing structure formation at late times. Since we essentially fixed the primordial amplitude by using CMB information in this analysis, the preference for a low late-time amplitude (i.e. the $S_8$ tension) is converted into a preference for a positive $A_{\rm ds}$. This effect is also evident in the anti-correlation of $A_{\rm ds}$ with the $S_8$ parameter displayed in \autoref{Fig:K1K_DS_CMB}. \\

\begin{figure}
\centering
\includegraphics[width=\textwidth]{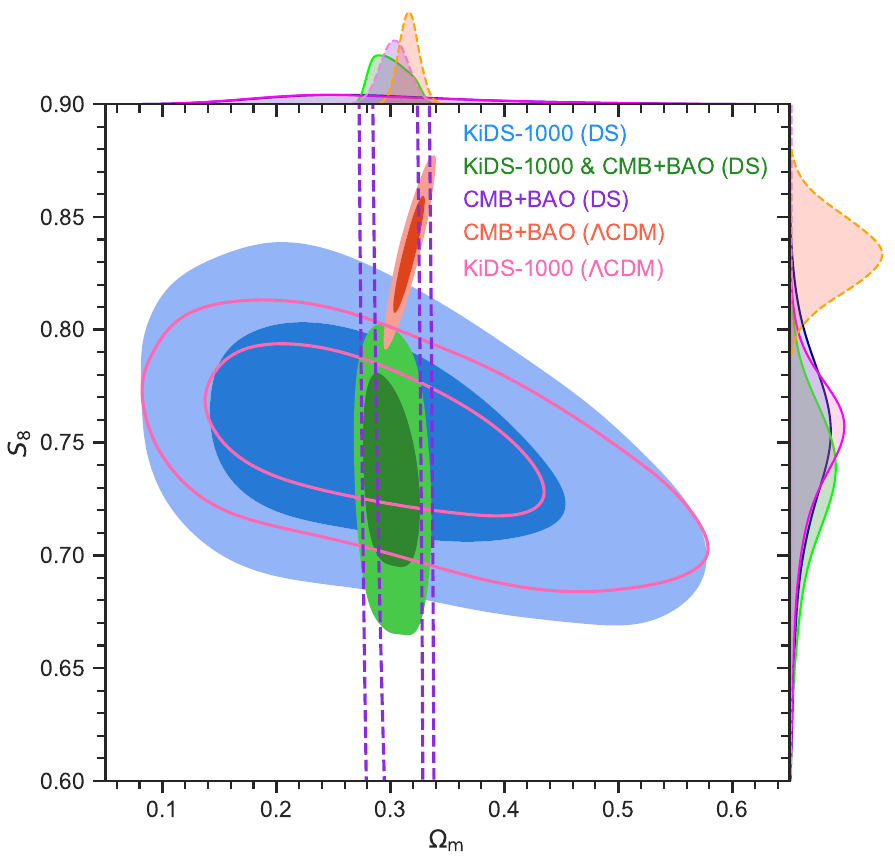}
\caption[Constraints projected on the $\Omega_{\rm m}$-$S_8$ plane]{Constraints projected on the $\Omega_{\rm m}$-$S_8$ plane ($68 \%$ and $95 \%$ C.L.). Firstly, we show in dashed violet an estimation of the constraints from the CMB+BAO analysis alone for the DS model, while in solid filled orange the full CMB+BAO analysis for the $\Lambda$CDM model. The filled blue contour represents DS constraints from the KiDS-1000 (COSEBIs only) analysis, while in solid pink lines we show the $\Lambda$CDM constraints. Finally, in green the constraints from KiDS-1000 \& CMB+BAO joint analysis for the DS model is displayed.}
\label{Fig:cosebis_all_models}
\end{figure}

\noindent Lastly, \autoref{Fig:cosebis_all_models} illustrates the comparison of contours on the $\Omega_{\rm m}$-$S_8$ plane for the DS analysis with KiDS-1000 data as well as the KiDS-1000+CMB+BAO joint analysis (both evaluating the COSEBIs statistics). An approximation of the analysis of CMB+BAO (excluding KiDS-1000) for DS is also included. To obtain this contour we re-scaled the $S_8$ values of the $w$CDM chains (TT+TE+EE+lowE+BAO) by the DS growth factor for a broad range of values of $A_{\rm ds}$ (again assuming no constraining power from CMB+BAO on $A_{\rm ds}$, as justified above). This results in a complete broadening of the constraints in the $S_8$ direction, illustrating that, in the DS model, the CMB does not constrain the late-time amplitude, since it is insensitive to one of the parameters controlling it -- the interaction amplitude $A_{\rm ds}$. The full contour plot can be seen in \autoref{Appendix_b}. Cosmic shear then constrains the late-time amplitude and together with CMB+BAO, determines the value of $A_{\rm ds}$ that resolves the tension between early and late Universe probes of the amplitude of density fluctuations. Note that the additional small difference between the CMB+BAO and the KiDS+CMB+BAO contours is due to the differences in the priors employed. More specifically, the flat priors employed in the KiDS-1000 scenarios, do not exactly correspond to the correlated near-Gaussian posteriors inherent in the CMB+BAO estimation. \\
For comparison, we include also in \autoref{Fig:cosebis_all_models} the results for the $\Lambda$CDM model from CMB+BAO as well as those from KiDS-1000. We can see that between $\Lambda$CDM and DS, the $S_8$ constraint is broadened, given the additional amplitude parameter being fitted. Additionally, when the CMB+BAO information is added, we can see that the $S_8$ constraint shifts to lower values. This is because $A_{\rm ds}$ is constrained to be positive, which can only lower the amplitude. This is clear when comparing with the $\Lambda$CDM case, as they have similar upper bounds on $S_8$ (corresponding to $A_{\rm ds}=0$), but differ in the lower bounds.

\begin{table}
\centering
\caption[Mean and marginalised $68\%$ constraints on key parameters from the combined analysis of KiDS-1000 with CMB+BAO measurements]{Mean and marginalised $68\%$ constraints on key parameters of the DS model and the baryonic parameter, from the combined analysis of all KiDS-1000 probes with CMB+BAO measurements. The results are also presented in \autoref{Fig:K1K_DS_CMB}. Note that the constraints on $w$ are also prior dominated for each probe.
}
\renewcommand{\arraystretch}{1.75}
\setlength{\tabcolsep}{3.5pt}
\begin{tabular}{|c|c|c|c|}
  \hline 
  & \multicolumn{1}{c}{\textbf{Band Powers}}                       & \multicolumn{1}{c}{\textbf{COSEBIs}}                           & \multicolumn{1}{c|}{\textbf{2PCFs}} \\
  & \multicolumn{1}{c}{\textbf{(CMB+BAO)}}                       & \multicolumn{1}{c}{\textbf{(CMB+BAO)}}                           & \multicolumn{1}{c|}{\textbf{(CMB+BAO)}} \\
  \hline 
    $S_{8}$  
    & $0.729 \pm 0.029$ 
    & $0.734 \pm 0.027$
    & $0.746^{0.029}_{-0.024}$
    \\    \hline
    $A_{\rm ds}$  
    & $12.5^{+5.7}_{-7.8}$ b/GeV 
    & $10.6^{+4.5}_{-7.3}$ b/GeV
    & $8.4^{+3.8}_{-6.7}$ b/GeV
    \\     \hline
    $w$  
    & $-0.969^{+0.015}_{-0.026}$
    & $-0.967^{+0.027}_{-0.015}$
    & $-0.968 \pm 0.019$
    \\ \hline
    $c_{\rm min}$  
    & $2.97 \pm 0.50$ 
    & $2.43^{+0.15}_{-0.39}$
    & $2.36^{+0.12}_{-0.34}$ \\
    \hline
    \end{tabular}
\label{tab:bestfit_K1K_params_CMB_priors}
\end{table}

\noindent In summary, \autoref{Fig:cosebis_all_models} shows that the DS model is consistent with early and late-Universe measurements, thus offering a viable approach to alleviate the $S_8$ tension linked to the measured value of $A_{\rm ds}$. However, it is important to emphasize the benefit of emulator-based methodologies employed in this analysis. Implementing emulators from \texttt{CosmoPower}, we efficiently obtained the contours for the KiDS-1000 pipeline within a mere few minutes of computation, in a $24$ CPU cores machine. This can also be easily extended to include photometric galaxy clustering in a 3x2pt analysis or Stage IV galaxy survey forecast, with the purpose of constraining the DS model. 
Moreover, the implications of adopting alternative models for intrinsic alignments (IA), baryonic feedback and other systematics can also be more efficiently explored in the future with these emulators. This would then allow for a thorough investigation of their impact on the inference of the DS parameter and its degeneracies with e.g. baryonic feedback, or massive neutrinos as explored in \autoref{Chapter4}.

\newpage
\thispagestyle{empty}

\chapter{Stage IV forecasts}\label{Chapter7}

\vspace{1cm}

As the inherent nonlinearity of cosmological models being more complex and parameter spaces expand, directly computing posterior distributions will be intractable. Standard MCMC techniques are commonly employed to sample from these distributions, but they can be computationally intensive and require careful tuning. In this chapter, we present the last inference analysis of this thesis and forms part of my third publication \citep{2024arXiv241010603C}. Here, we explore the power of Stage IV cosmic shear surveys to constrain dark energy. In a standard forecasting, actual observational data is absent. Instead, a fiducial model is used as a proxy to simulate potential observations. This is achieved using a novel, accelerated framework for parameter estimation and model comparison, where we extend previous work presented in \cite{2024MNRAS.532.3914C} and \cite{2024OJAp....7E..73P}. Our analysis remains centred on DS model, while exploring how baryonic feedback and massive neutrinos affect cosmological constraints. Once again we make use of our neural emulators, even though, now embedded them into a fully-differentiable pipeline for gradient-based cosmological inference. Consequently, the batch likelihood call running on a single GPU is up to $\mathcal{O}(10^5)$ times faster than traditional approaches. \\
To compare models, we apply the learnt harmonic mean estimator from the \texttt{harmonic} \citep{2021arXiv211112720M} software to posterior samples, that is process independently of the inference, as outlined in \autoref{subsec:harmonic}. These results show great promise for constraining DS with Stage IV data; furthermore, our methodology can be straightforwardly extended to a wide range of dark energy and modified gravity models.

\section{\texttt{JAX} framework}
\label{sec:jax_pipeline}

The entire pipeline of this chapter is powered by \texttt{JAX} \citep{jax2018github}, a software that benefits from Just-In-Time (JIT) compilation for rapid and efficient computations. This compilation works through translating Python packages to be highly optimised by the Accelerated Linear Algebra (XLA) compiler, enabling high-performance execution across various hardware platforms such as central processing units (CPUs), graphics and tensor processing units (GPUs and TPUs). \\
Nevertheless, the standout innovation of \texttt{JAX} is its automatic differentiation (autodiff) engine \citep{Wengert1964ASA,BARTHOLOMEWBIGGS2000171,2016arXiv161100712M,2021ascl.soft11002B}. This, combined with XLA compilation and batch evaluation into tracing functions from \texttt{Python} packages, like those from \texttt{NumPy} \citep{2020Natur.585..357H}, resulting in \texttt{JAX} achieving computational efficiency and memory usage comparable to standard GPU-accelerated methods. Moreover, \texttt{JAX} shares a \texttt{NumPy}-like syntax, making it easy to use for whom familiar with \texttt{NumPy} library. \\
Autodiff adopts a ``trace-and-transform", in which operates in computation of derivatives by applying differentiation rules to obtain functions composed by the underlying primitives with known derivatives. This eliminates the need for methods like finite differences or symbolic differentiation, which can be imprecise and computationally inefficient. Therefore, this design makes \texttt{JAX} more flexible\footnote{
Although, one concern about autodiff in complex models is controlling memory usage while ensuring optimal performance.} in large-scale computations\footnote{Going through further be adapted into methods like differentiation (\texttt{grad}), vectorization (\texttt{vmap}), and parallelisation (\texttt{pmap}).}. \\

\begin{figure}[t!]
\centering 
\includegraphics[width=.5\textwidth]{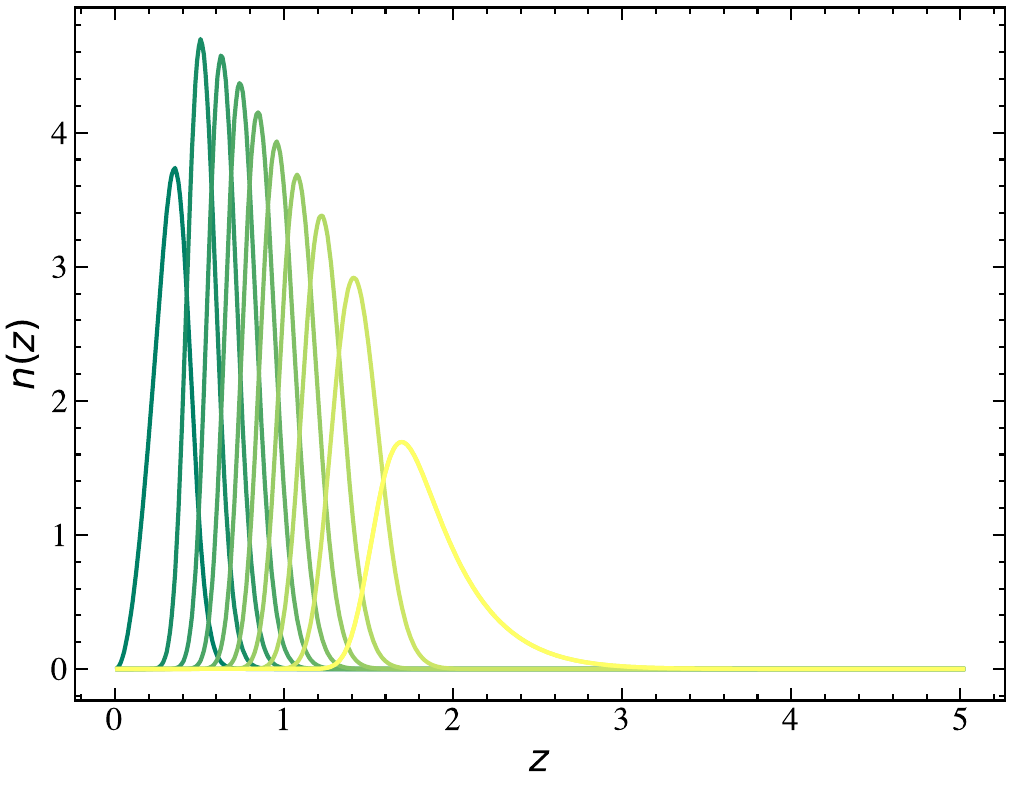}
\caption[Redshift bins for Stage IV forecast]{
We take $10$ equipopulated tomographic bins with galaxies for $n_{i}(z)$ between $z_{\textrm{min}}=0$ and $z_{\textrm{max}}=5$. Each tomographic bin $(i = 1, \dots, 10$) is modelled using a Gaussian kernel density estimation with a bandwidth of $0.01$. To account for photometric redshift uncertainties, we introduce a shift parameter $D_{z_i}$, shifting every $i$-bin distribution as $n_i'(z) = n_i(z - D_{z_i})$. Following \cite{2023OJAp....6E..40S}, we keep assuming a Gaussian prior $\mathcal{N}(0,10^{-4})$ for each of them.}
\label{fig:z_i_stage_iv}
\end{figure}

\noindent \texttt{JAX} has demonstrated to be an exploitable tool in cosmological context, across various studies \citep{2024ApJS..270...36L,2024OJAp....7E..11R,2024OJAp....7E..10B,2024A&A...686A..10B,2025arXiv250202294H}. In this thesis, we harness its capabilities to accelerate Bayesian cosmological inference, in which it is massively accelerated by orders-of-magnitude. In particular, we recur to the \texttt{JAX}-based library of \texttt{jax-cosmo} \citep{2023OJAp....6E..15C} pipeline to generate forecasts for the cosmic shear power spectrum $C_{ij}^{\hat{\gamma} \hat{\gamma}}(\ell)$ from Eq.~\eqref{eq:cell_shear_full} in the context of Stage IV galaxy surveys, as detailed in \autoref{Chapter3}. In order to load and manage our \texttt{CosmoPower} emulators, presented in \autoref{Chapter5}. We resort to \texttt{CosmoPower-JAX} \citep{2024OJAp....7E..73P} environment, as its name claims, this is a \texttt{JAX}-based implementation of the \texttt{CosmoPower} that  allows to have auto differentiable emulators.\\
We incorporate these emulators of DS nonlinear power spectrum with the baryonic correlation into \texttt{jax-cosmo}, replacing the \texttt{HaloFit} \citep{2012ApJ...761..152T} prescription for computing the nonlinear power spectrum in \texttt{jax-cosmo}. Once again, taking into account these emulators validity.  
Moreover, the cosmic shear is modelled including systematics contributions arising from uncertain redshift distributions and intrinsic alignments from Eq.~\eqref{eq:cell_shear_full}. \autoref{fig:z_i_stage_iv} shows the modelled redshift bins of galaxies that follows a Gaussian kernel (KDE) distribution \citep{1995MNRAS.273..277S,2010A&A...523A...1J} as:
\bea
n(z) \propto \dfrac{1}{\sqrt(2 \pi) \sigma_z} \exp \left[- \left( z - z_0 \right)^{2} / (2\sigma^2_z) \right] \, .
\label{Eq:nz}
\eea

\subsection{Inference method}
\label{subsec:inference}

To sample the posterior distribution, we employ the No-U-Turn Sampler (\texttt{NUTS}) \citep{JMLR:v15:hoffman14a} algorithm as implemented in the \texttt{NumPyro} \citep{2019arXiv191211554P} 
library. As mentioned in \autoref{subsec:sampling}, this sampler is an adaptive variant of the HMC algorithm, allowing for efficient and scalable sampling in high-dimensional spaces. \texttt{NumPyro} offers native compatibility with \texttt{JAX}, and thus with \texttt{jax-cosmo} and \texttt{CosmoPower-JAX}, enabling the development of a fully-differentiable inference pipeline. \\
The gradients for standard HMC algorithms are typically calculated using finite difference methods, which often lack precision and can be computationally expensive. In the \texttt{NUTS} framework, these inefficiencies are mitigated through adapting the sampling path to prevent it from looping back to regions of the parameter space that have already been sampled, hence optimizing the overall efficiency of the sampling process. In our case, with a differentiable pipeline in place, gradients can be efficiently computed using automatic differentiation. Thus, the key requirement is that the likelihood must be differentiable, which we achieve using \texttt{CosmoPower-JAX} and \texttt{jax-cosmo}.\\
This differentiable framework allows us to perform faster batch evaluations of the likelihood, taking $0.19$ seconds on a single GPU versus $\sim 8$ hours on a single CPU to produce $1000$ spectra. This results in an acceleration of up to $\mathcal{O}(10^5)$ with respect to traditional non-differentiable methods. This significant efficiency not only accelerates parameter inference but also facilitates the exploration of various scenarios based on Stage IV cosmic shear mock data. \\ 

\noindent Regarding the inference setup, our selection of \texttt{NUTS} hyperparameters ensures that each run achieves convergence, controlled by the \texttt{r\_hat} $< 1.05$ parameter, in accordance with the Gelman-Rubin criterion \citep{1992StaSc...7..457G,2018arXiv181209384V}. We consider the next configuration:

\begin{itemize}
    \item We use \texttt{chain\_method=`vectorized'}.
    \item We set \texttt{num\_chains} to $10$.
    \item Each chain consists of at least $10^3$ \texttt{num\_warmup} and \texttt{num\_samples}.
\end{itemize}

\noindent For each of our next forecasts, we assume noiseless data vectors, as typical in the literature for forecasts of this type \citep{2024arXiv240513491E}. We also consider a Gaussian likelihood throughout, with a Gaussian covariance \citep{2008A&A...477...43J}. This can be extended through including the impact of non-Gaussian contributions (see \cite{2013MNRAS.429..344K,2013PhRvD..87l3538S,2017MNRAS.470.2100K}) and super-sample covariance \citep{2014PhRvD..89h3519L, 2018JCAP...06..015B,2023A&A...671A.115L} terms. The priors used in the inference pipeline are listed in Table \ref{tab:priors_and_fiducial}. \\

\begin{table}[t!]
  \centering 
   \caption[All priors and fiducial values used for the Stage IV cosmic shear forecast]{Overview of the prior distributions applied to cosmological and nuisance parameters in the simulated analyses. The specified ranges for cosmological and baryonic feedback parameters ensure compatibility with the validity limits of the emulators used in this work. The last column presents the fiducial values of the cosmological parameters, where the values for the $\Lambda$CDM case are taken from the best-fit results of Planck 2018 TTEETE+BAO \citep{Planck:2018vyg}.
   In the DS model we allow for a non-zero dark sector interaction. In particular, the choice of $(w, A_{\rm ds})$ values are motivated by the best-fit findings from the previous results \citep{2024MNRAS.532.3914C} from \autoref{Chapter6}.
   The units of $m_\nu$ are [eV], while $A_{\rm ds}$ has units of [b/GeV].}
\label{tab:priors_and_fiducial}  \renewcommand{\arraystretch}{1.75}
  \setlength{\tabcolsep}{2.pt}
  \begin{tabular}{c c c c}
    &\textbf{Input Parameter}  &  \textbf{Prior} &  $\Lambda$CDM / DS fiducial \\
    \hline
    \hline
     \parbox[c]{5mm}{\multirow{8.5}{*}{\rotatebox[origin=c]{90}{\textbf{Cosmology}}}} 
     &$\omega_{\mathrm{b}} = \Omega_{\mathrm{b}} h^2$ & 
    $\mathcal{U}(0.01875, 0.02625)$ 
    & $0.02242$ \, / \, $0.02242$ \\
    &$\omega_{\mathrm{cdm}} = \Omega_{\mathrm{cdm}} h^2$ & $\mathcal{U}(0.05, 0.255)$   
    & $0.11933$ \, / \, $0.11933$ \\
    &$h$ & 
    $\mathcal{U}(0.64, 0.82)$  
    & $0.682$ \, / \, $0.682$\\
    &$n_{\mathrm{s}}$ & 
    $\mathcal{U}(0.84, 1.1)$  
    & $0.9665$ \, / \, $0.9665$\\
    &$S_8$ &  
    $\mathcal{U}(0.6, 0.9)$  
    & $0.825$ \, / \, $0.825$ \\
    &$m_\nu$ &  
    $\mathcal{U}(0,0.2)$  
    & $0.06$ \, / \, $0.06$ \\
    &$w$  & 
    $\mathcal{U}(-1.3, -0.7)$   
    & $-1$ \, / \, $-0.967$\\ 
    &$|A_{\mathrm{ds}}|$ & 
    $\mathcal{U}(0, 30)$  
    & $0$ \, / \, $10.6$ \\[1.ex]
    \hline
    \hline
    \\[-3.5ex] \parbox[c]{5mm}{\multirow{1.5}{*}{\rotatebox[origin=c]{90}{\textbf{Baryons}}}}
     &$c_{\mathrm{min}}$  & 
     $\mathcal{U}(2, 4)$   
     & $2.6$ \, / \, $2.6$  \\[3.ex]
    \hline
    \hline
    \\[-5ex] \parbox[c]{5mm}{\multirow{2.5}{*}{\rotatebox[origin=c]{90}{\textbf{IA}}}} 
     &$A_{\mathrm{IA}}$  & 
     $\mathcal{U}(-5, 5)$  
     & $0.8$ \, / \, $0.8$ \\
    &$\eta_{\mathrm{IA}}$   & 
    $\mathcal{U}(-5, 5)$   
    & $0$ \, / \, $0$ \\[1ex]
    \hline
    \hline
    \\[-3.5ex] \parbox[c]{5mm}{\multirow{1.5}{*}{\rotatebox[origin=c]{90}{\textbf{z-bins}}}} 
     & $D_{z_i, \rm source}$ \, $i = 1, \cdots 10$ &            $\mathcal{N}(0, 10^{-4})$ 
     & $0$ \, / \, $0$  \\[2.5ex]
    \hline
    \hline
    \end{tabular}
\end{table}

\noindent In our analysis, we use the \texttt{harmonic} package, which implements the learnt harmonic mean estimator (see \autoref{subsec:harmonic}) with normalizing flows \citep{2021arXiv211112720M, 2023arXiv230700048P} to estimate the Bayesian evidence. To estimate it for each model, we apply the \texttt{RQSplineModel} \citep{2019arXiv190604032D} provided within \texttt{harmonic}, which comprises $4$ layers and $128$ spline bins. In the flow training process (imposed through a temperature parameter $T\in [0,1]$), we separate $50\%$ of the chains as training samples, while the remaining chains are employed to compute the evidence; we set $T=0.8$. 

\section{Forecasting Stage IV surveys}
\label{sec:forecast_stage-iv}

Throughout our analysis we consider a survey with the following Stage IV configuration: a sky fraction $f_{\rm sky} = 0.36$, a surface density of galaxies $n_{\rm g}=30$~galaxies/arcmin$^2$ and an observed ellipticity dispersion $\sigma_{\epsilon} = 0.3$. In comparison, an LSST-like survey \citep{2018arXiv180901669T} covers $18,000$ sq. deg. ($44\%$ of the sky) with $27$ galaxies/arcmin$^2$ and $\sigma_{\epsilon} = 0.26$. Although our setup differs slightly, it remains well-suited for our analysis, by also incorporating redshift distributions with $10$ tomographic bins. We compare three cosmological models: $\Lambda$CDM, $w$CDM and DS. We generated two scenarios where the data vector is generated from a $\Lambda$CDM cosmology, while for the second the Universe has interaction $|A_{\rm ds}| > 0$. \\
Over each model, we first consider a conservative scenario with scales ranging from $\ell_{\rm{min}} = 30$ to $\ell_{\rm{max}} = 1500$.\footnote{\texttt{ReACT} and established methods like \texttt{HMcode} can reliably handle the $\ell_{\rm max} = 1500$ case.} For the second scale cut we extend the upper limit to $\ell_{\rm max} = 3000$. Finally, we also consider an upper limit of $\ell_{\rm max} = 5000$. As a larger $\ell_{\rm max}$ is considered for the scale cut, we expect to achieve greater sensitivity in the signal. 

\subsection{Mock data based on \texorpdfstring{$\Lambda$}CCDM}
\label{subsec:Fiducial_LCDM}

For our first set of forecasts we create a mock data vector described by a $\Lambda$CDM model with massive neutrinos, with parameter values taken from the Planck 2018+BAO \cite{Planck:2018vyg} best-fit analysis. As we mentioned above, we use the modelling described in \autoref{sec:shear}; the $C_{ij}^{\hat{\gamma} \hat{\gamma}}(\ell)$ spectra of Eq.~\eqref{eq:cell_shear_full} are binned over $30$-$\ell$ bins between $\ell_{\rm{min}} = 30$ and $\ell_{\rm{max}} = 5000$, and scale cuts\footnote{Keep in mind for surveys like Euclid $\ell \sim 3000$ is being conservative.} are subsequently applied. 

\begin{figure}[b!]
\centering 
\includegraphics[width=.75\textwidth]{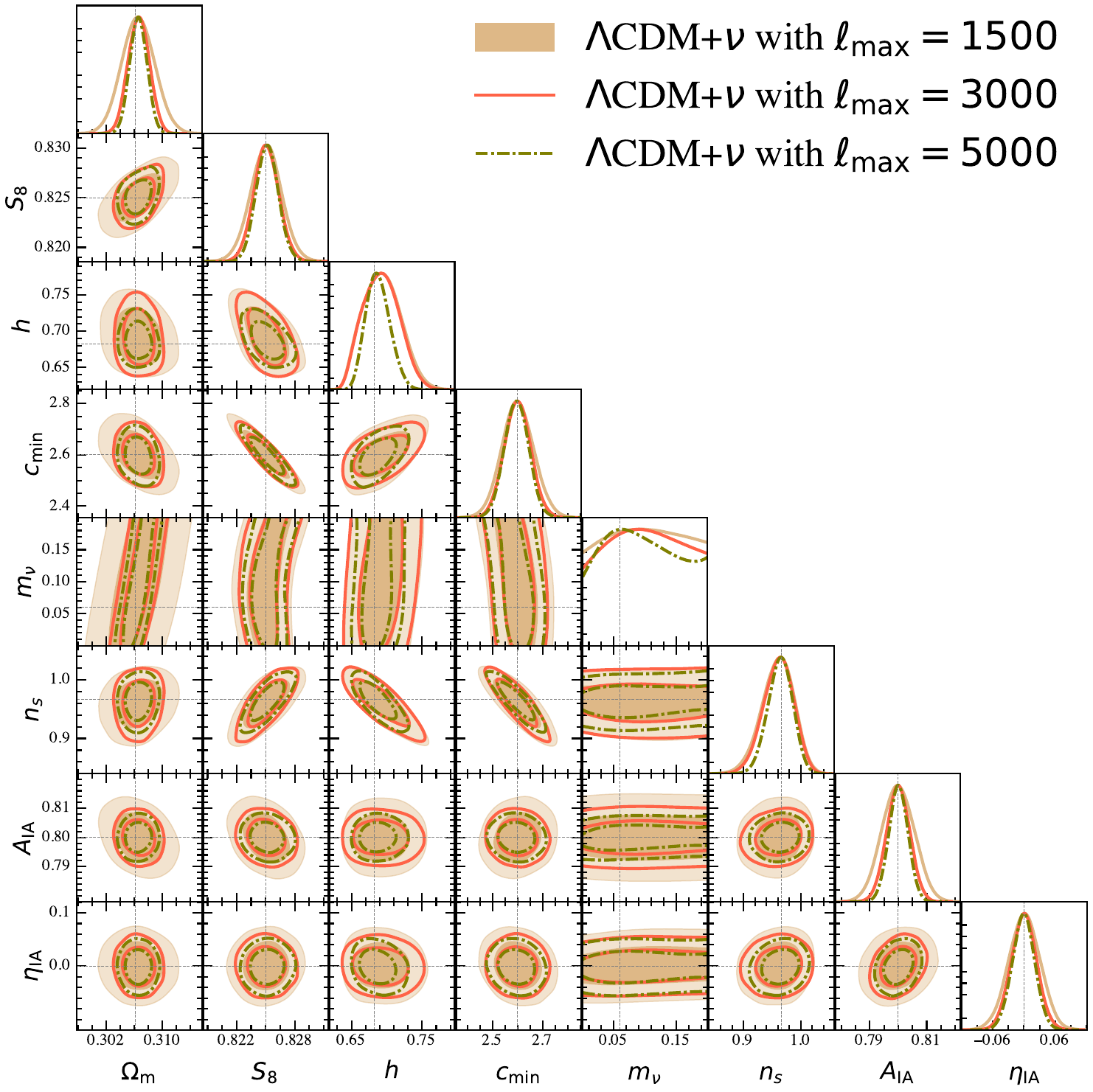}
\caption[$\Lambda$CDM constraints from $\Lambda$CDM data vector]{$68\%$ and $95\%$ 2D and 1D marginalised posterior contours for key cosmological and nuisance parameters for the $\Lambda$CDM model using three different scale cuts. The thin dashed-grey straight lines indicate the $\Lambda$CDM fiducial values used to generate the mock data vector.}
\label{fig:lcdm_comparison} 
\end{figure}

\noindent Our initial model for the analysis is $\Lambda$CDM, including contributions from baryonic feedback and massive neutrinos, i.e. the same model used to generate our mock data vector. The input parameters for the emulators are assigned uniform distributions, whereas the shifts in the redshift distribution follow a Gaussian prior (see \autoref{tab:priors_and_fiducial}). The second model we consider is $w$CDM, i.e. in practice we employ the same priors used for $\Lambda$CDM, but we also let $w$ free to vary uniformly between $[-1.3, -0.7]$. Finally, we also analyse the $\Lambda$CDM mock data assuming a DS model, with $A_{\rm ds}$ varying uniformly in $[0, 30]$ b/GeV, similar to our analysis of \autoref{Chapter6}.

\begin{figure}[b!]
\centering 
\includegraphics[width=.75\textwidth]{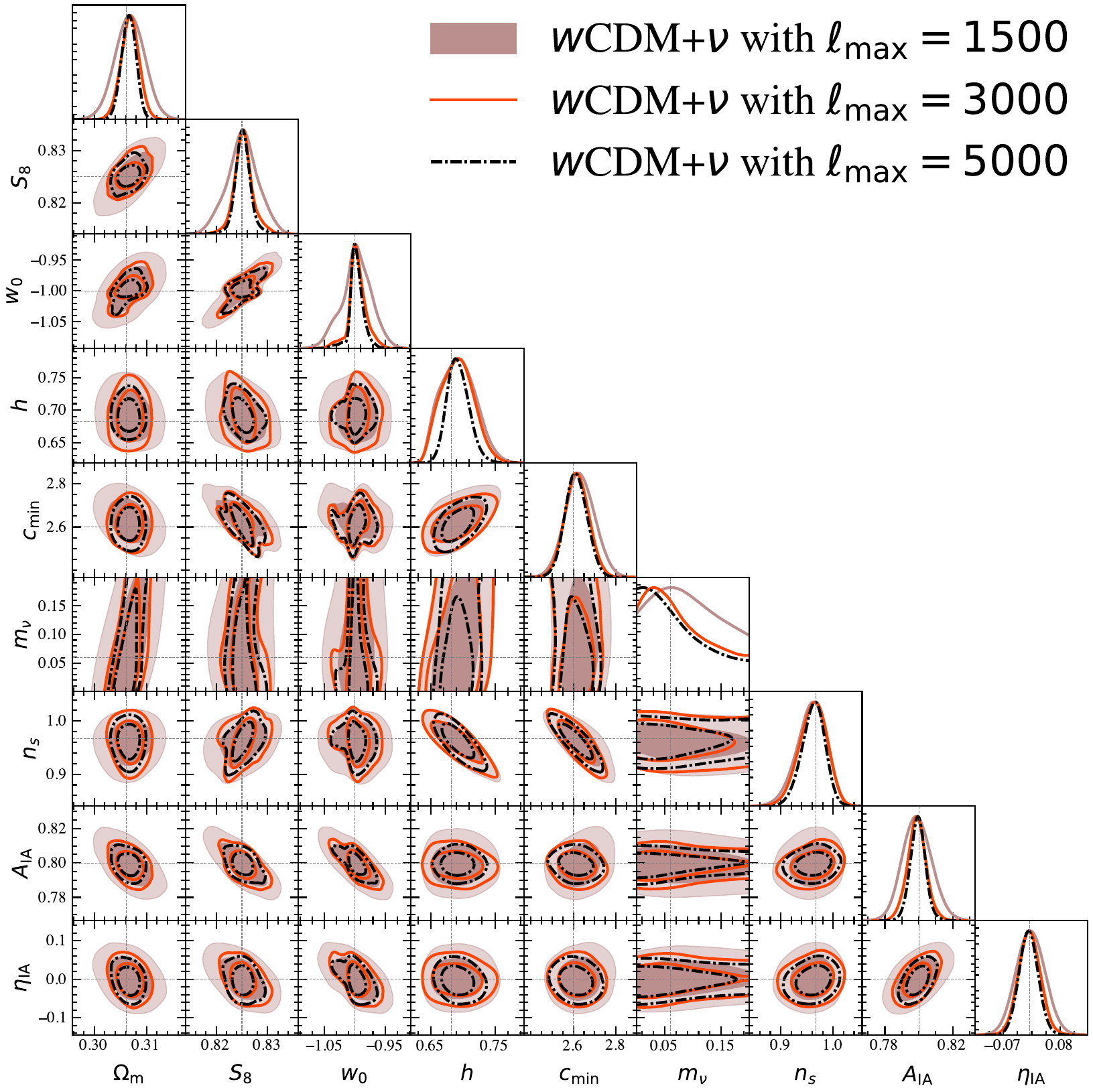}
\caption[$w$CDM constraints from $\Lambda$CDM data vector]{Same as in \autoref{fig:lcdm_comparison}, but for a $w$CDM model.}
\label{fig:wcdm_comparison} 
\end{figure}

\begin{figure}[b!]
\centering 
\includegraphics[width=.75\textwidth]{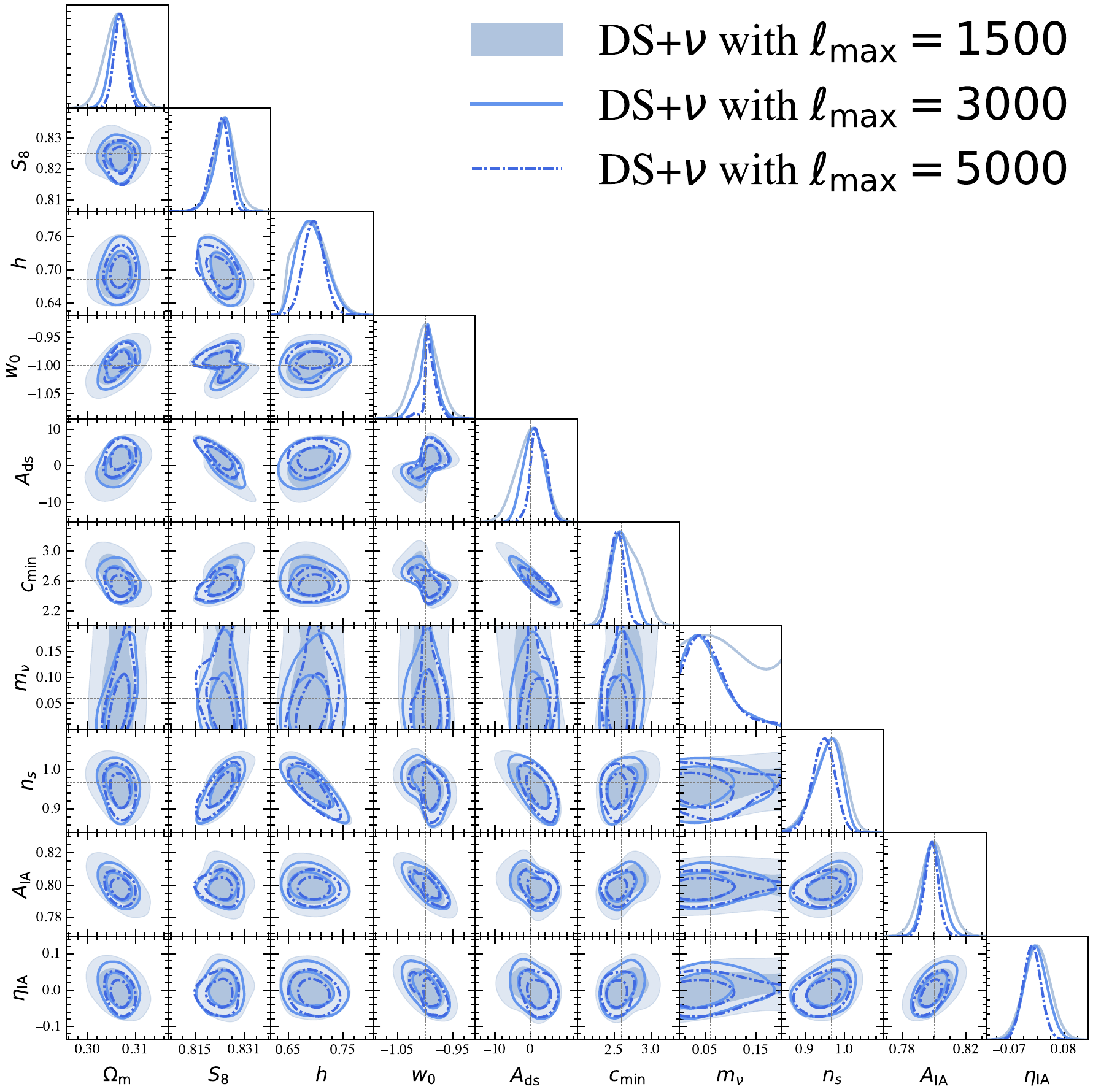}
\caption[DS constraints from $\Lambda$CDM data vector]{Same as in \autoref{fig:lcdm_comparison}, but for a DS model.}
\label{fig:ds_comparison} 
\end{figure}

\noindent \autoref{fig:lcdm_comparison}, \autoref{fig:wcdm_comparison} and \autoref{fig:ds_comparison} show results assuming a $\Lambda$CDM, $w$CDM and DS model, respectively. Once again, all contour plots in this chapter are obtained using \texttt{GetDist} \citep{2019arXiv191013970L}. In each corner plot, filled contours are used for the $\ell_{\mathrm{max}} = 1500$ case, solid contours are used for $\ell_{\mathrm{max}} = 3000$ and dashed contours for $\ell_{\mathrm{max}} = 5000$. In \autoref{Appendix_c}, we report two tables with marginalised $1\sigma$ constraints for key cosmological parameters and for all combinations of scale cuts and cosmological models. We do not include the final constraints on the redshift distribution shifts $D_{z_{i}}$ in the figures and table, as we find these are prior-dominated and do not show strong degeneracies with any other cosmological or nuisance parameter.\\
For all three models, we notice that increasing the scale cut from $\ell_{\rm max} = 1500$ to $\ell_{\rm max} = 3000$ leads to tighter constraints on $\Omega_{\rm m}$ and $S_8$, as expected, as these are the parameters best constrained by cosmic shear. However, going from $\ell_{\rm max} = 3000$ to $\ell_{\rm max} = 5000$ does not lead to major improvements in the constraints. Similarly, constraints on $m_{\nu}$ and $w_0$ reduce by a factor $2$ going from $\ell_{\mathrm{max}} = 1500$ to $3000$, but the inclusion of multipoles up to $\ell_{\mathrm{max}} = 5000$ does not lead to significant improvements. This is in line with the results of \cite{2023OJAp....6E..40S}, who found that including these highly nonlinear scales only mildly improves the constraining power on cosmological parameters if baryonic feedback parameters are \textit{a priori} unconstrained (as is the case here for $c_{\rm min}$), as baryonic parameters tend to absorb a large fraction of the constraining power. This motivates further research into ways to constrain the priors on these parameters (see e.g. for recent examples \cite{2024A&A...690A.188A,2024MNRAS.534..655B,2024PhRvL.133e1001F}). \\

\noindent When assuming a DS model, we observe that for all scale cuts, $\ell_{\rm max}$, the mean value is $w_0>-1$, leading to $A_{\rm ds}>0$, while the 1D marginalised posterior remains consistent with the fiducial value. We attribute the observed shift in $|A_{\rm ds}|$ primarily to parameter degeneracies, which may introduce biases, especially with baryonic feedback ($c_{\rm min}$), as shown in \autoref{fig:ds_comparison}. It is important to highlight the characteristic ``butterfly"-shaped contours in the $w_0$-$A_{\rm ds}$ plane associated with this model \citep{2021JCAP...10..004C,2024MNRAS.532.3914C,2025arXiv250203390T} due to the degeneracy between these two parameters -- cf. Eq.~\eqref{eq:Interaction_term_eff}. Overall, we find that $A_{\rm ds}$ is constrained with a marginalised $1\sigma$ uncertainty that is roughly an order of magnitude smaller than that obtained in our previous cosmic shear analysis of KiDS data \citep{2024MNRAS.532.3914C}, showcasing the promising constraining power of Stage IV configurations.

\noindent Interestingly, for all scale cuts the DS model produces uncertainties on the baryon feedback parameter $c_{\min}$ larger than both $\Lambda$CDM and $w$CDM. This can be attributed to the known degeneracy between the effects of baryonic feedback and the DS interaction on the matter power spectrum (see \autoref{fig:spectra_impact}). A similar degeneracy between the dark sector interaction and the effect of massive neutrinos on the matter power spectrum explains why the constraints lead to a well-defined 1D $m_\nu$ posterior distribution only for scale cuts at $\ell_{\mathrm{max}} = 3000$ and $5000$. Both of these degeneracies were already highlighted in \autoref{Chapter4} and \cite{2022MNRAS.512.3691C}, and we may confirm their impact on contours here.

\subsection{Mock data based on DS}
\label{subsec:Fiducial_DS}

The second set of forecasts we consider assumes a mock dataset generated under the assumption of a DS fiducial model; specifically, we assume an interaction value of $A_{\rm ds} = 10.6$ b/GeV for our mock data vector (see  \autoref{tab:priors_and_fiducial}). The choice of value is motivated by best-fit obtained from our previous analysis of \autoref{Chapter6}. 

\begin{figure}[b!]
\centering 
\includegraphics[width=.75\textwidth]{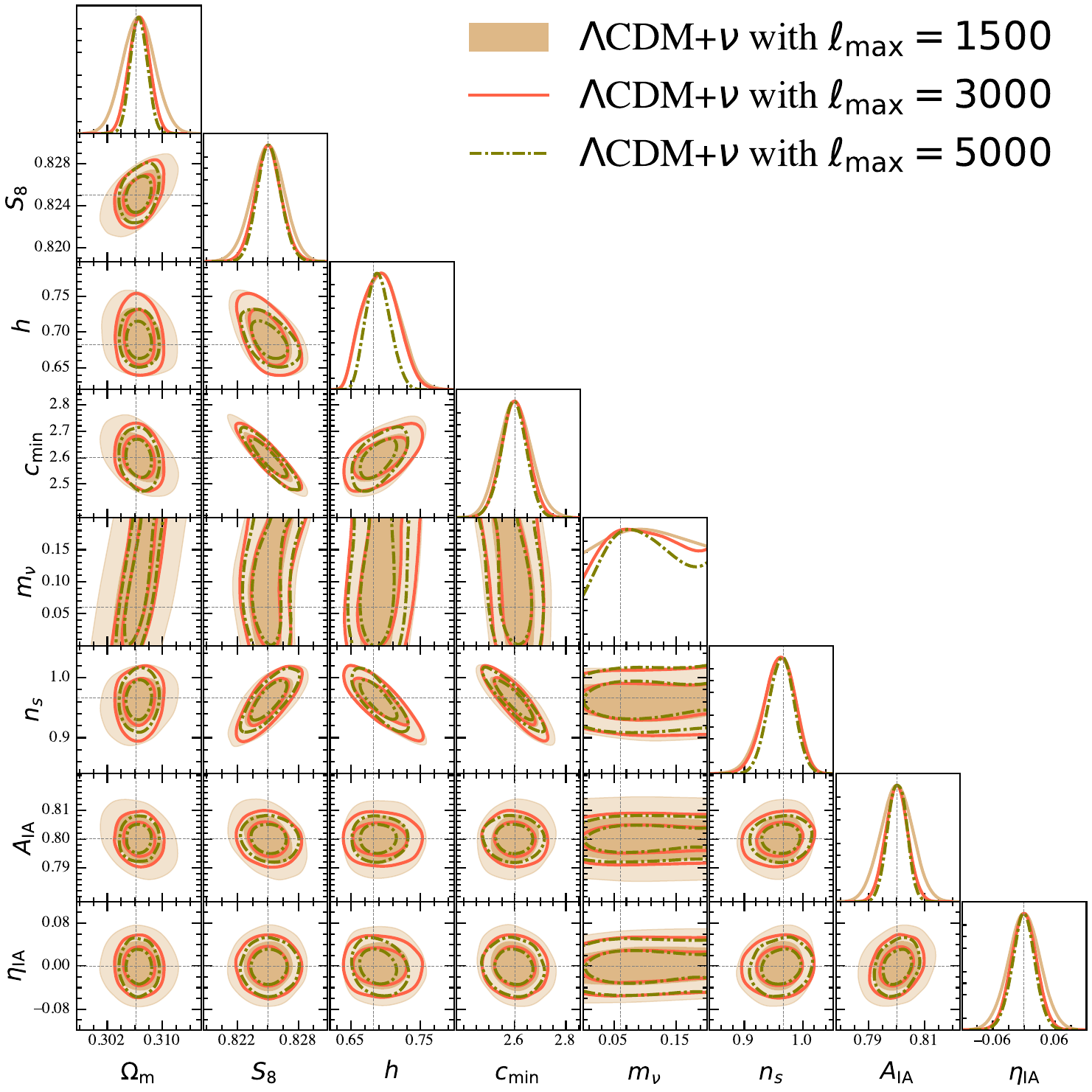}
\caption[$\Lambda$CDM constraints from DS data vector]{$68\%$ and $95\%$ 2D and 1D marginalised posterior contours for key cosmological and nuisance parameters for the $\Lambda$CDM model using three different scale cuts. The thin dashed-grey straight lines indicate the DS fiducial values used to generate the mock data vector.}
\label{fig:lcdm_comparison_f2} 
\end{figure} 

\begin{figure}[b!]
\centering 
\includegraphics[width=.75\textwidth]{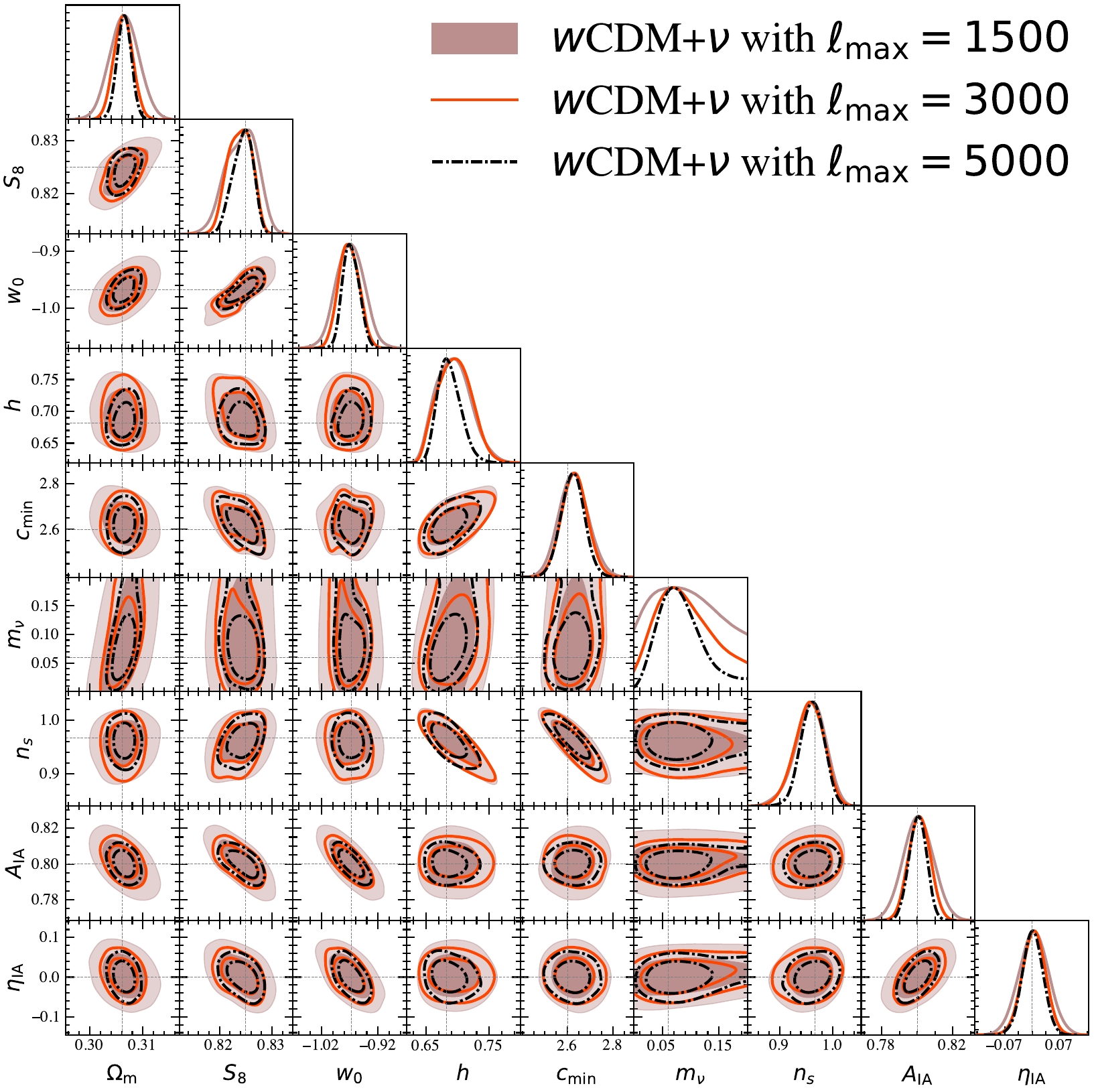}
\caption[$w$CDM constraints from DS data vector]{Same as in \autoref{fig:lcdm_comparison_f2}, but for a $w$CDM model.}
\label{fig:wcdm_comparison_f2} 
\end{figure}

\begin{figure}[t!]
\centering
\includegraphics[width=.75\textwidth]{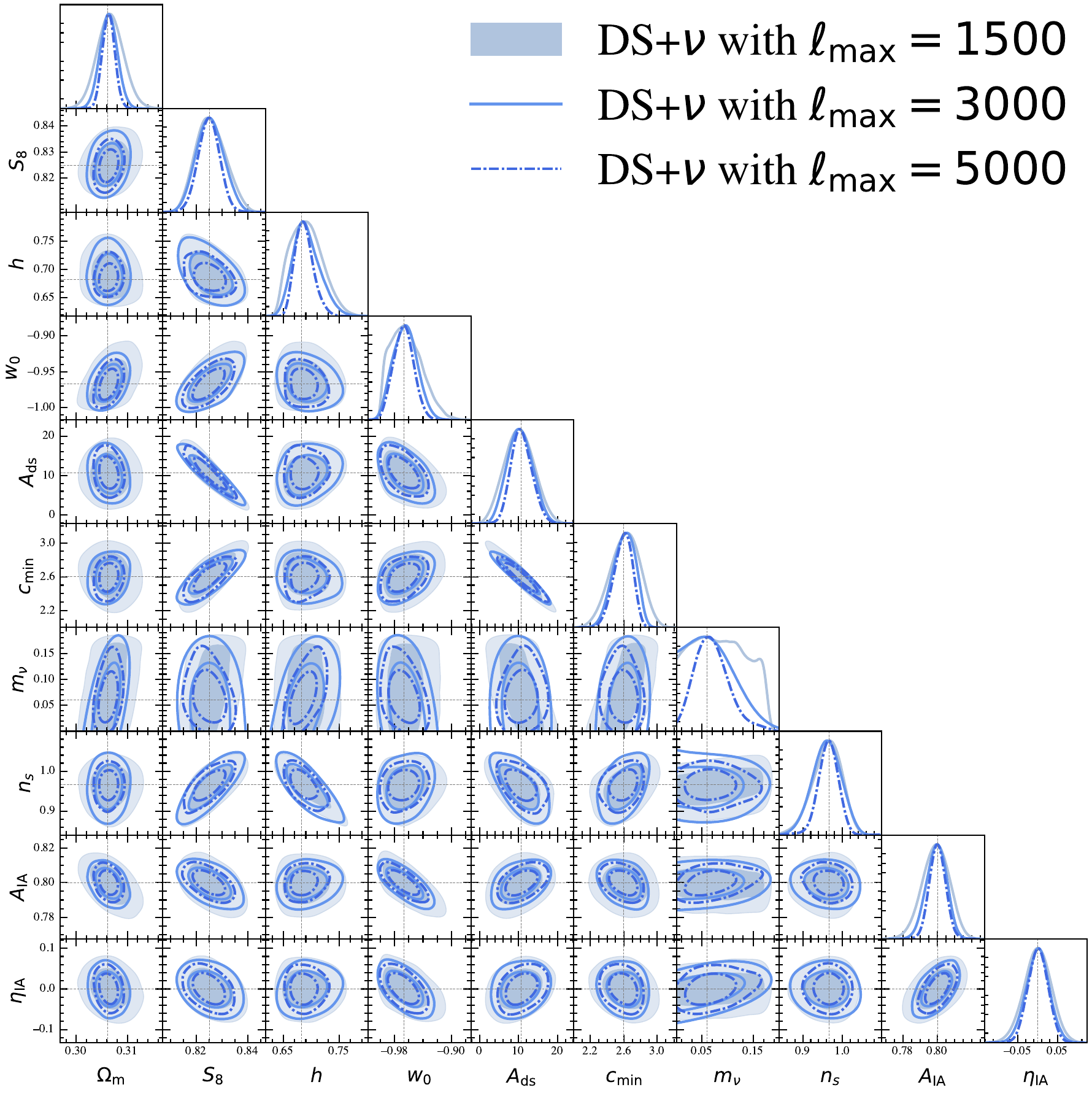} 
\caption[DS constraints from DS data vector]{Same as in \autoref{fig:lcdm_comparison_f2}, but for a DS model.}
\label{fig:ds_comparison_f2} 
\end{figure}

\noindent The results assuming a $\Lambda$CDM model are shown in \autoref{fig:lcdm_comparison_f2}, which indicates the best-fit values of the posteriors closely match the fiducial values. For the $w$CDM model, presented in \autoref{fig:wcdm_comparison_f2}, we note a very moderate bias in the baryonic feedback parameter $c_{\rm min}$. Moreover, in the $w$CDM case, the fiducial value of $w_0$ in the mock data imposes a better constraint than our first analysis and shows a correlation between $w_0$ and $S_8$. Finally, \autoref{fig:ds_comparison_f2} shows the posterior contours obtained by modelling our theory predictions with a DS model, with which we generated the mock data vector. The constraints on the DS amplitude parameter $A_{\rm ds}$ improve slightly when using larger values of the maximum angular scale: $A_{\rm ds}$ is constrained with a $1\sigma$ relative uncertainty $\sigma_{A_{\rm ds}}/A_{\rm ds} \sim 36\%,\, 30\%,\, 24\%$ for $\ell_{\rm max} = 1500, 3000, 5000$, respectively.

\noindent Importantly, the typical ``butterfly” structure is absent in this case, as the parameter $w$ is constrained to values $w>-1$. Additionally, an anti-correlation is observed between the parameter $A_{\rm ds}$ and both $S_8$ and $c_{\rm min}$. We refer to \autoref{Appendix_c} where is summarised the marginalised $1\sigma$ constraints for all parameters, reflecting the three combinations of scale cuts over all cosmological models.

\noindent Our forecasts demonstrate that several key parameters of the model are moderately sensitive to different $\ell_{\rm max}$, showing the impact of scale cuts on the final cosmological constraints. For example, considering a DS model for both the mock data and the modelling in the inference pipeline, we find that the relative $1\sigma$ uncertainty in the DS amplitude parameter $A_{\rm ds}$ goes from $36\%$ to $24\%$ as we increase $\ell_{\rm max}$ from $1500$ to $5000$. To put things into perspective, the constraints presented in this forecast show a promising future for constraining DS models with Stage IV cosmic shear data. These forecasts lead an increase in constraining power on $A_{\rm ds}$ of approximately an order of magnitude over Stage III cosmic shear configurations from \autoref{Chapter6}. As a potential future work, we will combine cosmic shear with galaxy clustering in a 3x2pt analysis \citep{2021A&A...646A.140H, DESY3:2022} 
to forecast the further improvement expected from a joint analysis of these probes. In addition, we will include a more complete modelling of baryonic feedback, e.g. using the \texttt{BCEmu} \citep{2021JCAP...12..046G} or \texttt{Bacco} \citep{2021MNRAS.506.4070A} emulators.

\subsection{Model comparison through harmonic estimator}
\label{subsec:model_comparison}

Regarding the model comparison results, the values of the logarithmic Bayes factor reported in \autoref{tab:harmonic_values}. As expected, given the $\Lambda$CDM mock data vector considered in this section, our results indicate a preference for the $\Lambda$CDM model; this is evident in the values of both the contours and the log-Bayes factors, $\log B$, the latter reported in \autoref{tab:harmonic_values}. For a given scale cut, $\Lambda$CDM generally shows tighter constraints compared to the other two models, particularly in $\Omega_{\rm m}$ and $S_8$, which is expected given the fewer degrees of freedom in $\Lambda$CDM, as compared to $w$CDM and DS. The values of the log-Bayes factors produced by \texttt{harmonic} present a coherent picture, confirming $w$CDM and DS to be disfavoured compared to $\Lambda$CDM. As expected, the DS model is even more disfavoured than the $w$CDM model due to the introduction of an additional parameter. We also note a trend in the log-Bayes factor computed between two models, namely that this value increases as $\ell_{\rm max}$ increases, for each pair of models compared in the analysis. This is also expected, since more information becomes available as we include more nonlinear scales. We note, however, that the uncertainty on the log-Bayes factor also increases with $\ell_{\rm max}$. This can be attributed to the shape of the posteriors shrinking at higher values of $\ell_{\rm max}$, which makes it more challenging for the normalising flow employed by \texttt{harmonic} to have tails narrower than its target distribution, i.e. the posterior. \\

\noindent We stress here that the reason for the particularly high values of the log-Bayes factors reported in this work is due to the noiseless data vectors considered for our forecasts, as is customary in the literature \citep{Euclid:2024yrr}. In realistic applications, noisy data would lead to lower values of the log-Bayes factor. However, even with noisy mock data we expect to see a similar trend in the impact of nonlinear scales on the values of the log-Bayes factor. This demonstrates the importance of including nonlinear information in cosmological inference pipelines not only to improve parameter estimation, but also model comparison. 

\begin{table}[t!]
\centering
\caption[Evidence values for different scale cuts $\ell_{\rm max}$, for $\Lambda$CDM and DS data vectors]{Evidence values for different scale cuts $\ell_{\rm max}$, for $\Lambda$CDM and DS mock data. We stress that, as typical in the literature for forecasts of this type, the fiducial data vector is assumed noiseless, which leads to high values of the logarithmic Bayes factor (see text for details).}
\label{tab:harmonic_values}
\renewcommand{\arraystretch}{2.0}
\setlength{\tabcolsep}{3.5pt}
\begin{tabular}{c||c c||c c}
\multicolumn{1}{c}{~} & \multicolumn{2}{c}{\textbf{$\Lambda$CDM mock data}} & \multicolumn{2}{c}{\textbf{DS mock data}} \\ 
$\ell_{\rm max}$ & $ \log \left( \dfrac{\mathcal{Z}_{\Lambda \rm CDM}}{\mathcal{Z}_{w\rm CDM}} \right)$ & $ \log \left( \dfrac{\mathcal{Z}_{\Lambda \rm CDM}}{\mathcal{Z}_{\rm DS}} \right)$ & $ \log \left( \dfrac{\mathcal{Z}_{\rm DS}}{\mathcal{Z}_{\rm \Lambda CDM}} \right)$ & $ \log \left( \dfrac{\mathcal{Z}_{\rm DS}}{\mathcal{Z}_{w \rm CDM}} \right)$\\ 
\hline \hline
$1500$ & $2.39^{\, +0.13}_{-0.14}$ & $4.22^{\, +0.41}_{-0.45}$ 
& $3.39^{\, +0.53}_{-0.62}$ & $1.40^{\, +0.34}_{-0.41}$ \\
$3000$ & $3.23^{\, +0.84}_{-1.13}$ & $5.30^{\, +0.45}_{-0.48}$ 
& $5.64^{\, +0.41}_{-0.35} $ & $4.20^{\, +0.32}_{-0.28}$  \\ 
$5000$ & $4.04^{\, +1.48}_{-1.18}$ & $
5.89^{\, +1.42}_{-1.86}$ 
& $6.10^{\, +1.15}_{-1.34}$ & $4.57^{\,+ 0.72}_{-0.81}$ \\ 
\end{tabular}
\end{table}

\noindent While DS is correctly identified by \texttt{harmonic} as the better fit to the data compared to both $w$CDM and $\Lambda$CDM, and values of the log-Bayes factors increase as one includes more nonlinear scales in the analysis.
Thus, we confirm for this set of forecasts the same trend reported in the \autoref{subsec:Fiducial_LCDM}. An important difference with respect to our previous analysis presented in \autoref{Chapter6} 
is the inclusion of varying massive neutrinos in the inference pipeline. Additionally, we perform model comparison using the learnt harmonic mean estimator, allowing us to straightforwardly estimate the evidence for the DS, $w$CDM and $\Lambda$CDM models through a process completely decoupled from the parameter estimation pipeline. \\

\noindent To sum up this chapter, we generate two forecasts through considering expected configurations for a Stage IV cosmic shear survey. We run our forecasts using a fully-differentiable \texttt{JAX} pipeline which allows us to use gradient-based methods for posterior sampling. Loading our neural emulators by \texttt{CosmoPower-JAX}, we incorporate the DS nonlinear spectrum, including the impact of massive neutrinos and baryon feedback. Notably, a few GPUs resources greatly accelerate computations in our pipeline, achieving orders-of-magnitude speedup comparable to consider hundreds of CPUs, especially for complex posteriors with more parameters and larger scale cuts. For instance, according to \cite{2023OJAp....6E..40S} an inference analysis of $14$-parameter with $\ell_{\rm max} = 5000$ took about a day on $512$ CPU cores. \\ 
Finally, we evaluate model evidence with the learnt harmonic mean estimator, that is fully decoupled from the inference process. This thus permits flexibility on the sampling method.
\chapter{Closing remarks \& Outlook}\label{Chapter8}

\vspace{1cm}

This thesis aimed to present a comprehensive and structured pipeline for exploring alternative cosmological models through refining nonlinear modelling, and using advanced tools to enhance the efficiency of Bayesian analysis for current and future galaxy surveys analyses. \\
The process involves modelling the nonlinearities inherent to the chosen cosmological model. Specifically, we have selected the halo model reaction formalism introduced in \autoref{Chapter2} to build a complete model for the nonlinear matter power spectrum of the model of choice:  the Dark Energy interacting model called Dark Scattering (DS), which is described in \autoref{Chapter4}. We did this by extending the halo model to include the additional force acting on dark matter particles, which required modifying the spherical collapse dynamics, as well as the virial theorem, in addition to the linear evolution. A graphical representation of the influence of the DS interaction amplitude, $A_{\rm ds}$, on the matter power spectrum is illustrated in \autoref{Fig:interaction_term_spectrum}. To our knowledge, this represented the first analytical model for the nonlinear power spectrum for an IDE model. We conjecture that many other models of momentum-exchange interactions could be readily described by this formalism, simply by modifying the time-dependence of the interaction function, $\Xi$, in Eq.~\eqref{Eq:Interaction_term}. Much of the formalism constructed here could be useful to generate nonlinear predictions for more complex interacting models. These extensions of DS model have been implemented into \texttt{ReACT}. As part of this effort, we conducted a validation test against high-resolution $N$-body simulations to assess the accuracy and reliability of our approach. We match the matter power spectrum output of simulations with the DS interaction presented in  \cite{2017MNRAS.465..653B}, using a pseudo spectrum generated with \texttt{HMCode2020}. In \autoref{Fig:react_w0} and \autoref{Fig:react_cpl}, we show that our predictions have $1\%$ agreement with simulations for scales up to $\mathcal{O}(1)~h/{\text{Mpc}}$ at redshift zero, improving beyond that at higher redshift. While at $z=1$, in the case of a constant $w$, the accuracy deviates by more than $1\%$ only beyond $\mathcal{O}(10)~h/{\text{Mpc}}$. For a scenario with significantly larger values ($\xi = 50$ b/GeV), the accuracy is only worse than $1\%$ up to $k = 2~h/{\text{Mpc}}$. In order to further extend the reach into even smaller scales, additional steps could be taken. This includes using a more accurate concentration-mass relation, fitted to simulations, which was shown by \cite{2020MNRAS.491.3101C} to improve the accuracy. In addition, improving the modelling of angular momentum loss during collapse could also enhance the accuracy, given that this is a crucial contributor to the effects of DS on the smallest scales \citep{2015MNRAS.449.2239B}. \\

\noindent We have also included the effects of baryonic feedback using \texttt{HMCode} \citep{2021MNRAS.502.1401M}; as well as massive neutrinos, using the full reaction formalism \citep{2021MNRAS.508.2479B}. We can thus generate predictions for the full power spectrum that will be directly probed by experiment. We have analysed the degeneracies between the DS interaction and those two ingredients by attempting to mimic the nonlinear effects of DS by varying the baryon feedback parameter or the neutrino mass. We find that limiting the scales to those for which our accuracy is within $1\%$ reveals a degeneracy in the case of baryon feedback. Extending beyond those scales, however, we expect that the degeneracy could be broken, given the stronger nonlinear effects generated in the interacting model. For the case of massive neutrinos, no significant degeneracy is found. \\

\noindent Afterwards in \autoref{Chapter5}, in view of the successful validation of the DS nonlinear spectrum against simulations with reaching percent-level accuracy, we show how to create a DS emulator trained with \texttt{CosmoPower}. As an initial step, we emulate the linear DS matter power spectrum to gain familiarity with \texttt{CosmoPower}, before extending this method to emulate the DS nonlinear spectra built upon the framework of the halo model reaction using \texttt{ReACT}. These emulators are parametrised by nine parameters, as specified in Eq.~\eqref{Eq:DS_parameters_training}, within a $5\sigma$ range of the Planck 2018 best-fit cosmological parameters \citep{Planck:2018vyg}, listed in \autoref{tab:range_of_emulators}. Additionally, we train an emulator for baryonic feedback, providing a correction to the matter power spectrum. During the training stage, the $k$-modes in the emulators are confined to the range $[10^{-3}, 10]~h/\text{Mpc}$, while the redshift $z$ spans from $0$ to $5$ and is treated as an additional input parameter for each emulator, as detailed in \autoref{tab:range_of_emulators}. \autoref{fig:acc_linear_ds}, \autoref{fig:acc_nonlinear_ds}, and \autoref{fig:acc_baryon} show the achieved accuracy of each emulator in comparison to the corresponding real predictions. The developed emulators are publicly accessible at \texttt{DS-emulators} repository. \\

\noindent With this computational tool at hand, in \autoref{Chapter6} we constrain the DS model using the approximately $1000~\text{deg}^2$ galaxy shear (the theoretical weak lensing aspects are covered and reviewed in \autoref{Chapter3} of the thesis) catalogue from the KiDS consortium. To enhance computational efficiency, we integrate our emulators directly into the inference pipeline, replacing calls to the Boltzmann code providing the linear and nonlinear spectra. Despite the emulation step is recommended; we emphasize the benefit of the emulator-based methodologies used in this thesis. By implementing emulators, we efficiently generated the contours, in this case for the KiDS-1000 pipeline within just a few minutes of computation on a $24$-core CPU machine (as shown in \autoref{Appendix_b}) -- an astonishing improvement in efficiency. \\

\noindent Our results show that the KiDS-1000 data constrains the DS parameter to be $\vert A_{\rm ds} \vert \lesssim 20$ b/GeV at $68\%$ C.L., as displayed in \autoref{Fig:K1K_DS_free_CMB}. Thus, we interpreted that the KiDS-1000 cosmic shear catalogue is sensitive to a combination of the growth history of this IDE model (i.e. the redshift evolution of $\sigma_8(z)$) and its specific nonlinear effects. \\
In the joint analysis of KiDS-1000 with CMB and BAO information we obtain a stronger constraint on the DS model parameter, finding now $A_{\rm ds} = 10.6^{+4.5}_{-7.3}$ b/GeV (for COSEBIs, see \autoref{Fig:K1K_DS_CMB} for all cases). We find that the combined analysis favours positive values of $A_{\rm ds}$, since these lead to a reduction of the amplitude of the matter power spectrum at late times, as illustrated in \autoref{fig:spectra_impact}. While in the KiDS-only case, this reduction could be compensated by an increase of the primordial amplitude, the CMB and BAO information essentially breaks that degeneracy, thus allowing for a clearer determination of the interaction amplitude $A_{\rm ds}$. Accordingly, \autoref{Fig:cosebis_all_models} shows that the DS model is consistent with early and late-Universe measurements, thus offering a viable approach to alleviate the $S_8$ tension linked to the measured value of $A_{\rm ds}$. \\

\noindent Our methodology has at reach the analysis of different weak lensing probes, such as DES-Y3, or the joint analysis of both of them (KiDS-1000 + DES-Y3), with the purpose of constraining the DS model or any other beyond $\Lambda$CDM model. These emulators can also be readily used for analyses including photometric galaxy clustering in a 3x2pt analysis or other probes \citep{2023PhRvD.107h3526T,2025arXiv250203390T}, and can also be easily extended to constrain the DS model with a time-dependent $w(z)$. Additionally, these emulators will allow for an accelerated exploitation of the Stage IV data that will become available from Euclid and Rubin's LSST in the forthcoming years. There, the importance of this fast analysis tool is even more critical, given the huge nuisance-parameter spaces that need to explored. Moreover, the implications of adopting alternative models for Intrinsic Alignment, baryonic feedback and other systematics, can be more efficiently explored in the future with these emulators. Thus allowing for a thorough investigation of their impact on the inference of the DS parameter and its degeneracies with e.g. baryonic feedback, or massive neutrinos.\\

\noindent To illustrate this, in \autoref{Chapter7} we evaluate the applicability of our methodology by testing the DS model through forecasting the expected power of the next-generation cosmological surveys. To this end, we generate synthetic cosmic shear data for a Stage IV survey configuration, employing the \texttt{jax-cosmo} library. By embedding our fast and precise \texttt{CosmoPower} emulators within a fully-differentiable inference pipeline in \texttt{JAX} that runs on GPUs, we drastically accelerate Bayesian inference with employing a gradient-based sampling by \texttt{NUTS}. Such enhancements are crucial for extracting precise and accurate insights from the extensive datasets, as those anticipated to be the product of Stage IV galaxy surveys; moreover, the accelerated analysis allows us to consider different fiducial models, thereby exploring the different possibilities that the data may present. \\ 
Our presented forecasts demonstrate that several key parameters of the model are moderately sensitive to different scale cuts ($\ell_{\rm max}$) in the final cosmological constraints. For example, considering a DS model for both the mock data and the modelling in the inference pipeline, we find that the relative $1\sigma$ uncertainty on the DS amplitude parameter $A_{\rm ds}$ goes from $36\%$ to $24\%$ as we increase the shear signal through $\ell_{\rm max}$ from $1500$ to $5000$. We also investigate degeneracies between dark energy and systematics parameters for those scale cuts, where we highlighted the importance of setting strong priors on baryonic feedback parameters \citep{2024A&A...690A.188A,2024MNRAS.534..655B,2024PhRvL.133e1001F}, to ensure that including more nonlinear scales in the analysis effectively leads to a significant increase in constraining power. \\

\noindent Whereas, assuming a DS model for the mock data vector, we find that a Stage IV survey cosmic shear analysis can constrain the DS amplitude parameter $A_{\mathrm{ds}}$ with an uncertainty roughly an order of magnitude smaller than current constraints from Stage III surveys presented in \autoref{Chapter6}, even after marginalising over baryonic feedback, intrinsic alignments and redshift distribution uncertainties. These results show great promise for constraining DS with Stage IV data. An important aspect over our previous work \citep{2024MNRAS.532.3914C} is the inclusion of varying massive neutrinos within the inference pipeline. 
Additionally, we perform model comparison using the learnt harmonic mean estimator, which allows us to straightforwardly estimate the evidence for the DS, $w$CDM, and $\Lambda$CDM models. This process is completely decoupled from the parameter estimation pipeline and is directly applied to the output chains from the inference process, using the harmonic mean estimator implemented (also mentioned in \autoref{Chapter1}) in the software \texttt{harmonic}. It is important to emphasize that our \texttt{JAX} ecosystem pipeline can be easily adapted to study a wide range of dark energy and modified gravity models or other cosmological models beyond $\Lambda$CDM. We are producing parameter constraints from simulated Stage IV cosmic shear data running on a single GPU, rather than relying on hundreds of CPUs, which may not be accessible to everyone. Furthermore, the pipeline can scale to even higher dimensional parameter spaces (see \cite{2024OJAp....7E..73P} for details), making it suitable for more complex cosmological analyses involving additional systematic parameters or new physics. 
In particular, the framework presented here can be integrated with complementary cosmological probes such as CMB lensing, the Sunyaev-Zel’dovich effects, and Redshift Space Distortions to place tighter constraints on extensions to the standard $\Lambda$CDM model. Such multi-probe analyses will be essential in the context of upcoming Stage IV surveys, as demonstrated by recent findings from DESI and KiDS consortium. \\

\noindent As we hope to have illustrated in the present  thesis, it is evident that the future of cosmology lies in the breakthroughs enabled by the vast potential of next-generation galaxy surveys, which are set to usher in an era of unparalleled statistical precision. As these surveys unfold, they will not only represent an opportunity to resolve existing cosmological tensions, but also they will fill the gaps in our current understanding regarding the dark matter and dark energy.\\
Nonetheless, by pushing the boundaries of observational cosmology in LSS, these efforts to strengthen the connection between observation and theory entail refining our modelling, driving the development of innovative methodologies and sophisticated tools, and opening new pathways for making efficient statistical inference approaches -- such like the effort presented in this work. \\

\noindent Looking ahead, as cosmological data continues to evolve, such new era promises to reshape the way we view the Universe. We hope that the results, methodologies, and perspectives developed in this thesis will serve as a foundation for future cosmological analyses, pushing closer to open doors to answers that remain just beyond our reach today. \vfill \textit{``For those who come after..."}

\newpage

\bibliography{references} 

\newpage
\thispagestyle{empty}

\newpage 

\begin{appendices}
\appendixtrue 

\renewcommand{\chaptername}{Appendix} 
\renewcommand{\thechapter}{\Alph{chapter}} 

\chapter{}
\label{Appendix_a}

\vspace{1cm}

Before recombination, it is well-known that baryon-photon perturbations experienced acoustic oscillations, while dark matter perturbations grew under gravity, seeding the formation of large-scale dark matter filaments and structures. Hence, here we focus solely on dark matter perturbations, treating them as a dust-like fluid. In this \autoref{Appendix_a}, we derive the Newtonian perturbations following the notation of \cite{2002PhR...367....1B}. The Lagrangian for a single particle in a gravitational potential $\phi$ is
\bea
L=\frac{1}{2} m \dot{r}^2 - m \phi \, .
\label{Eq:lagrangian_single}
\eea

\noindent After using the Euler-Lagrange equation, we derive the equation of motion (e.o.m.) given by, 
\bea
\dot{\boldsymbol{u}}= -H \boldsymbol{u} -\nabla_{\bm{r}}\Phi \, , \qquad \text{or in comoving coordinates} \qquad
\dfrac{\partial \boldsymbol{p}}{\partial \eta } = - ma \nabla_{\bm{x}} \Phi \, ,
\label{Eq:eom}
\eea

\noindent being $\boldsymbol{p} = m \, a \, \boldsymbol{u}$. Hereafter, we use comoving coordinates $\boldsymbol{r} = a(t)\boldsymbol{x}$ where $\boldsymbol{x}$ is referred to as the comoving distance, and conformal time $\eta$ with $d\eta = \frac{dt}{a(t)}$. In addition, allows us to express the conformal Hubble parameter as $\mathcal{H} = a H$.

\subsubsection{Vlasov equation} In physics the Vlasov equation is used to describe dissipative systems of interactive particles by defining the particle number density in the phase-space $f(\boldsymbol{x},\boldsymbol{p},\eta)$. Vlasov described an infinite chain of self-linking equations for the distribution functions of random variables (see \cite{2018arXiv181109424P}). Thus, the phase-space conservation implies,  

\bea
\dfrac{d f}{d\eta} =  \dfrac{\partial f}{ \partial \eta}
+ \dfrac{\partial f}{ \partial \boldsymbol{x}} \cdot \dfrac{\partial \boldsymbol{x}}{ \partial \eta}
+ \dfrac{\partial f}{ \partial \boldsymbol{p}} \cdot \dfrac{\partial \boldsymbol{p}}{ \partial \eta} = 0 \, .
\label{Eq:Vlasov_classical}
\eea

\noindent This is called the Vlasov equation. From Eq.~\eqref{Eq:eom}, we derive:
\bea
\dfrac{d f}{d\eta} =  \dfrac{\partial f}{ \partial \eta}
+ \dfrac{\boldsymbol{p}}{ma} \cdot  \dfrac{\partial f}{ \partial \boldsymbol{x}} 
- ma \nabla_{\bm{x}} \Phi \cdot \dfrac{\partial f}{ \partial \boldsymbol{p}} = 0 \, .
\label{Eq:Vlasov_classical_2}
\eea

\noindent The zeroth-order moment relates the phase-space density to the conformal density field through the next equation,
\bea
\int d^3 \boldsymbol{p} \  f(\boldsymbol{x},\boldsymbol{p}, \eta) 
\equiv \rho(\boldsymbol{x}, \eta) \, .
\label{Eq:zero_moment}
\eea

\noindent Moreover, the first moment is related to the velocity field of particles,
\bea
\int d^3 \boldsymbol{p} \ \dfrac{\boldsymbol{p}}{ma} f(\boldsymbol{x},\boldsymbol{p}, \eta) 
\equiv \rho(\boldsymbol{x}, \eta)\boldsymbol{u}(\boldsymbol{x}, \eta) \, .
\label{Eq:first_moment}
\eea

\noindent Finally, to derive the Euler equation, we integrate only up to the second moment,
\bea
\int d^3 \boldsymbol{p} \ \dfrac{p_i p_j}{m^2 a^2} f(\boldsymbol{x},\boldsymbol{p}, \eta) 
\equiv \rho(\boldsymbol{x}, \eta) u_i u_j + \sigma_{ij} \, .
\label{Eq:second_moment}
\eea

\noindent The term $\sigma_{ij}$ represents generalized pressure forces. In the absence of any pressure perturbation, as is the case for cold pressure-less matter, therefore, we have $\sigma_{ij} = 0$. \\
The continuity equation is derived by performing a $\boldsymbol{p}$-integration of Eq.~\eqref{Eq:Vlasov_classical_2} and applying the definition of both moments from Eq.~\eqref{Eq:zero_moment} and Eq.~\eqref{Eq:first_moment}. This results in the equation expressed in terms of the density contrast $\delta(\boldsymbol{x}, \eta)$:
\bea
\dfrac{\partial \delta}{\partial \eta} + \nabla_{\bm{x}} \cdot [(1 + \delta) \boldsymbol{u}] = 0 \, .
\label{Eq:Continuity_delta}
\eea

\noindent Subsequently, the Euler equation is derived by considering the first moment\footnote{Bear in mind that the distribution function $f$ evaluated at the boundary vanishes, since $f$ does not include particles with infinite momentum.} of Eq.~\eqref{Eq:Vlasov_classical_2}, simplifying the resulting terms, and arriving at the following expression:
\bea
\dfrac{\partial \boldsymbol{u}}{\partial \eta} 
+ (\boldsymbol{u} \cdot \nabla_{\bm{x}}) \boldsymbol{u}   
+ \mathcal{H} \boldsymbol{u} + \nabla_{\bm{x}} \Phi  = 0 \, .
\label{Eq:Euler}
\eea

\noindent Ultimately, applying the divergence and substituting the Poisson equation in comoving coordinates (see Eq.~\eqref{Eq:Poisson_comoving}), 
\bea
\dfrac{\partial \theta}{\partial \eta} 
+ \nabla_{\bm{x}} \cdot (\boldsymbol{u} \cdot \nabla_{\bm{x}}) \boldsymbol{u}   
+ \mathcal{H} \theta + \dfrac{3}{2} \Omega_{\rm m} \mathcal{H}^2 \delta  = 0 \, ,
\label{Eq:Euler_theta}
\eea

\noindent where we introduce the velocity divergence $\theta \equiv \nabla_{\bm{x}} \cdot  \boldsymbol{u}$, commonly referred to as the volume expansion rate (the 3D analogue of the Hubble parameter $H$). It is important to note that on the Euler equation the curl contribution of the velocity $\omega \equiv \nabla_{\bm{x}} \times \boldsymbol{u}$, known as the vorticity, has been neglected. This is because the vorticity decays as $\omega \propto 1/a$ with the expansion, and thus becomes negligible at late times.

\section*{Linear regime}
\label{sec:linear_regime}

The nonlinear Eq.~\eqref{Eq:Continuity_delta} and Eq.~\eqref{Eq:Euler}, can be linearised by removing the higher-orders. The continuity equation then becomes:
\bea
\dfrac{\partial \delta}{\partial \eta} + \theta = 0 \, ,
\label{Eq:Continuity_linear}
\eea

\noindent Next, we linearise the Euler equation in Eq.~\eqref{Eq:Euler_theta} and work in Fourier space, where the $k$-modes evolve independently at linear order, yielding:
\bea
\dfrac{\partial \tilde{\theta}_{\bm{k}}}{\partial \eta} 
+ \mathcal{H}  \tilde{\theta}_{\bm{k}} - k^2 \Phi = 0 \, .
\label{Eq:Euler_linear}
\eea

\noindent By combining Eq.~\eqref{Eq:Poisson_comoving}, Eq.~\eqref{Eq:Continuity_linear}, and Eq.~\eqref{Eq:Euler_linear}, we obtain an equation solely for the density contrast:
\bea
\dfrac{\partial^2 
\tilde{\delta}_{\bm{k}}(\eta)}{\partial \eta^2} + \mathcal{H} \dfrac{\partial 
\tilde{\delta}_{\bm{k}}(\eta)}{\partial \eta} - \dfrac{3}{2} \Omega_{\rm m} \mathcal{H}^2 
\tilde{\delta}_{\bm{k}}(\eta)  = 0 \, .
\label{Eq:delta_eq_linear}
\eea

\noindent The general solution is expressed in  Eq.~\eqref{Eq:delta_lin}. From such solution, we obtain $\theta$ via Eq.~\eqref{Eq:Continuity_linear}, using the relation $d/d \ln D_+ = (\mathcal{H} f)^{-1} d/d \eta$, therefore we obtain,
\bea
\tilde{\theta}_{\bm{k}}(\eta) =  - \mathcal{H}(\eta) f A_{\bm{k}} \, .
\label{Eq:theta_sol}
\eea

\noindent With $f$ being the growth rate function from Eq.~\eqref{Eq:f_growth}.

\section*{Nonlinear regime}
\label{sec:non-linear_regime}

\noindent In this context, the linear approximation holds on large scales, where the density contrast satisfies $\delta \ll 1$. However, starting from an initial value of $\delta \approx 10^{-5}$, these perturbations grow on sub horizon scales, and the dynamics become significantly different. The linear regime is no longer valid once $\delta$ approaches close to one. Our task now is to extend CPT to the nonlinear regime. To achieve this, the nonlinear terms in Eq.~\eqref{Eq:Continuity_delta} and Eq.~\eqref{Eq:Euler} are repositioned to the r.h.s., expressing the equations as follows:
\bea
\delta' + \theta &=& -\delta \theta - (\boldsymbol{u} \cdot \nabla_{\bm{x}} \delta) \, . 
\label{Eq:continuity_new} \\ 
\theta' + \mathcal{H}  \theta + \dfrac{3}{2} \Omega_{\rm m} \mathcal{H}^2 \delta   &=& -(\boldsymbol{u} \cdot \nabla_{\bm{x}}) \theta - \left[ \nabla_{\bm{x}} ( \boldsymbol{u} \cdot \nabla_{\bm{x}} )  \cdot \boldsymbol{u} \right] \, .
\label{Eq:euler_new}
\eea

\noindent As previously, it is beneficial to describe the matter fields in Fourier space. To extend the solutions beyond the linear order, fields will be expressed as infinite perturbative series, by setting them from configuration space into Fourier space as follows, 
\bea
\Theta(\boldsymbol{x}, \eta) =\sum^{\infty}_{n=1} \Theta^{(n)}(\boldsymbol{x}, \eta) \, , \quad \Rightarrow \quad \widetilde{\Theta}(\boldsymbol{k}, \eta) =\sum^{\infty}_{n=1} \widetilde{\Theta}^{(n)}(\boldsymbol{k}, \eta) \, . 
\label{Eq:Theta_field}    
\eea

\noindent The superscript $(n)$ establishes the perturbative order, e.g. the density field case $\delta^{(1)}$ and $\theta^{(1)}$ represents the fields at linear order. The Fourier mapping of Poisson equation holds at each perturbative order,\footnote{A similar relation is present for the peculiar velocity, but only at first order, where it is given by: \\ $\tilde{u}^{(1)}(\boldsymbol{k}, \eta) = i \dfrac{\boldsymbol{k}}{k^2} \mathcal{H} f(\eta) \tilde{\delta}^{(1)}(\boldsymbol{k}, \eta).$} as it provides a linear relation between the gravitational potential and the density contrast. Then, leading to,
\bea
\nabla^2_{\bf{x}} \Phi^{(n)}(\boldsymbol{x}, \eta) = \dfrac{3}{2} \Omega_{\rm m} \mathcal{H}^2 \delta^{(n)}(\boldsymbol{x}, \eta) \, , \quad \Rightarrow \quad
 \tilde{\Phi}^{(n)}(\boldsymbol{k}, \eta) = - \dfrac{3}{2} \Omega_{\rm m} \dfrac{\mathcal{H}^2}{k^2} \tilde{\delta}^{(n)}(\boldsymbol{k}, \eta) \, .
\label{Eq:Poisson_fourier}
\eea

\noindent As stated earlier, at very late times, the linear order can be rewritten as,
\bea
\delta^{(1)}(\boldsymbol{x}, \eta) = \delta_0(\boldsymbol{x}) D_+(\eta) \, . 
\label{Eq:delta_sol_1st}
\eea

\noindent We normalise the field at the present time $\eta_0$, such that $\delta_0(\boldsymbol{x}) = \delta(\boldsymbol{x}, \eta_0)$ and the velocity field is given by Eq.~\eqref{Eq:theta_sol}. To address the nonlinearities, we follow the standard approach \citep{2002PhR...367....1B,2020moco.book.....D}. Specifically, we focus on the nonlinear terms from Eq.~\eqref{Eq:continuity_new} and Eq.~\eqref{Eq:euler_new} by replacing the series approach of each perturbative quantity as follows:
\bea
\sum^{\infty}_{n=1} \delta'^{(n)} + \sum^{\infty}_{n=1} \theta^{(n)} &=& - \sum^{\infty}_{n=1} \sum^{\infty}_{k=1} \delta'^{(n)} \theta^{(k)} \notag\\ & & - \sum^{\infty}_{n=1} \sum^{\infty}_{k=1} \left(\boldsymbol{u}^{(n)} \cdot \nabla_{\bm{x}} \delta^{(k)} \right) \, . 
\label{Eq:continuity_new_n} \\ 
\sum^{\infty}_{n=1} \theta'^{(n)}  + \mathcal{H} \sum^{\infty}_{n=1} \theta^{(n)} + \dfrac{3}{2} \Omega_{\rm m} \mathcal{H}^2 \sum^{\infty}_{n=1} \delta^{(n)}  &=& - \sum^{\infty}_{n=1} \sum^{\infty}_{k=1} (\boldsymbol{u}^{(n)}  \cdot \nabla_{\bm{x}}) \theta^{(k)} \notag\\ & & - \sum^{\infty}_{n=1} \sum^{\infty}_{k=1} \left[ \nabla_{\bm{x}} ( \boldsymbol{u}^{(n)} \cdot \nabla_{\bm{x}} )  \cdot \boldsymbol{u}^{(k)} \right] \, .
\label{Eq:euler_new_n}
\eea

\noindent In accordance with the perturbative order hierarchy, each term in the equation will be truncated based on the imposed order, as outlined below:
\begin{align}
\text{1st order:} \qquad  \delta'^{(1)} + \theta^{(1)}  =& \ 0.  \\
\theta'^{(1)} + \mathcal{H}  \theta^{(1)} + \dfrac{3}{2} \Omega_{\rm m} \mathcal{H}^2 \delta^{(1)}  =& \ 0. \\ \notag  \\ 
\text{2nd order:} \qquad \delta'^{(2)} + \theta^{(2)} =& -\delta^{(1)} \theta^{(1)} - \boldsymbol{u}^{(1)} \cdot \nabla_{\bm{x}} \delta^{(1)}. \label{Eq:continuity_2nd} 
\\ \theta'^{(2)} + \mathcal{H}  \theta^{(2)} + \dfrac{3}{2} \Omega_{\rm m} \mathcal{H}^2 \delta^{(2)}   =& -(\boldsymbol{u}^{(1)} \cdot \nabla_{\bm{x}}) \theta^{(1)} - \left[ \nabla_{\bm{x}} ( \boldsymbol{u}^{(1)} \cdot \nabla_{\bm{x}} )  \cdot \boldsymbol{u}^{(1)} \right]. \label{Eq:Euler_2nd} \\ \notag \\
\text{3rd order:} \qquad \delta'^{(3)} + \theta^{(3)}  =& -\delta^{(2)} \theta^{(1)} -\delta^{(1)} \theta^{(2)} - \boldsymbol{u}^{(2)} \cdot \nabla_{\bm{x}} \delta^{(1)} - \boldsymbol{u}^{(1)} \cdot \nabla_{\bm{x}} \delta^{(2)}.  
\\ \theta'^{(3)} + \mathcal{H}  \theta^{(3)} + \dfrac{3}{2} \Omega_{\rm m} \mathcal{H}^2 \delta^{(3)}   =& -(\boldsymbol{u}^{(2)} \cdot \nabla_{\bm{x}}) \theta^{(1)} -(\boldsymbol{u}^{(1)} \cdot \nabla_{\bm{x}}) \theta^{(2)}  \notag\\ & - \left[ \nabla_{\bm{x}} ( \boldsymbol{u}^{(2)} \cdot \nabla_{\bm{x}} )  \cdot \boldsymbol{u}^{(1)} \right]    - \left[ \nabla_{\bm{x}} ( \boldsymbol{u}^{(1)} \cdot \nabla_{\bm{x}} )  \cdot \boldsymbol{u}^{(2)} \right].  \\ 
\vdotswithin{=}& \notag 
\end{align}

\noindent For illustrative purposes, we explore only to second-order perturbations. By transforming Eq.~\eqref{Eq:continuity_2nd} and Eq.~\eqref{Eq:Euler_2nd} to Fourier space, and employing the convolution theorem along with the properties of the Dirac delta function $\delta_{\rm D}$, we can derive the following results:
\bea
\tilde{\delta}'^{(2)}(\boldsymbol{k}, \eta) 
+ \tilde{\theta}^{(2)}(\boldsymbol{k}, \eta)  & = &  \int \dfrac{d^3 k_1}{(2\pi)^3}  \int d^3 k_2 \  \delta_{\rm D}(\boldsymbol{k} - \boldsymbol{k}_{1} - \boldsymbol{k}_{2}) \left[ 1  +  \dfrac{(\boldsymbol{k}_{1} \cdot \boldsymbol{k}_{2})}{k^2_1} \right] 
\notag \\ 
& & \times \left[ \mathcal{H} f(\eta) \tilde{\delta}^{(1)}(\boldsymbol{k}_{1}, \eta)\tilde{\delta}^{(1)}(\boldsymbol{k}_{2}, \eta) \right] \, ,
\label{Eq:Continuity_2nd_FT}
\\
\tilde{\theta}'^{(2)} + \mathcal{H}  \tilde{\theta}^{(2)} + \dfrac{3}{2} \Omega_{\rm m} \mathcal{H}^2 \tilde{\delta}^{(2)}   &=&  \int \dfrac{d^3 k_1}{(2\pi)^3}  \int d^3 k_2 \  \delta_{\rm D}(\boldsymbol{k} - \boldsymbol{k}_{1} - \boldsymbol{k}_{2}) \left[ \dfrac{(\boldsymbol{k}_{1} \cdot \boldsymbol{k}_{2})}{k^2_1} + \dfrac{(\boldsymbol{k}_{1} \cdot \boldsymbol{k}_{2})^2}{k^2_1 k^2_2} \right] 
\notag \\ 
& & \times \left[ (\mathcal{H} f(\eta))^2 \tilde{\delta}^{(1)}(\boldsymbol{k}_{1}, \eta)\tilde{\delta}^{(1)}(\boldsymbol{k}_{2}, \eta) \right] \, .
\label{Eq:Euler_2nd_FT}
\eea

\noindent By rewriting the derivatives as $\partial/\partial \ln D_+ = (\mathcal{H} f)^{-1} \partial/\partial \eta $ and rescaling the velocity field as $\tilde\vartheta^{(2)} \equiv \tilde{\theta}^{(2)}/(\mathcal{H} f)$ to simplify the equations, we obtain the following:
\bea
\dfrac{\partial\tilde{\delta}^{(2)}(\boldsymbol{k}, D_+)}{\partial \ln D_+} 
+ \tilde{\vartheta}^{(2)}(\boldsymbol{k}, D_+) & = & D^2_+ I_1(\boldsymbol{k}) \, , 
\label{Eq:Continuity_2nd_FT_D+} \\
\dfrac{\partial\tilde{\vartheta}^{(2)}(\boldsymbol{k}, D_+)}{\partial \ln D_+} + \left[ \dfrac{3}{2} \dfrac{\Omega_{\rm m}}{f^2} - 1 \right] \tilde{\vartheta}^{(2)}(\boldsymbol{k}, D_+) + \dfrac{3}{2} \dfrac{\Omega_{\rm m}}{f^2} \tilde{\delta}^{(2)}(\boldsymbol{k}, D_+) & = & D^2_+ I_2(\boldsymbol{k}) \, .
\label{Eq:Euler_2nd_FT_D+}
\eea
 
\noindent The terms $I_1$ and $I_2$ represent the nonlinear integrals, which depend solely on $\boldsymbol{k}$. For dark energy cosmologies with similar expansion histories, it is found that the quantity $\Omega_{\rm m}(\eta)/f^2(\eta)$ remains close to $1$ throughout time. Consequently, the only time-dependent terms in Eq.~\eqref{Eq:Continuity_2nd_FT_D+} and Eq.~\eqref{Eq:Euler_2nd_FT_D+} are the source terms (via $D_+$). This leads us to make the following power-law ansatz:
\bea
\tilde{\delta}^{(2)} = C_1(\boldsymbol{k})   D^m_+ \, , \quad \text{and} \quad \tilde{\vartheta}^{(2)} = C_2(\boldsymbol{k})   D^m_+ \, .
\label{Eq:sol_anz}    
\eea

\noindent Imposing this, we then obtain:
\bea
m C_1 D_+^m +  C_2 D^m_+ & = & D^2_+ I_1(\boldsymbol{k}) \, ,
\label{Eq:Continuity_2nd_FT_ansat} \\
m C_2 D^m_+ + \dfrac{1}{2} C_2 D^m_+ + \dfrac{3}{2} C_1 D_+^m  & = & D_+^2 I_2(\boldsymbol{k}) \, .
\label{Eq:Euler_2nd_FT_ansat}
\eea

\noindent  Clearly, the only solution explored in this case is $m = 2$, such that:
\bea
C_1 & = &  + \dfrac{5}{7} I_1(\boldsymbol{k}) - \dfrac{2}{7} I_2(\boldsymbol{k}) \, ,
\label{Eq:C_1} \\
C_2 & = & - \dfrac{3}{7} I_1(\boldsymbol{k}) + \dfrac{4}{7} I_2(\boldsymbol{k}) \, .
\label{Eq:C_2}
\eea

\noindent Finally, we can express the solutions at second order in terms of $\eta$, as follows:
\bea
\tilde{\delta}^{(2)}(\boldsymbol{k}, \eta) & = & \ D^2_+  \int \dfrac{d^3 k_1}{(2\pi)^3}  \int d^3 k_2 \  \delta_{\rm D}(\boldsymbol{k} - \boldsymbol{k}_{1} - \boldsymbol{k}_{2}) F_2(\boldsymbol{k}_{1}, \boldsymbol{k}_{2}) \delta_0(\boldsymbol{k}_{1}) \delta_0(\boldsymbol{k}_{1}) \, , \label{Eq:delta_2nd}
\\
\tilde{\vartheta}^{(2)}(\boldsymbol{k}, \eta) & = & - D^2_+  \int \dfrac{d^3 k_1}{(2\pi)^3}  \int d^3 k_2 \  \delta_{\rm D}(\boldsymbol{k} - \boldsymbol{k}_{1} - \boldsymbol{k}_{2}) G_2(\boldsymbol{k}_{1}, \boldsymbol{k}_{2}) \delta_0(\boldsymbol{k}_{1}) \delta_0(\boldsymbol{k}_{1}) \, .
\label{Eq:theta_2nd}
\eea

\noindent In which $F_2$ and $G_2$ are the symmetrized kernels at second order, and they are calculated as, 
\bea
F_2(\boldsymbol{k}_{1}, \boldsymbol{k}_{2}) &=& \dfrac{5}{7} + \dfrac{2}{7} \dfrac{(\boldsymbol{k}_{1} \cdot \boldsymbol{k}_{2})^2}{k^2_1 k^2_2} + \dfrac{1}{2} \left(\boldsymbol{k}_{1} \cdot \boldsymbol{k}_{2} \right) \left( \dfrac{k_1}{k_2} + \dfrac{k_2}{k_1} \right) \, ,
\label{Eq:F_2} \\
G_2(\boldsymbol{k}_{1}, \boldsymbol{k}_{2}) &=& \dfrac{3}{7} + \dfrac{4}{7} \dfrac{(\boldsymbol{k}_{1} \cdot \boldsymbol{k}_{2})^2}{k^2_1 k^2_2} + \dfrac{1}{2} \left(\boldsymbol{k}_{1} \cdot \boldsymbol{k}_{2} \right) \left( \dfrac{k_1}{k_2} + \dfrac{k_2}{k_1} \right) \, .
\label{Eq:G_2}
\eea

\noindent Building upon this methodology, the solution at the $n$-th order can be formulated as,
\bea
\tilde{\delta}^{(n)}(\boldsymbol{k}, \eta) =  \ D^n_+   \left[ \prod_{i=1}^{n}  \int  \dfrac{d^3 k_i}{(2\pi)^3} \right] (2\pi)^3 \delta_{\rm D}\left(\boldsymbol{k} - \sum_{i=1}^{n} \boldsymbol{k}_{i} \right) F_n(\boldsymbol{k}_{1}, \dots ,\boldsymbol{k}_{n}) \delta_0(\boldsymbol{k}_{1}) \cdots \delta_0(\boldsymbol{k}_{n}) \, ,
\label{Eq:delta_nth} \\
\tilde{\vartheta}^{(n)}(\boldsymbol{k}, \eta)  =   - D^n_+  \left[ \prod_{i=1}^{n}  \int  \dfrac{d^3 k_i}{(2\pi)^3} \right] (2\pi)^3 \delta_{\rm D}\left(\boldsymbol{k} - \sum_{i=1}^{n} \boldsymbol{k}_{i} \right) G_n(\boldsymbol{k}_{1}, \dots ,\boldsymbol{k}_{n}) \delta_0(\boldsymbol{k}_{1}) \cdots \delta_0(\boldsymbol{k}_n)\, .
\label{Eq:theta_nth}
\eea

\noindent Certainly, at first order, the kernels simplify to $F_1 = 1$ and $G_1 = 1$. A strategy to determine the higher-order kernels $F_n$ and $G_n$ is by resorting to recurrence formulae or solve them numerically. This result allows us to explicitly calculate how structure in the Universe evolves nonlinearly. This however, is limited to only dark matter.
As a result, while CPT provides a powerful framework for modelling LSS formation, it is insufficient for a complete description of the nonlinear regime, especially at very smaller scales or when including the contributions of other components like baryons and radiation.
\chapter{}
\label{Appendix_b}

\vspace{1cm}

\begin{figure*}[b!]
    \centering
    \includegraphics[width=0.75\textwidth]{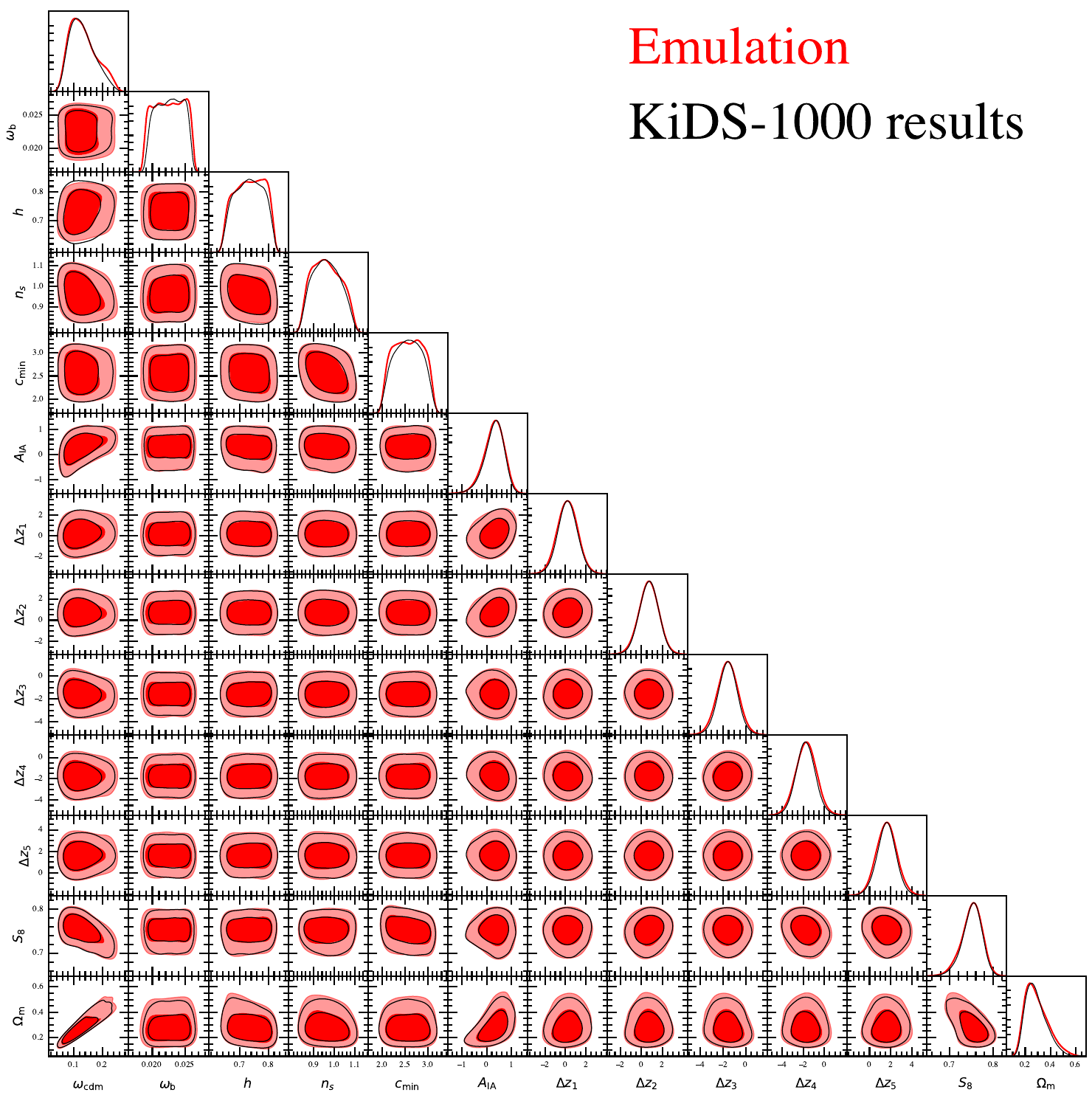}
  \caption[Validation between the posterios from KiDS-1000 and emulators with reflecting the $\Lambda$CDM case]{Full COSEBIs statistics posteriors from the KiDS-1000 chains \citep{KiDS:2020suj} (black lines), and from emulators with parameters fixed to reflect the $\Lambda$CDM case (solid red). The plot clearly shows the desired consistency of the emulated data with the official set. Note that the full emulator chains were calculate in a few minutes, while the sole Boltzmann solver (\texttt{CAMB}) takes days to process this data.}
\label{Fig:LCDM_emu_vs_k1k_data}
\end{figure*}

In this appendix we present important supplementary contour plots of this work. The first validity test for emulators involved assessing whether they are capable to reproduce the KiDS-1000 $\Lambda$CDM official analysis \citep{KiDS:2020suj}. 
This is shown in \autoref{Fig:LCDM_emu_vs_k1k_data}, which displays a full comparison between the publicly available KiDS-1000 results for $\Lambda$CDM and posteriors obtained through our emulators in a few a minutes.

\newpage 

\begin{figure*}[t!]
    \centering
\includegraphics[width=0.75\textwidth]{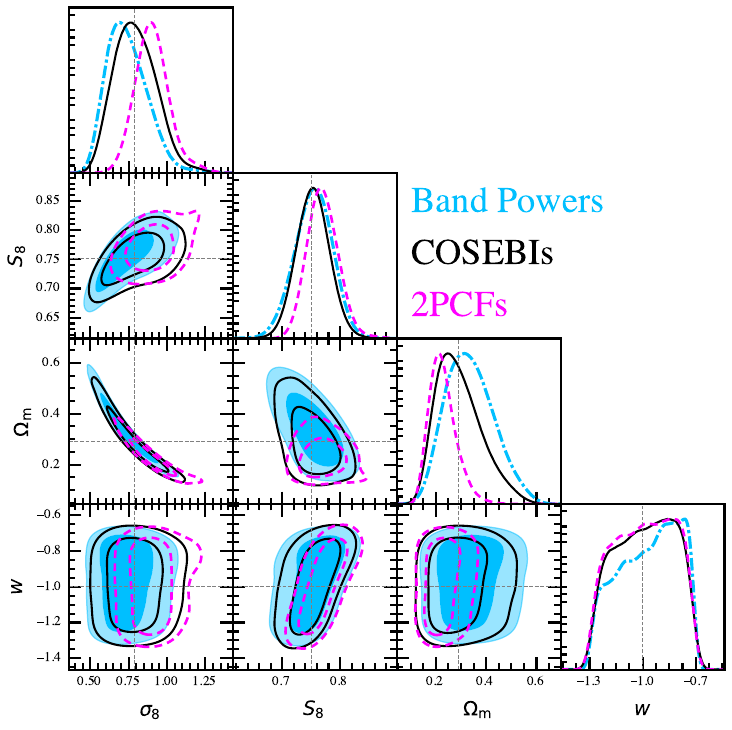}
  \caption[Posteriors from the emulator with reflecting the $w$CDM case with KiDS-1000]{Posteriors from the emulator, for the $w$CDM cosmology, using Band Powers (blue), COSEBIs (black) and Two point Correlation Functions (magenta) statistics. The dashed lines represent COSEBIs mean values from $\Lambda$CDM case. }
\label{Fig:wCDM_emu_all_sets}
\end{figure*}

\noindent Additionally, we conducted a second test by evaluating the emulators within the context of the $w$CDM cosmology, as detailed in \cite{2021A&A...649A..88T}. The results of this validation is presented in \autoref{Fig:wCDM_emu_all_sets}, where we used the same \texttt{MultiNest} settings for consistency.

\newpage 

\begin{figure}[b!]
    \centering
\includegraphics[width=\textwidth]{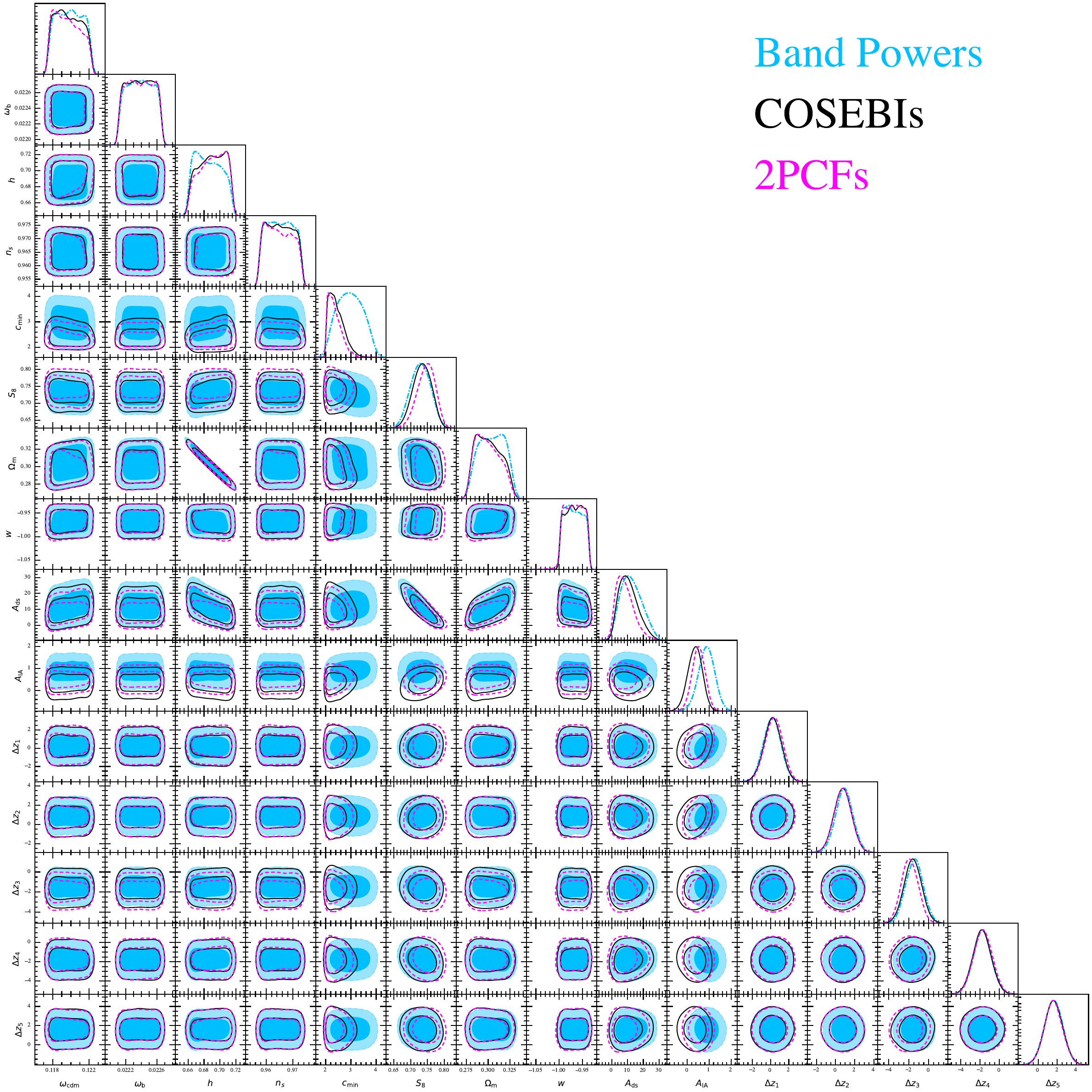}
  \caption[Full DS posteriors from KiDS-1000 and CMB+BAO combined analysis]{Full posteriors of DS model and also photo-z errors from KiDS-1000 and CMB+BAO combined analysis. The units of $A_{\rm ds}$ are [b/GeV].}
\label{Fig:all_k1k_sets}
\end{figure}

\noindent Lastly, \autoref{Fig:all_k1k_sets} illustrates the posterior distributions of nine cosmological parameters and five nuisance parameters associated with the redshift distributions ($\Delta z_i$). These distributions are derived from our combined analysis using CMB+BAO data and considering the three distinct KiDS-1000 statistical approaches: Band Powers, COSEBIs, and 2PCFs.

\newpage
\thispagestyle{empty}

\chapter{}
\label{Appendix_c}

\vspace{1cm}

We report mean and $68\%$ credible intervals for the two fiducial cases considered in this work in \autoref{tab:lcdm_fiducial_data} and \autoref{tab:ds_fiducial_data}.

\vspace{3cm}

\begin{table*}[h]
\centering
\caption[Mean and $68\%$ marginalised credible intervals for key parameters from the $\Lambda$CDM mock data vector]{Mean and $68\%$ marginalised credible intervals for key parameters characterising the $\Lambda$CDM, $w$CDM and DS models, for scale cuts $\ell_{\rm max} = 1500, 3000$, and $5000$. The mock data vector is generated assuming a $\Lambda$CDM fiducial.}
\label{tab:lcdm_fiducial_data}
    \renewcommand{\arraystretch}{1.75}
    \setlength{\tabcolsep}{3.5pt}
    \begin{adjustbox}{max width=\textwidth}
    \begin{tabular}{|c|c|c|c||c|c|c||c|c|c|}
        \multicolumn{10}{c}{{\Large\textbf{$\Lambda$CDM mock data}}} \\ \hline
        \multirow{2}{*}{\textbf{Parameter}} & \multicolumn{3}{c||}{\textbf{$\ell_{\rm max} = 1500$}} & \multicolumn{3}{c||}{\textbf{$\ell_{\rm max} = 3000$}} & \multicolumn{3}{c|}{\textbf{$\ell_{\rm max} = 5000$}} \\ 
        \cline{2-10} \cline{2-10}                                    
        & \textbf{$\Lambda$CDM} & \textbf{$w$CDM} & \textbf{DS} & \textbf{$\Lambda$CDM} & \textbf{$w$CDM} & \textbf{DS} & \textbf{$\Lambda$CDM} & \textbf{$w$CDM} & \textbf{DS} \\ \hline
        
        $\omega_{\rm b}$  
        & $0.023 \pm 0.002$ & $0.023 \pm 0.002$ & $0.023 \pm 0.002$ 
        & $0.023 \pm 0.002$ & $0.023 \pm 0.002$&  $0.023 \pm 0.002$ 
        & $0.022 \pm 0.002$ & $0.023 \pm 0.002$ & $0.023 \pm 0.002$ \\ \hline
        
        $\omega_{\rm cdm}$  
        & $0.124 \pm 0.010$ & $0.124 \pm  0.010$ & $0.123 \pm 0.011$ 
        & $0.123 \pm 0.009$ & $0.123 \pm 0.009$ & $0.124 \pm 0.010$ 
        & $0.122 \pm 0.006$ & $0.123 \pm 0.006$ & $0.125 \pm 0.007$ \\ \hline
        
        $h$  
        & $0.69 \pm 0.03$ & $0.69 \pm 0.03$ & $0.69 \pm 0.03$ 
        & $0.69 \pm 0.03$ & $0.69 \pm 0.025$ & $0.69 \pm 0.03$ 
        & $0.69 \pm 0.02$ & $0.69 \pm 0.02$ & $0.70 \pm 0.02$ \\ \hline
        
        $n_s$  
        & $0.96 \pm 0.03$ & $0.96 \pm 0.03$ & $0.96 \pm 0.04$ 
        & $0.96 \pm 0.03$ & $0.96 \pm 0.03$ & $0.96 \pm 0.03$ 
        & $0.96 \pm 0.02$ & $0.96 \pm 0.02$ & $0.95 \pm 0.03$ \\ \hline

        $\Omega_{\rm m}$  
        & $0.306 \pm 0.002$ & $0.306 \pm 0.002$ &  $0.306 \pm 0.002$
        & $0.306 \pm 0.001$ & $0.306 \pm 0.001$ &  $0.306 \pm 0.001$
        & $0.306 \pm 0.001$ & $0.306 \pm 0.001$ & $0.306 \pm 0.001$ \\ \hline

        $S_8$  
        & $0.825 \pm 0.002$ & $0.825 \pm 0.003$ & $0.825 \pm 0.004$ 
        & $0.825 \pm 0.001$ & $0.825 \pm 0.002$ & $0.824 \pm 0.003$ 
        & $0.825 \pm  0.001$ & $0.825 \pm 0.002$ & $0.823 \pm 0.003$ \\ \hline
        
        $m_\nu$  
        & $0.10 \pm 0.06$ & $0.09 \pm 0.05$ & $0.09 \pm 0.06$ 
        & $0.10 \pm 0.06$ & $0.08 \pm 0.05$ & $0.06 \pm 0.04$ 
        & $0.10 \pm 0.06$ & $0.08 \pm 0.05$ & $0.06 \pm 0.04$ \\ \hline
        
        $w_0$  
        & $-$ & $-0.997 \pm 0.025$ & $-1.003 \pm 0.023$ 
        & $-$ & $-0.996 \pm 0.015$ & $-0.997 \pm 0.017$ 
        & $-$ & $-0.997 \pm 0.012$ & $-0.991 \pm 0.012$ \\ \hline
        
        $|A_{\rm ds}|$  
        & $-$ & $-$ & $3.02 \pm 2.17$ 
        & $-$ & $-$ & $2.33 \pm 1.73$ 
        & $-$ & $-$ & $2.47 \pm 1.81$ \\ \hline
        
        $c_{\rm min}$  
        & $2.60 \pm 0.06$ & $2.63 \pm 0.07$ & $2.69 \pm 0.19$ 
        & $2.60 \pm 0.05$ & $2.62 \pm 0.06$ & $2.60 \pm 0.13$ 
        & $2.60 \pm 0.05$ & $2.61 \pm 0.05$ & $2.54 \pm 0.09$ \\ \hline
        
        
        \hline
    \end{tabular}
    \end{adjustbox}
\end{table*}

\begin{table*}[b!]
    \centering
    \caption[Mean and $68\%$ marginalised credible intervals for key parameters from the DS mock data vector]{Same as \autoref{tab:lcdm_fiducial_data}, but for a DS fiducial.}
    \label{tab:ds_fiducial_data}
    \renewcommand{\arraystretch}{1.75}
    \setlength{\tabcolsep}{3.5pt}
    \begin{adjustbox}{max width=\textwidth}
    \begin{tabular}{|c|c|c|c||c|c|c||c|c|c|}
        \multicolumn{10}{c}{{\Large\textbf{DS mock data}}} \\ \hline
        \multirow{2}{*}{\textbf{Parameter}} & \multicolumn{3}{c||}{\textbf{$\ell_{\rm max} = 1500$}} & \multicolumn{3}{c||}{\textbf{$\ell_{\rm max} = 3000$}} & \multicolumn{3}{c|}{\textbf{$\ell_{\rm max} = 5000$}} \\ 
        \cline{2-10} \cline{2-10}                                    
        & \textbf{$\Lambda$CDM} & \textbf{$w$CDM} & \textbf{DS} & \textbf{$\Lambda$CDM} & \textbf{$w$CDM} & \textbf{DS} & \textbf{$\Lambda$CDM} & \textbf{$w$CDM} & \textbf{DS} \\ \hline
        
        $\omega_{\rm b}$  
        & $0.023 \pm 0.002$ & $0.023 \pm 0.002$ & $0.023 \pm 0.002$
        & $0.023 \pm 0.002$ & $0.023 \pm 0.002$ & $0.023 \pm 0.002$
        & $0.023 \pm 0.002$ & $0.022 \pm 0.002$ & $0.023 \pm 0.002$ \\ \hline
        
        $\omega_{\rm cdm}$  
        & $0.124 \pm 0.009$ & $0.124 \pm 0.010$ & $0.123 \pm 0.011$ 
        & $0.123 \pm 0.009$ & $0.124 \pm 0.010$ & $0.123 \pm 0.009$
        & $0.122 \pm 0.006$ & $0.121 \pm 0.007$ & $0.121\pm 0.006$ \\ \hline 
        
        $h$  
        & $0.69 \pm 0.03$ & $0.69 \pm 0.03$ & $0.69 \pm 0.03$
        & $0.69 \pm 0.03$ & $0.70 \pm 0.03$ & $0.69 \pm 0.02$
        & $0.69 \pm 0.02$ & $0.69 \pm 0.02$ & $0.69 \pm 0.02$ \\ \hline
        
        $n_s$  
        & $0.96 \pm 0.03$ & $0.96 \pm 0.03$ & $0.96 \pm 0.04$
        & $0.96 \pm 0.03$ & $0.96 \pm 0.03$ & $0.96 \pm 0.04$
        & $0.96 \pm 0.02$ & $0.96 \pm 0.02$ & $0.96 \pm 0.03$ \\ \hline

        $\Omega_{\rm m}$  
        & $0.306 \pm 0.002$ & $0.306 \pm 0.002$ &  $0.306 \pm 0.002$
        & $0.306 \pm 0.001$ & $0.306 \pm 0.001$ &  $0.306 \pm 0.001$
        & $0.306 \pm 0.001$ & $0.306 \pm 0.001$ & $0.306 \pm 0.001$ \\ \hline
        
        $S_8$  
        & $0.825 \pm 0.002$ & $0.824 \pm 0.003$ & $0.826 \pm 0.006$
        & $0.825 \pm 0.001$ & $0.824 \pm 0.002$ & $0.825 \pm 0.005$
        & $0.825 \pm 0.001$ & $0.824 \pm 0.002$ & $0.825 \pm 0.004$ \\ \hline
        
        $m_\nu$  
        & $0.10 \pm 0.06$ & $0.09 \pm 0.05$ & $0.09 \pm 0.05$
        & $0.10 \pm 0.06$ & $0.09 \pm 0.05$ & $0.07 \pm 0.04$
        & $0.10 \pm 0.05$ & $0.08 \pm 0.04$ & $0.07 \pm 0.03$ \\ \hline
        
        $w_0$  
        & $-$ & $-0.969 \pm 0.025$ & $-0.963 \pm 0.022$
        & $-$ & $-0.972 \pm 0.018$ & $-0.965 \pm 0.017$
        & $-$ & $-0.968 \pm 0.014$ & $ -0.967 \pm 0.014$ \\ \hline
        
        $|A_{\rm ds}|$  
        & $-$ & $-$ & $10.20  \pm 3.77$
        & $-$ & $-$ & $10.36 \pm 3.16$
        & $-$ & $-$ & $10.74 \pm 2.67$ \\ \hline
        
        $c_{\rm min}$  
        & $2.60 \pm 0.06$ & $2.63 \pm 0.06$ & $2.62 \pm 0.18$
        & $2.60 \pm 0.05$ & $2.63 \pm 0.06$ & $2.61 \pm 0.13$
        & $2.60 \pm 0.05$ & $2.62 \pm 0.05$ & $2.59 \pm 0.11$ \\ \hline
        
        
        \hline

    \end{tabular}
    \end{adjustbox}
\end{table*}

\end{appendices}

\end{document}